\documentclass[11pt]{amsart}
\usepackage{amsmath}
\usepackage{amsxtra}
\usepackage{amscd}
\usepackage{amsthm}
\usepackage{amsfonts}
\usepackage{amssymb}
\usepackage{eucal}
\usepackage{epsfig}

\usepackage{latexsym,amsthm}

\DeclareFontEncoding{OT2}{}{}

\input epsf

\def\figin{\epsfcheck\figin}\def\figins{\epsfcheck\figins}
\def\epsfcheck{\ifx\epsfbox\UnDeFiNeD
\message{(NO epsf.tex, FIGURES WILL BE IGNORED)}
\gdef\figin##1{\vskip2in}\gdef\figins##1{\hskip.5in}
\else\message{(FIGURES WILL BE INCLUDED)}%
\gdef\figin##1{##1}\gdef\figins##1{##1}\fi}
\def\DefWarn#1{}
\def\figinsert{\goodbreak\topinsert}
\def\ifig#1#2#3#4{\DefWarn#1\xdef#1{fig.~\the\figno}
\writedef{#1\leftbracket fig.\noexpand~\the\figno}%
\figinsert\figin{\centerline{\epsfxsize=#3mm \epsfbox{#2}}}
\bigskip\medskip\centerline{\vbox{\baselineskip12pt
\advance\hsize by -1truein\noindent\footnotefont{\sl Fig.~\the\figno:}\sl\ #4}}
\bigskip\endinsert\noindent\global\advance\figno by1}

\def\Figx#1#2#3{
\bigskip
\vbox{\centerline{\epsfxsize=#1 cm \epsfbox{#2.eps}}
\centerline{ #3}}\bigskip}

\def\half{\frac{1}{2}}

\newtheorem{thm}{Theorem}[section]

\theoremstyle{remark}
\newtheorem{remark}[thm]{Remark}

\theoremstyle{definition}

\numberwithin{equation}{section}

\newcommand{\Ref}[1]{{$($\ref{#1}$)$}}
\newcommand{\bean}{\begin{eqnarray}}
\newcommand{\eean}{\end{eqnarray}}
\newcommand{\be}{\begin{displaymath}}
\newcommand{\ee}{\end{displaymath}}
\newcommand{\bea}{\begin{eqnarray*}}
\newcommand{\eea}{\end{eqnarray*}}

\newcommand{\secref}[1]{Section~\ref{#1}}

\newcommand{\nc}{\newcommand}
\nc{\on}{\operatorname}
\nc{\p}{{\partial}}
\nc{\pa}{{\p}}
\nc{\ch}{\mbox{ch}}
\nc{\Z}{{\mathbb Z}}
\nc{\C}{{\mathbb C}}
\nc{\pone}{{\mathbb P}^1}

\nc{\dlx}[1]{{\CD}^{#1}{\bf X}}
\nc{\odl}[1]{{\ol{\CD}}^{#1}{\bf \ol{X}}}
\nc{\dlpsi}[1]{{\CD}^{#1}{\bf \Psi}}
\nc{\odlpsi}[1]{{\ol{\CD}}^{#1}{\bf \ol{\Psi}}}

\nc{\CA}{{\mathcal A}}
\nc{\CB}{{\mathcal B}}
\nc{\CC}{{\mathcal C}}
\renewcommand{\CD}{{\mathcal D}}
\nc{\CE}{{\mathcal E}}
\nc{\CF}{{\mathcal F}}
\nc{\CG}{{\mathcal G}}
\nc{\CH}{{\mathcal H}}
\nc{\CI}{{\mathcal I}}
\nc{\CJ}{{\mathcal J}}
\nc{\CK}{{\mathcal K}}
\nc{\CL}{{\mathcal L}}
\nc{\CM}{{\mathcal M}}
\nc{\CN}{{\mathcal N}}
\nc{\CO}{{\mathcal O}}
\nc{\CP}{{\mathcal P}}
\nc{\CQ}{{\mathcal Q}}
\nc{\CR}{{\mathcal R}}
\nc{\CS}{{\mathcal S}}
\nc{\CT}{{\mathcal T}}
\nc{\CU}{{\mathcal U}}
\nc{\CV}{{\mathcal V}}
\nc{\CW}{{\mathcal W}}
\nc{\CX}{{\mathcal X}}
\nc{\CY}{{\mathcal Y}}
\nc{\CZ}{{\mathcal Z}}

\nc{\zb}{\ol{z}}
\nc{\jb}{\ol{j}}
\nc{\ib}{\ol{i}}
\nc{\xb}{\ol{x}}
\nc{\yb}{\ol{y}}
\nc{\ub}{\ol{u}}
\nc{\pb}{\bar\partial}
\nc{\wb}{\ol{w}}
\nc{\qb}{\ol{q}}
\nc{\nb}{\ol{n}}
\nc{\phb}{\ol\ph}
\nc{\Xb}{\ol{X}}

\nc{\ph}{p}

\nc{\bi}{{\bf i}}
\nc{\bj}{{\bf j}}
\nc{\bk}{{\bf k}}
\nc{\bq}{{\bf q}}
\nc{\bv}{{\bf v}}

\nc{\dirac}{D\hspace*{-2.5mm}\slash}

\nc{\al}{\alpha}
\nc{\bt}{{\beta}}
\nc{\dl}{{\delta}}
\nc{\la}{\lambda}
\nc{\m}{\mu}
\nc{\ep}{\epsilon}
\nc{\si}{\sigma}
\nc{\om}{\omega}

\nc{\De}{\Delta}
\nc{\Ga}{\Gamma}
\nc{\La}{\Lambda}

\nc{\el}{\ell}

\nc{\arr}{\rightarrow}
\nc{\larr}{\longrightarrow}
\nc{\ri}{\rangle}
\nc{\lef}{\langle}
\nc{\su}{\widehat{{\mathfrak s}{\mathfrak l}}_2}
\nc{\sw}{{\mathfrak s}{\mathfrak l}}
\nc{\g}{{\mathfrak g}}
\nc{\h}{{\mathfrak h}}
\nc{\n}{{\mathfrak n}}
\nc{\N}{\widehat{\n}}
\nc{\G}{\widehat{\g}}
\nc{\gt}{\widetilde{\g}}
\nc{\one}{{\mathbf 1}}
\nc{\z}{{\mathfrak Z}}
\nc{\wt}{\widetilde}
\nc{\wh}{\widehat}
\nc{\cri}{_{\kappa_c}}
\nc{\kk}{h^\vee}
\nc{\sun}{\widehat{\sw}_N}
\nc{\ol}{\overline}
\nc{\ds}{\displaystyle}
\nc{\dzz}{\frac{dz}{z}}
\nc{\Res}{\on{Res}}
\nc{\mc}{\mathcal}
\nc{\Cal}{\mathcal}
\nc{\bb}{{\mathfrak b}}
\nc{\ot}{\otimes}
\nc{\R}{{\mathbb R}}
\nc{\yy}{{\mc Y}}
\nc{\ga}{\gamma}

\nc{\us}{\underset}
\nc{\opl}{\oplus}
\nc{\beq}{\begin{equation}}
\nc{\Rep}{\on{Rep}}
\nc{\sssec}{\subsubsection}
\nc{\ssec}{\subsection}
\nc{\lan}{\langle}
\nc{\ran}{\rangle}

\nc{\Vect}{\on{Vect}}
\nc{\ghat}{\G}
\nc{\T}{\mc T}
\nc{\Tloc}{\T^\g_{\on{loc}}}
\nc{\vac}{|0\ran}
\nc{\Wick}{{\mb :}}
\nc{\mb}{\mathbf}
\nc{\delz}{\partial_z}
\nc{\cali}{\mathcal}
\nc{\li}{\mathfrak l}
\nc{\lt}{\widetilde{\li}}
\nc{\astar}{a^*}
\nc{\cA}{{\mc A}}
\nc{\ka}{\kappa}

\nc{\OO}{{\mc O}}
\nc{\AutO}{\on{Aut} O}
\nc{\DerO}{\on{Der} O}
\nc{\DerpO}{\on{Der}_+ O}
\nc{\Au}{{\mc A}ut}
\nc{\mf}{\mathfrak}
\nc{\V}{{\mc V}}
\nc{\hh}{\wh{\h}}

\nc{\pp}{{\mathfrak p}}
\nc{\mm}{{\mathfrak m}}
\nc{\rr}{{\mathfrak r}}
\nc{\ket}{\rangle}
\nc{\zz}{{\mathfrak z}}
\nc{\gr}{\on{gr}}
\nc{\Spe}{\on{Spec}}
\nc{\rv}{\crho}
\nc{\can}{\on{can}}
\nc{\MOp}{\on{MOp}_G(D)}
\nc{\Db}{{\mathbb D}}
\nc{\ww}{w}

\nc{\Con}{\on{Conn}(\Omega^{\crho})_D}
\nc{\ConD}{\on{Conn}(\Omega^{\crho})_{\Db}}
\nc{\ConDL}{\on{Conn}(\Omega^{\rho})_{\Db}}
\nc{\ConDtL}{\on{Conn}(\Omega^{\rho})_{\Db^\times}}
\nc{\OpD}{\on{Op}_G(\Db)}
\nc{\crho}{\check{\rho}}
\nc{\chal}{\check{\al}}
\nc{\cchi}{\check{\chi}}
\nc{\cLa}{\check\Lambda}
\nc{\cla}{\check\la}
\nc{\cmu}{\check\mu}

\nc{\PP}{{\mathbb P}}
\nc{\TT}{{\mathbb T}}

\nc{\bone}{{\mb 1}}

\nc{\bs}{\backslash}

\def\tr{{\rm tr}}

\nc{\zzb}{z \zb}

\nc{\pf}{\int\hspace*{-3.5mm}\bs}

\nc{\inn}{\on{in}}
\nc{\out}{\on{out}}
\nc{\covac}{\langle 0|}
\nc{\ptwo}{{\mathbb C}{\mathbb P}^2}
\nc{\BS}{{\mathbb S}}
\nc{\BT}{{\mathbb T}}

\begin{document}

\vspace*{-20mm}

\title{Instantons beyond topological theory II}\thanks{Supported by
DARPA and AFOSR through the grant FA9550-07-1-0543}

\author{E. Frenkel}

\address{Department of Mathematics, University of California,
       Berkeley, CA 94720, USA}

\author{A. Losev}

\address{Institute of Theoretical and Experimental Physics,
       B. Cheremushkinskaya 25, Moscow 117259, Russia}

\author{N. Nekrasov}

\address{Institut des Hautes \'Etudes Scientifiques, 35, Route de
Chartres, Bures-sur-Yvette, F-91440, France}

\centerline{\hfill MSRI-08/01 \ ITEP-TH-05/08 \ IHES-P-08/15}
\centerline{\hphantom{void}}
\centerline{\hphantom{void}}

\date{March 2008}

\begin{abstract}
The present paper is the second part of our project in which we
describe quantum field theories with instantons in a novel way by
using the ``infinite radius limit'' (rather than the limit of free
field theory) as the starting point. The theory dramatically
simplifies in this limit, because the correlation functions of all,
not only topological (or BPS), observables may be computed explicitly
in terms of integrals over finite-dimensional moduli spaces of
instanton configurations. In Part I we discussed in detail the
one-dimensional (that is, quantum mechanical) models of this
type. Here we analyze the supersymmetric two-dimensional sigma models
and four-dimensional Yang--Mills theory, using the one-dimensional
models as a prototype. We go beyond the topological (or BPS) sectors
of these models and consider them as full-fledged quantum field
theories. We study in detail the space of states and find that the
Hamiltonian is not diagonalizable, but has Jordan blocks. This leads
to the appearance of logarithms in the correlation functions. We find
that our theories are in fact logarithmic conformal field theories
(theories of this type are of interest in condensed matter
physics). We define jet-evaluation observables and consider in detail
their correlation functions.  They are given by integrals over the
moduli spaces of holomorphic maps, which generalize the Gromov--Witten
invariants. These integrals generally diverge and require
regularization, leading to an intricate logarithmic mixing of the
operators of the sigma model. A similar structure arises in the
four-dimensional Yang--Mills theory as well.
\end{abstract}

\maketitle


\tableofcontents

\newpage


\section{Introduction}

Many two- and four-dimensional quantum field theories have two kinds
of coupling constants: the actual coupling $g$, which in particular
counts the loops in the perturbative calculations, and the topological
coupling, ${\vartheta}$, the theta-angle, which is the chemical
potential for the topological sectors in the path integral. These
couplings can be combined into the complex coupling ${\tau}$ and its
complex conjugate $\ol{\tau}$.

For example, in the four-dimensional gauge theory one combines the
Yang--Mills coupling $g$ and the theta-angle $\vartheta$ as follows:
\begin{equation}
{\tau} = \frac{\vartheta}{2\pi} + \frac{4{\pi}i}{g^{2}}, \qquad
\ol{\tau} = \frac{\vartheta}{2\pi} - \frac{4{\pi}i}{g^{2}}.
\label{tauym}
\end{equation}
For the two-dimensional sigma model with the complex target space $X$,
endowed with a Hermitian metric $g_{i\jb}$ and a $(1,1)$ type
two-form $B_{i\jb}$ one defines
\begin{equation}
{\tau}_{i\jb} = B_{i\jb} + i g_{i\jb}, \qquad \ol{\tau} =
B_{i\jb} - i g_{i\jb}
\label{tausigma}
\end{equation}
If $dB = 0$, then the two-form $B$ plays the role of the theta-angle.

\ssec{Weak coupling limit}

We wish to study the dependence of the theory on $\tau$, $\ol\tau$ as
if they were two separate couplings, not necessarily complex conjugate
to each other. The resulting theory should greatly simplify in the
limit when
\begin{equation}
\ol{\tau} \to - i \infty \ , \ \qquad {\tau} \ \on{is} \ \on{fixed}.
\label{ourlimit}
\end{equation}
This is the weak coupling limit, in which the theta-angle has a large
imaginary part. In two-dimensional sigma models, it is also known as the
``infinite radius limit''. This limit has been studied in the
literature since the early days of the theory of instantons, see,
e.g., \cite{David}.

The reason for this simplification is that the path integral of the
theory in this limit, as described by a first-order Lagrangian,
represents the ``delta-form'' supported on the instanton moduli space,
which is a union of {\em finite-dimensional} components labeled by
``instanton numbers''. Therefore the correlation functions are
expressed as linear combinations of integrals over these
finite-dimensional components.

The idea is to use the theory in this weak coupling limit as the {\em
starting point} for investigating more realistic models corresponding
the finite values of coupling constants. In other words, we first
describe the theory in this limit, and then develop a {\em
perturbation theory} around this point in the space of couplings in
order to reach the theories defined for other values of the coupling
constants. In the framework of this perturbation theory we could, in
principle, compute all correlation functions in terms of
finite-dimensional integrals. A more detailed discussion of this
point, along with a summary of our project, may be found in
\cite{Cargese}.

We view this as a viable alternative to the conventional approach in
quantum field theory of using the perturbation theory around a
Gaussian point describing a free field theory. The advantage of our
approach is that we do not need to impose from the beginning a linear
structure on the space of fields. On the contrary, the non-linearity
is preserved and is reflected in the geometry of the moduli space of
instantons. This is why we believe that our approach may be beneficial
for understanding some of the hard dynamical questions, such as
confinement, that have proved elusive in the conventional formalism.

\medskip

The 4D $S$-{\em duality} and its 2D analogue, the {\em mirror
symmetry}, give us another tool for connecting the theory at our
special limit to the theories in the physical range of coupling
constants. In a physical theory, in which $\ol\tau$ is complex
conjugate to $\tau$, the $S$-duality sends $\tau \mapsto -1/\tau$. It
is reasonable to expect that $S$-duality still holds when we
complexify the coupling constants $\tau, \ol\tau$. It should then act
as follows:
$$
\tau \mapsto -1/\tau, \qquad \ol\tau \mapsto -1/\ol\tau.
$$
Now observe that applying this transformation to $\ol\tau = -i \infty$
and finite $\tau$, we obtain $\tau' = -1/\tau$ and $\ol\tau' =
0$. These coupling constants are already within the range of physical
values, in the sense that both the coupling constant $g$ and the
theta-angle $\vartheta$ are finite! Therefore we hope that our
calculations in the theory with $\ol\tau = -i \infty$ could be
linked by $S$-duality (or mirror symmetry) to exact non-perturbative
results in a physical theory beyond the topological sector.

\ssec{Topological sector}

In supersymmetric models there is an important class of observables,
called the {\em topological}, or {\em BPS observables}, whose
correlation functions are independent of $\ol\tau$. They commute with
the supersymmetry charge $\CQ$ of the theory and comprise the {\em
topological sector} of the theory. Their correlation functions are
closely related to the Gromov--Witten and Donaldson invariants, in
two-dimensional sigma models and four-dimensional Yang--Mills theory,
respectively. The perturbation away from the point $\ol\tau = -i
\infty$ is given by a $\CQ$-exact operator, and therefore the
correlation functions of the BPS observables (which are $\CQ$-closed)
remain unchanged when we move away from the special point. This is the
secret of success of the computation of the correlation functions of
the BPS observables achieved in recent years in the framework of
topological field theory: the computation is actually done in the
theory at $\ol\tau = -i \infty$, but because of the special properties
of the BPS observables the answer remains the same for other values of
the coupling constant. But for general observables the correlation
functions do change in a rather complicated way when we move away from
the special point.

We would like to go {\em beyond the topological sector} and consider
more general correlation functions of non-BPS observables. Multiple
reasons for this are described in detail in the Introductions to our
earlier papers \cite{PartI,Cargese}, and we will not repeat them
here. Clearly, if we wish to use the model in the limit $\ol\tau = - i
\infty$ as a launchpad for studying the models at more general values
of the coupling constants, we must first understand it as a
full-fledged quantum field theory, beyond its topological sector.

\medskip

In \cite{PartI}, to which we will henceforth refer as ``Part I'', we
have launched a program of systematic study of the $\ol\tau \to -i
\infty$ limit of the instanton models in one, two and four
dimensions. In Part I we have described in detail the one-dimensional
models. These are supersymmetric quantum mechanical models on K\"ahler
manifolds $X$ equipped with a Morse function $f$.  Here we analyze the
supersymmetric two-dimensional sigma models and four-dimensional
Yang--Mills theory, using the one-dimensional models as a prototype.

\ssec{Sigma models}

Let us describe briefly our results concerning two-dimensional
supersymmetric sigma models. The first step is to recast these models
in the framework of the quantum mechanical models that we have studied
in Part I. For a fixed Riemann surface $\Sigma$ the space of bosonic
fields in the supersymmetric sigma model with the target manifold $X$
is $\on{Maps}(\Sigma,X)$, the space of maps $\Sigma \to X$. If we
choose $\Sigma$ to be the cylinder $I \times \BS^1$, then we may
interpret $\on{Maps}(\Sigma,X)$ as the space of maps from the interval
$I$ to the loop space $LX = \on{Maps}(\BS^1,X)$. Thus, we may think of
the two-dimensional sigma model on the cylinder with the target $X$ as
the quantum mechanical model on the loop space $LX$. Hence it is
natural to try to write the Lagrangian of the sigma model in such a
way that it looks exactly like the Lagrangian of the quantum
mechanical model on $LX$ with a Morse function $f$.

It turns out that if $X$ is a K\"ahler manifold, this is ``almost''
possible. However, there are two caveats. First of all, our function
$f$ is the so-called Floer function \cite{Floer} which has
non-isolated critical points corresponding to the constant loops in
$LX$, so is in fact a Morse--Bott function. We can deal with this
problem by deforming this function so that it only has isolated
critical points, corresponding to the constant loops whose values are
critical points of a Morse function on $X$ (this amounts to
considering the sigma model in the background of a gauge field, as we
explain in \secref{gauged sm}). The second, and more serious, issue is
that the Floer function $f$ is not single-valued on $LX$, but only on
the universal (abelian) cover $\wt{LX}$. In other words, it is an
example of a Morse--Novikov function, or, more properly,
Morse--Bott--Novikov function. Because of that, the instantons are
identified with gradient trajectories of the pull-back of $f$ to
$\wt{LX}$.

Suppose that $I = \R$ with a coordinate $t$. In the limits $t \to \pm
\infty$ a gradient trajectory tends to the critical points of $f$ on
$\wt{LX}$, which are the preimages of constant maps in $LX$. Therefore
the gradient trajectory may be interpreted as the map of the cylinder
compactified by two points at $\pm \infty$ to $X$, or equivalently, a
map $\C\pone \to X$. Moreover, the condition that it corresponds to a
gradient trajectory of the Floer function $f$ simply means that this
map is {\em holomorphic}. Thus, we obtain that the instantons of the
two-dimensional sigma model are holomorphic maps $\C\pone \to X$, and
more generally, $\Sigma \to X$, where $\Sigma$ is an arbitrary compact
Riemann surface.

In order to describe the structure of the space of states of the
two-dimensional sigma model in the infinite radius limit, we first
need to generalize our results obtained in Part I to the case of
Morse--Novikov and Morse--Bott--Novikov functions. (Actually, examples
of such functions arise already for finite-dimensional manifolds with
non-trivial fundamental groups.) Essentially, this amounts to
considering the universal (abelian) cover (which is $\wt{LX}$ in the
case of sigma model). One also needs to impose an equivariance
condition on the states of the model corresponding to the action of
the (abelianized) fundamental group $H_2(X,\Z)$ on $\wt{LX}$.  Besides
those changes, the structure of the space of states in the limit
$\ol\tau \to \infty$ is similar to the one that we have observed in
our analysis of quantum mechanical models in Part I.

There are spaces of ``in'' and ``out'' states with a canonical pairing
between them. Each of them is isomorphic to the space of semi-infinite
``delta-forms'' on the loop space $LX$ supported on the subset of
boundary values of holomorphic maps from the disc (which is the
interior of the circle for the ``in'' states and the exterior for the
``out'' states). These spaces of delta-forms may be described in terms
of the familiar Fock representations of the chiral and anti-chiral
$\beta\gamma,bc$-systems (the chiral-anti-chiral de Rham complex of
$X$).

However, this ismorphism is not canonical because of the instanton
effects which cause non-trivial self-extensions between the spaces of
semi-infinite delta-forms. These extensions are similar to the
extensions arising in quantum mechanics which we have studied in
detail in Part I. One consequence of this is that the Hamiltonian is
not diagonalizable, but has Jordan blocks. Another consequence is that
our models are logarithmic conformal field theories.

\subsection{Logarithmic structure of the correlation functions}

The logarithmic features of our model are revealed upon the
computation of the correlation functions. As we discussed above, in
the infinite radius limit the path integral localizes on the moduli
space of instantons, which are holomorphic maps $\Sigma \to
X$. Because we are dealing here with a Morse--Bott--Novikov function,
this moduli space now has infinitely many connected components labeled
by $\beta \in H_2(X,\Z)$ (the moduli space is non-empty only if the
integral of the K\"ahler class $\omega$ of $X$ over $\beta$ is
non-negative). All of them are finite-dimensional.

The simplest class of observables of this model consists of the
evaluation observables corresponding to differential forms on
$X$. Their correlation functions are given by integrals of their
pull-backs to the moduli spaces of holomorphic maps, or, more
precisely, their Kontsevich compactifications, under the evaluation
maps.

Such correlation functions have been studied extensively in the
literature for the BPS observables, corresponding to closed
differential forms on $X$. They are expressed in terms of the
Gromov-Witten invariants of $X$. However, in order to describe the
structure of the sigma model as a full-fledged quantum field theory
(and, for instance, observe that the Hamiltonian is
non-diagonalizable), we must go beyond the topological sector of the
model and study correlation function of more general, non-BPS,
observables. These observables include evaluation observables
corresponding to differential forms on $X$ that are not closed.

In this paper we introduce an even larger class of non-BPS observables
which we call the {\em jet-evaluation observables}. They keep track of
not only the value of a map $\Sigma \to X$ at a point $p \in \Sigma$,
but also its derivatives with respect to a local coordinate at
$p$. These observables correspond to differential forms not on $X$,
but on its jet space $JX$, the space of jets of holomorphic maps from
a small disc to $X$. Their correlation functions are signficantly more
complicated than those of the evaluation observables. They are again
given by integrals over the (compactified) moduli spaces of
holomorphic maps $\Sigma \to X$, but now these integrals may diverge
at the boundary divisors and require regularization. We have
encountered the necessity of the regularization of instanton integrals
in Part I when studying one-dimensional quantum-mechanical models. An
immediate consequence of this regularization is the {\em logarithmic
mixing of operators}: the naive perturbative jet-evaluation operators
are not well-defined as operators of the full quantum field
theory. They are only well-defined up to the addition of their
``logarithmic partners''. This ambiguity is reflected in the ambiguity
of the regularization of our integrals. Roughly speaking, to obtain a
true operator of the non-perturbative sigma model, we need to take a
jet-evaluation observable in its perturbative definition
together with a consistent set of regularization rules for all of
these integrals.

Together, operators and their logarithmic partners span
finite-dimensional subspaces in the space of operators (or,
equivalently, states), on which the Hamiltonian acts as a Jordan
block. The logarithmic mixing is also responsible for the appearance
of logarithmic terms in the operator product expansion (OPE) of these
operators, as we will see in explicit examples below.

We note that the upper-triangular entries in the Jordan blocks are
proportional to the instanton parameters of the model, so their
appearance is caused by the instanton effects. When we move away from
the point $\ol\tau = -i \infty$ (that is, to finite radius),
anti-instantons contributions arise which cause the appearance of
non-zero lower-triangular entries, and the Hamiltonian becomes
diagonalizable. In other words, Jordan blocks appear because there are
instantons, but no anti-instantons, in the infinite radius limit.

Thus, we find that the two-dimensional supersymmetric sigma model in
the infinite radius limit is a {\em logarithmic conformal field
theory} with central charge $0$. Theories of this kind have been
studied extensively recently (see, e.g.,
\cite{Gurarie,Kogan,GL,Schomerus,DF,MR} and references therein), in
part because of their applications to condensed matter physics. The
theory has a large chiral algebra, which is nothing but the space of
global sections of the chiral de Rham complex of $X$ introduced in
\cite{MSV}. (The chiral algebra is not affected by logarithms; only
operators depending on both holomorphic and anti-holomorphic
coordinates are so affected.)

To illustrate these phenomena, we consider explicitly two important
examples: the first is the sigma model with an elliptic curve as the
target. This is essentially a free field theory which may be described
at both finite and infinite radii. We explain what the passage to the
infinite radius means for these models. The second case of interest is
when the target manifold is $\pone$. In \cite{AiB} this model was
described as a deformation of the free field theory by {\em holomortex
operators}. Here we revisit and rederive this description and present
numerous examples of correlation functions, OPE and logarithmic mixing
in this model.

\medskip

In Part III of this paper we will consider similar structures in the
${\mc N} = (0,2)$ supersymmetric, as well as purely bosonic, sigma
models. One of the motivations for the study of these models is that
when the target manifold is the flag variety of a simple Lie group
$G$, this model has an affine Kac--Moody algebra symmetry of critical
level (see \cite{FF:si,F:rev}), even though the theory is not
conformally invariant. The correlation functions in this model,
coupled to a gauge field, may be viewed as solutions of systems of
linear differential equations on the moduli space of $G$-bundles on
the worldsheet, which arise naturally in the geometric Langlands
correspondence.

\ssec{Yang--Mills theory}

The next step in our program is the investigation of the
four-dimensional gauge theory. Again, we start with the supersymmetric
model, which is a twisted ${\CN}=2$ super-Yang--Mills theory with
gauge group $G$ on a four-dimensional manifold ${\mathbf M}^4$, in the
limit $\ol\tau \to -i\infty$ with fixed
$$
{\tau} = \frac{\vartheta}{2\pi} + \frac{4\pi i}{g^{2}}.
$$
Suppose that ${\bf M}^{4} = {\R} \times M^3$,
where $M^3$ is a compact three-dimensional manifold. Let $t$ denote
the coordinate along the $\R$ factor. Then the Yang-Mills theory may
be interpreted as quantum mechanics on the space ${\CA}/{\CG}$ of
gauge equivalence classes of $G$-connections on $M^3$, with the
Morse--Novikov function being the Chern--Simons functional
\cite{Atiyah,W:tft}. However, there is again a new element, compared
to the previously discussed theories, and that is the appearance of
gauge symmetry. The quotient ${\CA}/{\CG}$ has complicated
singularities because the gauge group ${\CG}$ has non-trivial
stabilizers in the space ${\CA}$. For this reason we should consider
the {\em gauged Morse theory} on the space ${\CA}$ of connections
itself.

This theory is defined as follows. Let $X$ be a manifold equipped with
an action of a group $G$ and a $G$-invariant Morse function $f$. Then
the gradient vector field $v^{\m} \pa_{x^{\m}}$, where $v^{\m} =
h^{\m\nu} {\p}_{x^\nu}f$ commutes with the action of $G$. Denote by
$\CV_a^{\m} \pa_{x^{\m}}$ the vector fields on $X$ corresponding to
basis elements $J^a$ of the Lie algebra $\g = \on{Lie}(G)$. We define
a gauge theory generalization of the gradient trajectory: it is a pair
$(x(t): \R \to X,A_t(t)dt \in \Omega^1(\R,\g^*))$, which is a solution
of the equation
\begin{equation}    \label{gauged Morse}
\frac{d{x}^{\m}}{dt} = v^{\m} (x(t))+ {\CV}^{\m}_{a}(x(t))
A^{a}_{t}(t).
\end{equation}
The group of maps $g(t): \R \to G$ acts on the space of solutions by
the formula 
$$
g: \left( x(t),A_t(t)dt \right) \mapsto \left( g(t) \cdot x(t), \
g^{-1}(t) {\p}_{t} g (t)+ g^{-1}(t) A_{t}(t) g(t) \right),
$$
and the moduli space of gradient trajectories is, naively, the
quotient of the space of solutions of \eqref{gauged Morse} by this
action. However, because this action has non-trivial stabilizers and
the ensuing singularities of the quotient, it is better to work
equivariantly with the moduli space of solutions of the equations
\eqref{gauged Morse}.

In \secref{ym} we develop a suitable formalism of equivariant
integration on the moduli space of gradient trajectories of the gauged
Morse theory. We then apply this formalism to the case when $X = \CA$,
the space of connections on a three-manifold $M^3$ and $f$ is the
Chern--Simons functional (note that this formalism may also used to
define gauged sigma models in two dimensions, see \secref{gauging
away}). The corresponding equivariant integrals give us the
correlation functions of evaluation observables of the Yang-Mills
theory in our weak coupling limit $\ol\tau \to -i \infty$. In the case
of the BPS observables these correlation functions are the {\em
Donaldson invariants} \cite{W:tft}. They comprise the topological (or
BPS) sector of the theory.

We obtain more general (off-shell) correlation functions by
considering more general, i.e., non-BPS, observables. We will present
some sample computations of these off-shell correlation functions
which exhibit the same effects as in one- and two-dimensional models
considered above. In particular, we will observe the appearance of the
logarithm function in the correlation functions. Moreover, we find the
same kind of logarithmic mixing that we have observed in
two-dimensional sigma models. We also present an example of instanton
corrections to the OPE of some natural observables in the Yang--Mills
theory which involve logarithm. Thus, we conclude that the
supersymmetric Yang--Mills theory in the weak coupling limit is a
logarithmic CFT in four dimensions.

\ssec{Contents}

The paper is organized as follows. In \secref{2D sigma} we give the
definition of the two-dimensional sigma models and its infinite radius
limit. We then interpret it as quantum mechanics on the the loop space
of the target manifold $X$, equipped with the Floer function. The
multi-valuedness of the Floer function requires certain modifications
in the analysis of quantum mechanical models from Part I. Hence we
discuss in \secref{qm with mn} the general structure of quantum
mechanics on non-simply connected manifolds. Having developed the
general theory, we go back to our main example, the sigma models, in
\secref{back}. We describe its space of states in terms of certain
spaces of delta-forms supported on semi-infinite ascending manifolds
of the Floer function inside the universal covering of the loop space
of $X$. Algebraically, these are given in terms of Fock modules over
the $\beta\gamma$-$bc$-systems. In particular, the chiral algebra of
the sigma model may be identified with the chiral de Rham complex
introduced in \cite{MSV}. We isolate an important subspace in the
space of states: the space of differential forms on the jet space of
$X$. The corresponding operators are the jet-evaluation observables
which we study in detail in the subsequent sections. We also consider
two examples: when $X$ is an elliptic curve and $\pone$.

In \secref{cor fns} we take up the correlation functions. We first
recall the definition of a simplest ones, which are the Gromov--Witten
invariants. Then we explain the difference between the BPS (or
topological) and non-BPS observables and explain why it is important
to go beyond the topological sector of the model. We give examples of
correlation functions of non-BPS observables and show how to extract
from them some unusual features of our model, such as the
non-diagonalizability of the Hamiltonian. The next section,
\secref{log mixing}, plays a special role. Here we look closely at the
correlation functions and the OPE of the jet-evaluation observables
introduced in \secref{back}. We give examples of logarithmic mixing of
observables and explain the underlying reasons for this phenomenon
using the geometry of the moduli spaces of stable maps and the
description of the space of states in terms of consecutive extensions
of spaces of delta-forms.

In \secref{gauged sm} we consider the sigma models in the background
of a $\C^\times$-gauge field. These models correspond to a deformation
of the Morse--Bott--Novikov function to a Morse--Novikov function,
with isolated critical points. This leads to simplifications in the
description of the space of states of the model and allows us to make
a closer contact with the results of Part I. In particular, we
consider the case of the sigma model with the target $\pone$ and
relate the formulas for the Hamiltonian involving the so-called
Cousin--Grothendieck operators obtained in Part I to the description
of the sigma model of $\pone$ as a deformation of the free theory by
holomortex operators.

In \secref{ym} we consider the four-dimensional supersymmetric
Yang--Mills theory. We explain how to modify the formalism developed
in the previous sections in the presence of gauge invariance. We then
compute explicitly non-BPS correlation functions of the $SU(2)$ model
in the one instanton sector using the ADHM construction of the moduli
space of anti-self-dual connections. We show that these correlation
functions exhibit the same kind of logarithmic behavior that we have
seen the one- and two-dimensional models.

Finally, in \secref{conclusions} we present our conclusions and
outlook.

\ssec{Acknowledgments}

We are grateful to DARPA and AFOSR for their generous support through
the FAThM Program and the grant FA9550-07-1-0543 which enabled us to
carry out this project.

\medskip

In addition, the research of AL was supported by the grants RFBR
07-02-01161, INTAS 03-51-6346, NWO-RFBR-047.0112004.026
(05-02-89000-NWOa) and NSh- \linebreak 8065.2006.2 (for the support of
Russian scientific schools). The research of NN was partially
supported by the European commission via the RTN network
``Constituents, Fundamental Forces and Symmetries of the Universe'',
by the ANR grants for the programs ``Structure of vacuum, topological
strings and black holes'', and ``Geometry and integrability in
Mathematical Physics'' and by the grants RFFI 06-02-17382 and
NSh-3036.2008.2 (for the support of Russian scientific schools).

Part of the research was done while NN visited Princeton University
during the Spring semester of 2007. He thanks Departments of
Mathematics and Physics for hospitality. EF thanks the Institute for
Advanced Study, IHES and MSRI for hospitality. AL thanks IHES
for hospitality. 

Some of the results of the paper were reported by EF at the Erwin
Schr\"odinger Institute in Vienna in June of 2006, at the Mathematics
Department of MIT in October of 2006, and at Kyoto University in March
of 2007; by NN at the Cargese Summer School in 2006, at Strings '06 in
Beijing, at the New High Energy Theory Center at Rutgers University in
2006, in the Simons Lecture Series in Stony Brook University in 2007,
at the Physics Department at Princeton University in 2007, and at
DAMTP, Cambridge University in 2008. We thank all these institutions
for the opportunity to present this paper and our audience for helpful
remarks.

\section{Two-dimensional sigma models}    \label{2D sigma}

In this section we give the definition of the supersymmetric
two-dimensional sigma model in the infinite radius limit. We then
interpret it as quantum mechanics on the loop space, equipped with a
Morse--Bott--Novikov function.

\ssec{Definition of the supersymmetric sigma model}

Let $X$ be a compact K\"ahler manifold. We will denote by $X^a,
a=1,\ldots,N=\dim X$, local holomorphic coordinates on $X$, and by
$X^{\ol{a}} = \ol{X^a}$ their complex conjugates. We write the metric
as $g_{a\ol{b}} dX^a dX^{\ol{b}}$ and the K\"ahler form as
\begin{equation}    \label{kahler}
\omega_K =\frac{i}{2} g_{a\ol{b}} dX^a \wedge dX^{\ol{b}}.
\end{equation}

We consider the type A twisted ${\CN}=(2,2)$ supersymmetric sigma
model on a Riemann surface $\Sigma$ with the target manifold
$X$. Given a map $\Phi: \Sigma \to X$, we consider the pull-backs of
$X^a$ and $X^{\ol{a}}$ as functions on $\Sigma$, denoted by the same
symbols.  We also have fermionic fields $\psi^a$ $\psi^{\ol{a}},
a=1,\ldots,N$, which are sections of $\Phi^*(T^{1,0} X)$ and
$\Phi^*(T^{0,1} X)$, respectively, and and $\pi_a$ and $\pi_{\ol{a}}$,
which are sections of $\Phi^*(\Omega^{1,0} X) \otimes \Omega^{1,0}
\Sigma$ and $\Phi^*(\Omega^{0,1} X) \otimes \Omega^{0,1} \Sigma$,
respectively. The Levi-Civita connection on $T X$ corresponding to the
metric $g_{a\ol{b}}$ induces a connection on $\Phi^*(T X)$. The
corresponding covariant derivatives have the form
\begin{align*}
D_{\ol{z}} \psi^a &= \pa_{\ol{z}} \psi^a + \pa_{\ol{z}}
X^b \cdot \Gamma^a_{bc} \psi^c, \\
D_z \psi^{\ol{a}} &= \pa_z \psi^{\ol{a}} +  \pa_z X^{\ol{b}} \cdot
\Gamma^{\ol{a}}_{\ol{b}\ol{c}} \psi^{\ol{c}},
\end{align*}
where $\Gamma^a_{bc} = g^{a \ol{b}} \pa_{b} g_{c \ol{b}}$.

The standard action of the supersymmetric sigma model is
\begin{multline}    \label{2D first action}
\int_{\Sigma} \; \left( \frac{1}{2} \la (g_{a\ol{b}}
(\pa_{\ol{z}} X^a \pa_z X^{\ol{b}} + \pa_z X^a \pa_{\ol{z}}
X^{\ol{b}}) \right. \\ \left. + i \pi_a D_{\ol{z}} \psi^{a} + i
\pi_{\ol{a}} D_{z} \psi^{\ol{a}} + \frac{1}{2} \la^{-1}
R^{a\ol{b}}{}_{c\ol{d}} \pi_a \pi_{\ol{b}} \psi^c \psi^{\ol{d}}
\right) d^2 z,
\end{multline}
We are now in the same position in which we were in quantum mechanics
(see formula (2.6) in Part I), and our analysis will proceed along the
same lines.

We start by describing the instantons and the anti-instantons of this
model. Applying the ``Bogomolny trick'' as in the quantum mechanical
model of (see Part I, Section 2.3), we may rewrite the bosonic part of
the action as
$$
\la \int_{\Sigma} d^2 z \; |\pa_{\ol{z}} X|^2 +
\la \int_\Sigma \Phi^*(\omega_K) = \la \int_{\Sigma} g_{a\ol{b}}
\pa_{\ol{z}} X^a \pa_z X^{\ol{b}} d^2 z + \la \int_\Sigma
\frac{i}{2} g_{a\ol{b}} dX^a \wedge dX^{\ol{b}},
$$
or as
$$
\la \int_{\Sigma} d^2 z \; |\pa_{z} X|^2 - \la \int_\Sigma
\Phi^*(\omega_K) = \la \int_{\Sigma} g_{a\ol{b}} \pa_{z} X^a
\pa_{\ol{z}} X^{\ol{b}} d^2 z - \la
\int_\Sigma \frac{i}{2} g_{a\ol{b}} dX^a \wedge dX^{\ol{b}},
$$
(here $\omega_K$ is the K\"ahler form given by formula
\eqref{kahler}). Thus, we see that in a given topological sector,
where $\int_{\Sigma} {\Phi}^{*}{\om}_{K}$ is fixed, the absolute
minima of the action are given by the holomorphic maps (satisfying
$\pa_{\ol{z}} X^a = 0$) or the anti-holomorphic maps (satisfying
$\pa_{z} X^a = 0$). These are the instantons and the anti-instantons
of the sigma models, just like the gradient trajectories and the
anti-gradient trajectories were the instantons and the anti-instantons
in quantum mechanics. Similarly, to their quantum mechanical
counterparts, the contributions of the instantons and anti-instantons
to the path integral are suppressed by the exponential
factor\footnote{For holomorphic maps $\int_\Sigma \Phi^*(\omega_K)
\geq 0$ and for anti-holomorphic maps $\int_\Sigma \Phi^*(\omega_K)
\leq 0$.} $e^{-\la \left| \int_\Sigma \Phi^*(\omega_K) \right|}$.

The next step is to add to the action \eqref{2D first action} the term
$$
\int_{\Sigma} \Phi^*(B) = \int_\Sigma B_{a\ol{b}} dX^a \wedge
dX^{\ol{b}},
$$
where
$$
B = B_{a\ol{b}} dX^a \wedge dX^{\ol{b}}
$$
is a closed two-form on $X$, called the $B$-{\em field}. Our goal
is to enhance the effect of the instantons and further suppress the
anti-instantons. To this end we choose the $B$-field to be of the form
$$
B = - \la \omega_K + \tau,
$$
where
$$
\tau = \tau_{a\ol{b}} dX^a \wedge dX^{\ol{b}}
$$
is another closed two-form.\footnote{The action is CPT invariant if
$\ol{B} = -B$, i.e., if $B$ is purely imaginary. However, like in
quantum mechanics, we break CPT invariance by considering a complex
$B$-field with the real part equal to $-\la \omega_K$.} Note that it
is analogous to the term $(- \la - i \tau) \int_I df$ that we
added to the Lagrangian of the quantum mechanical model (see
Sections 2.3--2.4 of Part I). The resulting action reads
\begin{multline}    \label{2D second action}
\int_{\Sigma} d^2 z \; \left( \la |\pa_{\ol{z}} X|^2 + i \pi_a
D_{\ol{z}} \psi^{a} + i \pi_{\ol{a}} D_{z} \psi^{\ol{a}} + \frac{1}{2}
\la^{-1} R^{a\ol{b}}{}_{c\ol{d}} \pi_a \pi_{\ol{b}} \psi^c
\psi^{\ol{d}} \right) \\ + \int_\Sigma \tau_{a\ol{b}}
dX^a \wedge dX^{\ol{b}}.
\end{multline}
The holomorphic maps (i.e., the instantons of our model) are the
absolute minima of this action, and they are no longer suppressed in
the path integral, whereas the anti-holomorphic maps (anti-instantons)
are now doubly suppressed by the exponential factor $e^{-2 \la \left|
\int_\Sigma \Phi^*(\omega_K) \right|}$.

\ssec{Infinite radius limit}

We now wish to take the limit $\la \to \infty$, in which the metric on
$X$ becomes very large (hence the name ``infinite radius limit'').  In
this limit the instantons survive, but the anti-instantons disappear.
In terms of the coupling constants
$$
\tau_{a\ol{b}} = B_{a\ol{b}} + \frac{i}{2} \la g_{a\ol{b}}, \qquad
\ol\tau_{a\ol{b}} = B_{a\ol{b}} - \frac{i}{2} \la g_{a\ol{b}}
$$
it is the limit in which $\ol\tau_{a\ol{b}} \to - i \infty$, but
the $\tau_{a\ol{b}}$'s are kept finite.

As in quantum mechanics, in order to implement this limit we first
pass to the first order Langrangian (see \cite{W:tsm,BS}):
\begin{multline}    \label{2D second action prime}
\int_{\Sigma} \left( -i p_a \pa_{\ol{z}} X^{a} - i p_{\ol{a}}
\pa_z \ol{X^a} + \la^{-1} g^{a\ol{b}} p_a p_{\ol{b}} \right. \\ +
\left.  i \pi_a D_{\ol{z}} \psi^{a} + i \pi_{\ol{a}} D_{z} \psi^{\ol{a}}
+ \frac{1}{2} \la^{-1} R^{a\ol{b}}{}_{c\ol{d}} \pi_a
\pi_{\ol{b}} \psi^c \psi^{\ol{d}} \right) d^2 z + \int_\Sigma
\tau_{a\ol{b}} dX^a \wedge d\ol{X^b}
\end{multline}
(compare with the action (2.12) in Part I). For finite
values of $\la$, by eliminating the momenta variables $p_a,
p_{\ol{a}}$ using the equations of motion, we obtain precisely the
action \eqref{2D
 second action}. Therefore the two actions are
equivalent for finite values of $\la$. But now we can take the limit
$\la \to \infty$ in the new action. The resulting action is
\begin{equation}    \label{2D third action}
- i \int_{\Sigma} \left( p_a \pa_{\ol{z}} X^{a} + p_{\ol{a}} \pa_z
X^{\ol{a}} - \pi_a D_{\ol{z}} \psi^{a} - \pi_{\ol{a}} D_{z}
\psi^{\ol{a}} \right) d^2 z + \int_\Sigma \tau_{a\ol{b}} dX^a \wedge
dX^{\ol{b}}
\end{equation}
(compare with formula (2.13) of Part I).

As in the quantum mechanical model, we may redefine the momentum
variables as follows:
\begin{equation}    \label{prime}
p_a \mapsto p'_a = p_a + \Gamma^b_{ac} \pi_b \psi^c, \qquad p_{\ol{a}}
\mapsto p'_{\ol{a}} = p_{\ol{a}} + \Gamma^{\ol{b}}_{\ol{a}\ol{c}}
\pi_{\ol{b}} \psi^{\ol{c}}.
\end{equation}
Then the action \eqref{2D third action} becomes
\begin{equation}    \label{2D fourth action}
- i \int_{\Sigma} \left( p'_a \pa_{\ol{z}} X^{a} + p'_{\ol{a}} \pa_z
X^{\ol{a}} - \pi_a \pa_{\ol{z}} \psi^{a} - \pi_{\ol{a}} \pa_{z}
\psi^{\ol{a}} \right) d^2 z + \int_\Sigma \tau_{a\ol{b}} dX^a \wedge
dX^{\ol{b}}
\end{equation}
(compare with formula (2.16) in Part I). However, the transformation
formulas for the new momenta variables under the changes of
coordinates now become more complicated (see
\secref{delta-forms}). They are similar to the quantum mechanical
formulas (2.17) in Part I.

\ssec{Sigma model in the infinite radius limit as a CFT}

The theory with the action \Ref{2D fourth action} appears to be a
conformal field theory (CFT). Indeed, up to a topological term $\int
{\Phi}^{*}{\tau}$, the action is a curved version of the beta-gamma-bc
system. In particular, the stress-energy tensor is given by the
formula:
\begin{equation} 
T = i ( p^{\prime}_{a} {\pa}_{z}X^{a} + 
{\pi}_{a}{\pa}_{z}{\psi}^{a} ) 
\label{tzz}
\end{equation}
which is invariant under the coordinate transformations. In the
perturbative limit, where we consider the sigma model with the target
space a single coordinate patch $U \approx {\C}^{n}$, the
stress-tensor \Ref{tzz} forms a Virasoro algebra with the vanishing
central charge. At the same time, the following OPEs are obtained from
the action \eqref{2D fourth action}:
\begin{align} \notag
p_a(z) X^b(w) &\sim - \frac{i \delta_{a}^{b}}{z-w}, \qquad \pi_a(z)
\psi^b(w) \sim - \frac{i \delta_{a}^{b}}{z-w}, \\ p_{\ol{a}}(\zb)
X^{\ol{b}}(\wb) &\sim - \frac{i \delta_{\ol{a}}^{\ol{b}}}{\zb-\wb},
\qquad \pi_{\ol{a}}(\zb) \psi^{\ol{b}}(\wb) \sim - \frac{i
\delta_{\ol{a}}^{\ol{b}}}{\zb-\wb}.
  \label{OPEfrom}
\end{align}
However, including non-perturbative effects makes the theory a
logarithmic CFT, and induces interesting corrections to the operator
product expansions derived from \Ref{OPEfrom}.

\ssec{Sigma model as quantum mechanics on the loop space}    \label{as
  qm}

We now interpret the supersymmetric sigma model in the infinite radius
limit as a quantum mechanical model on the loop space $LX$. Let
$\Sigma$ be a cylinder $I \times \BS^1$, where $I$ is an interval. We
will view a map $(I \times \BS^1) \to X$ as a map $I \to LX$, where
$LX = \on{Maps}(\BS^1,X)$ is the loop space of $X$.

We introduce coordinates $t$ on $I$ and $\sigma$ on $\BS^1$, so that
$\sigma$ is periodic with the period $2\pi$. Then the holomorphic
coordinate on $\Sigma$ is $z=t+i\sigma$. Using local holomorphic and
anti-holomorphic coordinates $X^a$ and $X^{\ol{a}}$ on an open subset
$U \subset X$, we may represent a map $\BS^1 \to U$ by the Fourier
series
\begin{equation}    \label{Xa exp}
X^a(\sigma) = \sum_{n \in \Z} X^a_n e^{-in\sigma}, \qquad
X^{\ol{a}}(\sigma) = \sum_{n \in \Z} X^{\ol{a}}_{n} e^{in\sigma}.
\end{equation}
Therefore we may use $X^a_n, X^{\ol{a}}_n$ as local holomorphic and
anti-holomorphic coordinates on the loop space $LX$. Let $p'_{a,n}$
and $p'_{\ol{a},n}$ be the Fourier coefficients of the corresponding
expansions of the momenta variables $p'_a$ and $p'_{\ol{a}}$, dual to
$X^a_{-n}$ and $X^{\ol{a}}_{-n}$, respectively. Note that we have
$$
\ol{X^a_n} = X^{\ol{a}}_{n}, \qquad \ol{p'_{a,n}} = p'_{\ol{a},n}.
$$

Similarly, expanding in Fourier series, we obtain the fermionic
variables $$\psi^a_n, \psi^{\ol{a}}_n, \pi_{a,n}, \pi_{\ol{a},n},
\qquad n \in \Z.$$ In terms of these variables, the action \eqref{2D
fourth action} on the cylinder may be rewritten as follows (we ignore
for a moment the last term $\int_\Sigma \Phi^*(\tau)$):
\begin{multline*}
- i \int_I \left( \sum_{n \in \Z} p'_{a,n} \left(
\frac{dX^a_n}{dt} + n X^a_n \right) + \sum_{n \in \Z} p'_{\ol{a},n}
\left( \frac{dX^{\ol{a}}_n}{dt} + n X^{\ol{a}}_n \right) \right. \\ -
\left. \sum_{n \in \Z} \pi_{a,n} \left( \frac{d\psi^a_n}{dt} + n
\psi^a_n \right) - \sum_{n \in \Z} \pi_{\ol{a},n} \left(
\frac{d\psi^{\ol{a}}_n}{dt} + n \psi^{\ol{a}}_n \right) \right) dt
\end{multline*}
(up to the inessential factor of $\pi$). We recognize in this formula
the action (see formula (1.1) of Part I)
\begin{multline}    \label{1D action}
S = -i \int_{I} \left( p'_A \left( \frac{dX^A}{dt} - v^A \right) +
p'_{\ol{A}} \left(\frac{d{X^{\ol{A}}}}{dt} - \ol{v^A} \right) +
\right. \\ \left. - \pi_A \left( \frac{d \psi^A}{dt} - \frac{\pa
v^A}{\pa X^B} \psi^B \right) - \pi_{\ol{A}} \left(
\frac{d{\psi^{\ol{A}}}}{dt} - \frac{\pa \ol{v^A}}{\pa X^{\ol{B}}}
{\psi^{\ol{B}}} \right) \right) dt - i \tau \int df,
\end{multline}
of the quantum mechanical model associated to the loop space $LX$ and
the vector field
\begin{equation}    \label{vector field v}
v = - \sum_{n \in \Z} n X^a_n \frac{\pa}{\pa
X^a_n} - \sum_{n \in \Z} n X^{\ol{a}}_n \frac{\pa}{\pa X^{\ol{a}}_n}
\end{equation}
In other words, we have written
$$
p'_a \pa_{\ol{z}} X^a = \frac{1}{2} p'_a(\pa_t X^a + i \pa_\sigma
X^a), \qquad p'_{\ol{a}} \pa_z X^{\ol{a}} = \frac{1}{2} p'_a(\pa_t
X^{\ol{a}} - i \pa_\sigma X^{\ol{a}}).
$$

Thus, the above vector field $v$ (which is real) corresponds to the
vector field $i\pa_\sigma$ when acting on holomorphic coordinates and
to $-i\pa_\sigma$ when acting on anti-holomorphic coordinates. Note
that $\pa_\sigma$ is the natural vector field on $LX$ corresponding to
infinitesimal rotation of the loop. Therefore this vector field comes
from the natural $U(1)$-action on $LX$ corresponding to the loop
rotation. Since $X$ is a complex manifold, we may complexify this
action to a $\C^\times$-action. The vector field $v$ then comes from
the action of the subgroup $\R^\times \subset \C^\times$. The gradient
flow is therefore given by the equations
\begin{align}    \label{cr1}
\pa_t X^a + i \pa_\sigma X^a = \pa_{\ol{z}} X^a &= 0, \\
\pa_t X^{\ol{a}} - i \pa_\sigma X^{\ol{a}} = \pa_z X^{\ol{a}} &= 0.
\label{cr2}
\end{align}
These are the Cauchy--Riemann equations for the map $\Phi: \Sigma = I
\times \BS^1 \to X$, and so the gradient trajectories are the
holomorphic maps with given boundary conditions.

Thus, we have interpreted the two-dimensional sigma model on the
cylinder as a quantum mechanical model of the type that we have
studied previously. However, in order to apply our results we need to
represent the vector field $v$ given by formula \eqref{vector field v}
as the {\em gradient vector field} of a function $f$ on $LX$: $v =
\nabla f$.\footnote{Most importantly, our construction involves
multiplication of the wave-functions by $e^{\la f}$, so we do need to
have $f$.} Here we consider a natural metric on $LX$ induced by the
metric $g$ on $X$ (and the measure $d\sigma$ on $\BS^1$). A tangent
vector at a point $\ga: \BS^1 \to X$ is a section of the vector bundle
$\ga^*(TX)$. Given two such sections $\eta_1,\eta_2$, we define their
scalar product as $$\int_{\BS^1} \langle \eta_1,\eta_2 \rangle_g
d\sigma.$$ In order to construct the function $f$, we take the
contraction of the above vector field $v$ with this metric on
$LX$. This gives us a one-form $\beta$ on $LX$, and if $v = \nabla f$,
then $\beta = df$. If $H_1(LX,{\mathbb Q})=0$, then such $f$ would
exist. When we considered above quantum mechanical models on a
K\"ahler manifold $X$ we had assumed that the set of zeros of the
vector field $v$ on $X$ was non-empty. It is known \cite{Frankel} that
in this case $H_1(X,{\mathbb Q}) = 0$ (actually, even $H_1(X,\Z) =
0$), and so the function $f$ exists.

However, for a K\"ahler manifold $X$, the group $H_1(LX,{\mathbb Q})$
is always non-trivial. The best we can do in this case is to construct
a {\em multi-valued} Morse function, also known as {\em Morse--Novikov
function} \cite{novikov}, $f$ on $LX$ whose gradient is the vector
field $v$. This function becomes single valued when pulled back to a
covering $\wt{LX}$ of $LX$. The critical points of this function are
the preimages of constant loops in $\wt{LX}$, so this is strictly
speaking not a Morse--Novikov function, but a {\em
Morse--Bott--Novikov function}. Another important phenomenon is that
at the critical points the Hessian of the function $f$ has infinitely
many positive and negative eigenvalues, so only the relative index of
two critical points is well-defined. This indicates that the
corresponding Morse--Novikov complex computes ``semi-infinite''
cohomology of $LX$. These properties require that we make some
adjustments in our construction.

Before explaining how the theory changes when we take into account all
of these phenomena, we recall how to construct this function in our
case. This construction goes back to the work of Floer \cite{Floer}
(for the connection to two-dimensional sigma models, see, e.g.,
\cite{Sadov}). Let us assume for simplicity that $X$ is
simply-connected, and so $LX$ is connected. Then any loop $\gamma:
\BS^1 \to X$ can be contracted to a point, and hence $\gamma$ may be
extended to a map $\wt\gamma: D \to X$, where $D$ is a two-dimensional
disc with the boundary $\BS^1$. Now set
\begin{equation}    \label{f on LX}
f(\wt\gamma) = \int_D \wt\ga^*(\omega_K).
\end{equation}
How does $f$ depend on the choice of $\wt\ga$ extending a fixed loop
$\ga$? Let $\wt\ga'$ be another such extension. Then gluing $\wt\ga$
and $\wt\ga'$ together along the boundary we obtain a map
$\Phi_{\wt\ga,\wt\ga'}: \BS^2 \to X$. It is clear from the definition
that
$$
f(\wt\gamma) = f(\wt\gamma') + \int_{\BS^2}
\Phi_{\wt\ga,\wt\ga'}^*(\omega_K).
$$

Thus, as a function of $\ga \in LX$, the function $f$ is defined up to
the addition of an integral of the K\"ahler form $\omega_K$ over
cycles in $H_2(X,\Z)$ represented by two-dimensional spheres. Let
$\wt{LX}$ be the space of equivalence classes of maps $\wt\ga: D \to
X$ modulo the following equivalence relation: we say that $\wt\ga \sim
\wt\ga'$ if $\wt\ga|_{\pa D} = \wt\ga'|_{\pa D}$ and $\wt\ga$ is
homotopically equivalent to $\wt\ga'$ in the space of all maps $D \to
X$ which coincide with $\wt\ga$ and $\wt\ga'$ on the boundary circle
$\pa D$. We have the obvious map $\wt{LX} \to LX$, which realizes
$\wt{LX}$ as a covering of $LX$. The group of deck transformations is
naturally identified with $H_2(X,\Z)$. Under our assumption that
$\pi_1(X)$ is trivial, we have $H_2(X,\Z) = \pi_2(X) = \pi_1(LX)$, and
$\wt{LX}$ is the universal cover of $LX$.

The function $f$ defined by formula \eqref{f on LX} is a single valued
function on $\wt{LX}$. It is easy to see that its differential $df$ is
the pull-back of the one-form $\beta$ obtained by contracting the
metric on $LX$ with the vector field $v$. Indeed, consider the
evaluation map $\on{ev}: LX \times \BS^1 \to X$ and the projection
$\on{pr}: LX \times \BS^1 \to LX$. Then $df$ is the pull-back to
$\wt{LX}$ of the one-form on $LX$ which is the push-forward of
$\on{ev}^*(\omega_K)$ with respect to $\on{pr}$. The value of this
one-form on $LX$ on a tangent vector $\eta \in T_\ga(LX) =
\Gamma(\BS^1,\ga^*(TX))$ is the integral
$$
\int_{\BS^1} \langle \ga_*\eta,\omega_K \rangle.
$$
Using formula \eqref{kahler}, we find that in terms of the local
holomorphic coordinates $X^a_n$ on the loop space this one-form is
equal to
$$
- \sum_{n \in \Z} n g_{a\ol{b}} (X^a_n d X^{\ol{b}}_n + n
X^{\ol{b}}_n dX^{a}_n),
$$
which is precisely the contraction of the metric on $LX$ and $v$.
Thus, the vector field $v$ is indeed the gradient of the (multi-valued)
function $f$.

Now we can identify the action \eqref{2D fourth action} with the
quantum mechanical action \eqref{1D action} on the loop space $LX$
equipped with the multi-valued function $f$ given by formula \eqref{f
on LX}. In the next section we will discuss how the multi-valuedness
of this function changes the structure of the model.

\section{Quantum mechanics with Morse--Novikov functions}    \label{qm
with mn}

In this section $X$ denotes a compact finite-dimensional manifold,
which is the target space of a quantum mechanical model. We will focus
on the effects in this model caused by non-simply connectedness of
$X$.  For compact K\"ahler manifolds considered in Part I the
(abelianized) fundamental group is always trivial. Therefore we will
consider here the more general case of a smooth real Riemannian
manifold $X$.

The quantum mechanics on a non-simply connected manifold has some
interesting new features. They stem from the fact that the space of
paths on $X$ is disconnected. Thus the path integral, computing the
amplitudes of propagation from one point on $X$ to another involves a
sum over the connected components, which may be labeled by the
fundamental group $\pi_1(X)$. We will discuss how the instanton
calculus developed in Part I should be adjusted in this more general
setting, with the aim of applying the results to the case of the loop
spaces that is relevant to the two-dimensional sigma models (where the
role of $X$ is played by $LX$, the loop space).

\ssec{Path integral analysis on non-simply connected manifolds}

We study supersymmetric quantum mechanics on a smooth real compact
Riemannian manifold $X$, with a non-trivial fundamental group
${\pi}_{1}(X) \neq 0$. In designing the path integral describing the
kernel $G( x_{i}, x_{f})$ of the evolution operator of getting from a
point $x_{i} \in X$ to a point $x_{f} \in X$ in this situation, a
physicist faces a choice: either to fix a homotopy class of paths
connecting the points $x_{i}$ and $x_{f}$, or to sum over all homotopy
classes with some weights. It is a well-known fact (see, e.g.,
\cite{Coleman}) that one should sum over all homotopy classes in
order to get a unitary theory. However, there are different ways of
summing over them, which correspond to different sectors of the
theory.

In order to define these sectors, it is convenient to introduce a
closed one-form $b \in Z^{1}(X)$ and change the action of the theory
as follows:
\begin{equation}
S \longrightarrow S_{b} = S + 2{\pi}i \int x^{*} b
\label{thetaction}
\end{equation}
The modification \Ref{thetaction} does not modify the equations of
motion.  In fact, it only knows about the homotopy class of a
path. Moreover, up to boundary terms, it only depends on the class
$[b] \in Z^{1}(X)/Z^{1}_{\Z}(X)$.  The space of states becomes a
direct sum, (or rather a direct integral
\begin{equation}
\int_{[b] \in Z^{1}(X)/Z^{1}_{\Z}(X)} {\mathcal H}_{[b]}
\label{thetavac}
\end{equation}
of orthogonal spaces, known as {\it $\vartheta$-vacuum superselection
sectors}). Each of these spaces is isomorphic to the standard space of
$L^2$ differential forms on $X$, but the action of the Hamiltonian on
${\mathcal H}_{b}$ depends on $b$:
$$
H_b = \{ \CQ_b,\CQ_b^* \},
$$
where $\CQ_b = \CQ_0 + 2 \pi i b \wedge$ and $\CQ_b^*$ is its adjoint.
We will analyze this in more detail below.

Note that changing the one-form $b$ by the {\it real} exact form $b
\to b + d \alpha$, does not change the physics of the problem, as this
change can be compensated by the unitary transformation
\begin{equation}
{\Psi} \longrightarrow e^{- 2\pi i \alpha} \cdot {\Psi} 
\label{gauge}
\end{equation}
of the wave-functions. In particular, the spectrum of the Hamiltonian
does not change. This is why in \Ref{thetavac} we have only the
finite-dimensional space
\begin{equation}
Z^{1}(X)/Z^{1}_{\Z}(X) = H^{1}(X, {\R}) / H^{1}(X, {\Z})
\label{thetav}
\end{equation}
which labels different $\vartheta$-sectors. 

Now we shall generalize the standard discussion and allow {\em
complex-valued} closed one-forms $b \in Z^{1}_{\C}(X)$. We will then
discuss the corresponding modification of the space of states.

Let us recall the set-up of Part I, Section 2.3. We start with the
Lagrangian of a quantum mechanical model on $X$ given by formula (2.6)
of Part I:
\begin{multline}   \label{first action}
S = \int_I \left( \frac{1}{2} \la g_{\mu\nu} \frac{dx^\mu}{dt}
\frac{dx^\nu}{dt} + \frac{1}{2} \la g^{\mu\nu} \frac{\pa f}{\pa x^\mu}
\frac{\pa f}{\pa x^\nu} \right. + \\ \left. i \pi_\mu D_t \psi^\mu - i
g^{\mu\nu} \frac{D^2 f}{D x^\nu D x^\al} \pi_\mu \psi^\al +
\frac{1}{2} \la^{-1} R^{\mu\nu}_{\al\beta} \pi_\mu \pi_\nu \psi^\al
\psi^\beta \right) dt.
\end{multline}
Next, we modify it by adding the term $-i \vartheta \int_I df$, where
$f$ is a Morse function. We write $\vartheta = \tau - i \la$, where
$\la$ is the (real) parameter in front of the metric in the action
\eqref{first action}. Then we wish to take the limit $\la \to +\infty$
with $\tau$ finite and fixed. We observe (see Part I, Section 3.2)
that for finite $\la$ the correlation functions of this model are
equivalent to the correlation functions of the model described by the
Lagrangian \eqref{first action} with the additional term $-i \tau
\int_I df$. However, there is a price to pay: we need to 
rescale the wave-functions of the 
``in'' states by $e^{\la f}$, rescale the wave-functions of
the ``out'' states by $e^{-\la
f}$ and conjugate the observables by $e^{\la f}$. By making these
transformations we relate the correlation functions of the theory with
the term $-i \vartheta \int_I df$, where $\vartheta$ has a large
imaginary part, to those of the theory with the term $-i \tau \int_I
df$, where $\tau$ is finite.

Let us revisit our formulas from Part I in the case of
simply-connected $X$. The supercharges and the Hamiltonian of the
theory with the action \eqref{first action}, acting on the space of
$L^{2}$ differential forms on $X$, read as follows (see formulas
(2.4)--(2.4) of Part I)
\begin{align*}
{\mathcal Q} &= d_\la = e^{-\la f} d e^{\la f} = d + {\la} df \wedge
\\ {\mathcal Q}^{*} &= (d_\la)^* = \frac{1}{\la} e^{\la
f} d^* e^{-\la f} = \frac{1}{\la} d^{*} + \iota_{v}, \\
H &= \frac{1}{2} \{ {\mathcal Q} , {\mathcal
Q}^{*} \} = \frac{1}{2} \left( - {\la}^{-1} {\Delta} + \la \Vert df
\Vert^2 + ({\mathcal L}_{v} + {\mathcal L}_{v}^{*}) \right),
\end{align*}
where $v = \nabla f$ (recall that for a vector field $\xi$ we denote
by ${\mc L}_\xi$ its Lie derivative acting on differential forms).

Next, consider the model with the action \eqref{first action} plus the
term $- \la \int_I df$. Then the supercharges and the Hamiltonian take
the form (see formulas (3.4)--(3.6) of Part I):
\begin{align}
Q_\la &= \wt{\CQ} = e^{\la f} \CQ e^{-\la f} = d, \\
Q^*_\la &= \wt{\CQ}^* = e^{\la f} {\CQ}^{*} e^{-\la f} = 2 \iota_{v} +
\frac{1}{\la} d^{*}, \\    \label{laham}
H_\la &= e^{\la f} H e^{-\la f} = \frac{1}{2} \{ Q_\la,Q^*_\la \} =
{\mc L}_v - \frac{1}{2\la} \Delta.
\end{align}

Now let us also add the term $- i \tau \int_I df$ to the previous
action (so the net result is the action \eqref{first action} plus the
term $- i \vartheta \int_I df$). The corresponding supercharges and
Hamiltonian read
\begin{align} \notag
Q_{\la,\tau} &= e^{i \tau f} Q_\la e^{- i \tau f} = d - i \tau df
\wedge, \\ Q^*_{\la,\tau} &= e^{i \tau f} Q_\la^* e^{- i \tau f} = 2
\iota_{v} + \frac{1}{\la} (d^{*} + i \tau i_v), \notag \\ H_{\la,\tau}
&= e^{i \tau f} H_\la e^{- i \tau f} = {\CL}_{v} - i {\tau} \
\Vert v \Vert^2 + \frac{1}{2\la} \left( \ -
{\Delta} + {\tau} \left( {\CL}_{v} + {\CL}_{v}^{*} \right) + i
{\tau}^2 \ \Vert v \Vert^2 \right).
\label{tauham}
\end{align}
In the case of simply-connected $X$, the Hamiltonians \Ref{tauham} and
\eqref{laham} are related by conjugation with $e^{i \tau f}$ (and
likewise for the supercharges). Therefore in the case of
simply-connected manifold $X$, this finite $\tau$-term is not
important (see the discussion at the end of of Section 2.4 of Part
I). That is why in most of the discussion of Part I we had dropped
this term. But on non-simply connected manifolds this term is
important as it corresponds to the choice of the ``$\vartheta$-vacuum''
sector. Therefore we need to include it in our formulas in this case.

\ssec{Maximal abelian cover}    \label{abelian}

In all of the above formulas the Hamiltonian and the supercharges
depend only on the vector field $v$. If $H_1(X,\Z) = 0$ (in
particular, for simply-connected $X$) there exists a function $f$ on
$X$ such that $v = \nabla f$, and we have used this function in the
above formulas. Let us suppose now that $H_1(X,\Z) \neq 0$. Then $v$
may still be written as $\nabla f$, but $f$ may be multi-valued, i.e.,
defined on the $H_1(X,\Z)$-cover $\wt{X}$ of $X$. This is the maximal
abelian cover of $X$.\footnote{We will assume that $H_1(X,\Z)$ has no
torsion; otherwise, we have to take the quotient of $H_1(X,\Z)$ by its
torsion subgroup.} The points of $\wt{X}$ may be described as pairs
$(x,[I])$, where $x \in X$ and $[I]$ is an equivalence class of a path
$I$ connecting $x$ with another, fixed, reference point $x_0 \in
X$. The equivalence relation identifies two paths $I$ and $I'$ if the
image of their difference in $H_1(X,\Z)$ is equal to $0$. The group
$H_1(X,\Z)$ naturally acts on $\wt{X}$.

Let $b_v$ be the closed one-form obtained by contracting $v$ and the
metric $g$ on $X$ (thus, $b_v = df$ in the simply-connected
case). Then the sought-after function $f$ on $\wt{X}$ is constructed
as follows: its value at the point $(x,[I]) \in \wt{X}$ is equal to
$\int_I b_v$. Then we have $v = \nabla f$, in the sense that the
vector field $\nabla f$ on $\wt{X}$ is $H_1(X,\Z)$-invariant and
corresponds to the vector field $v$ on $X$.  Note also that $df$ is
the pull-back of $b_v$ to $\wt{X}$. In what follows, by a slight
abuse of notation, we will sometimes write $df$ for $b_v$.

Let us consider the case when $f$ is a function on $\wt{X}$ that is
not $H_1(X,\Z)$-invariant, and hence gives rise to a multi-valued
function on $X$. In this case conjugation by $e^{i \tau f}$ is
problematic, since it maps differential forms on $X$ (i.e.,
$H_1(X,\Z)$-invariant forms on ${\wt X}$) to differential forms on
$\wt X$, which are $\tau$-equivariant. These are the forms
${\omega} \in {\Omega}^{\bullet}({\wt X})$ such that
\begin{equation}
{\mu}^{*}{\omega} = {\exp} \ \left(  i \tau \int_{\mu}df \right)
\ {\omega}, \qquad \mu \in H_1(X,\Z).
\label{tauequiv}
\end{equation}
We denote the space of such forms by ${\Omega}_{\tau}^{\bullet}(X)$.

Thus, we find that there are two possible pictures for the Hamiltonian
formalism with non-zero $\tau$ (at finite $\la$ for now): we may
either consider the space ${\Omega}^{\bullet}(X)$ with the Hamiltonian
$H_{\la,\tau}$ (which tends to ${\CL}_{v} + i {\tau} \Vert v
\Vert^2$ in the limit $\la \to \infty$), or the space
${\Omega}_{\tau}^{\bullet}(X)$ with the Hamiltonian $H_\la$ (which
tends to $\CL_v$ in the limit $\la \to \infty$). Given an eigenstate
$\Psi$ of $H_{\la,\tau}$ in ${\Omega}^{\bullet}(X)$, we obtain an
eigenstate $e^{-i \tau f} \Psi$ of $H_\la$ in
${\Omega}_{\tau}^{\bullet}(X)$.

However, as in the simply-connected case (see Part I, Section 3), in the
limit $\la \to \infty$ both spaces decouple into spaces of "in" and
"out" states and undergo a violent transformation which results in
certain spaces of delta-forms. In the first picture, those should be
defined on $X$, and in the second -- on $\wt{X}$, with the two
pictures again related by the operator of multiplication by $e^{i \tau
f}$.

\ssec{Ground states}

The first question to consider is the structure of the ground
states. Here we encounter the following puzzle. In the case of
simply-connected $X$ there were two obvious ground states (for all
finite values of $\la$): the zero form $1$ and the top form $e^{2\la
f} {\rm vol}_{g}$ (see Part I, Section 3.5). But now the analogue of
the "in" ground state $1$ in ${\Omega}^{\bullet}(X)$ is the
eigenfunction $e^{i {\tau} f}$ of $H_{\la,\tau}$. However, this
function is not periodic, and hence it does not belong to the space of
states (the corresponding function in ${\Omega}_{\tau}^{\bullet}(X)$
would be $1$, which is not $\tau$-periodic, and hence also not
allowed). So, naively, we could conclude that the ground state which
is of degree $0$ as a differential form is absent in the spectrum of
the theory. A similar analysis shows that the top degree form would be
absent as well.

However, it may be easily shown by the path integral analysis that
these non-periodic wave-functions are in fact the correct ground states
of the $\la = \infty$ theory.

The point is that we should view the function $e^{-i \tau f}$ on
$X$ not as a function, but as a {\em generalized function} (or
{\em distribution}) on $X$. As we explained in Part I, Section 4,
including such generalized functions into the spectrum is inevitable
for the theory in the limit $\la \to \infty$. Now we see that
for non-simply laced manifolds $X$ one has to allow distributions with
mild singularities (such as the jump of $f$ across some hypersurface)
even for finite values of $\la$ in order to have a consistent
Hamiltonian formulation of the theory. This type of singularity is
similar, and actually milder, than the delta-form type singularities we
have encountered in the study of the excited states in the limit $\la
\to \infty$ in case of simply-connected $X$, and which we will
encounter below for non-simple connected $X$ as well. It might lead,
in the worst case, to the Jordan block form of the Hamiltonian
\Ref{tauham} at $\la = \infty$.

Since it is not {\em a priori} obvious whether to allow such jumps
of the eigenfunctions or not, it is instructive to analyze this
phenomenon in detail in the simplest possible case, that of a circle
$\BS^1$. This will be done in the next section and will help us to
confirm that sometimes such discontinuous periodic functions are
indeed bona fide eigenfunctions of the Hamiltonian.

\ssec{Example: $X = {\mathbb S}^{1}$}    \label{BS1}

As an example, take $X = {\mathbb S}^{1}$, with the coordinate 
$x \sim x + 2{\pi}$, and the Morse--Novikov function 
\begin{equation}
f(x) = {\mu} x + {\rm cos}(x).
\label{fofx}
\end{equation}

\begin{center}
\Figx{10}{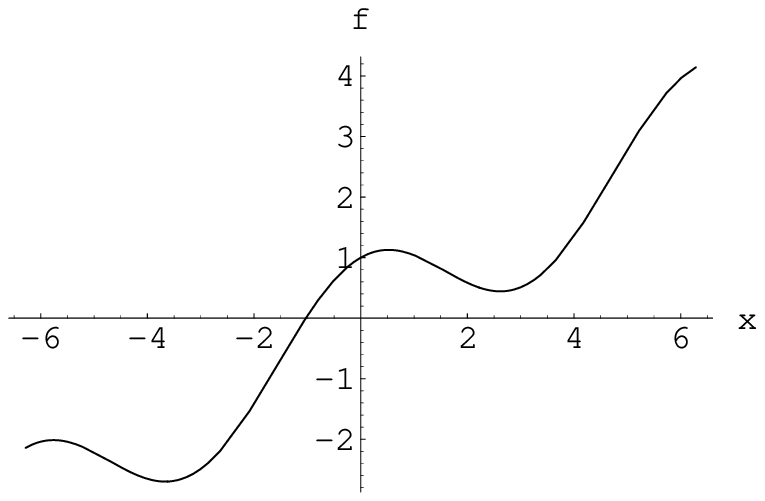}{{\bf Figure 1. Morse--Novikov function on ${\mathbb
S}^{1}$}}
\end{center}

\sssec{Act I}    \label{actI}

We start with the model described by the action \eqref{first action}
with the additional term $(- i\tau - \la) \int_I df$ in the limit $\la
\to \infty$. We rewrite the corresponding action in the first order
formalism, following Part I, Section 2.3, like this:
$$
S = - i \int_I p( \dot x - V(x)) dt - i \int_I V(x) \dot x dt,
$$
where our vector field is $v = V(x) \pa_x$, with
$$
V(x) = {\mu} - {\rm sin}(x).
$$

Suppose first that $0 < | {\mu} | < 1$. Then it has two critical
points: $x_{+}$ and $x_{-}$, ${\rm sin}x_{\pm} = {\mu}$, with ${\rm
cos}x_{+} < 0$, ${\rm cos}x_{-} > 0$.  Let us choose the branch
$f_{+}(x)$ of $f(x)$ such that it is smooth at $x_{+}$ and has a jump
at $x_{-}$:
$$
f(x_{-} - 0) = f ( x_{-} + 0 ) + 2{\pi}{\mu}.
$$
Let us also define the branch $f_{-}(x)$, which is smooth at $x_{-}$,
and jumps by $ 2{\pi}{\mu}$ across $x_{+}$.

The Hamiltonian at $\la = \infty$ is given by the first order
operator:
\begin{equation}
H_{\infty,\tau} = V(x) \left( {\pa}_{x} - i {\tau} V (x) \right).
\label{cirham}
\end{equation}
acting on the degree zero differential forms, i.e., functions on
${\BS}^{1}$.  It annihilates the ground state wave-function:
$$
{\Psi}^{(0)\rm in} (x) = e^{ - i \tau f_{+} (x)}.
$$
Indeed, the ${\delta}( x - x_{-})$ term in
$H_{\infty, \tau} {\Psi}^{(0)\rm in}$, due to the jump of $f(x)$, is
multiplied by $V(x_{-}) = 0$. The second ground state, a one-form, is
given by the delta-form supported at the attractive critical point
$x_{-}$:
$$
{\Psi}^{(1)\rm in}(x) = e^{i {\tau} f_{-}(x_{-})} \ {\delta}( x -
x_{-}) dx.
$$
Analogously, the "out" states are built using the branch $f_{-}(x)$:
\begin{align}
& {\Psi}^{(0)\rm out} (x) = e^{ i \tau f_{-}(x)} \\
& {\Psi}^{(1)\rm out} (x) = e^{i {\tau} f_{+}(x_{+})} {\delta}( x -
  x_{+}) dx.
\label{outstates}
\end{align}
This is in complete analogy with the analysis on $\C\pone$ presented
in Part I, Section 3.5.

Now let us briefly discuss the excited states. By analogy with the
analysis of Part I, Section 3.7, we have two series of (generalized)
"in" eigenstates of the operator $H_{\infty,\tau}$ on the zero degree
forms corresponding to the two critical points (and similarly for
one-forms). The first of them is found to be proportional to
$$
{\Psi}^{(0){\rm in}}_{n,-} = {1\over (n-1)!} {\pa}^{n-1}_{w} \left(
e^{- i \tau f_{-}(x)} {\delta}(w) \right), \qquad n>0,
$$ 
where $w$ is a new coordinate on ${\BS}^{1}$ near $x = x_{-}$,
such that
\begin{equation}
w = e^{{\kappa}_{-} y}, \quad y =   \int_{0}^{x} {dx \over V(x)},
\label{linear}
\end{equation}
and the constant $\kappa_{-}$ is chosen so that the map $x \mapsto w$
is a local diffeomorphism near $x = x_{-}$:
\begin{equation}
{\kappa}_{-} = V^{\prime}(x_{-})
\label{adju}
\end{equation}
In our example ${\kappa}_{-} = \sqrt{1 - {\mu}^{2}}$. 

The corresponding eigenvalue of the Hamiltonian \eqref{cirham} is
given by:
\begin{equation}
E_{n,-} = \sqrt{1 - {\mu}^{2}} \ n \ , \ \quad n > 0
\label{energ}
\end{equation} 
(the zero energy state is given by the wave-function ${\Psi}^{(0){\rm
in}}$).

The second series of excited states on the the zero-forms are
generalized eigenstates of the Hamiltonian \eqref{cirham} given by the
formula
$$
{\Psi}^{(0){\rm in}}_{n,+} = z^{n} e^{-i {\tau}f_{+}(x)},
$$
where $z$ is another coordinate on $\BS^{1}$, which is related by the
local diffeomorphism to $x$ near $x_{+}$, where it vanishes:
\begin{equation}
z = e^{{\kappa}_{+} y} , \qquad {\kappa}_{+} = V^{\prime} (x_{+}).
\label{adjui}
\end{equation} 
In our example ${\kappa}_{+} = - {\kappa}_{-}$ and
$$
z = {1\over w}.
$$
The Hamiltonian has Jordan blocks mixing the states ${\Psi}^{(0){\rm
in}}_{n,+}$ and ${\Psi}^{(0){\rm in}}_{n+1,-}$.

There is also a similar picture for excited "in" states among
one-forms, and also for the "out" states.

\sssec{Intermezzo}

Consider a more general Morse--Novikov function
$$
f(x) = {\mu} x  + {\varphi}(x)
$$
with sufficiently small $\mu$ so that
the vector field $V(x) = f^{\prime}(x)$ has multiple zeroes $x_{\al}$
on ${\BS}^{1}$:
$$
{\mu} + {\varphi}^{\prime}(x_{\al}) = 0.
$$ 
The spectrum of the theory depends crucially on the "weights"
$$
{\kappa}_{\al} = V^{\prime} (x_{\al}).
$$
The spectrum can be either simple or degenerate. If the weights have
"resonances", as in the above example, the Hamiltonian has Jordan
blocks, familiar from our analysis in Part I. But for generic weights
the spectrum is simple, and hence no Jordan blocks arise. Note that we
may set $\mu = 0$.

\sssec{Act II} \label{actII} 

Next, we increase $\vert \mu \vert$. Once $| {\mu} | > 1$, the vector
field $V(x) {\pa}_{x}$ has no zeroes, and hence it can be brought to
the normal form
\begin{equation}
V(x) {\pa}_{x} = {2\pi\over T} {\pa}_{y},
\label{norml}
\end{equation}
where $T$ is the period of revolution:
\begin{equation}
T = \int_{0}^{2\pi} {dx \over V(x)},
\label{periodt}
\end{equation}
and the periodic coordinate $y \sim y + 2{\pi}$ is given by the integral 
\Ref{linear}. The period $T$ in our case is given by:
\begin{equation}
T = {2\pi \over \sqrt{{\mu}^{2} - 1}}.
\label{periodtt}
\end{equation}
Miraculously enough, we can now find perfectly smooth eigenfunctions
of the Hamiltonian $H_{\infty,\tau}$:
\begin{equation}
{\Psi}^{(0)\rm in}_{n} = e^{- i {\tau} \left( f(x) - {\mu} {y\over T}
  \right) + i n y }, \qquad n \in {\Z},
\label{psin}
\end{equation}
with the eigenvalues
\begin{equation}
E_{n} = i \sqrt{{\mu}^{2}-1} \left( n - {\tau}{\mu} \right).
\label{energn}
\end{equation}
Thus, we find that the spectrum is discrete and has a non-zero
imaginary part.

\sssec{Act III} \label{actIII}

Let us now compare our results with the standard example of the free
particle on the circle, which can be solved for finite $\la$. This is
somewhat similar to testing the $\la = \infty$ theory against the
harmonic oscillator (see Part I, Sects. 3.3--3.4).

So, we take again $X = {\BS}^{1}$, with the metric ${\la} {\rm d}x^2$,
$x \sim x + 2{\pi}$. We consider the function $f = {\mu}x$. In a
sense, this function corresponds to the limit ${\mu} \to \infty$ of
the function from \secref{actII}, where we neglect the contribution of
the cosine function ${\rm cos}(x)$. The Lagrangian, with the
$\vartheta$-term, is given by:
\begin{equation}
S_{\vartheta} = \int {\la\over 2} \left( \left( {\dot x}^{2} - {\mu}^2
\right) + i {\vartheta}{\mu} {\dot x} \right) dt,
\label{theth}
\end{equation}
where, as before $\vartheta = \tau - i \la$. The quantization of the
model \Ref{theth} is done in a standard way, leading to the
eigenstates, labeled by the integer $ n \in \Z$,
$$
{\Psi}_{n}(x) = \frac{1}{\sqrt{2{\pi}}}\ {\exp} \ i  n x,
$$
of the Hamiltonian $H_{\la,\tau}$ with the energy
\begin{equation}
   E_{n}  = \frac{\left( n - {\vartheta} {\mu} \right)^2 }{ 2 \la}
+ {1\over 2} {\la} {\mu}^{2} \ , \qquad n \in {\Z}.
\label{perspec}
\end{equation}

For real $\vartheta$ the spectrum is real, positive, discrete, and
bounded from below. Now let us rewrite \Ref{perspec} in terms of
${\tau} = {\vartheta} + i {\la}$:
$$
E_{n, \la} = i {\mu} ( n + {\tau}{\mu} ) + \frac{( n+ {\tau}{\mu}
  )^2}{2\la}.
$$
Clearly, the ${\la} \to \infty$ limit of this expression with $\tau$
kept finite gives \Ref{energn} up to the replacement $${\mu} \to
\sqrt{{\mu}^2 -1} \ . $$

\sssec{Corrections in $\frac{1}{\la}, \frac{1}{\mu}$}

The Hamiltonian for $ f (x) = {\mu}x + {\rm cos}(x)$ can be obtained
from the one for $f(x) = {\mu}x$ by perturbing it with the term ${\rm
  sin}(x){\p}_{x}$ (after the conjugation by $e^{i {\tau}f(x)}$). The
second order perturbation theory gives:
\begin{equation}
E_{n} = ( n + {\tau}{\mu} ) \left( i {\mu} +{\epsilon}_{n} \right) 
\left( 1 + \frac{2}{ (2 i {\mu} + 4 {\epsilon}_{n} )^2 - {\la}^{-2}}
\right) \ , \label{scndord}
\end{equation}
where ${\epsilon}_{n} = \frac{n + {\tau}{\mu}}{2\la}$. In the limit
$\la \to \infty$ we get
$$
E_{n} \sim i {\mu} ( n  + {\tau}{\mu} ) ( 1 - \frac{1}{2{\mu}^{2}} ),
$$
which is the expansion of $\sqrt{{\mu}^{2}-1}$. If we expand
\Ref{scndord} in $\frac{1}{\la}$, the first correction is given by:
$$
E_{n} \sim ( n + {\tau}{\mu} ) \left(  i {\mu} ( 1 -
\frac{1}{2{\mu}^2}) + {\epsilon}_{n} ( 1 + \frac{1}{2{\mu}^{2}} )
\right).
$$ 

Note that there are two ways to describe the
situation here.  In one approach, we deal with $L^2$ functions on $X =
{\BS}^{1}$, and the Hamiltonian is $\vartheta$-dependent:
$$
H_{\la,\tau} = {1\over 2 \la} \left( -i {\pa}_{x} + (\tau - i \la) {\mu}
\right)^2 + {1\over 2} {\la} {\mu}^{2}.
$$
In the second approach, we lift the wave-functions to the covering
space ${\wt X} = {\R}$ and perform the gauge transformation
\begin{equation}
{\Psi} (x) = e^{ i {\vartheta} f} {\wt\Psi}(x),
\label{thetavv}
\end{equation}
which maps the Hamiltonian to
$$
H_{\la} = -{1\over 2 \la} {\pa}_{x}^{2} + {1\over 2}
{\la} {\mu}^{2},
$$
making it $\tau$-independent.  The price to pay for the simpler form
of the Hamiltonian is the non-trivial ${\pi}_{1}(X) =
{\Z}$-equivariance condition on the resulting wave-functions $\wt\Psi$:
\begin{equation}
{\wt\Psi}(x + 2{\pi}) = e^{-2{\pi} i {\vartheta}{\mu}}
{\wt\Psi}(x),
\label{thetavvv}
\end{equation}
which leads to the same spectrum \Ref{perspec} if the appropriate
analytic conditions are imposed.

\sssec{Act IV} \label{actIV} 

Let us now discuss a puzzle. It is {\sl well-known} that in the
presence of the periodic potential, the energy levels split to form
continuous bands \cite{Coleman}, \cite{Abrikosov}:
\begin{equation}
E ( {\vartheta} ) = E_{0} + K e^{-S_{0}} {\rm cos} {\vartheta},
\label{Colspec}
\end{equation}
where one assumes a generic periodic potential $U(x)$:

\Figx{10}{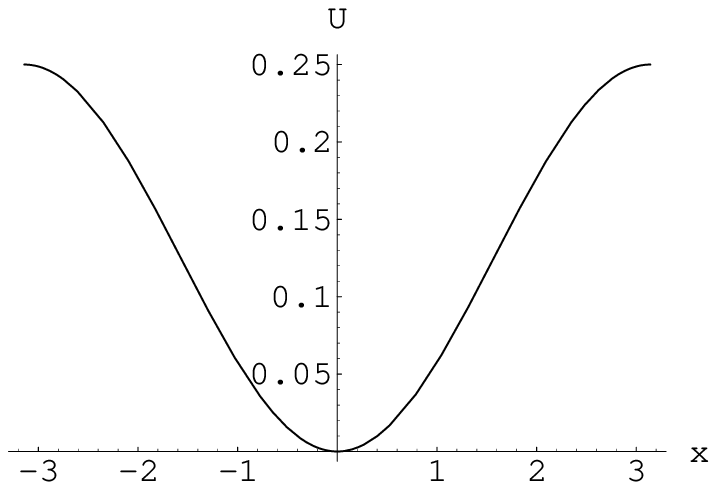}{{\bf Figure 2. A typical periodic potential, i.e.,
a potential on ${\mathbb S}^{1}$.}}

and $S_{0}$ is the instanton action, i.e.
$$
S_{0} = \int \sqrt{U(x)} dx
$$
Now let us write
$$
U(x) = {\la} f^{\prime}(x)^2
$$
and take our limit: ${\la} \to \infty$, ${\vartheta} \to -i\infty$, 
${\tau}$ finite. The spectrum \Ref{Colspec} becomes complex:
\begin{equation}
E ( {\tau} ) \sim E_{0} + K e^{- {\tau}}
\label{complspec}
\end{equation}
This result should surprise us, since we are not in the situation of
act III (see \secref{actIII}) where the function $f(x)$ had no
critical points and the spectrum of Hamiltonian turned out to be
imaginary. In the case at hands the function $f(x)$ clearly has the
critical point(s) (which correspond(s) to the minima of $U(x)$).  What
is going on here?

The resolution of the paradox is quite instructive. The superpotential
$f(x)$ in this example is not a Morse function! Indeed, if we want to
insist on the periodicity of $df$, not just $\Vert df \Vert^2$ then we
have to admit that near the minimum of $U(x)$ (in our example it is
unique) the function $f(x)$ behaves as:
$$
f(x) \sim x | x | \ , 
$$
or, if we want to achieve the same qualitative behavior with
$C^{\infty}$-functions, $f(x) \sim x^3$, a $A_2$ singularity.  Once we
perturb it a little bit, two Morse critical points will occur, and the
spectrum will change dramatically -- the eigenvectors will be replaced
by the adjoint eigenvectors as in the act I (\secref{actI}).

\ssec{From path integral to the space of states}

Having analyzed in detail the case $X = \BS^1$, we turn to a general
non-simply connected target manifold $X$.

Let $\theta$ be a (complex) one-form on $X$. Consider the quantum
mechanical model described by the action \eqref{first action} plus the
term $-i \int_I \theta$. Let us write
\begin{equation}    \label{theta}
\theta = \tau - i \la df,
\end{equation}
where $\tau$ is another one-form. Recall that $f$ is {\em a priori} a
multi-valued function. In this formula, $df$ really stands for
$b_v$, the contraction of the vector field $v$ with the metric
$g$. Formula \eqref{theta} is a small generalization of the formulas
above in which we had $\tau df, \tau \in \C$, for the one-form
$\tau$, so that $\theta = \vartheta df$. In fact, the one-form $\tau$
does not have to be proportional to $df$.

The Hilbert space of states of this model is the space of $L^{2}$
differential forms on $X$. The question before us is to describe what
happens with this space in the limit when $\la \to +\infty$, and
$\tau$ is fixed.

\medskip

As discussed in Part I, Section 3.2, the values of the ``in'' states
of the theory at points of $X$ may be constructed as path integrals
over maps from the half-line $I = (-\infty,0]$ to $X$. The value of
the wave-function corresponding to a state $\Psi$ at $x \in X$ is
given by the path integral over those trajectories for which the end
point $0 \in I$ goes to $x$. Now, as explained in Part I, Section 3.2,
the theory with the term $- i \int_I \theta$ is equivalent to the
theory without this term, but in which the ``in'' states are
multiplied by $e^{i \int_{x_-}^x \theta}$ and the ``out'' states -- by
$e^{-i \int_{x_-}^x \theta}$ (here $x_-$ is the boundary condition at
$-\infty \in I$ corresponding to the choice of the vacuum). If
$\theta$ were exact, these states would be well-defined functions on
$X$ (as in our analysis in Part I). But we are now interested in the
case when $\theta$ is not exact. Then this integral depends on the
choice of the integration contour $I$ going from $x_-$ to $x$.

In fact, if we replace $I$ by another path $I'$ starting and ending at
the same points, the action will change by the factor $e^{i
\int_{[I]-[I']} \theta}$, where we integrate $\theta$ over the closed
loop obtained by gluing together $I$ and $-I'$. Thus, the
corresponding path integral is naturally a function not on $X$, but on
$\wt X$, the $H_1(X,\Z)$-covering $\wt{X}$ of $X$ (as before, we
assume here that $H_1(X,\Z)$ has no torsion). As explained above, in
\secref{abelian}, the points of $\wt{X}$ are pairs $(x,[I])$, where $x
\in X$ and $[I]$ is a homotopy class of a path $I$ connecting $x$ with
another, fixed, point $x_0 \in X$ (corresponding to the boundary
condition at $-\infty$). The equivalence relation identifies two paths
$I$ and $I'$ if the image of their difference in $H_1(X,\Z)$ is equal
to $0$. Note that the group $H_1(X,\Z)$ naturally acts on $\wt{X}$.

The ``in'' states of the model with the term $- i \int_I \theta$ are
therefore the differential forms $\Psi$ on $\wt{X}$ satisfying the
equivariance condition
\begin{equation}    \label{eq cond}
\gamma^*(\Psi) = e^{i \int_\gamma \theta} \Psi, \qquad \ga \in
H_1(X,\Z).
\end{equation}
In the physical model $\theta$ is a real one-form. In this case the
space of such states carries a natural structure of a Hilbert space
with the hermitean inner product given by the formula
$$
\langle \Phi \vert \Psi \rangle = \int_{X} (\star {\overline{\Phi}})
\wedge {\Psi}.
$$
The equivariance conditions on $\Psi$ and $\Phi$ give rise to factors
that are inverse to each other in this case, and so their product is a
differential form on $\wt{X}$ that is a pull-back of a form on $X$
which we can integrate over $X$ (this is the appropriate version of
the $L^2$ condition in the non-simply connected case). In other words,
the operation of complex conjugation and Hodge star identifies the
spaces of ``in'' and ``out'' states in this case.

The existence of a hermitean inner product is expected in a model
described by a CPT invariant action, as explained in Part I, Section
3.2. Our action is CPT invariant precisely when $ \theta$ is real.
Note that there is a special case of this construction when the
lattice of periods $\int_\ga \theta$ in $\C$ is equal to
$2\pi\Z$. Then the equivariance condition \eqref{eq cond} means that
$\Psi$ is a differential form on $X$, so the space of states is not
affected. Therefore we obtain a family of hermitean theories
parametrized by $H^1(X,{\R})/2\pi H^1(X,\Z)$.

\medskip

For our purposes, however, we wish to consider a complex one-form
$\theta = -i \la df + \tau$, where $\la$ is a real parameter which
coincides with the factor in front of the metric in the action
\eqref{first action}. Let us assume that $\tau$ is a real one-form. We
wish to take $\la$ to $+\infty$, while keeping $\tau$ fixed. Adding
this $\theta$-term breaks the CPT invariance of the action, so that
we have, as before, separate spaces of ``in'' and ``out'' states. What
are these spaces?

Let $f$ be the function on $\wt{X}$ defined by the formula $f(x,[I]) =
\int_I \beta$. Let us assume that $f$ has only isolated non-degenerate
critical points on $\wt{X}$. In this case we will say that $f$ is a
{\em Morse--Novikov function on} $X$ (see \cite{novikov}).  Note that
while $f$ is not well-defined on $X$, the gradient vector field
$\nabla f$ and the corresponding $\C^\times$-action $\phi$ descend to
$X$, by our assumption. Therefore the critical points of $f$ on
$\wt{X}$ are the preimages of the fixed points $x_\al, \al \in A$, of
$\phi$ on $X$. Let $S_\al$ be the set of preimages of $x_\al$ in
$\wt{X}$. For each $\al \in A$, $S_\al$ is a torsor over $H_1(X,\Z)$;
in other words, $H_1(X,\Z)$ acts on $S_\al$ simply transitively, even
though $S_\al$ cannot be canonically identified with $H_1(X,\Z)$.

As in Part I, we obtain the vectors in the ``in'' space of states by
multiplying the differential forms $\Psi$ on $\wt{X}$ satisfying
equivariance condition \eqref{eq cond} by $e^{\la f}$. These newly
rescaled states $\wt\Psi$ then satisfy the equivariance condition
\begin{equation}   \label{eq cond1}
\gamma^*(\wt\Psi) = e^{i  \int_\gamma \tau} \wt\Psi, \qquad
\gamma \in H_1(X,\Z).
\end{equation}

Likewise, the vectors in the ``out'' space are obtained by multiplying
the differential forms $\Phi$ on $\wt{X}$ satisfying the equivariance
condition
$$
\gamma^*(\Phi) = e^{- i  \int_\gamma \theta} \Phi, \qquad
\gamma \in H_1(X,\Z),
$$
by $e^{-\la f}$. These new states $\wt\Phi$ satisfy the equivariance
condition
\begin{equation}   \label{eq cond2}
\gamma^*(\wt\Phi) = e^{- i  \int_\gamma \tau} \wt\Phi, \qquad
\gamma \in H_1(X,\Z).
\end{equation}
Because the equivariance conditions \eqref{eq cond1} and \eqref{eq
cond2} are opposite to each other, the wedge product $\Phi \wedge
\Psi$ is the pull-back of a differential form on $X$, which we
integrate over $X$. Hence we obtain a well-defined pairing between the
spaces of ``in'' and ``out'' states.

\ssec{Spaces of delta-forms}    \label{spaces of delta}

The limit $\la \to \infty$ is now described in the same way as in the
case when $f$ is a Morse function on $X$. Quasi-classical analysis,
as in Part I, Section 3.8, shows that before multiplication by $e^{\la
f}$, the wave-functions are concentrated near the critical points of
the Morse--Novikov function on $\wt{X}$, where they are approximately
given by Gaussian type distributions. However, after we multiply them
by $e^{\la f}$, the terms in the Gaussian distribution corresponding
to the positive eigenvalues of the Hessian of $f$ at the critical
point disappear, whereas the ones corresponding to the negative
eigenvalues get doubled. As the result, we obtain delta-forms
supported at the strata of the decomposition of $\wt{X}$ into the
ascending manifolds with respect to $f$ (this analysis applies to the
``in'' states; for the ``out'' states we obtain delta-forms supported
at the descending manifolds).

Let $\wt{X}_{\al,\mu}, \al \in A, \mu \in S_\al$, (resp.,
$\wt{X}^{\al,\mu}$) be the strata of the decomposition of $\wt{X}$
into the union of ascending (resp., descending) manifolds of the
function $f$. The strata $\wt{X}_{\al,\mu}, \mu \in S_\al$, are the
inverse images in $\wt{X}$ of the strata $X_\al \subset X$ of the
decomposition
\begin{equation}    \label{stratification}
X = \bigsqcup_{\al \in A} X_\al
\end{equation}
coming from the $\C^\times$-action $\phi$ on $X$, and similarly for
the strata $\wt{X}^{\al,\mu}$. Let $\wt\CH^{\inn}_{\al,\mu}$ be the
space of ``delta-forms'' supported on $\wt{X}_{\al,\mu} \subset
\wt{X}$, defined as in Part I, Section 3.8. Set
$$
\wt\CH^{\inn}_{\al} = \prod_{\mu \in S_\al}
\wt\CH^{\inn}_{\al,\mu}.
$$
The group $H_1(X,\Z)$ naturally acts on $\wt\CH^{\inn}_{\al,\mu}$ by
shifting the index $\mu$ by $\ga \in H_1(X,\Z)$ (recall that $S_\al$
is an $H_1(X,\Z)$-torsor).

The space of ``in'' states of our model in the limit $\la \to \infty$
is then the subspace of those vectors $(\wt\Psi^{\inn}_{\al,\mu})$ in
$\bigoplus_{\al \in A} \wt\CH^{\inn}_{\al}$ that satisfy the
$\tau$-equivariance condition
\begin{equation}    \label{tau eq}
\wt\Psi^{\inn}_{\al,\mu+\ga} = e^{i \int_\gamma \tau}
\wt\Psi^{\inn}_{\al,\mu}, \qquad \ga \in H_1(X,\Z).
\end{equation}
Therefore for a fixed $\al$ all vectors $\wt\Psi^{\inn}_{\al,\mu},
S_\al$, are determined once we know one of them (for a particular
$\mu$). Thus, we find that the space of states is non-canonically
isomorphic to $\bigoplus_{\al \in A} \CH^{\inn}_\al$, where
$\CH^{\inn}_\al$ is the space of delta-forms on $X_\al$ defined in
Part I, Section 3.8. The space of ``out'' spaces is defined similarly.

Both spaces should be viewed as subspaces of the spaces of equivariant
distributions on $\wt{X}$, as explained in Part I. The definition of
these distributions requires regularization. The effect of this
regularization is that the ``big'' space of states is non-canonically
isomorphic to the direct sum of the subspaces $\wt\CH^{\inn}_{\al}$,
but has a canonical filtration whose successive quotients are
isomorphic to $\wt\CH^{\inn}_{\al}$. In particular, the Hamiltonian is
not diagonalizable, but has Jordan blocks, by the same mechanism as
the one described in Part I.

The space of ``out'' states is defined similarly, as a successive
extension of spaces of delta-forms supported at the strata
$\wt{X}^{\al,\mu}, \al \in A, \mu \in S_\al$.

\begin{remark} In the case when $X$ is a finite-dimensional K\"ahler
manifold, equipped with a $\C^\times$-action with a non-empty set of
fixed points (which has been our assumption in this paper), we always
have $H_1(X,\Z) = 0$ and so {\em any} closed one-form is exact (see
\cite{Frankel}). Therefore we do not need to consider the possibility
that $\beta$ is not exact, and so the above discussion appears to be
superfluous. But for {\em infinite-dimensional} manifolds, such as the
loop space $LX$ and the space of connections on $\BS^3$ which we
consider below in the context of four-dimensional Yang--Mills theory,
this is no longer true, so the above discussion will be useful in
these cases.

Note, however, that ``toy models'' of Morse--Novikov quantum mechanics
may already be constructed on finite-dimensional manifolds. But we
need to consider a real manifold $X$ with $H_1(X,{\mathbb Q}) \neq
0$. We have already considered above the new phenomena which occur in
the case of Morse--Novikov quantum mechanics in the example of $X =
{\BS}^{1}$. In this more general situation we may still take the limit
$\la \to \infty$ of the corresponding quantum mechanical model, and
most of our results obtained in the case of K\"ahler manifolds will
carry over to this case. In particular, the space of ``in'' states of
such a model will be isomorphic to the subspace of equivariant states
in $\prod_{\al,\mu} \wt\CH_{\al,\mu}$, where $\wt\CH_{\al,\mu}$ is
the space of delta-forms on the descending manifolds $\wt{X}_{\al,\mu}
\subset \wt{X}$ of the Morse--Novikov function. Note however that these
strata are no longer isomorphic to $\C^n$, and so we obviously do not
have a holomorphic factorization of the corresponding spaces of
delta-forms. In realistic non-simply connected finite-dimensional
situation one encounters various combinations of the two basic
situations we have seen in the case of the circle: the discrete real
spectrum with Jordan blocks and the discrete imaginary spectrum with
periodic functions.

\end{remark}


\subsection{Non-isolated critical points}    \label{non isolated}

Up to now we have considered quantum mechanics on a non-simply
connected manifold $X$ with a Morse--Novikov function $f$ defined on a
covering $\wt{X}$ of $X$. The assumption that $f$ is a Morse--Novikov
function means that its critical points are isolated and
non-degenerate. In this section we discuss what happens when the fixed
points are not isolated and $f$ is a {\em Morse--Bott--Novikov
function}. This is precisely the situation we encounter in
two-dimensional sigma models (where $X$ is a loop space).

We have already discussed such quantum mechanical models in the case
when $X$ is simply-connected in Part I, Section 6.5. Let us recall
briefly the results of our analysis. Thus, we are given a compact
K\"ahler manifold $X$ with a holomorphic $\C^\times$-action preserving
the K\"ahler structure. We will assume that the set of fixed points of
this action is non-empty. Then, according to \cite{Frankel}, there
exists a Morse--Bott function $f$ such that its critical points are
the fixed points of the $\C^\times$-action. Let $C_\al, \al \in A$, be
the components of the fixed point set of the $\C^\times$-action on $X$
(under our old assumptions, each $C_\al$ consisted of a single
point). According to the results of \cite{BB,CS,Wu}, in this case $X$
still has decompositions
\begin{equation}    \label{strat}
X = \bigsqcup_{\al \in A} X_\al = \bigsqcup_{\al \in A} X^\al,
\end{equation}
with the ascending and descending manifolds $X_\al$ and $X^\al$
defined in the same way as before. Each $X_\al$ is a
$\C^\times$-equivariant holomorphic fibration over $C_\al$. Each fiber
is isomorphic to $\C^{n_\al}$, where $n_\al$ is the number of positive
eigenvalues of the Hessian of $f$ at the points of $C_\al$. Moreover,
locally over $C_\al$, the bundle $X_\al$ is isomorphic to the
subbundle $N^+_\al$ of the normal bundle to $C_\al \subset X$ spanned
by the eigenspaces of the Hessian of the function $f$ with positive
eigenvalues. In general, $X_\al$ does not have a natural structure of
a vector bundle; in other words, the transition functions between
local identifications of $X_\al$ with $N^+_\al$ (or any other vector
bundle) over open subsets of $C_\al$ may not be linear. However, we
will assume in what follows that they are algebraic (that is,
polynomial). This is the case when $X$ is a projective algebraic
variety (see \cite{BB}). Thus, we can speak of functions on $X_\al$
that are polynomial along the fibers of the projection $X_\al \to
C_\al$.

Likewise, $X^\al$ is also a $\C^\times$-equivariant holomorphic bundle
over $C_\al$. Each fiber is isomorphic to $\C^{n-n_\al-\dim
C_\al}$. Locally, the corresponding bundle over $C_\al$ is isomorphic
to the subbundle $N^-_\al$ of the normal bundle to $C_\al \subset X$
spanned by the eigenspaces of the Hessian of the function $f$ with
negative eigenvalues. Again, we will assume that the transition
functions between local identifications of $X_\al$ with $N^-_\al$ over
open subsets of $C_\al$ are algebraic.

Consider, for example, the case of $X=\ptwo$ with the
$\C^\times$-action corresponding to $f$ given by the formula
$(z_1:z_2:z_3) \mapsto (qz_1:z_2:z_3)$. Then the fixed point set has
two components: the point $C_1=(1:0:0)$ and the one-dimensional
component $C_2 = \{ (0:z_2:z_3) \}$ isomorphic to $\C\pone$. The
corresponding strata $X_1$ and $X_2$ are the point $(1:0:0)$ and its
complement, respectively. Note that $X_2$ is a line bundle over
$C_2=\C\pone$ isomorphic to ${\mc O}(1)$, which is also isomorphic to
the normal bundle of $C_2 \subset \ptwo$. The strata $X^1$ and $X^2$
are the plane $\{ (1:u_1:u_2) \}$ and $C_2 = \C\pone$, respectively.

The description of the spaces of ``in'' and ``out'' states of this
model is similar to the one obtained previously in the Morse function
case. Namely, $\CH^{\inn}$ is isomorphic to the direct sum of the
spaces $\CH^{\inn}_\al, \al \in A$. Roughly speaking, each space
$\CH^{\inn}_\al$ is the space of $L^{2}$ differential forms on $C_\al$
extended in two ways: by polynomial differential forms in the bundle
directions of $X_\al$ and by polynomials in the derivatives in
the transversal directions to $X_\al$ in $X$. More precisely, the
states in $\CH_\al$ are $L^{2}$ sections of vector bundles over
$C_\al$. For example, in the case when $X = \ptwo$ we have $X_2 \simeq
N^+_2$, which is the line bundle ${\mc O}(1)$ over $C_2 =
\C\pone$. Then the purely bosonic part of $\CH_2$ is the direct sum of
the spaces of $L^{2}$ sections of the line bundles ${\mc
O}(-1)^{\otimes n} \otimes \ol{{\mc O}(-1)}^{\otimes \ol{n}}, n,
\ol{n} \geq 0$. To include the fermions, we need to add the
corresponding spaces of differential forms, which are defined
similarly.

In particular, the ground states, on which the Hamiltonian ${\mc
L}_v$, where $v = \nabla f$, acts by zero, correspond to just the
ordinary $L^{2}$ differential forms on $C_\al$. Given such a form
$\omega_\al$, let $\wt\omega_\al$ be its pull-back to $X_\al$
under the projection $X_\al \to C_\al$. Then $\wt\omega_\al$
defines a ``delta-like'' distribution supported on $X_\al$, whose
value on $\eta \in \Omega^\bullet(X)$ is equal to
$$
\int_{X_\al} \wt\omega_\al \wedge \eta|_{X_\al}.
$$
While these are the ground states of the model at $\la=\infty$, only
those of them which correspond to harmonic differential forms
$\omega_\al \in \Omega^\bullet(C_\al), \al \in A$, may be deformed to
ground states for finite values of $\la$.

Other elements of $\CH_\al$ are distributions obtained by applying to
the distributions $\wt\omega_\al$ Lie derivatives in the transversal
directions to $X_\al$ as well as multiplying them by differential
forms on $X_\al$ which are polynomial along the fibers of the
projection $X_\al \to C_\al$. The definition of these distributions
requires a regularization similar to the one we used in the case of
isolated critical points in Part I. Because of this regularization, we
obtain non-trivial extensions between different spaces
$\CH^{\inn}_\al$, and the action of the Hamiltonian is not
diagonalizable.

The space of "out'' states is defined in a similar fashion. As in the
case of isolated critical points, we have a canonical pairing between
the two spaces.

Now we consider the case of non-simply connected (but connected) $X$
and a {\em Morse--Bott--Novikov function} $f$. The resulting picture
is a combination of the case of Morse--Novikov function discussed in
\secref{spaces of delta} and the case of Morse--Bott function
discussed in this section. Namely, we have the decomposition
\eqref{strat} defined using the $\C^\times$-action corresponding to
the vector field $\xi$, which is the holomorphic part of the gradient
vector field $v = \nabla f = \xi + \ol{\xi}$. Let $\wt{X}_{\al,\mu},
\al \in A, \mu \in S_\al$ (resp., $\wt{X}^{\al,\mu}$), be the strata
of the decomposition of the $H_1(X,\Z)$-cover $\wt{X}$ into the union
of ascending (resp., descending) manifolds of the function $f$. Here,
as above, $S_\al$ is the $H_1(X,\Z)$-torsor of components of the
preimage of $C_\al$ in $\wt{X}$. The strata $\wt{X}_{\al,\mu}, \mu \in
S_\al$, are the inverse images in $\wt{X}$ of the strata $X_\al
\subset X$, and similarly for the strata $\wt{X}^{\al,\mu}$. Let
$\wt\CH^{\inn}_{\al,\mu}$ be the space of ``delta-forms'' supported on
$\wt{X}_{\al,\mu} \subset \wt{X}$, defined as in Part I, Section
3.8. We set
$$
\wt\CH^{\inn}_{\al} = \prod_{\mu \in S_\al}
\wt\CH^{\inn}_{\al,\mu}.
$$
The space of states of the Morse--Bott--Novikov quantum mechanical
model is the space of vectors
$$
(\wt\Psi^{\inn}_{\al,\mu}) \in \prod_{\al \in A, \mu \in S_\al}
\wt\CH^{\inn}_{\al,\mu}
$$
satisfying the $\tau$-equivariance condition \eqref{tau eq}. As
before, the spaces $\wt\CH^{\inn}_{\al,\mu}$ are defined as certain
spaces of distributions on $\wt{X}$. Because of the regularization
involved in the definition of these distributions, there are
non-trivial extensions between them, and the ``big'' space of states
$\bigoplus_{\al \in A} \wt\CH^{\inn}_{\al}$ is really a successive
extension of the spaces $\wt\CH^{\inn}_{\al,\mu}$ (rather than a
direct sum). Furthermore, the Hamiltonian is not diagonalizable, but
has Jordan blocks, as before.

Thus, the structure we obtain is very similar to the case of isolated
critical points. One essential difference is that we observe
holomorphic factorization only in the fiber directions of the maps
$X_\al \to C_\al$, but not along the manifolds $C_\al$ themselves.

\section{Back to sigma models}    \label{back}

We now apply the results of the previous section to the (type A
twisted) two-dimensional sigma models in the limit $\ol\tau \to
\infty$. As we explained in \secref{2D sigma}, such a model, with the
target being a K\"ahler manifold $X$, may be viewed as a quantum
mechanical model on the loop space $X$, corresponding to the
Morse--Bott--Novikov function $f$ given by formula \eqref{f on
LX}. These are precisely the types of models considered in \secref{qm
with mn}.

\subsection{Space of states}
\label{delta-forms}

According to the analysis of \secref{qm with mn}, in order to describe
the space of states of our model we need to find the set of critical
points of our Morse--Bott--Novikov function $f$ on the loop space $LX$
of a K\"ahler manifold $X$. This function is the Floer function given
by formula \eqref{f on LX}.

In order to simplify the exposition, we will assume throughout this
section that $X$ is simply-connected, so that $LX$ is connected (we
will consider the case of a non-simply connected $X$ -- namely, a
torus -- in \secref{torus1}). Then the critical points of $f$ are the
constant maps. Thus, the set of critical points of $f$ is the
submanifold $X \subset LX$, so it has only one component. We need to
describe the corresponding spaces $\wt\CH_{\al,\mu} =
\wt\CH^{\inn}_{\al,\mu}$, in the notation of \secref{non isolated}
(the spaces of ``out'' states may be described similarly). Here $\al$
takes only one value (since the manifold of critical points has only
one component), so we will suppress this index. The other index $\mu$
runs over a torsor over $H_1(LX,\Z) = H_2(X,\Z)$, which is by
definition the set of components of the preimage $p^{-1}(X)$ of $X
\subset LX$ in the $H_1(LX,\Z)$-cover $p: \wt{LX} \to LX$. Since
$\wt{LX}$ is described in terms of maps from a two-dimensional disc
$D$ to $X$ (see \secref{as qm}), we actually have a canonical
component in $p^{-1}(X)$. It consists of the (equivalence classes of)
constant maps $D \to X$. Therefore this $H_2(X,\Z)$-torsor has a
canonical trivialization, and so the index $\mu$ really takes values
in $H_2(X,\Z)$. We therefore have the spaces $\wt\CH_{\mu}, \mu \in
H_2(X,\Z)$. All of them are isomorphic canonically to each other, via
the action of the group $H_2(X,\Z)$ on the covering. Let us describe
the structure of one of them; namely, $\wt\CH_0$.

By analogy with the results in the finite-dimensional case (which were
based on the semi-classical analysis of the wave-functions),
$\wt\CH_0$ is the space of ``delta-forms'' on the ascending
submanifold $\wt{LX}_0 \subset \wt{LX}$. (More precisely, we will use
a particular model for this space described below.) In terms of the
description of $\wt{LX}$ as homotopy classes of maps $D \to X$, where
$D$ is a unit complex disc, given in \secref{as qm}, points of
$\wt{LX}_0$ correspond to {\em holomorphic} maps $D \to X$. Indeed,
the gradient flows of our Morse--Bott--Novikov function $f$ correspond
to the Cauchy--Riemann equations \eqref{cr1}, \eqref{cr2} for the maps
$D \to X$.

To give a more concrete description of the space $\wt\CH_0$, we recall
the explicit (local) coordinates $X^a_n, X^{\ol{a}}_n, n \in \Z$, on
the loop space $LX$, the momenta $p'_{a,n}, p'_{\ol{a},n}, n \in \Z$,
and the corresponding fermionic variables
$$
\psi^a_n, \psi^{\ol{a}}_n, \pi_{a,n}, \pi_{\ol{a},n},
\qquad n \in \Z,
$$
introduced in \secref{as qm}.

\medskip

{\em Warning}: to simplify notation, from now on we will denote
$p'_{a,n}, p'_{\ol{a},n}$, by $p_{a,n}, p_{\ol{a},n}$, and similarly
for the corresponding fields.

\medskip

The OPE \Ref{OPEfrom} give rise to the usual commutation relations
\begin{align}    \label{betagamma}
[p_{a,n},X^b_m] &= - i \delta_{a}^{b} \delta_{n,-m}, \qquad
[\pi_{a,n},\psi^b_m] = - i \delta_{a}^{b} \delta_{n,-m}, \\
\label{betagamma1}
[p_{\ol{a},n},X^{\ol{b}}_m] &= - i \delta_{\ol{a}}^{\ol{b}} \delta_{n,-m},
\qquad [\pi_{\ol{a},n},\psi^{\ol{b}}_m] = - i
\delta_{\ol{a}}^{\ol{b}} \delta_{n,-m}.
\end{align}

\subsubsection{The case of flat space}    \label{flat}

Consider first the case when $X = \C^N$. Technically, it does not fall
in the category of manifolds we consider, since it is not compact, but
it is instructive to consider it because it provides a useful local
model for the compact target manifolds. In this case holomorphic maps
$D \to X$ are simply described by their Taylor series
\begin{equation}    \label{Xa}
\sum_{n \leq 0} X^a_n z^{-n}, \qquad a=1,\ldots,N.
\end{equation}
Hence we obtain a natural set of coordinates on the ascending manifold
$\wt{LX}_0$ (actually, in this case there is no covering, but we will
keep the tilde in the notation); namely, $X^a_n$ and $X^{\ol{a}}_n$,
where $a=1,\ldots,N; n \leq 0$. Therefore, following the discussion in
\secref{non isolated}, it is natural to define the space $\wt\CH_0$ of
``delta-forms'' supported on $\wt{LX}_0$ as the tensor product of
three spaces: the space of $L^{2}$ differential forms on $X = \C^N$
realized in terms of the zero modes $X^a_0, X^{\ol{a}}_0$ and
$\psi^a_0, \psi^{\ol{a}}_0$; the space of polynomial functions in the
remaining coordinates on $\wt{LX}_0$, $X^a_n, \psi^a_n, n<0$, and
their complex conjugates; and the space of polynomial functions in the
``derivatives'' in the transversal directions, $p_{a,n}, \pi_{a,n},
n<0$, and their complex conjugates.

The space $\wt\CH_0$ should be compared with the usual Fock
representation of the free $\beta\gamma$-$bc$ system described by the
superalgebra with the commutation relations \eqref{betagamma} and
\eqref{betagamma1}. It is the tensor product of the chiral Fock
representation ${\mc F}_0$ generated by the vacuum vector $\vac$
satisfying
$$
X^a_n \vac = \psi^a_n \vac = 0, \quad n > 0, \qquad p_{a,m} \vac =
\pi_{a,m} \vac = 0, \quad m \geq 0,
$$
and its anti-chiral counterpart $\ol{\mc F}_0$ generated by the
vector $\ol\vac$ satisfying analogous relations with respect to
half of the the anti-holomorphic generators. Therefore
\begin{align}    \label{fock1}
{\mc F}_0 &= \C[X^a_n,p_{a,m}]_{n\leq 0,m<0} \otimes
\Lambda[\psi^a_n,\pi_{a,m}]_{n\leq 0,m<0} \cdot \vac, \\ \label{fock2}
\ol{\mc F}_0 &= \C[X^{\ol{a}}_n,p_{\ol{a},m}]_{n\leq 0,m<0} \otimes
\Lambda[\psi^{\ol{a}}_n,\pi_{\ol{a},m}]_{n\leq 0,m<0} \cdot \ol\vac.
\end{align}

The tensor product ${\mc F}_0 \otimes \ol{\mc F}_0$ coincides with
$\wt\CH_0$, except for the zero mode part. In the case of $\wt\CH_0$
we have the space of $L^{2}$ differential forms in the zero modes,
whereas in the case of ${\mc F}_0 \otimes \ol{\mc F}_0$ we have the
space of polynomials in the zero modes. Thus, $\wt\CH_0$ may be viewed
as an $L^{2}$ version of the tensor product of the Fock
representations of the chiral and anti-chiral free $\beta\gamma$-$bc$
systems.

\subsubsection{General K\"ahler manifolds}

For a general K\"ahler target manifold $X$, the space $\wt\CH_0$ may
be described in similar terms. Morally, this should be the space of
``semi-infinite'' delta-forms supported on the stratum $\wt{LX}_0
\subset \wt{LX}$, which consists of holomorphic maps from the unit
disc $D$ to $X$. Let us choose a covering of $X$ by coordinate patches
$X = \cup_{\beta} U_{\beta}$, which each patch $U_\beta$ isomorphic to
an open analytic subset of $\C^N$, providing us with complex
coordinates $X^a, a=1,\ldots,N$. Suppose that a map $\Phi: D \to X$
sends $0 \in D$ to a point in $U_\beta$. Then the restriction of
$\Phi$ to a smaller disc $D' \subset D$ will also land in
$U_\beta$. Therefore we obtain a map $D' \to \C^N$, which we may
expand in Taylor series, using the coordinates $X^a$, as in formula
\eqref{Xa}. Thus, we obtain a set of coordinates $X^a_n, n \leq 0$,
on the space of maps $\Phi: D \to X$ such that $0 \mapsto
U_\beta$. These really capture the jet of the holomorphic map $\Phi$
at the origin $0 \in D$.\footnote{It is certainly possible that under
the map $\Phi$ the entire disc $D$ does not land in any given patch
$U_\beta$. However, the values of the coordinates $X^a_n$,
corresponding to a given patch $U_\beta$, on such maps are
well-defined as long as $\Phi(0) \in U_\beta$.}

On the other hand, a general loop $\gamma: \BS^1 \to X$, which is not
the boundary value of a holomorphic map $D \to X$, may be expanded in
the Fourier series in both positive and negative directions, as in
formula \eqref{Xa exp},
$$
X^a(\sigma) = \sum_{n \in \Z} X^a_n e^{-in\sigma},
$$
where $\sigma$ is the coordinate on the circle to which the coordinate
$z$ on the disc used in formula \eqref{Xa} is related by the formula
$e^{i\sigma}$ on the disc boundary $|z|=1$. Therefore $X^a_n, n>0$,
give us coordinates in the transversal directions to $\wt{LX}_0
\subset \wt{LX}$.

Note that the differentials of the coordinates $X^a_n, n \neq 0$, and
their complex conjugates may be viewed as the fiberwise coordinates on
the normal bundle to $X \subset \wt{LX}$ restricted to $U_\beta
\subset X$. According to formula \eqref{vector field v}, the
eigenspaces of the Hessian of the Floer function with positive
eigenvalues correspond to $X^a_n, n<0$, and the eigenspaces of the
Hessian of the Floer function with negative eigenvalues correspond to
$X^a_n, n>0$.

\medskip

This discussion suggests the following model for the space of
semi-infinite delta-forms supported on $\wt{LX}_0 \subset \wt{LX}$.

Let $\psi^a_n$ be the fermionic counterparts of the coordinates
$X^a_n$ corresponding to the patch $U_\beta$, and let $p_{a,n},
\pi_{a,n}$ be the corresponding momenta variables. For each patch
$U_\beta$ we have the space of states of the free field theory
described in the same way as in \secref{flat}, except that we need to
replace the $L^{2}$ condition by the smoothness condition. The reason
for this is the following. When we considered the sigma model with the
target $\C^N$, the $L^{2}$ condition was imposed as the condition
specifying the behavior of the wave-functions at infinity. In the case
of a compact K\"ahler manifold $X$ the $L^{2}$ condition is is
replaced by the gluing condition on overlaps of different patches.

These spaces should be glued on the overlaps $U_\beta \cap
U_\gamma$. In other words, on the overlap of any two patches $U_\beta
\cap U_\gamma$ we must define a transition function between the spaces
of states attached to them. Then these spaces would form a vector
bundle over $X$, and a state in $\wt\CH_0$ would by definition be a
global section of this vector bundle. More concretely, it would be
represented by a collection of local states on the patches, which
agree on the overlaps.

\medskip

Describing the transition functions for this vector bundle amounts to
describing how the generating fields $X^a(z), p_a(z)$, etc., transform
under general changes of coordinates on the target manifold $X^a =
f^a(\{ \wt{X}^b \})$. The existence of consistent transition
functions is by no means obvious. In fact, in a purely bosonic version
of the theory they may not exist. This has to do with an anomaly which
has been studied extensively in recent years (see
\cite{MSV,GMS,BD,Witten:cdo,N:curved}). However, in the
supersymmetric models that we are considering now the anomaly cancels
and the transformation formulas do exist. They are given by the
following formulas from \cite{MSV}:
\begin{align}    \notag
\wt{X}^\mu(z) &= g^\mu(X^\mu(z)), \\ \notag \wt{p}_\mu(z) &= \Wick
\frac{\pa f^\nu}{\pa \wt{X}^\mu} p_\nu \Wick \; + \; \frac{\pa^2
f^\la}{\pa \wt{X}^\mu \pa \wt{X}^\al} \frac{\pa g^\al}{\pa X^\nu} \;
\Wick \pi_\la \psi^\nu \Wick \, , \\
\label{trans} \wt\psi^\mu &= \frac{\pa g^\mu}{\pa X^\nu} \psi^\nu, \\
\notag \wt\pi_\mu &= \frac{\pa f^\nu}{\pa \wt{X}^\mu} \pi_\nu.
\end{align}
Here we let the Greek indices $\mu,\nu,...$ take both holomorphic and
anti-holomorphic values, e.g., $\mu=a, \ol{a}$, etc., and we set
$f^{\ol{a}} = \ol{f^a}$. We denote $\wt{X}^\mu = g^\mu(\{ X^\nu \})$
the inverse change of variables to $\wt{X}^\nu = f^\nu(\{ \wt{X}^\mu
\})$. These formulas are checked by an explicit computation (see
\cite{MSV}).

Thus, we obtain here the smooth (rather than holomorphic) version of
the chiral de Rham complex, as introduced in \cite{MSV} and reviewed
more recently in \cite{BHS}. To avoid the terminological confusion
with the actual chiral de Rham complex of $X$ involving only
holomorphic variables (which we consider below in \secref{chiral}), we
will call $\wt\CH_0$ the {\em chiral-anti-chiral de Rham complex}.

The transformation formulas for the fermions $\psi^\mu, \pi_\mu$
simply mean that they transform as sections of the tangent and
cotangent bundles to $X$, respectively, as expected. The only surprise
is the second term in the formula for the transformation of the
momenta $p_\mu$. This is due to the fact that the momenta variables we
use here are the transformed variables denoted $p'_\mu$ at the end of
\secref{as qm}. Had we used the original momenta $p_\mu$, we would not
have this term. But then the action would be given by formula
\eqref{2D third action} rather than \eqref{2D fourth action}, which
involves the covariant derivatives (with respect to the Levi-Civita
connection) rather than the ordinary derivative, with respect to the
local coordinates $X^\mu$. Therefore there would be non-trivial OPEs
between the momenta variables themselves. We have avoided this by
redefining the momenta $p_\mu \mapsto p'_\mu$ in formula \eqref{prime}
(and changing the notation for $p'_\mu$ back to $p_\mu$) at the cost
of introducing an inhomogeneous second term in the transformation
formula \eqref{trans} for the $p_\mu$'s (see \cite{Kapustin} for a
discussion of this point).

Formulas \eqref{trans}, when rewritten in terms of the Fourier
coefficients of the fields, give rise to the transition functions of
the bundles of spaces of states. It is clear from these formulas that
this vector bundle is a successive extensions of tensor products of
symmetric (for bosonic variables) and exterior (for fermionic
variables) powers of the tangent and cotangent bundles on $X$ and
their complex conjugates (see \cite{MSV} for more details). This
completes the definition of $\wt{\mc H}_0$.

\medskip

We now define, as in \secref{non isolated}, the ``big'' space of
states as the direct product
$$
\wt\CH = \prod_{\mu \in H_2(X,\Z)} \wt\CH_\mu,
$$
with $\wt\CH_\mu$ being the space of ``delta-forms'' on $\wt{LX}$
supported on the ascending manifold $\wt{LX}_\mu$ of the component of
the set of critical points of the Floer function $f$ labeled by
$\mu$. These spaces are canonically isomorphic to $\wt\CH_0$,
described above. The big space of states should be realized as a space
of distributions on $\wt{LX}$. For this reason, as we have explained
in detail in Part I and in \secref{qm with mn} in the quantum
mechanical setting, this space is non-canonically isomorphic to the
above direct product. Canonically, we only have a filtration
corresponding to the closure relations among the strata $\wt{LX}_\mu
\subset \wt{LX}$. The associated graded pieces are isomorphic to
$\wt\CH_\mu$ as in the quantum mechanical models discussed above.

\subsubsection{Definition of the space of states}

We can now describe the space of states of our sigma model (in the
infinite radius limit $\ol\tau \to -i \infty$, as defined in
\secref{2D sigma}). Such a state is, by definition, a collection
$$
(\wt\Psi_{\mu}) \in \prod_{\mu \in H_2(X,\Z)}
\wt\CH_{\mu}
$$
satisfying the $\tau$-equivariance condition
\begin{equation}    \label{tau eq1}
\wt\Psi_{\mu+\gamma} = e^{\int_\gamma \tau} \; \;
\wt\Psi_{\mu}, \qquad \ga \in H_2(X,\Z).
\end{equation}
Therefore we see that $\Psi_0$ determines the remaining $\Psi_\mu$,
and so the space of states is isomorphic to $\wt\CH_0$. However, this
isomorphism is non-canonical because the direct product decomposition
of the ``big'' space of states is non-canonical.

The fact that the space of states of the sigma model in the infinite
radius limit is not canonically isomorphic to chiral-anti-chiral de
Rham complex $\wt\CH_0$ is due to the instanton effects. What it really
means is that the identification of $\wt\CH_0$ with the space of
states is only valid {\em perturbatively}, that is, without the
instantons. As we will see below (and as we have already seen in
quantum mechanics in Part I), the {\em instanton corrections} occur
precisely because of the intricate structure of extensions between the
spaces of delta-forms $\wt\CH_\mu$. This leads to the
non-diagonalizability of the Hamiltonian, divergence of the
correlation functions, logarithms in the operator product expansion
and other interesting phenomena. We will discuss this in more detail
in \secref{log mixing}.

\subsection{The sigma model in the infinite radius limit as a
  logarithmic CFT}

As in the quantum mechanical case discussed in detail in Part I, the
action of the Hamiltonian of the two-dimensional sigma model can be
read off the correlation functions. We will now use this information
to show that this Hamiltonian is non-diagonalizable and that the sigma
model in the infinite radius limit is in fact a {\em logarithmic}
conformal field theory.

The Hamiltonian on each graded piece $\wt\CH_\mu$ of the ``big'' space
of states is equal to $L_0 + \ol{L}_0$, where $L_0$ is the $0$th mode
of the stress tensor $T(z)$, generating the Virasoro algebra with
central charge $0$, and $\ol{L}_0$ is its anti-holomorphic
counterpart. In the free $\beta\gamma$-$bc$ conformal field theory
corresponding to each patch $U_\beta$ these fields are given by the
usual formulas
\begin{align}    \label{Tz}
T(z) &= i \ \Wick \left( p_a(z) {\pa}_zX^a(z) + \pi_a(z) {\pa}_z
\psi^a(z) \right) \Wick \, , \\ \label{barTz} \ol{T}(\ol{z}) &= i \
\Wick \left( p_{\ol{a}}(\ol{z}) {\pa}_{\zb} X^{\ol{a}}(\ol{z}) +
\pi_{\ol{a}}(\ol{z}) {\pa}_{\zb} \psi^{\ol{a}}(\ol{z}) \right) \Wick
\, \ .
\end{align}
Thus, the Hamiltonian is diagonalizable on each graded piece
$\wt\CH_\mu$. However, the fact that the big space of states is a
successive extension of these pieces, rather than a direct product,
opens the door to potential non-diagonalizability of the Hamiltonian
of the sigma model. In other words, the true Hamiltonian may have
Jordan blocks, as we saw in the quantum mechanical models in Part
I. In fact, we will compute sample correlation functions below, and
this will confirm that this is the case in sigma models as well. In
addition, we will compute explicitly the nilpotent corrections to the
Hamiltonian causing the Jordan blocks in the case of the target
manifold $\pone$ in \secref{Morse def} (more precisely, we will do
this for the sigma model on $\pone$ in a background gauge field).

What kind of Jordan block structure should we expect to see in the
action of the Hamiltonian on the space of states of the sigma model?
Let us identify the space of states of our model with $\wt\CH_0$ using
the $\tau$-equivariance condition \eqref{tau eq1}.  The nilpotent
entries of the Hamiltonian would bump $\Psi_0 \in \wt\CH_0$ to some
$\Psi' \in \wt\CH_\ga \simeq \wt\CH_0$ corresponding to a stratum
$\wt{LX}_\ga, \ga \in H_2(X,\Z)$ in the closure of $\wt{LX}_0$. We
will call such $\gamma$ negative, because they have negative relative
dimension compared to $\wt{LX}_0$. By the $\tau$-equivariance
condition, this $\Psi'$ corresponds to $\Psi'' = e^{- \int_\gamma
\tau} \Psi' \in \wt\CH_0$. Thus, the nilpotent entries of the
Hamiltonian will necessarily contain factors of the form $e^{-
\int_\gamma \tau}$ for negative $\gamma$.

The Jordan block nature of the Hamiltonian implies that the sigma
model in the limit $\ol\tau \to \infty$ is a logarithmic conformal
field theory (LCFT). Note that the logarithmic corrections to the
Virasoro generators have non-perturbative character: they are caused
directly by the instantons!

The reason why we get Jordan blocks is the absence of anti-instantons
in our model at $\ol\tau = -i \infty$. If they were present, the
anti-instantons would contribute a small matrix element under the
diagonal in the Hamiltonian making it diagonalizable.

\subsection{The observables}

The observables of our model are obtained by the state--operator
correspondence from vectors in the space of states. Consider as a toy
model the case of $X=\C^N$.  Then, as we discussed above, the space of
states is essentially the tensor product of the Fock representations
of the chiral and anti-chiral $\beta\gamma$-$bc$ systems. The
corresponding fields are therefore normally ordered products of the
basic fields $X^i(z), p_i(z)$, etc., their derivatives and complex
conjugates. For a general K\"ahler target manifold $X$ we have a sheaf
of spaces of states which locally look like the spaces of states of
the free theory. Therefore the states, as well as the fields, may be
described as collections of states (or fields) in these free theories
which agree on overlaps.

For example, the zero modes generate a subspace in the space of states
which is isomorphic to the space $\Omega^\bullet(X)$ of differential
forms on $X$. The corresponding fields are locally generated by the
fields $X^a(z)$ and $\psi^a(z)$: a globally defined smooth
differential form which locally, on a particular chart in $X$, is
given by formula
$$
\omega =
\omega_{b_1,\ldots,b_m;\ol{b}_1,\ldots,\ol{b}_{\ol{m}}}(X^a,X^{\ol{a}})
dX^{b_1} \wedge \ldots \wedge dX^{b_m} \wedge dX^{\ol{b}_1} \wedge
\ldots \wedge dX^{\ol{b}_{\ol{m}}}
$$
gives rise to the observable, whose restriction to this patch is
$$
{\mc O}_\omega =
\omega_{b_1,\ldots,b_m;\ol{b}_1,\ldots,\ol{b}_{\ol{m}}}\left( X^a(z),
X^{\ol{a}}(\zb)\right) \psi^{b_1}(z) \ldots \psi^{b_m}(z)
\psi^{\ol{b}_1}(\zb) \ldots \psi^{\ol{b}_{\ol{m}}}(\zb).
$$
These are called the {\em evaluation observables}. However, there are
many more observables in the theory.

\ssec{Jet-evaluation observables}    \label{jet-ev}

First of all, we have an obvious generalization of the evaluation
observables involving higher derivatives of the fields $X^\mu,
\psi^\mu$. To define them, let $J X$ be the space of $\infty$-jets of
holomorphic maps from a small complex disc $D$ to $X$. Such a map is
defined by its Taylor series
$$
X^a(z) = \sum_{n \leq 0} X_n z^{-n},
$$
or equivalently, by the values of its derivatives $\pa_z^n X^a(z)$ at
the origin in $D$ (with respect to some coordinates on an open subset
$U_\beta$ in $X$ which contains the image of origin). These are formal
power series, with no convergence condition assumed. We have a natural
forgetful map $JX \to X$, whose fibers are complex affine spaces with
the coordinates $X^a_n, n<0$. We will refer to $JX$ as the {\em jet
space} of $X$. It also goes by the name {\em jet scheme} (see, e.g.,
\cite{vertex}, Sections 9.4.4 and 11.3.3).

Next, let $\Sigma$ be a smooth algebraic curve. We introduce the
bundle ${\mc J} X$ of jet spaces over $\Sigma$ whose fiber at $p \in
\Sigma$ is the space $J_p X$ of jets of holomorphic maps to $X$ from a
small complex disc $D_p$ {\em around the point} $p$. A more precise
definition is as follows: let $\AutO$ be the group of jets coordinate
changes
$$
z \mapsto \rho(z) = a_1 z + a_2 z^2 + \ldots,
$$
where $a_1 \neq 0$, and $a_n, n>1$, are arbitrary complex
numbers. (These are formal coordinate changes, so we do not assume
that the series converges anywhere.) This group acts on $JX$ as
follows: $X^a(z) \mapsto X^a(\rho(z))$. On the other hand, we have a
natural principal $\AutO$-bundle ${\mc A}ut_\Sigma$ on $\Sigma$, whose
fiber ${\mc A}ut_p$ at $p \in X$ is the space of jets of holomorphic
local coordinates at $p$ (see, e.g., \cite{vertex}, Ch. 6, for the
precise definition). If $t_p$ is one such jet of coordinates, then
$\rho(t_x)$ is another, for any $\rho \in \AutO$, and for any two jets
of coordinates, $t'_p, t''_p$, there exists a unique $\rho \in AutO$
such that $t''_p = \rho(t'_p)$ (so that ${\mc A}ut_p$ is an
$\AutO$-torsor). Now we define ${\mc J} X$ as the associated bundle
$$
{\mc J} X := {\mc A}ut_\Sigma \underset{\AutO}\times JX.
$$
Then the fiber of ${\mc J} X$ at $p \in \Sigma$ is indeed the space
of jets of maps $D_p \to X$. If we choose a particular coordinate $z$
at $p \in \Sigma$, we identify $D_p$ with $D$ (the ``coordinatized''
disc), and hence the fiber $J_p X$ of ${\mc J} X$ at $p$
with $JX$. But the above definition of ${\mc J} X$ is independent of
any such choices.

Next, let $\Omega^\bullet_{\on{vert}}({\mc J} X)$ be the sheaf of
vertical differential forms on the bundle ${\mc J} X \to X$ (here
``vertical'' means that we consider differential forms only in the
fiber directions of this bundle). By definition, a {\em jet-evaluation
observable} is a smooth global section of
$\Omega^\bullet_{\on{vert}}({\mc J} X)$. Locally, over an open subset
${\mc U}$ in $\Sigma$ with a coordinate $z$, the bundle ${\mc J} X$
may be trivialized: ${\mc J} X|_{\mc U} \simeq {\mc U} \times JX$ and
so a vertical differential form is just a function on ${\mc U}$ with
values in differential forms on $JX$. Choosing an open subset $U_\beta
\subset X$ with holomorphic coordinates $\{ X^a \}$, we can write such
a form of degree $(p,\ol{p})$ as follows
\begin{multline}
\Omega = {\mathbb{A}}_{IJKL\ldots} (X^a(z), \Xb^{\ol{a}}(\ol{z}))
dX^{i_{1}}_{0}\wedge \ldots dX^{i_{m_{1}}}_{0} \wedge dX^{j_{1}}_{-1}
\wedge \ldots \wedge dX^{j_{m_{2}}}_{-1} \\ \wedge dX^{k_{1}}_{-2} \wedge
\ldots \wedge dX^{k_{m_{3}}}_{-2} \wedge \ldots \wedge
d\ol{X}^{\ol{i}_{1}} \wedge \ldots \wedge
d\ol{X}^{\ol{k}_{\ol{m}_{3}}}_{-2} \ldots,
\label{aform}
\end{multline}
where the jet-scheme $JX$ is coordinatized by:
$$
X^{a}(z) = X^{a}_{0} + X^{a}_{-1}z + X^{a}_{-2} z^2 + \ldots,
$$
and $m = m_{1}+m_{2}+\ldots \ , \ \ol{m}=
\ol{m}_{1}+\ol{m}_2+\ldots$.

To the form $\Omega$ we then associate the operator
\begin{equation}
\mathcal{O}_{{\mathbb A}} = \mathbb{A}_{IJKL\ldots } ( X^a (t),
\ol{X}^{\ol{a}} (\ol{t}) ) {\psi}^{i_{1}} \ldots \psi^{i_{m_{1}}}
{\pa}{\psi}^{j_{1}} \ldots {\pa} {\psi}^{j_{m_{2}}} {{\pa}^{2}
\psi}^{k_{1}} \ldots \ol{\psi}^{\ol{i}_{1}} \ldots ,
\label{oopera}
\end{equation}
where $I = i_{1}i_{2} \ldots i_{m_{1}}$, $J = j_{1}j_{2}\ldots
j_{m_{2}}$, $\ldots$ , $K = k_{1}k_{2}\ldots k_{m_{3}}$, etc.

Introduce the following notation. For a partition
${\la} = (\la_1 \geq \ldots \geq \la_n)$ we denote $n=\ell(\la)$, 
$$
| {\la} | = \sum_{i} {\la}_{i},
$$ 
and write
$$
\dlx{\la} =
\prod_{i=1}^{{\ell}({\la})} \frac{1}{{\la}_{i}!} {\pa}_z^{{\la}_{i}}X,
\qquad \odl{\la} =
\prod_{i=1}^{{\ell}({\la})} \frac{1}{{\la}_{i}!}
     {\pa}_{\ol{z}}^{{\la}_{i}}\ol{X}.
$$
For instance,
$$
\dlx{1^{3}} = ({\pa}X)^{3} \ , \qquad \dlx{2,1} = \frac{1}{2}
    {\pa}^{2}X {\pa}X \ , \qquad \dlx{3} = \frac{1}{6}{\pa}^{3}X,
$$
$$
\odl{1^{3}} = ({\ol{\pa}}\ol{X})^{3} \ , \qquad
\odl{2,1} = \frac{1}{2}{\ol{\pa}}^{2}\ol{X} {\ol{\pa}}\ol{X} \ , \qquad
\odl{3} = \frac{1}{6}{\ol{\pa}}^{3}\ol{X},
$$
and so on. For the target space of complex dimension $d$ we introduce
operators labeled by the colored partitions: $\vec\la = (
{\la}^{(1)}, {\la}^{(2)}, \ldots , {\la}^{(d)} )$: the bosonic ones
$$
\dlx{\vec\la} = \prod_{a=1}^{d} \prod_{i=1}^{{\ell}({\la}^{(a)})}
\frac{1}{{\la}^{(a)}_{i}!} {\pa}_{z}^{{\la}^{(a)}_{i}}X^{a}\ , 
$$
and the fermionic ones:
$$
\dlpsi{\vec\la} = \prod_{a=1}^{d} \prod_{i=1}^{n_{a}} \frac{1}{\left(
  {\la}^{(a)}_{i} - i + n_{a}-1\right)!}
      {\pa}_{z}^{{\la}^{(a)}_{i}-i+n_{a}-1}{\psi}^{a}\ , 
$$
where $n_{a} = {\ell}({\la}^{(a)})$ is the number of rows in the Young
diagram of the partition ${\la}^{(a)}$. Analogously, we define the
complex conjugate operators $\odl{\vec\la}, \odlpsi{\vec\la}$.

Then we formally expand: 
\begin{equation}
\mathbb{A}_{IJKL\ldots } ( X^a (t), \ol{X}^{\ol{a}} (\ol{t}) ) =
\sum_{\vec\la , \vec\mu} \mathbb{A}_{IJKL\ldots | \vec{\la}\vec{\mu}}
( X^a_{0}, \ol{X}^{\ol{a}}_{0}) \dlx{\vec\la} \odl{\vec\mu}.
\label{lamuexp}
\end{equation}
Similarly, the fermionic content of the operator \Ref{oopera} may be
repackaged using colored partitions:
\begin{equation}
\mathcal{O}_{\mathbb{A}} = \sum_{\vec\la, \vec\mu, \vec\nu, \vec\rho}
\mathbb{A}_{\vec\la\vec\mu \vec\nu\vec\rho}(X_{0}^{a}, \ol{X}_{0}^{a})
\dlx{\vec\la}\odl{\vec\nu}\dlpsi{\vec\mu}\odlpsi{\vec\rho}.
\label{ooopera}
\end{equation}

On the overlaps $U_\beta \cap U_\gamma$ we have transition functions
defined explicitly by formulas \eqref{trans}. It is easy to see that
these are precisely the (classical) transition functions on the jet
bundle ${\mc J} X$ defined above.

Since these observables do not depend on the momenta variables
$p_a(z), \pi_a(z)$ and their complex conjugates, they are
``classical'' in the sense that in perturbation theory no normal
ordering (or any other kind of regularization) is necessary to define
them. (We will see below that non-perturbatively even these
observables require regularization.) They transform in the same
way as the classical jets (see formula \eqref{trans}), without any
quantum correction terms. Hence this class of observables is the
easiest to study (beyond the ordinary evaluation observables).

\begin{remark}

Note that we have a tautological map $JX \to X$ corresponding to
evaluating a jet of maps $D \to X$ at $0 \in D$, $X^a(z) \mapsto
X^a(0)$. Likewise, we have a map ${\mc J} X \to \Sigma \times X$,
defined in the same way pointwise. Any differential form on $X$ gives
rise to a differential form on the product $\Sigma \times X$ (constant
along the first factor), and, by pull-back, on ${\mc J} X$. The
corresponding observables are the ordinary evaluation observables;
they depend on the fields $X^\mu, \psi^\mu$, but not on their
derivatives. Likewise, we may consider the space $J^N X$ of $N$-jets
of maps $D \to X$; these are determined by the first $(N-1)$
derivatives of $X^a(z)$ at the origin. Let ${\mc J}^N X$ be the
corresponding bundle over $\Sigma$. We have natural forgetful maps $JX
\to J^N X$ and ${\mc J} X \to {\mc J}^N X$ (in fact, $JX$ and ${\mc J}
X$ are the inverse limits of $J^N X$ and ${\mc J}^N X$, respectively,
as $N \to \infty$). The jet-evaluation observables that depend only on
the first $(N-1)$ derivatives of $X^\mu$ and $\psi^\mu$ correspond
precisely to the differential forms on ${\mc J} X$ obtained by pull-back
from ${\mc J}^N X$.\qed

\end{remark}

In addition to the jet-evaluation observables, which do not depend on
the momenta variables $p_a, \pi_a$ and their complex conjugates, there
are also observables that do depend on them. For instance, each global
vector field on $X$ which locally reads as
$$
v = v^b(X^a,X^{\ol{a}}) \frac{\pa}{\pa X^b} +
v^{\ol{b}}(X^a,X^{\ol{a}}) \frac{\pa}{\pa X^{\ol{b}}}
$$
gives rise to the observable
\begin{align} \notag
v(z) &= \Wick v^b(X^a(z),X^{\ol{a}}(\zb)) p_b(z) \Wick \\ \notag &+
\Wick \frac{\pa v^b}{\pa X^c}(X^a(z),X^{\ol{a}}(\zb)) \pi_b(z)
\psi^c(z) \Wick + \Wick \frac{\pa v^b}{\pa
X^{\ol{c}}}(X^a(z),X^{\ol{a}}(\zb)) \pi_b(z) \psi^{\ol{c}}(z) \Wick \\
\label{vz}
&+ \Wick v^{\ol{b}}(X^a(z),X^{\ol{a}}(\zb)) p_{\ol{b}}(\zb) \Wick \\
\notag
&+ \Wick \frac{\pa v^{\ol{b}}}{\pa X^c}(X^a(z),X^{\ol{a}}(\zb))
\pi_{\ol{b}}(z) \psi^c(z) \Wick + \Wick \frac{\pa v^{\ol{b}}}{\pa
X^{\ol{c}}}(X^a(z),X^{\ol{a}}(\zb)) \pi_{\ol{b}}(z) \psi^{\ol{c}}(z)
\Wick \, ,
\end{align}
which corresponds to the action of the Lie derivative ${\mc L}_v$ on
the differential forms.

In particular, to purely holomorphic differential forms or vector
fields correspond purely chiral fields (i.e., the ones annihilated by
$\pa_{\ol{z}}$). Yet more general fields may be obtained from global
differential operators on $X$. If these operators are holomorphic,
then we obtain fields from the chiral algebra of our model, also known
as (the space of global section of) the chiral de Rham complex. We
will discuss it in more detail in \secref{chiral}.

\subsection{Chiral algebra of the sigma model and chiral de Rham
  complex}    \label{chiral}

Now consider the chiral algebra of the sigma model in the infinite
radius limit. Because of the state--operator correspondence, it is
isomorphic to the space of states of the model which are annihilated
by the operator $\ol{L}_{-1}$ (a Fourier coefficient of the field
$\ol{T}(\ol{z})$ given by formula \eqref{barTz}) which corresponds to
the derivative $\pa_{\ol{z}}$ (the chiral states). Locally, for each
coordinate patch $U_\beta \subset X$ isomorphic to $\C^N$, we have the
space of states of the free theory on $\C^N$ (more precisely, its
version in which the $L^{2}$ condition on the zero modes is replaced by
the smoothness condition, see \secref{delta-forms}). Its subspace of
states annihilated by the operator $\ol{L}_{-1}$ is therefore
isomorphic to the Fock module ${\mc F}$ of the free {\em chiral}
$\beta\gamma$-$bc$ system on $U_\beta$, as defined in
\secref{delta-forms}. A global chiral state is therefore a collection
of local chiral states $\Psi_\beta$, that is, elements of the chiral
Fock module ${\mc F}$ corresponding to the patch $U_\beta$, which are
compatible on the overlaps $U_\beta \cap U_\gamma$.

Formulas \eqref{trans} for the transformation of chiral fields under
holomorphic changes of coordinates coincide with the formulas given in
\cite{MSV}. Therefore we arrive at the definition of the {\em chiral
de Rham complex} from \cite{MSV} (see also \cite{vertex},
Sect. 18.5). Thus, we find that {\em the chiral algebra of the sigma
model in the infinite radius limit is the space of global sections
over $X$ of the chiral de Rham complex}, as was previously observed in
\cite{AiB} (see also \cite{Kapustin,Witten:cdo}).

However, in contrast to most of the mathematical literature, we are
not interested in the chiral algebra {\it per se}. Rather, we are
interested in the full quantum field theory in the infinite radius
limit $\ol\tau \to \infty$, in which the chiral and anti-chiral
sectors are combined in a non-trivial way. Perturbatively, the space
of states is isomorphic to the chiral-anti-chiral de Rham complex
discussed above. The inclusion of the instantons requires that we take
into account the non-trivial self-gluings of this space, realized as
the space of delta-forms on semi-infinite strata in $\wt{LX}$. This
changes the structure of the space of states and the correlation
functions and leads to the logarithmic mixing of states and operators
discussed in \secref{cor fns} below.

\bigskip

We now consider some explicit examples of sigma models. As our first
example we take up the case of an elliptic curve $\BT = \C/(\Z +
\Z\tau)$. Though this is essentially a free field theory, it is
instructive to see how the familiar description of the space of states
at finite radius changes when we take the infinite radius limit
$\ol\tau \to \infty$. (In \secref{from to} we will compute the bosonic
part of the partition function of this model and show how it can be
obtained as a limit of the well-known partition function of the free
bosonic field theory compactified on a torus.) We will then consider
the target manifold $\C\pone$ and relate the description of the space
of states given above with the one obtained in \cite{AiB} using
``holomortex operators''.

\ssec{The case of a torus}    \label{torus1}

Consider the supersymmetric sigma model with the target manifold
elliptic curve $\BT = \C/2\pi(\Z + \Z T)$, where $T$ is in the
upper-half plane. Note that in our discussion above we had assumed
that the target manifold is simply-connected. The case at hand is
different, as $\pi_1(\BT) = \Z^2$. This means that the loop space
$L\BT$ has components labeled by $\Z^2$, which we will denote by
$L\BT_\la, \la \in \Z^2$. In addition, the Floer function is not
single-valued on any of these components, and so it is necessary to
consider an $H_{2} ({\BT},{\Z}) = {\Z}$-covering of each of them. This
covering is defined as follows: for each component $L\BT_\la$ we have
to choose a loop $\ga^0_\la: {\BS}^1 \to {\BT}$ which is in the
homotopy class $\la$. Then points of the covering $\wt{L\BT}_\la$ are
by definition pairs $(\ga,[\wt\ga])$, where $\ga \in L\BT_\la$ and
$[\wt\ga]$ is the equivalence class of maps $\wt\ga: \BS \times [0,1]
\to \BT$ such that
$$
\wt\ga(\BS^1 \times 0) = \ga^\la_0\, , \qquad \wt\ga(\BS^1 \times 1) =
\ga,
$$
modulo the equivalence relation identifying any two maps $\wt\ga_1,
\wt\ga_2$ satisfying these boundary conditions whose difference in
$H_2(\BT,\Z)$ is equal to $0$. Then the Floer function \eqref{f on LX}
lifts to a single-valued function on $\wt{L\BT}_\la$ (compare with the
general construction in \secref{abelian}).

Note that for $\la=0$ we may choose as the initial loop $\ga^0_0$ any
constant map and then the definition of $\wt{L\BT}_0$ is the same as
discussed above: we consider equivalence classes of maps from the disc
$D$ to $\BT$. But for $\la \neq 0$ there are no maps $D \to \BT$ which
are in the homotopy class $\la$ on the boundary. Instead, points of
the covering are represented by maps from the cylinder $\BS^1 \times
[0,1]$ with prescribed image on one boundary circle. This description
enables us to identify each component $L\BT_\la$ with the product of
$\BT$, corresponding to the zero mode, and an infinite-dimensional
vector space corresponding to other modes (see formula \eqref{Xz}
below). For $\la=0$, our Floer function becomes a Bott-Morse function
on $\wt{L\BT}_0$; its critical points are the constant maps $D \to
\BT$ and their translates by the group $\Z$ of deck transformations of
the covering. But on all other components this function has {\em no
critical points}. We are therefore rather in the situation of a circle
${\mathbb R}/2\pi \Z$, equipped with the multi-valued function $x
\mapsto \mu x$ considered in \secref{BS1}. More precisely, we have
that kind of function along the first factor of the decomposition $\BT
\times \C^\infty$, and along the second factor we are in the situation
of the target manifold $X=\C$. Therefore the space of states also
decomposes into the tensor product of the space corresponding to the
zero mode, which exhibits the same kind of spectrum as in
\secref{BS1}, and the space of delta-forms, or, more concretely, the
tensor product of Fock representations of the chiral and anti-chiral
$\beta\gamma$-$bc$ systems (see \secref{delta-forms}), without the
zero modes.

Let us describe this space of states more explicitly. First of all,
the states supported on different components of $L\BT$ are
distinguished by different winding numbers, as in the case of the
sigma model at the finite radius. Therefore we obtain a decomposition
of the space of states\footnote{Here, as before, we discuss the space
of ``in'' states; the structure of the space of ``out'' states is
similar.} ${\mc H}$ into sectors corresponding to different winding
numbers
\begin{equation}    \label{dec winding}
{\mc H} = \bigoplus_{m,n \in \Z} {\mc H}_{m,n}.
\end{equation}
To describe the sectors ${\mc H}_{m,n}$, we expand a general
smooth map from the circle with coordinate $\sigma$, where $\sigma
\sim \sigma + 2\pi$, to $\BT$ in Fourier series
\begin{equation}    \label{Xz}
X(\sigma) = \omega \sigma + X_0 + \sum_{n \neq 0} X_n e^{- i n \sigma},
\end{equation}
where $\omega$ is the {\em winding operator}, which takes values in
the lattice $\Z + \Z T$. It acts on ${\mc H}_{m,n}$ by multiplication
by $m + n T$. The zero mode $X_0$ is a periodic variable taking values
in $\BT = \C/2\pi(\Z + \Z T)$, and the remaining modes take
arbitrary complex values.

The space ${\mc H}_{m,n}$ is generated by a vector $|m,n\rangle$,
which is annihilated by $X_k, \psi_k, k>0$, $p_{k}, \pi_{k}, k
\geq 0$, and their complex conjugates. Thus, we have
\begin{equation}    \label{Hmn}
{\mc H}_{m,n} = \Omega(X_0,\ol{X}_0,\psi_0,\ol{\psi}_0) \otimes
\C[X_k,p_k,\ol{X}_k,\ol{p}_k]_{k<0} \otimes
\Lambda[\psi_k,\pi_k,\ol{\psi}_k,\ol{\pi}_k]_{k<0} \cdot |m,n\rangle,
\end{equation}
where $\Omega(X_0,\ol{X}_0,\psi_0,\ol{\psi}_0)$ is the space of
differential forms on $\BT$ realized in terms of the zero modes
$X_0,\ol{X}_0,\psi_0,\ol{\psi}_0$. It is spanned by the monomials
\begin{equation}    \label{spanned}
e^{i r (X_0 \ol{T} - \ol{X}_0 T)/(\ol{T}-T) + i
    s (X_0-\ol{X}_0)/(T - \ol{T})} \psi_0^{\ph} \ol{\psi}_0^{\phb},
    \qquad r,s \in \Z, \quad \ph,\phb = 0,1.
\end{equation}

Let us discuss the state--operator correspondence. The fields
$\Psi_{m,n}(z,\zb)$ corresponding to the winding states $|m,n\rangle$
are versions of the {\em holomortex operators}\footnote{This
terminology is a shorthand for ``holomorphic vortex''.} introduced in
\cite{AiB} in the case of the target manifold $X=\C^\times$. These are
analogues of the well-known vortex operators responsible for the
winding in the sigma model at finite radius. They satisfy the OPE
\begin{align*}
X(z) \Psi_{m,n}(w,\wb) &= (m + nT) \log(z-w) \Psi_{m,n}(w,\wb) +
\ldots, \\ \ol{X}(\zb) \Psi_{m,n}(w,\wb) &= (m + n\ol{T})
\log(\zb-\wb) \Psi_{m,n}(w,\wb) + \ldots
\end{align*}
Using this OPE, we find the fields $\Psi_{m,n}(z,\zb)$ explicitly:
\begin{equation}    \label{Psi}
\Psi_{m,n}(z,\zb) = e^{i \int^z \left( (m+nT) P + (m+n\ol{T}) \ol{P}
  \right)},
\end{equation}
where
$$
e^{\int^z (\al P + \beta \ol{P})} := \exp \left( \al \int^z_{z_0} p(w)
dw + \beta \int^z_{z_0} \ol{p}(\wb) d\wb \right).
$$
This formula {\em a priori} depends on the choice of the point $z_0$
and the contour of integration. However, under a correlation function
this ambiguity disappears because of the ``charge conservation''
condition: the total winding should be equal to zero. If this
condition is satisfied, then we can pair the integrals in the exponent
to obtain a linear combination of integrals between the points of
insertion of these holomortex operators (see \cite{AiB}, Section 2.2,
for more details). A more precise algebraic construction of these
fields as operators acting on ${\mc H}$ is obtained in the same way as
in \cite{AiB}, Section 5.1.

To obtain the fields corresponding to other states in ${\mc H}$, we
use the description of ${\mc H}$ given in formulas \eqref{Hmn} and
\eqref{spanned}. This description gives us a natural basis of
monomials. We then replace the variables $X_k$ and $\ol{X}_k$ by
$\frac{1}{(-k)!}  \pa_z^{-k} X(z)$ and $\frac{1}{(-k)!} \pa_{\zb}^{-k}
\ol{X}(\zb)$, respectively, $p_k$ and $\ol{p}_k$ by $\frac{1}{(-k-1)!}
\pa_z^{-k-1} p(z)$ and $\frac{1}{(-k-1)!} \pa_{\zb}^{-k-1}
\ol{p}(\zb)$, respectively, and similarly for the fermionic
variables. The normally ordered product of these fields is the field
associated to a given monomial in the space of states.

\ssec{The case of $\pone$}    \label{target pone}

Now we consider the supersymmetric sigma model with the target
manifold $\pone = \C\pone$. We can describe this model (in the
infinite radius limit) explicitly in the case of a special $B$-field;
namely,
\begin{equation}    \label{B-field}
\frac{1}{2} \tau (\delta^{(2)}_0 + \delta^{(2)}_\infty),
\end{equation}
where $\delta^{(2)}_0$ and $\delta^{(2)}_\infty$ are the delta-like
two-forms on $\pone$ supported at two fixed points, $0$ and $\infty$,
and $\tau$ is a complex parameter. We choose the overall factor
$\frac{1}{2}$ so that the integral of the $B$-field is equal to
$\tau$. More generally, we could take $\tau_1 \delta^{(2)}_0 + \tau_2
\delta^{(2)}_\infty$, where $\tau_1+\tau_2=\tau$.

The states of our model have a path integral interpretation
where we integrate over maps from the two-dimensional disc $D$ of
radius $1$ to $\pone$. More precisely, the states are represented by
the path integrals of this form
\begin{equation}    \label{in st}
\Phi(\ga) = \underset{\Phi: D \to \pone, \Phi|_{\BS^1} = \ga} \int
{\mc O}_1(z_1) \ldots {\mc O}_n(z_n) e^{-S},
\end{equation}
where ${\mc O}_1,\ldots,{\mc O}_n$ are some observables, and
$z_1,\ldots,z_n$ are their positions on $D$. As we have stressed
before, in the infinite radius limit the path integral localizes on
the holomorphic maps $D \to \pone$. Such a map may be described as a
meromorphic function on $D$. Let $w_1^+,\ldots,w_m^+$ be the zeros of
this map and $w_1^-,\ldots,w^-_n$ the poles. Each zero and pole give
us a factor of $q^{1/2} = e^{\tau/2}$ in our path integral from the
$B$-field \eqref{B-field}.

Therefore we obtain that the result may be written as the following
state of the theory with the target $\C^\times = \pone \bs \{ 0,\infty
\}$:
\begin{multline}    \label{sum}
\sum_{m,n} \frac{q^{m/2}}{m!} \frac{q^{n/2}}{n!} \int \Psi^+(w^+_1)
\ldots \Psi^+(w^+_m) d^2w^+_1 \ldots d^2w^+_m \\ \cdot \int
\Psi^-(w^-_1) \ldots \Psi^-(w^-_n) d^2w^-_1 \ldots d^2w^-_n \cdot
|A\rangle,
\end{multline}
where $|A\rangle$ is a state in the theory on $\C^\times$ and
$\Psi^+(w), \Psi^-(w)$ are vertex operators in this theory
corresponding to the insertion of zeros and poles.

Summing up \eqref{sum}, we obtain
$$
\exp \left( q^{1/2} \int \left( \Psi^+(w) + \Psi^-(w) \right) d^2 w
\right) \cdot |A\rangle.
$$
This is a state in the theory with the target manifold $\C^\times$
which is {\em deformed} by the operators $q^{1/2} \Psi^+(w)$ and
$q^{1/2} \Psi^-(w)$. Thus, we obtain a free field realization of the
two-dimensional sigma model with the target manifold $\pone$ and with
the $B$-field \eqref{B-field}, as a deformation of the
free field theory with the target $\C^\times$ by these operators.
This realization was found in \cite{AiB}. Now we see that this
deformation naturally arises in the framework of the general formalism
developed in the present paper.

The vertex operators $\Psi^\pm(w)$ have been determined explicitly in
\cite{AiB}, Section 3.1: these are the {\em holomortex operators}
\begin{equation}    \label{Psi pm}
\Psi^\pm(z) = e^{\pm i \int^z(p(w) dw + \ol{p}(\ol{w}) d\ol{w})}
\pi(z) \ol\pi(\zb),
\end{equation}
where $p(z), \ol{p}(\ol{z})$ are the bosonic momenta variables
corresponding to the realization of $\C^\times$ as $\C/2\pi i \Z$, and
$\pi(z)$ and $\ol\pi(\ol{z})$ are the corresponding fermionic
variables (a different sign in the exponential, as compared with
\cite{AiB}, is due to the fact that the action we use here, and hence
the corresponding OPE, differ by a sign from those used in
\cite{AiB}). Using these formulas, we can express correlation
functions of the sigma model with the target $\pone$ as multiple
integrals of correlation functions of the free theory on
$\C^\times$. Some examples of such integrals may be found in
\cite{AiB}, and more examples are presented below in \secref{OPE}.

Since the sigma model on $\pone$ may be realized as a deformation of a
free conformal field theory by strictly marginal operators, we obtain
that the Hamiltonian and the supercharges in the theory on $\pone$ may
be obtained from those of the free theory by some deformation. This
deformation turns out to be ``nilpotent'', in the sense that the
diagonalizable Hamiltonian of the free theory is deformed by an upper
triangular matrix giving rise to Jordan blocks. In quantum mechanics
this mechanism was discussed in detail in Part I. In the case of the
sigma model on $\pone$ in the background of a non-trivial
$\C^\times$-gauge field an analogous derivation of the Hamiltonian
will be given below in \secref{Morse def}.

These results may be generalized from $\pone$ to other toric
varieties, along the lines of \cite{AiB}.

\section{Correlation functions}    \label{cor fns}

{}From the Lagrangian point of view, the correlation functions are
represented by the path integral
$$
\langle {\mc O}_1(p_1) \ldots {\mc O}_n(p_n) \rangle =
\underset{\Phi: \Sigma \to X}\int {\mc O}_1(p_1) \ldots {\mc
O}_n(p_n) e^{-S},
$$
where $S$ is the action \eqref{2D fourth action}. As discussed above
(see Part I, Sections 2.4--2.6, for the quantum mechanical version),
this path integral localizes on {\em holomorphic maps} $\Phi: \Sigma
\to X$, and we obtain a sum over the topological types of such maps,
which correspond to the homology class of the image of $\Sigma$ under
$\Phi$ in $H_2(X,\Z)$,
\begin{equation}    \label{sum over beta}
\sum_{\beta \in H_2(X,\Z)} e^{- \int_\beta \tau} \; \; \underset{{\mc
M}_{\Sigma}(X,\beta)}\int {\mc O}_1(p_1) \ldots {\mc O}_n(p_n)
\end{equation}
(of course, only positive elements of $H_2(X,\Z)$ with respect to the
K\"ahler form will give rise to non-trivial contributions). Here ${\mc
M}_{\Sigma}(X,\beta)$ denotes the moduli space of holomorphic maps
$\Phi$ from the {\em parametrized} Riemann surface $\Sigma$ to $X$ of
degree $\beta$, i.e., such that $[\Phi(\Sigma)] = \beta$. In what
follows we will often assume that there are sufficiently many marked
points $p_1,\ldots,p_n$ on $\Sigma$, so that the marked curve has no
continuous automorphisms (this means that $n\geq 3$ if $\Sigma$ has
genus $0$ and $n\geq 1$ if $\Sigma$ has genus $1$). In this case we
can interpret ${\mc M}_{\Sigma,(p_i)}(X,\beta)$ as the fiber of the
map $\pi_{g,n}$ discussed below, which will imply that the integral
converges.

It is customary to choose a basis $\beta_1,\ldots,\beta_N$ of
$H_2(X,\Z)$. Then this sum becomes a power series in $q_i =
e^{-\int_{\beta_i} \tau}, i=1,\ldots,N$.

In this section we consider these correlation functions in more
details and compute some explicit examples.

\ssec{Gromov--Witten invariants}    \label{GW}

In the case when ${\mc O}_i, i=1,\ldots,n$ are evaluation observables
corresponding to closed smooth differential forms $\omega_i,
i=1,\ldots,n$, on $X$, these integrals are special cases of the {\em
Gromov--Witten invariants}. More precisely, for each point $p_i \in
\Sigma, i=1,\ldots,n$, we have the evaluation map
\begin{equation}    \label{eval}
\on{ev}_{p_i}: {\mathcal M}_{\Sigma}(X,\beta) \to X.
\end{equation}
Now, for a given collection of differential forms
$\omega_1,\ldots,\omega_n$ on $X$, the $\beta$-term in the sum
\eqref{sum over beta} is given by the integral
\begin{equation}    \label{fixed Sigma}
\underset{{\mathcal M}_{\Sigma}(X,\beta)}\int
\on{ev}_{p_1}^*(\omega_1) \wedge \ldots \wedge
\on{ev}^*_{p_n}(\omega_n).
\end{equation}
Then, even though the moduli space ${\mathcal M}_{\Sigma}(X,\beta)$ is
not compact, the integral \eqref{fixed Sigma} is convergent for smooth
differential forms $\omega_i$ on $X$, which is assumed to be
compact. This follows from the fact that the evaluation maps extend to
the Kontsevich compactification (the moduli space of stable maps), as
we discuss below.

More general Gromov--Witten invariants are constructed as
follows. Suppose that $(\Sigma,(p_i))$ does not admit any continuous
automorphisms. Let ${\mathcal M}_{g,n}(X,\beta)$ be the moduli space
(more precisely, Deligne--Mumford stack) of the data
$(\Sigma,(p_i),\Phi)$, where $\Sigma$ is a genus $g$ Riemann surface,
$p_1,\ldots,p_n$ are distinct marked points on $\Sigma$, and $\Phi:
\Sigma \to X$ is a holomorphic map. Then we have a projection
$\pi_{g,n}: {\mathcal M}_{g,n}(X,\beta) \to {\mathcal M}_{g,n}$, where
${\mathcal M}_{g,n}$ is the moduli space (or Deligne--Mumford stack)
of genus $g$ curves with $n$ marked points.  ${\mathcal
M}_{\Sigma}(X,\beta)$ is nothing but the fiber of $\pi_{g,n}$ at
$(\Sigma,(p_i)) \in {\mathcal M}_{g,n}$. We have the evaluation maps
\begin{equation}    \label{evali}
\on{ev}_i: {\mathcal M}_{g,n}(X,\beta) \to X,
\end{equation}
corresponding to evaluation at the $i$th point $p_i$ (so that
$\on{ev}_{p_i} = \on{ev}_i|_{{\mathcal M}_{g,n} \times p_i}$). The
general Gromov-Witten invariants are obtained by taking the
push-forward
\begin{equation}    \label{GW inv}
\pi_{g,n*}(\on{ev}_{1}^*(\omega_1) \wedge \ldots \wedge
\on{ev}^*_{n}(\omega_n)),
\end{equation}
which is a differential form on ${\mathcal M}_{g,n}$. In particular,
\eqref{fixed Sigma} occurs as a special case when the degree of this
differential form is equal to zero, so we obtain a function on
${\mathcal M}_{g,n}$. Then its value at $(\Sigma,(p_i)) \in {\mathcal
M}_{g,n}$ is given by \eqref{fixed Sigma}. More general observables
give rise to differential forms of positive degree on ${\mathcal
M}_{g,n}$.

The fibers of the map $\pi_{g,n}$ are non-compact, hence {\em a
priori} the integrals obtained via this push-forward are not
well-defined. To show that they are, we replace ${\mathcal
M}_{g,n}(X,\beta)$ by its compactification, the Kontsevich moduli
space of stable maps $\ol{\mathcal M}_{g,n}(X,\beta)$ and ${\mathcal
M}_{g,n}$ by its Deligne-Mumford compactification $\ol{\mathcal
M}_{g,n}$. The map $\pi_{g,n}$ extends to a map
\begin{equation}    \label{olpi}
\ol\pi_{g,n}: \ol{\mathcal M}_{g,n}(X,\beta) \to \ol{\mc M}_{g,n}.
\end{equation}
which is already proper (has compact fibers). The evaluation maps
$\on{ev}_i$ also extend to $\ol{\mathcal M}_{g,n}(X,\beta)$. Then the
Gromov--Witten invariants may be defined by formula \eqref{GW inv}
with $\pi_{g,n}$ replaced by $\ol\pi_{g,n}$. Since $\ol\pi_{g,n}$ is
proper, we see that these invariants are well-defined. Hence the
original integrals \eqref{GW} are also well-defined (for smooth
differential forms $\omega_i$).

These are the correlation functions of what is often referred to as
the ``sigma model coupled to gravity'', with the observables being the
``cohomological descendants'' of the evaluation observables.

\ssec{BPS vs. non-BPS}

Among the observables, an important role is played by the {\em
topological}, or {\em BPS observables}. These are the observables
annihilated by the total supercharge ${\CQ} + \ol{\CQ}$ of the model. The
supercharges $\CQ$ and $\ol{\CQ}$ locally act on fields by the formulas
\begin{align*}
{\CQ} \cdot {\mc O}(w,\wb) &= \left[ \int i p_a(z) \psi^a(z) dz,{\mc
  O}(w,\wb) \right], \\
\ol{\CQ} \cdot {\mc O}(w,\wb) &= \left[ \int i p_{\ol{a}}(\zb)
  \psi^{\ol{a}}(\zb) d\zb,{\mc O}(w,\wb) \right].
\end{align*}
In particular, the supercharge acts on evaluation observables
${\mc O}_\omega$ corresponding to the differential forms $\omega$ as
the de Rham differential $d$:
$$
({\CQ}+\ol{\CQ}) \cdot {\mc O}_\omega = {\mc O}_{d\omega}.
$$
Therefore the BPS evaluation observables correspond to the closed
differential forms on $X$.

In the above definition of the Gromov--Witten invariants we considered
the BPS observables corresponding to closed differential forms
$\omega_i, i=1,\ldots,n$. However, {\em any} differential form
$\omega$ on $X$ gives rise to a legitimate observable in our theory,
and the correlation functions of such observables are still given by
the same integrals \eqref{fixed Sigma}. The difference is, of course,
that unlike the correlation functions of the BPS observables, the
correlation functions of more general observables depend of $\ol\tau$,
so this answer is correct only at $\ol\tau=-i \infty$.

Our goal in this paper is to go beyond the topological sector of the
sigma model and consider the correlation functions of non-BPS
observables. The reasons for doing this have already been explained in
Part I and in the Introduction to this Part. Here we want to stress
that if we only consider the BPS observables, we will not be able to
gain any insights into the structure of the space of states of our
theory beyond the ground states.

Indeed, from the Hamiltonian perspective, when $\Sigma$ is the
cylinder $\BS \times \R$, the correlation function is given by the
formula
\begin{equation}    \label{path int1}
\langle {\mc O}_1(p_1) \ldots {\mc O}_n(p_n) \rangle = \langle
         0 | \OO_n e^{(t_{n-1}-t_n)H} \ldots
         e^{(t_1-t_2)H} \OO_1 \vac,
\end{equation}
Here $t_i$ denotes the time coordinate of the point $p_i$ (along the
$\R$ factor of $\Sigma$), so that $|z_i| = e^{t_i}$, and we assume
that $t_1 < t_2 < \ldots < t_n$. Thus, in principle, we could derive
information about the spectrum of the Hamiltonian and its
diagonalizability by analyzing the correlation functions. For example,
the appearance of terms of the form $(t_i-t_{i+1}) e^{N(t_i-t_{i+1})}$
(which we will observe below), but not $(t_i-t_{i+1})^m
e^{N(t_i-t_{i+1})}, m>1$, means that the Hamiltonian $H$ has a Jordan
block of size $2$ with the generalized eigenvalue $N$.

However, BPS observable are not suitable for this purpose. Indeed,
since the vacuum state is annihilated by $\CQ$, any correlation
function of BPS observables (which commute with $\CQ$) is
automatically equal to $0$ as soon as one of the observables is
$\CQ$-exact, that is, equal to the commutator of another observable
and $\CQ$. This means that the correlation functions of BPS
observables only depend on their $\CQ$-cohomology classes. One can
modify any BPS observable by $\CQ$-exact terms so as to make it
commute with $\CQ$ and ${\CQ}^*$, where $[{\CQ},{\CQ}^*]_+ = H$. Such
a representative transforms a vacuum state, which is annihilated by
${\CQ}, {\CQ}^*$ to a state, which is again annihilated by $\CQ$ and
${\CQ}^*$, and hence by their anti-commutator $H$. Therefore no
excited states on which $H$ acts non-trivially, can arise in formula
\eqref{path int1}. Hence we do not learn anything about the spectrum
of the model. In contrast, non-BPS observables transform ground states
to excited states, and, as we will see below, we can learn a lot about
the structure of the space of states from their correlation functions.

In addition, considering non-BPS observables allows us to bring into
play some important $\CQ$-exact observables, which are ``invisible'' in
the BPS sector.

Examples are the observables corresponding to Lie derivatives with
respect to vector fields on $X$ given by formula \eqref{vz}. These
observables are $\CQ$-exact, as follows from the Cartan formula ${\mc
L}_v = \{ d,\imath_v \}$ for the Lie derivative ${\mc L}_v$.

This means that if we insert the observable $v(z)$ into a correlation
function of BPS observables, then the result will always be zero. But
these observables, and more general observables of this type
corresponding to differential operators on $X$, play a very important
role in the full theory. Indeed, on a K\"ahler manifold we often have
a large Lie algebra of global holomorphic vector fields, and the
corresponding Lie derivatives will belong to the chiral algebra of our
theory (see \secref{chiral} below). Hence they give rise to
non-trivial Ward identities which impose relations between correlation
functions in our model. But in order to obtain non-trivial correlation
functions involving these operators we must include non-BPS
observables.

Let us summarize: the correlation functions of the supersymmetric
sigma model in the infinite radius limit are expressed in terms of
integrals over finite-dimensional moduli spaces of holomorphic
maps. Our goal is to use them to obtain information about the space
of states of the theory and the action of the Hamiltonian. We are
particularly interested in the appearance of logarithms in the
correlation functions (when they are written in terms of coordinates
$z_i$ on the worldsheet $\pone$), which indicate that the Hamiltonian
is not diagonalizable.

However, the correlation functions of the BPS observables which have
been almost exclusively studied in the literature up to now (and which
correspond to the Gromov-Witten invariants) do not contain
logarithms. In order to observe the appearance of logarithms, we must
consider non-BPS observables. We note that the Hamiltonian is also
diagonalizable on all purely chiral (and anti-chiral) states; thus,
the chiral algebra of the theory, which we will discuss in more detail
in \secref{chiral} below, is free of logarithms.

\ssec{A simple example of non-BPS correlation function}

As our first example, we consider the target manifold $X = {\C\mathbb
P}^{1}$ and the worldsheet $\Sigma=\pone$. Then the instantons are
holomorphic maps from the parametrized $\pone$ to $\pone$. The moduli
spaces are labeled by non-negative integers in this case
corresponding to the degree of such a map. The moduli space of degree
$d$ instantons ${\mathcal M}_{d} = {\mathcal M}_{\pone}(\pone,1)$ (in
the notation of \secref{GW}) has complex dimension $2d+1$. We consider
the case when $d=1$.  Then the corresponding moduli space ${\mathcal
M}_{1}$ is isomorphic to $PGL_{2}({\C})$. Consider the correlator of
the following evaluation observables:
\begin{equation}
\langle {\CO}_{{\omega}_{0}}(0) {\CO}_{{\omega}_{\infty}}({\infty})
	{\CO}_{{\omega}_{\rm FS}}(1) {\CO}_{h}(z , \ol{z})
	\rangle_{d=1},
\label{corrsm}
\end{equation}
where
$$
{\omega}_{0} = {\delta}^{(2)}(x) d^{2}x \ , \quad
{\omega}_{\infty} = {\delta}^{(2)}\left({1\over x}\right)
\frac{d^{2}x}{\vert x \vert^{4}} \ , \quad
{\omega}_{\rm FS} = \frac{d^{2}x}{( 1+ \vert x \vert^{2})^{2}},
$$
\begin{equation}
h = \frac{1}{1+ \vert x \vert^{2}}.
\label{omf}
\end{equation}
The two-forms are closed, and hence correspond to BPS observables, but
the function $h$ is not. Its inclusion allows us to observe the
logarithmic effects.

The delta-function two-forms ${\om}_{0}$, ${\om}_{\infty}$, supported
at $x=0$ and $x=\infty$, respectively, reduce the integration over
${\mathcal M}_{1}$ to that over the locus consisting of the
holomorphic maps of the form
\begin{equation}
x(w) = A w.
\label{redm}
\end{equation}

Thus, \Ref{corrsm} is equal to:
\begin{equation}
q \int \ \frac{d^{2}A}{( 1 + | A |^{2})^{2}} \frac{1}{1+ z\ol{z}
|A|^{2}} \propto - q \left( \frac{1}{1 - z\ol{z}} + \frac{
z\ol{z}\, {\log}(z\zb)}{(1-z\ol{z})^{2}} \right)
\label{corrt}
\end{equation}
(see Part I for details on the computation of integrals of this type)
where $q = e^{-\tau}$ is the instanton factor.  The $z,
\ol{z}$-dependence in \Ref{corrt} implies the logarithmic nature of
the two-dimensional conformal theory, in the same way as in the case
of the quantum mechanical models analyzed in Part I. We recall that
from the Hamiltonian point of view the correlation function is
represented as the matrix element of the form \eqref{path
int1}. Therefore the monomials $(z\zb)^n$ correspond to eigenstates of
the Hamiltonian with the eigenvalue $n$, whereas the terms
$\log(z\zb)\cdot (z\zb)^n$ correspond to a Jordan block of size $2$
with the generalized eigenvalue $n$. Therefore, expanding the right
hand side of \eqref{corrt} in powers of $z\zb$ (here we assume that
$|z|<1$), we find that the spectrum of the Hamiltonian contains all
non-negative integers as well as Jordan blocks of size two with the
generalized eigenvalues equal to all positive integers.

When we move away from the point $\ol\tau = - i \infty$ (back to
finite radius), anti-instantons start making contributions. As the
result, the Jordan blocks deform to matrices with small non-zero
entries below the diagonal, and the Hamiltonian becomes
diagonalizable.

\ssec{Spectrum from the genus one correlation functions}

Next, we discuss the genus one correlation functions in the sigma
model with the target $\pone$. We want to use them to probe the
spectrum of our theory, the way we probed the spectrum in the
analogous quantum mechanical models in Part I (using factorization
over intermediate states). Again, in order to detect the spectrum on
the excited states, we need to throw in some non-BPS observables.

The simplest correlation function of this type is that of the
evaluation observables with the periodic boundary conditions on the
fermions. We keep the target space $X = {\pone}$, and the worldsheet
${\Sigma}$ is the elliptic curve, with the modular parameter $q =
{\exp} ( 2 {\pi} i {\tau})$.  Let $z$ denote the linear coordinate on
$\Sigma$, so that
$$
z \sim z + m + n {\tau}, \qquad m ,n \in \Z.
$$
Let ${\mc O}_\omega$ denote the evaluation observable corresponding to
a differential form $\omega$ on the target $\pone$.

Let ${\om} = {\om}(X, {\Xb})dXd{\Xb} \in {\Omega}^{2}
({\mathbb{P}}^{1})$, $f \in C^{\infty}({\mathbb{P}}^{1})$. We start
with the five-point function
$$
\Biggl\langle {\mathcal{O}}_{{\om}_{0}}(z_{1})
{\mathcal{O}}_{{\om}_{\infty}}(z_{2})
{\mathcal{O}}_{{\om}_{\infty}}(z_{3}) {\mathcal{O}}_{\om} (z_{4})
{\mathcal{O}}_{f}(z_{5}) \Biggr\rangle_{q} =
$$
\begin{multline}
\int_{{\mathbb{P}}^{1}} {\om} \left( A {{\vartheta}_{11}( z_{4} -
z_{1}) {\vartheta} ( z_{4} - z_{0} ) \over {\vartheta}_{11} ( z_{4} -
z_{2} ) {\vartheta} ( z_{4} - z_{3} )}, {\bar A}
\overline{{\vartheta}_{11}( z_{4} - z_{1}) {\vartheta} ( z_{4} - z_{0}
) \over {\vartheta}_{11} ( z_{4} - z_{2} ) {\vartheta} ( z_{4} - z_{3}
)} \right) \\ \qquad\qquad \times \left\vert {{\vartheta}_{11}( z_{4}
- z_{1}) {\vartheta} ( z_{4} - z_{0} ) \over {\vartheta}_{11} ( z_{4}
- z_{2} ) {\vartheta} ( z_{4} - z_{3} )} \right\vert^{2} \ d^{2}A
\; \cdot \; f \left( A {{\vartheta}_{11}( z_{5} - z_{1}) {\vartheta} (
z_{5} - z_{0} ) \over {\vartheta}_{11} ( z_{5} - z_{2} ) {\vartheta} (
z_{5} - z_{3} )} \right),
\label{fivepoint}
\end{multline}
where we chose for simplicity some of the observables to correspond to
the delta-forms supported at $0, \infty \in X$, as indicated by the
subscripts in \eqref{fivepoint} (compare with formula
\Ref{corrt}). These delta-forms force the holomorphic maps from
$\Sigma$ to $X$ to be of the special form: \be X(z) = A
{{\vartheta}_{11}( z - z_{1}) {\vartheta} ( z - z_{0} ) \over
{\vartheta}_{11} ( z - z_{2} ) {\vartheta} ( z - z_{3} )},
\label{specf}
\ee
where $A \in {\pone}$ is the remaining modulus, and $z_{0} = z_{2}
+ z_{3} - z_{1}$. To illustrate our point about the logarithmic nature
of the sigma model Hamiltonian we take the further simplifying limit,
where $z_{2}, z_{3} \to 0$. Let us also take for $\om$ the usual
Fubini-Study form.  In this case the integral \Ref{fivepoint}
simplifies to \be \int_{\pone} {d^{2}x \over ( 1 + |x|^{2} |q_{\rm
eff}|^{2} )^{2}} f (x)
\label{reducedfp}
\ee where \be q_{\rm eff} = \frac{{\wp}(z_{4}) -
{\wp}(z_{1})}{{\wp}(z_{5}) - {\wp}(z_{1})},
\label{effq}
\ee
where
$$
x = A (\wp(z_5) - \wp(z_1))
$$ and $\wp$ is the Weierstrass' function.  The result can be now
expanded in powers of $q$, as $q_{\rm eff}$ has such an expansion,
following from the $q$-expansion of the $\wp$-function, and will be of
the form \Ref{corrt}, i.e., with integer powers of $q$ or with ${\rm
log}(q)$ multiplying the integer powers of $q$.

On the other hand, the correlation function \Ref{fivepoint} can be
written as a trace:
\be {\rm Tr}_{\mathcal{H}} \left( (-1)^{F} \,
q^{H} e^{-z_{5} H} f e^{(z_{5} - z_{4}) H} {\mathcal{O}}_{\om}
e^{(z_{4}-z_{1})H} {\mathcal{O}}_{{\om}_{0}} e^{z_{1}H}
{\mathcal{O}}_{{\om}_{\infty}} {\mathcal{O}}_{{\om}_{\infty}} \right).
\label{trce}
\ee
Thus the $q$-dependence of \Ref{reducedfp} gives us the information
about the spectrum of conformal dimensions in our theory.
It confirms the integrality of the spectrum and the existence of
Jordan blocks in the action of the Hamiltonian in the sigma model with
the target $\pone$.

\subsection{Sigma model on the torus: from finite to infinite
  radius}    \label{from to}

Consider the sigma model with the torus $\BT$ as the target
manifold discussed in \secref{torus1}. An attractive feature of
this model is that the correlation functions may be computed exactly
at both finite and infinite radius, and this can help us learn how the
correlation functions behave in the limit of infinite radius, $\ol\tau
\to \infty$. After all, eventually we would like to compute the
(non-BPS) correlation functions in the sigma models at the finite
radius using those in the infinite radius limit as the starting point,
by some sort of perturbation theory.

The sigma model on the torus is also useful in that it allows us to
see explicitly how the spectrum of a well-defined unitary theory
acquires an imaginary part or an unbounded branch, in the $\ol{\tau}
\to \infty$ limit, as we have argued in \secref{BS1} in the context of
quantum mechanics on the non-simply connected target manifolds.

\subsubsection{Finite radius Hamiltonian}    \label{fin radius}

For the sake of generality consider the sigma model on the
$2d$-dimensional torus ${\BT}^{2d}$, with the translation invariant
metric $G_{mn}$ and the antisymmetric two-form $B_{mn}$.\footnote{Our
notation in this section is slightly different from that of the main
body of the paper. In particular, what we called previously the field
$B$ is now $B/4\pi \al^{\prime}$. This is done in order to make the
notation consistent with the traditional physics notation; see e.g.,
\cite{Pol}.} In Sections \ref{fin radius}--\ref{comparison} we will
consider the purely bosonic theory. We will be interested in its
partition function, which coincides with the partition function of the
bosonic sector of the supersymmetric sigma model. Then in
\secref{fermions} we will add fermions.

The bosonic theory has the following action on $\Sigma = {\BS}^{1}
\times I$:
\be
S = \frac{1}{4{\pi}{\alpha}^{\prime}} \int_{\Sigma}
G_{mn} \left( {\pa}_{t} X^{m} {\pa}_{t} X^{n} + {\pa}_{\si}
X^{m}{\pa}_{\si} X^{n} \right) + i B_{mn} \left( {\pa}_{t} X^{m}
{\pa}_{\si} X^{n} - {\pa}_{\si} X^{m}{\pa}_{t} X^{n} \right) d\sigma
dt
\label{torus}
\ee
where $X^{m}$, $m=1, \ldots , 2d$, correspond to the linear
coordinates on ${\BT}^{2d}$ with the period $2\pi$. We assume the
worldsheet to have Euclidean metric $dt^{2}+ d{\si}^{2}\ , \ {\si}
\sim {\si} + 2{\pi}$. Thus the path integral measure is
$$
DX \ e^{-S}
$$
The momentum conjugate to $X^{m}$ is then given by
\be
P_{m} = i \frac{{\delta}L}{{\delta}{\p}_{t}X^{m}} =
\frac{1}{2{\pi}{\alpha}^{\prime}} \left( i G_{mn} {\pa}_{t}
X^{n} - B_{mn} {\pa}_{\si} X^{n} \right),
\label{mom}
\ee
and the Hamiltonian is
\begin{multline}
H = \frac{1}{4\pi {\alpha}^{\prime}}\int_{{\BS}^{1}} {\rm d}{\si} \
G^{mn} \left( 2{\pi} {\alpha}^{\prime} P_{m} + B_{mk} {\pa}_{\si}X^{k}
\right)\left( 2{\pi} {\alpha}^{\prime} P_{n} + B_{nl} {\pa}_{\si}X^{l}
\right) + \\ + G_{mn}{\pa}_{\si} X^{m} {\pa}_{\si} X^{n} =
\int_{{\BS}^{1}}\ {\pi}{\alpha}^{\prime} G^{mn}P_{m}P_{n} + J^{m}_{n}
P_{m} {\pa}_{\si}X^{n} + \frac{g_{mn}}{4\pi {\alpha}^{\prime}}
{\pa}_{\si} X^{m} {\pa}_{\si}X^{n},
\label{ham}
\end{multline}
where
\begin{align*}
J_{m}^{n} \ &= \ G^{kn} B_{km}\ , \\ 
g_{mn} \ &= \ G_{mn} + B_{am}B_{bn}G^{ab}\ , \\
   & \qquad g \ = \  (1 - J^{2} ) G .
\end{align*}
At the level of zero modes, we have
$$
X^{m} \sim X_{0}^{m} +  w^{m}{\si} 
\ , 
$$
and the Hamiltonian \Ref{ham} reduces to
\be
H_{0} = \frac{1}{4{\pi}} \left(  {\alpha}^{\prime}
G^{mn} k_{m} k_{n} + 
2 J^{m}_{n} k_{m} w^{n} + \frac{g_{mn}}{{\alpha}^{\prime}} w^{m} w^{n}
\right),
\label{zrmd}
\ee
where $k_{m} \in {\Z}$ is the integer eigenvalue of the momentum zero
mode,
$$
- i \frac{\pa}{\pa X^{m}_{0}} = k_{m} \ , 
$$
while $w^{m} \in {\Z}$ is the winding number of the $X^{m}$
coordinate.

The inclusion of the oscillators, that is, the harmonics $X^{m}_{l}
e^{i l {\si}}$ and $P_{m,l}e^{- i {\si}l}$ with $l \in {\Z}_{\neq 0}$,
simply adds non-negative integers to the eigenvalues of the
Hamiltonian (modulo the zero point energies). Indeed, the
$B$-dependence of the Hamiltonian on the $X_{l}^{m}$, $P_{m,l} \sim -
i \frac{\pa}{\pa X_{m,l}}$ modes can be eliminated by the gauge
transformation of the wave-function
$$
{\Psi} (X) \longrightarrow \prod_{l \neq 0} e^{\frac{1}{4\pi
\al^{\prime}} il B_{mn} X_{l}^{m}X_{-l}^{n}} \; \cdot \; {\Psi} (X),
$$
mapping the Hamiltonian on the non-zero modes to that of a system of
harmonic oscillators
\be {\alpha}^{\prime}G^{mn}P_{m,l}P_{n,-l} +
\frac{l^{2}}{{\alpha}^{\prime}} G_{mn} X^{m}_{l}X^{n}_{-l}
\label{nzm}
\ee
with the eigenvalues 
\begin{equation}
| l | \left(N_{l} + \frac{1}{2} \right) \ , \ N_{l} \in {\Z}_{\geq
  0}.
\label{eq:oscse}
\end{equation}

\subsubsection{The limit ${\alpha}^{\prime} \to 0$} The infinite
radius limit that we are interested in is $\al^\prime \to
0$.\footnote{This is the limit denoted by $\ol\tau \to - i \infty$ in
the main body of the paper, but in this section and the next, $\tau$
denotes the complex modulus of the worldsheet torus rather than the
coupling constant.} The non-zero modes are not sensitive to the value
of ${\alpha}^{\prime}$, as we see in \Ref{eq:oscse}. However, the zero
mode Hamiltonian \Ref{zrmd} has a finite limit when ${\alpha}^{\prime}
\to 0$ only if
$$
g_{mn} \; \underset{{\alpha}^{\prime} \to 0}\longrightarrow \; 0
\quad \Longleftrightarrow \quad J^{2} \; \underset{{\alpha}^{\prime}
  \to   0}\longrightarrow \; 1.
$$
For a positive-definite metric $G_{mn}$ and a real two-form $B_{mn}$
the condition $J^{2}=1$ is impossible to fulfill.

However, we know that we should allow complex-valued $B_{mn}$ for
our limit to exist. In this case $J = i {\mathcal{J}} +
{\alpha}^{\prime} {\delta}J$, where $\mathcal{J}$ is a complex
structure on ${\R}^{2n}$.  Let us write,
\begin{align}
& B_{mn} = i\, {\omega}_{mn} + {\alpha}^{\prime} {\mc T}_{mn}, \\
& {\mathcal{J}} = G^{-1}{\omega}\ , \ {\mathcal{J}}^{2} = - 1.
\label{compls}
\end{align}
Then we obtain 
\begin{equation}
H_{0} = i {\mathcal{J}}_{m}^{n} \left( k_{n}w^{m} + i {\mc T}_{mn'}
w^{m}w^{n'} \right) \ , 
\end{equation}
with the characteristic imaginary part, exactly like in formula
\Ref{energn} for the energy levels in the quantum mechanical model on
the circle.

\subsubsection{The partition function}

Let us compute the partition function of the bosonic sector of the
supersymmetric sigma model on the torus,\footnote{The full
supersymmetric partition function is in the topological sector and is
equal to $q$ times the Euler characteristic of $\BT^{2d}$, that is,
$0$.} in the limit ${\alpha}^{\prime} \to 0$,
$$
{\rm Tr}_{\CH_{\on{bos}}} q^{L_{0}} \ol{q}^{\ol{L_{0}}}.
$$
It factorizes as a product of the zero mode part and the oscillator
contribution. We have:
$$
H = L_{0} + \ol{L}_{0},
$$
and
$$
L_{0} - \ol{L}_{0} = 2\int_{{\BS}^{1}} P_{m} {\p}_{\si} X^{m} = 2k_{m}
w^{m} + \on{oscillators}.
$$
The partition function becomes:
\begin{equation}
Z (q, \ol{q}) = {1\over | {\eta}(q) |^{2d}} \sum_{\vec k , \vec w}
q^{\frac{1}{2}(1+i\mathcal{J})\vec k \cdot \vec w}
\ol{q}^{\frac{1}{2}(1-i\mathcal{J})\vec k \cdot \vec w}\left( q\ol{q}
\right)^{i\mathcal{J}{\mc T}(\vec w , \vec w )}
\label{zfun}
\end{equation}

{}From the path integral point of view, we expect that the correlation
functions may be written as a sum over holomorphic maps. In the case
at hand those are the maps from the worldsheet torus $E_\tau =
\C/(\Z+\Z\tau)$ to the target torus ${\BT}^{2d}$, endowed with the
complex structure $\mathcal{J}$. A holomorphic map $E_\tau \to
\BT^{2d}$ exists if and only if there exist integral vectors $\vec
w^{\vee} , \vec w \in \Z^d$ such that
\begin{align*}
{\vec w}^{\vee} \mathcal{T} - \frac{1}{2}(1+ i
{\mathcal{J}}) {\vec w} &= 0 , \\ 
{\vec w}^{\vee} \mathcal{\ol{T}} - \frac{1}{2}(1- i
{\mathcal{J}}) {\vec w} &= 0.
\end{align*}
Hence we should be able to represent our partition function $Z(q,
\ol{q}), q = {\exp} 2{\pi} i \tau$, as a sum of delta-functions
corresponding to these constraints. We indeed obtain such an
expression using the Poisson resummation formula:
\begin{multline}
Z(q, \ol{q}) \sim \\ {1\over | {\eta}(q) |^{2d}} \sum_{\vec w^{\vee} ,
  \vec w \in \Z^d} {\delta}^{(2d)} \left( {\vec w}^{\vee} \mathcal{T}
  - \frac{1}{2}(1+ i {\mathcal{J}}) {\vec w} , {\vec w}^{\vee}
  \mathcal{\ol{T}} - \frac{1}{2}(1- i {\mathcal{J}}) {\vec w} \right)
  e^{- 2{\pi}{\rm Im}\mathcal{T} i\mathcal{J}{\mc T}(\vec w , \vec w
  )}\ .
\label{toruspf}
\end{multline}

\subsubsection{Comparison with the computation in the infinite radius
  limit}    \label{comparison}

Let us compare the limit \eqref{zfun} with the trace
$$
{\rm Tr}_{\CH_{\on{bos}}} q^{L_{0}} \ol{q}^{\ol{L_{0}}},
$$
computed directly in the infinite radius limit $\al^{\prime} \to
0$. Here $\CH_{\on{bos}}$ is the bosonic part of the space of states
in the infinite radius limit that we have described in
\secref{torus1}. According to formulas \eqref{dec winding} and
\eqref{Hmn}, $\CH_{\on{bos}}$ is the tensor product of the oscillator
part and the momentum/winding part. Formula \eqref{Tz} for the stress
tensor shows that the oscillators contribute the factor
$1/\eta(q)^2$. The momentum/winding part is spanned by the monomials
of the form
\begin{equation}    \label{hw states}
e^{i r (X_0 \ol{T} - \ol{X}_0 T)/(\ol{T}-T) + i
    s (X_0-\ol{X}_0)/(T - \ol{T})} \cdot e^{i \int \left( (m+nT) P +
(m+n\ol{T}) \ol{P} \right)}, \qquad r,s,m,n \in \Z.
\end{equation}
(the first factor corresponds to momentum, see formula
\eqref{spanned}, and the second factor corresponds to winding, see
formula \eqref{Psi}). Formulas \eqref{Tz} and \eqref{Xz} show that
$L_0$ and $\ol{L}_0$ act on this state by the formulas
$$
L_0 = i p_0 \omega, \qquad \ol{L}_0 = i \ol{p}_0 \ol\omega.
$$
In the simplest case when the $B$-field is zero, we have $p_0 = - i
\frac{\pa}{\pa X_0}, \ol{p}_0 = - i \frac{\pa}{\pa \ol{X}_0}$, and so
they act on these states by the formulas
$$
L_0 = \frac{(s-r\ol{T})(m+nT)}{T-\ol{T}}, \qquad \ol{L}_0 = -
\frac{(s-rT)(m+n\ol{T})}{T-\ol{T}}.
$$
Therefore we find that
\begin{equation}    \label{B zero}
\on{Tr}_{\CH_{\on{bos}}} q^{L_{0}} \ol{q}^{\ol{L}_{0}} =
\frac{1}{\eta(q)^2} \sum_{m,n,r,s \in \Z}
  q^{(s-r\ol{T})(m+nT)/(T-\ol{T})}
  \ol{q}^{-(s-rT)(m+n\ol{T})/(T-\ol{T})}.
\end{equation}

This agrees with formula \eqref{zfun} in the case when $d=1$ with the
zero $B$-field. Indeed, in this case the complex structure of the
torus is uniquely determined by the metric: $T = T_{1} + i T_{2}$,
\begin{equation}
G_{11} = \frac{1}{T_{2}} \sqrt{G}\, , \, G_{12} = \frac{T_{1}}{T_{2}}
\sqrt{G}\, , \, G_{22} = \frac{|T|^{2}}{T_{2}}\sqrt{G}.
\label{cmplx}
\end{equation} 
The remaining moduli of the torus are the volume and the $B$-field,
which combine into:
\begin{equation} 
U = \frac{1}{{\al}^{\prime}} \left( B_{12} + i \sqrt{G} \right).
\label{ubg}
\end{equation}
We take a limit ${\al}^{\prime} \to 0$ while keeping $U$ fixed. 
It means that $B_{12}$ is taken to be complex and that
\begin{equation}
\ol{U} = U  - \frac{2i}{{\al}^{\prime}}\sqrt{G} \to \infty.
\label{ubar}
\end{equation}
When $U=0$ we obtain formula \eqref{B zero}. For non-zero $U$ there is
an additional factor $(q \ol{q})^{U|m+nT|^2/(T-\ol{T})}$ on the right
hand side. This corresponds to a shift of the eigenvalues of $L_0$ and
$\ol{L}_0$ on the states \eqref{hw states} that is caused by the
following redefinition of $p_0, \ol{p}_0$ for non-zero $B$-field:
$$
p_0 \mapsto p_0 - i \frac{U}{T-\ol{T}} \ol\omega, \qquad \ol{p}_0
\mapsto \ol{p}_0 - i \frac{U}{T-\ol{T}} \omega.
$$

\subsubsection{Fermions}    \label{fermions}

The fermionic contribution is universal, it does not depend on the
choice of the complex structure on ${\BT}^{2d}$. The fermionic part of
the action is
$$
S_{\on{ferm}} = \int \, ( {\pi}_{a} {\ol{\pa}}_{\ol{z}} {\psi}^{a} +
{\ol{\pi}_{\ol{a}}} {\pa}_{z} \ol{\psi}^{\ol{a}} ) d^2 z. 
$$
The $\mathcal{Q}$-symmetry depends on the choice of the complex
structure. In other words, the sigma model has extended fermionic
symmetry on ${\BT}^{2d}$ for $d >1$. We shall not discuss this any
further.

\section{Logarithmic mixing of jet-evaluation observables}
\label{log mixing}

The most striking illustration of the logarithmic nature of the
two-dimensional sigma model is the behavior of the correlation
functions of the jet-evaluation observables introduced in
\secref{jet-ev}. In this section we discuss these correlation
functions. They are also given by integrals over the moduli spaces of
stable maps, but we will see that in general they diverge at the
boundary divisors corresponding to ``bubbles'' on the
worldsheet. These integrals require regularization, and the ambiguity
of the regularization scheme introduces a non-trivial mixing of
operators.

In other words, we find that each jet-evaluation observable ${\mc
O}_{\mathbb A}$ introduced in \secref{jet-ev} is only well-defined
perturbatively. The definition of a true operator of the sigma model
corresponding to it -- and its correlation functions -- requires
regularization. Depending on the regularization scheme, we obtain {\em
a priori} different operators, which differ from each other by a
linear combination of other operators. Those are the logarithmic
partners of ${\mc O}_{\mathbb A}$. Together with ${\mc O}_{\mathbb
A}$, the logarithmic partners span a subspace in the space of
operators -- or the space of states, via the state--operator
correspondence -- on which the Hamiltonian acts as a Jordan block.

This logarithmic mixing is in agreement with the description of the
space of states given in \secref{delta-forms} as an extension of the
spaces of delta-forms on the ascending manifolds in $\wt{LX}$. It is
also analogous to the similar logarithmic phenomena that we have
observed and explored in quantum mechanical models in Part I.

\ssec{Jet Gromov--Witten invariants}

Recall the jet bundle ${\mc J} X$ over $\Sigma$, whose fiber $J_p X$
over $p \in \Sigma$ consists of jets of holomorphic maps from a disc
$D_p$ around the point $p$ to $X$. Given a holomorphic map $\Phi:
\Sigma \to X$, we obtain a point in $J_p X$; namely, the restriction
$\Phi|_{D_p}$ of $\Phi$ to $D_p \subset \Sigma$. Thus, we obtain the
following generalization of the map $\on{ev}_p$,
\begin{equation}    \label{Jeval}
{\mc J}\on{ev}_p: {\mathcal M}_{\Sigma}(X,\beta) \to {\mc J}X, \\
\end{equation}
Now recall that jet-evaluation observables are sections of the bundle
$\Omega^\bullet_{\on{vert}}({\mc J} X)$ of vertical differential forms
on ${\mc J} X$. Given $n$ such observables
$\wh\omega_1,\ldots,\wh\omega_n$, we define their correlation function
as the following integral generalizing \eqref{fixed Sigma}:
\begin{equation}    \label{J fixed Sigma}
\int_{{\mathcal M}_{\Sigma}(X,\beta)} {\mc
  J}\on{ev}_{p_1}^*(\wh\omega_1) \wedge \ldots \wedge {\mc
  J}\on{ev}^*_{p_n}(\wh\omega_n).
\end{equation}
These are the correlation functions of the observables of the sigma
model involving the fields $X^a(z), \psi^a(z)$ and their complex
conjugates and all of their derivatives, but not the momenta variables
$p_a(z), \pi_a(z)$ and their complex conjugates and
derivatives.\footnote{For a discussion of the correlation functions of
the observables involving the momenta variables, see \cite{Cargese},
Section 5.} These integrals are compatible
with the integrals \eqref{fixed Sigma} in the following sense. Recall
the tautological map ${\mc J} X \to X \times \Sigma$ from
\secref{jet-ev}. If $\omega_1,\ldots,\omega_n$ are differential forms
on $X$ and $\wh\omega_1,\ldots,\wh\omega_n$ are their pull-backs to
${\mc J} X$ via the composite map ${\mc J} X \to X \times \Sigma \to
X$, then the integral \eqref{J fixed Sigma} is equal to \eqref{fixed
Sigma}. In other words, if the observables depend only on $X^a(z),
\psi^a(z)$ and their complex conjugates, but not on their derivatives,
then we obtain the same answer as before.

To make this formula more concrete, let us choose local holomorphic
coordinates $z_1,\ldots,z_n$ at the points $p_1,\ldots,p_n$. Then the
fiber of ${\mc J} X$ at $p_i$ may be identified with the space $JX$ of
jets of holomorphic maps from the coordinatized disc $D$ to
$X$. Therefore we obtain a map
\begin{equation}    \label{Jeval1}
\on{Jev}_{p_i}: {\mathcal M}_{\Sigma}(X,\beta) \to JX, \\
\end{equation}
sending $\Phi: \Sigma \to X$ to $\Phi|_{D_{p_i}}$, written as a power
series with respect to the coordinate $z_i$. Then we may define the
corresponding correlation function by the same formula as \eqref{J
fixed Sigma}, but with ${\mc J}\on{ev}_{p_i}$ replaced by
$\on{Jev}_{p_i}$ for all $i=1,\ldots,n$.

As in the case of the ordinary Gromov--Witten invariants, there is a
problem in the definition of these integrals as the moduli spaces
${\mathcal M}_{\Sigma}(X,\beta)$ are non-compact. In the case of the
Gromov--Witten invariants this problem is cured by using their
Kontsevich compactification. The boundary strata of this
compactification are not maps from $\Sigma$ to $X$, but maps to $X$
from stable singular curves which are obtained by attaching to
$\Sigma$ additional components of genus $0$ (``bubbles'') with fewer
than three marked points, so that they are collapsed under the map to
the moduli space of stable pointed curves (see Figure 3). One can
easily extend the evaluation maps $\on{ev}_{p_i}$ to this
compactification. But can we extend the jet-evaluation maps
$\on{Jev}_{p_i}$?

\begin{center}
\epsfig{file=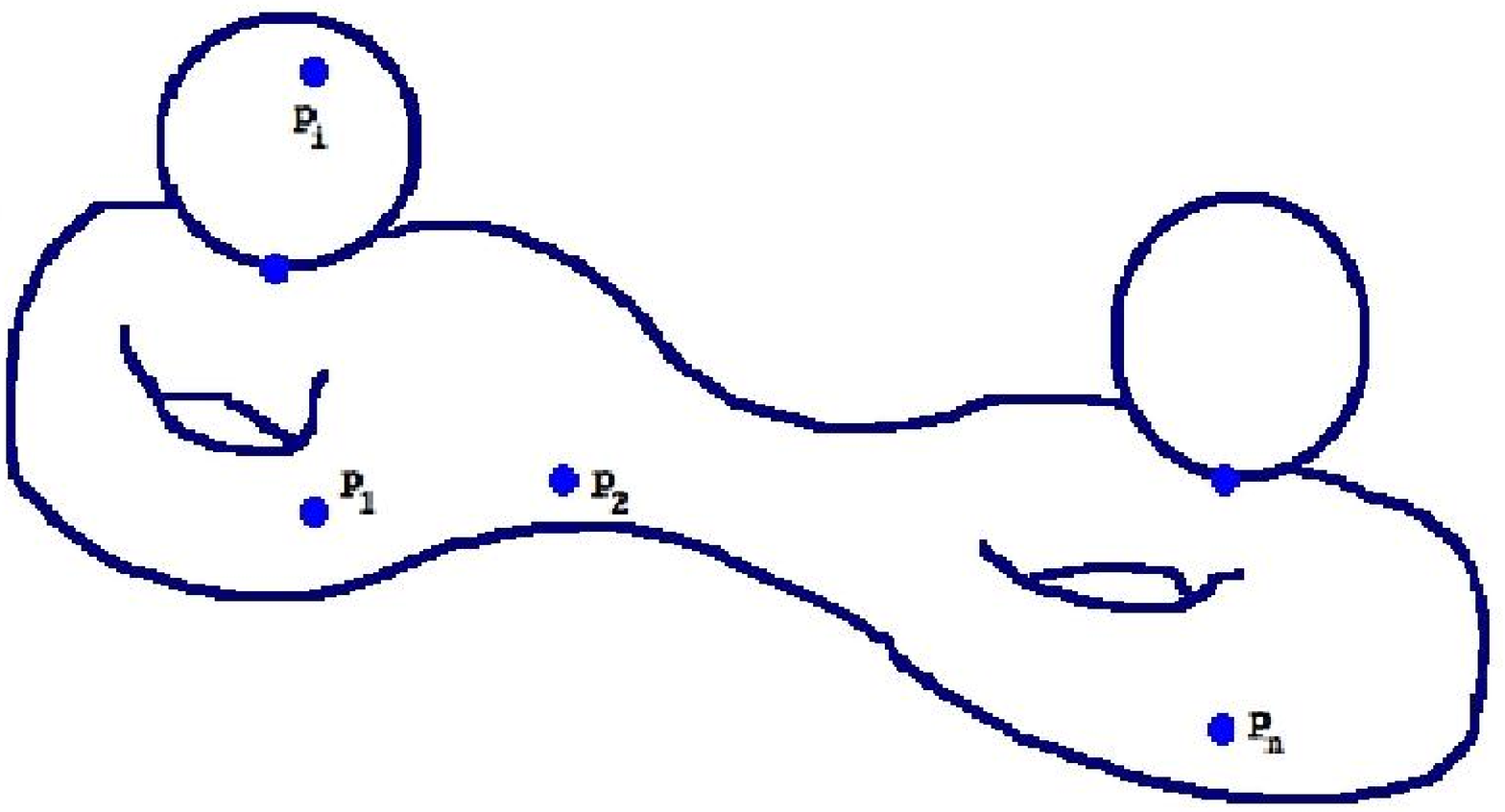,width=120mm}

{\bf Figure 3. Stable curves obtained by attaching ``bubbles'' to the
original curve.}
\end{center}

\ssec{A genus zero example}    \label{genus zero ex}

Let us consider the simplest example, the moduli space ${\mc
M}_{\pone}(\pone,1)$ of degree one maps from a genus $0$
curve $\Sigma = \pone$ with $3$ marked points $0,1,\infty$ to
$\pone$. In this case we have ${\mc M}_{\pone}(\pone,1)
= {\mc M}_{0,3}(\pone,1)$, in the notation of \secref{GW}. This moduli
space is isomorphic to the group $PGL_2$ of M\"obius transformations
on $\pone$:
\begin{equation}    \label{moebius}
\Phi: z \mapsto \frac{\al z + \beta}{\ga z + \delta}.
\end{equation}
We have an injective map
$$
{\mc M}_{\pone}(\pone,1) \to \pone \times \pone \times
\pone
$$
corresponding to evaluating $\Phi$ at the three points $0,1,\infty$,
whose image is the complement of the diagonals. Let $X_1,X_2,X_3$ be
the three $\pone$-valued functions on the moduli space ${\mc
M}_{\pone}(\pone,1)$, which are just the three evaluation
maps. The Kontsevich compactification (the moduli space of stable
maps)
$$
\ol{\mc M}_{\pone}(\pone,1) = \ol{\mc M}_{0,3}(\pone,1)
$$
is just the the blow-up of $(\pone)^{3}$ along the principal diagonal
$X_1=X_2=X_3$. 

In terms of the M\"obius coordinates \eqref{moebius} on ${\mc
M}_{\pone}(\pone,1) \simeq PGL_2$, we have
$$
X_1 = \frac{\beta}{\delta}, \qquad X_2 = \frac{\al+\beta}{\ga+\delta},
\qquad X_3 = \frac{\al}{\ga}.
$$
These maps extend, tautologically, to the compactification. In
addition to these evaluation maps $\on{ev}_{p_i} = X_i$ on ${\mc
M}_{\pone}(\pone,1)$, we have their jet analogues
$\on{Jev}_{p_i}$, which pick out not just the values, but also all the
derivatives of $\Phi$ at $0,1,\infty$ with respect to the local
coordinates $z,z-1,z^{-1}$. 
We have:
\begin{align}
{\Phi}(z) = & \, X_{1} + X_{31} \frac{z X_{12}/X_{23}}{1+ z
X_{12}/X_{23}} \\ &= \, X_{3} + X_{13} \frac{z^{-1} X_{23}/X_{12}}{1+
z^{-1}X_{23}/X_{12}}
\label{phx}
\end{align}
Therefore the first derivative of $\Phi$ at $z=\infty$ (with respect
to $z^{-1}$) is equal to
\begin{equation}    \label{divergence}
\left. -z^2 \frac{d\Phi}{dz} \right|_{z=\infty} =
\frac{\al\delta - \beta\ga}{\ga^2} =
\frac{(X_1-X_3)(X_2-X_3)}{X_1-X_2}.
\end{equation}
Thus, we see that this is a well-defined function on ${\mc
M}_{\pone}(\pone,1)$, but it {\em cannot} be extended to the divisor
in the Kontsevich compactification $\ol{\mc M}_{0,3}(\pone,1)$ where
$X_1=X_2$! This divisor consists of stable maps to $\pone$ from an
unstable curve that is a union of two genus zero curves, one
containing the points labeled $0,1$, and the other containing one
point $\infty$ (see Figure 4).  Thus, the second component has two
marked points ($\infty$ and the connecting points between the two
components), and therefore is unstable as a curve. As such, it has a
continuous group of automorphisms, namely $\C^\times$. Because of
this, the tangent space at the marked point $\infty$ is not
well-defined as a vector space; only its quotient by $\C^\times$ is
well-defined. Therefore the derivative of $\Phi$ at this point is not
well-defined either, leading to the divergence in formula
\eqref{divergence}.

\begin{center}
\epsfig{file=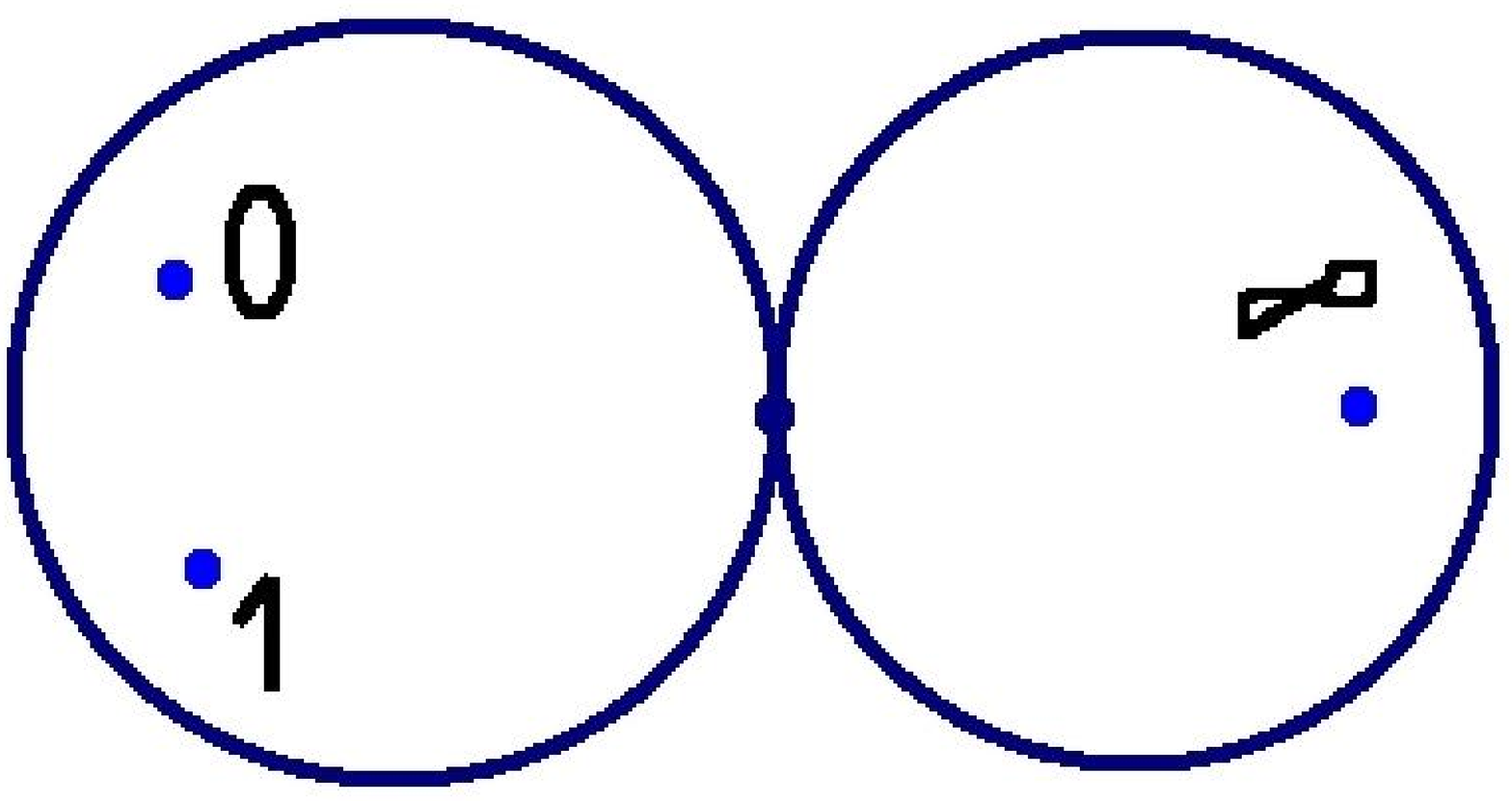,width=80mm}

{\bf Figure 4. Degeneration of $\pone$ with three points.}
\end{center}

\bigskip

Likewise, we find that the first derivative of $\Phi$ at $z=0$ (with
respect to $z$) is equal to
$$
\frac{(X_2-X_1)(X_3-X_1)}{X_3-X_2},
$$
and at $z=1$ (also with respect to $z$) it is
$$
\frac{(X_3-X_2)(X_1-X_2)}{X_3-X_1}.
$$
Thus, we observe the same kind of divergence of the derivative of
$\Phi$ at a given point on the worldsheet $\pone$ in the limit when
the other two points are coming together. (Note that, on the contrary,
the derivative becomes zero when our point of evaluation itself comes
close to another point. In this case the two appear together on a
stable component, on which the map $\Phi$ becomes constant, and hence
has zero derivative.)

Thus, we encounter an important geometric phenomenon, which, as we
will see in Sections \ref{reg} and \ref{OPE}, is responsible for the
appearance of the logarithmic partners of operators of the sigma model
and for the logarithmic terms in the OPE of jet-evaluation observables
of the sigma model. The logarithmic nature of the sigma models in the
infinite radius limit may therefore be traced to the fact that the jet
evaluation map cannot be extended to the Kontsevich moduli space of
stable maps.

\subsubsection{Higher derivatives}

For future use let us record here the expression for higher
derivatives $\dlx{\la} = \CD^\la \Phi$ at $z=0$ (we use the notation
of \secref{jet-ev})
\begin{equation}
\dlx{\la} = (X_{3} - X_{1})^{{\ell}({\la})} \left( \frac{X_{1} -
X_{2}}{X_{2} - X_{3}} \right)^{|{\la}|}
\label{dlzzx}
\end{equation}

\ssec{Generalization to higher genus}

To understand better what is going on here, it is instructive to
consider the more general setting of Gromov--Witten invariants
described in \secref{GW}, in which we are allowed to vary the pointed
curve $(\Sigma,(p_i))$. Thus, we have the moduli space ${\mathcal
M}_{g,n}(X,\beta)$ of triples $(\Sigma,(p_i),\Phi)$
and a morphism
$$
\pi_{g,n}: {\mathcal M}_{g,n}(X,\beta) \to {\mc M}_{g,n},
$$
as in \secref{GW}. Now we define the moduli space ${\mc
J}_{g,n}(X)$. This is a (non-trivial) bundle over ${\mathcal
M}_{g,n}$, whose fiber over $(\Sigma,(p_i))$ is $J_{p_1} X \times
\ldots \times J_{p_n} X$ (recall that $J_{p_i} X$ is the space of jets
of holomorphic maps $D_{p_i} \to X$, where $D_{p_i}$ is a small disc
around $p_i$). We have the jet analogues of the maps \eqref{evali},
\begin{equation}    \label{Jev DM}
{\mc J}\on{ev}_i: {\mathcal M}_{g,n}(X,\beta) \to {\mc
  J}_{g,n}(X),
\end{equation}
sending
$$
(\Sigma,(p_i),\Phi) \mapsto
\left(\Sigma,(p_i),\left(\Phi|_{D_{p_i}}\right) \right).
$$
Using these maps, we
define the {\em jet Gromov--Witten invariants} (which may be viewed as
the correlation functions of the jet-evaluation observables of the
sigma model with the target $X$ coupled to gravity) by the formula
\begin{equation}    \label{J GW inv}
\pi_{g,n*}({\mc J}\on{ev}_{1}^*(\wh\omega_1) \wedge \ldots \wedge
{\mc J}\on{ev}^*_{n}(\wh\omega_n)),
\end{equation}
of which \eqref{J fixed Sigma} is a special case (when the degree of
the resulting differential form on ${\mathcal M}_{g,n}$ is equal to
zero). Here $\wh\omega_i$ are vertical differential forms on ${\mc
J}_{g,n}(X)$, with respect to the map ${\mc J}_{g,n}(X) \to {\mathcal
M}_{g,n}$ (as defined in \secref{jet-ev}).

But here we again face the problem that the fibers of the map
$\pi_{g,n}$ are not compact. To cure this problem, it is natural to
use the Kontsevich moduli space of stable maps $\ol{\mathcal
M}_{g,n}(X,\beta)$, as in \secref{GW}. We also have the
Deligne--Mumford moduli space $\ol{\mc M}_{g,n}$ of stable pointed
curves $(\Sigma,(p_i))$ of genus $g$ with $n$ marked points, which is
a compactification of ${\mc M}_{g,n}$. The map $\pi_{g,n}$ extends to
a map \eqref{olpi} which has compact fibers. We then try to define the
regularized integrals by the formula
\begin{equation}    \label{J GW inv1}
\ol\pi_{g,n*}(\ol{\mc J}\on{ev}_{1}^*(\wh\omega_1) \wedge \ldots
\wedge \ol{\mc J}\on{ev}^*_{n}(\wh\omega_n)).
\end{equation}
However, to make sense of this formula, we need to extend the maps
${\mc J}\on{ev}_{i}$ to the compactification $\ol{\mathcal
M}_{g,n}(X,\beta)$.

It is natural to try to extend ${\mc J}_{g,n}(X)$, which is a bundle
over ${\mathcal M}_{g,n}$, to a bundle $\ol{\mc J}_{g,n}(X)$ over the
Deligne--Mumford compactification $\ol{\mathcal M}_{g,n}$, and then
extend the maps ${\mc J}\on{ev}_{i}$.

In $\ol{\mc M}_{g,n}$ the marked points $p_1,\ldots,p_n$ are always
smooth, and therefore the jet space $J_{p_i} X$ makes sense even if
$\Sigma$ becomes a stable singular curve. Hence the bundle ${\mc
J}_{g,n}(X)$ over ${\mathcal M}_{g,n}$ extends to a bundle over
$\ol{\mc M}_{g,n}$, which we denote by $\ol{\mc J}_{g,n}(X)$. Its
fiber over a stable pointed curve $(\Sigma,(p_i))$ is $J_{p_1} X
\times \ldots \times J_{p_n} X$ is again $(\Sigma,(p_i))$ is $J_{p_1}
X \times \ldots \times J_{p_n} X$. We now wish to extend the maps
\eqref{Jev DM} to maps
\begin{equation}    \label{olJ}
\ol{\mc J}\on{ev}_i: \ol{\mathcal M}_{g,n}(X,\beta) \to \ol{\mc
  J}_{g,n}(X).
\end{equation}
However, here we encounter a problem. Namely, the points
$(\Sigma,(p_i),\Phi)$ in $\ol{\mathcal M}_{g,n}(X,\beta)$ may well
correspond to unstable pointed curves $(\Sigma,(p_i))$. An example is
a curve $\wt\Sigma$ which has two components, $\Sigma_0$ and
$\Sigma'$, where $\Sigma_0$ is a genus zero component containing
exactly one marked point $p_i$. We denote by $\wt{p}_i$ the point of
intersection of the two components (see Figure 5). In fact, curves of
this type constitute one of the boundary divisors in $\ol{\mathcal
  M}_{g,n}(X,\beta)$.

\bigskip
\begin{center}
\epsfig{file=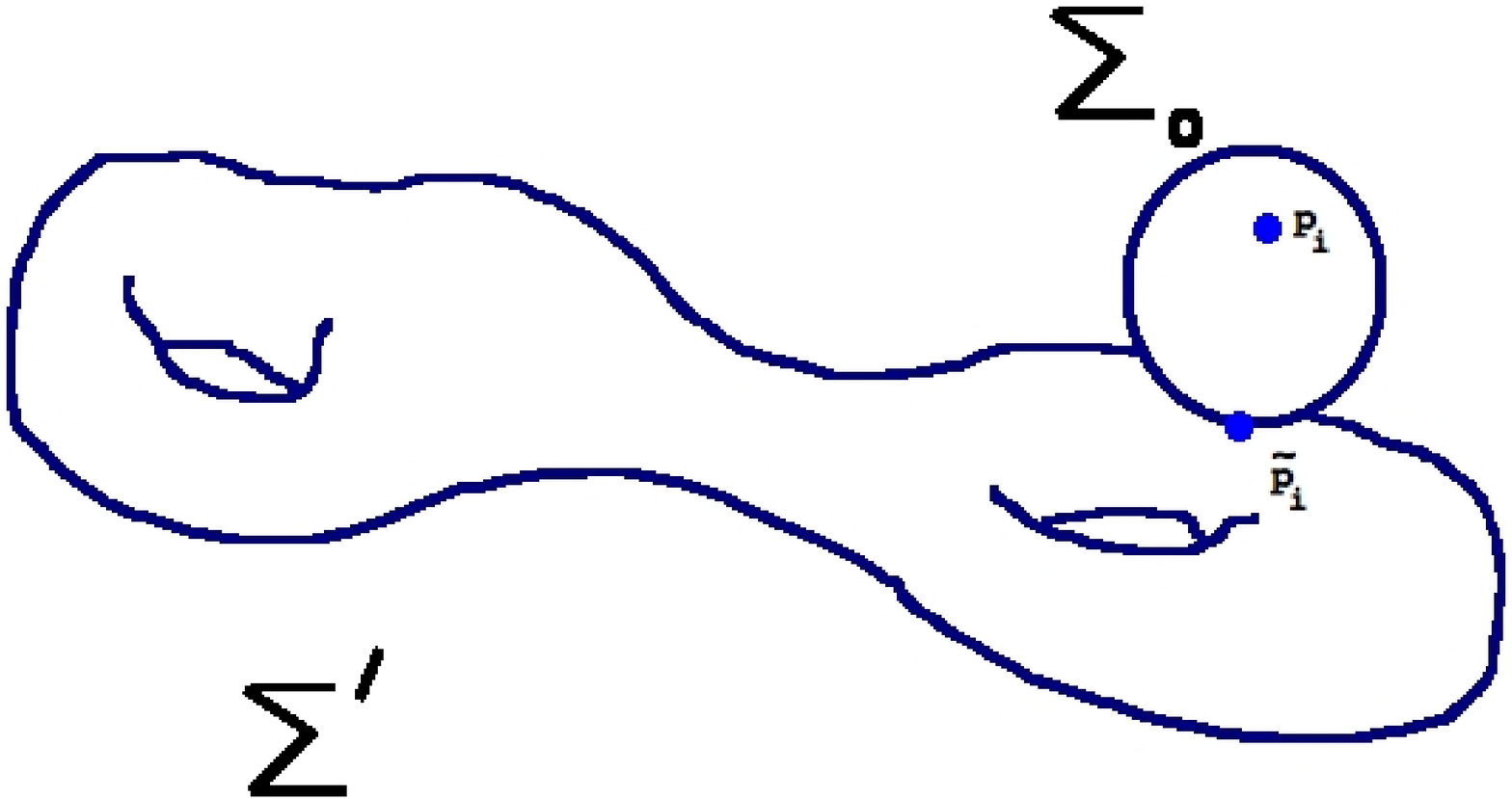,width=110mm}

\vspace*{5mm}

{\bf Figure 5. Stable curves corresponding to one of the divisors in
  $\ol{\mathcal M}_{g,n}(X,\beta)$.}
\end{center}

\vspace*{5mm}

The restriction of $\Phi$ to $\Sigma_0$ should have a non-zero degree
$\beta_0$; then the restriction of $\Phi$ to $\Sigma'$ has degree
$\beta' = \beta - \beta_0$. Now, the component $\Sigma_0$ is unstable,
and therefore it is collapsed under the map $\ol\pi_{g,n}$. In other
words, the image of $(\wt\Sigma,(p_i),\Phi)$ is
$(\Sigma',(p_1,\ldots,\wt{p}_i,\ldots,p_n)$, with the point $\wt{p}_i
\in \Sigma'$ replacing $p_i$.

Recall first the situation with the ordinary evaluation maps
$\on{ev}_i$. They take values in $X$, and hence they extend easily to
$\ol{\mathcal M}_{g,n}(X,\beta)$: in the case of a singular curve
$\wt\Sigma$ described above we just evaluate $\Phi$ at $p_i \in
\Sigma_0$. We do not care that $p_i$ ``disappears'' under the map
$\ol\pi_{g,n}$ and is replaced by $\wt{p}_i$ (where the value of the
map $\Phi$ would certainly be different in general), because we do not
use the Deligne-Mumford moduli space $\ol{\mathcal M}_{g,n}(X,\beta)$
in the definition of $\on{ev}_i$ (only in the definition of the
integrals \eqref{J GW inv1}). Since the evaluation maps extend to
$\ol{\mathcal M}_{g,n}(X,\beta)$ and the morphism $\ol\pi_{g,n}$ is
proper, we see that the corresponding push-forward
$$
\ol\pi_{g,n*}(\on{ev}_{1}^*(\omega_1) \wedge \ldots \wedge
\on{ev}^*_{n}(\omega_n))
$$
to $\ol{\mathcal M}_{g,n}$ is well-defined. Hence the integrals
\eqref{J fixed Sigma} are also well-defined.

The situation is different for the jet-evaluation maps. They take
values not in $X$, but in a jet bundle ${\mc J}_{g,n}(X)$ over
${\mathcal M}_{g,n}$. Note that we have a forgetful map $t: {\mc
J}_{g,n}(X) \to X^n \times {\mathcal M}_{g,n}$, which truncates the
jets of maps $D_{p_i} \to X$ to their values at $p_i$, and $\on{ev}_i
= \on{pr}_1 \circ t_i \circ {\mc J}\on{ev}_i$; this factorization is
the reason why the dependence on ${\mathcal M}_{g,n}$ disappears when
we consider the ordinary evaluation maps. The jet bundle ${\mc
J}_{g,n}(X)$ may be trivialized locally over ${\mathcal M}_{g,n}$ if
we choose a family of local coordinates at the points $p_i$ on
$\Sigma$. Indeed, by definition, the value of the map ${\mc
J}\on{ev}_i$ on $(\Sigma,(p_i),\Phi)$ for a smooth $\Sigma$ is the jet
of the restriction of $\Phi$ to the disc $D_{p_i}$. Once we pick a
local coordinate at $p_i$, we may view it as a point of $JX = \{ D \to
X \}$. Hence we may pull back differential forms on $JX$ to ${\mathcal
M}_{g,n}$.

Now, when $\Sigma$ degenerates to a singular curve described above, we
would like to consider the restriction of $\Phi$ to the disc
$D_{p_i}$, where the point $p_i$ is now on the unstable bubble
$\Sigma_0$. In order to be able to view it as a point of $JX$, we need
to choose a coordinate at $p_i$. But the image of $(\Sigma_0 \cup
\Sigma',(p_i),\Phi)$ in $\ol{\mathcal M}_{g,n}$ under $\ol\pi_{g,n}$
is $(\Sigma',(p_1,\ldots,\wt{p}_i,\ldots,p_n)$. It does not know
anything about $p_i$! The point $p_i$ has disappeared under the map
$\ol\pi_{g,n}$ and instead we now have the point $\wt{p}_i$ on the
other component $\Sigma'$. A local coordinate at $p_i$ has nothing to
do with a local coordinate at $\wt{p}_i$, and in any case the jet of
$\Phi$ at $p_i$ is completely independent of the jet at $\wt{p}_i$,
viewed as a point of $\Sigma'$.

Thus, we see that we cannot extend the map ${\mc J}\on{ev}_i$ to a map
\eqref{olJ}. Of course, we could try to use instead of $\ol{\mathcal
M}_{g,n}$ another moduli space (or stack), $\wt{\mathcal M}_{g,n}$, in
which the component $\Sigma_0$ and the point $p_i$ are included. For
instance, we could take as $\wt{\mathcal M}_{g,n}$ the moduli stack of
prestable curves. Then unstable bubbles such as $\Sigma_0$, with two
marked points, would be allowed. The problem is that this unstable
component $\Sigma_0$ has the group of automorphisms $\C^\times$. This
group naturally acts on the corresponding space $J_{p_i} X$ of jets of
maps $D_{p_i} \to X$, and as the result we are only able to identify
$J_{p_i} X$ with $JX$ up to this action. In other words, the
restriction of $\Phi$ to $D_{p_i}$ only gives us a well-defined
element $JX$ modulo the action of $\C^\times$. Thus, the only
observables that make sense in this case are the differential forms
that are invariant under this $\C^\times$-action. These are just the
evaluation observables. They may be pulled back to $\ol{\mathcal
M}_{g,n}(X,\beta)$ and we obtain the usual Gromov--Witten
invariants. But the pull-back of more general differential forms to
$\ol{\mathcal M}_{g,n}(X,\beta)$ is not well-defined.

This indicates that the differential forms ${\mc
J}\on{ev}_{i}^*(\wh\omega_i)$ on ${\mathcal M}_{g,n}(X,\beta)$ (which
are well-defined before the compactification) may have singularities
when we try to extend them to the boundary divisors in $\ol{\mathcal
M}_{g,n}(X,\beta)$. This is indeed the case for $g=0, n=3$, as we have
seen above. In general, the result will be similar. Thus, we see that
these differential forms have {\em poles} along the boundary divisors
in $\ol{\mathcal M}_{g,n}(X,\beta)$. (Their residues could potentially
have poles on the codimension two boundary strata in $\ol{\mathcal
M}_{g,n}(X,\beta)$, and so on.)

\medskip

Let us summarize: in order to define the correlation functions of
jet-evaluation observables, we need to regularize integrals of the
form \eqref{J GW inv1}. However, the standard prescription, which
works well for the ordinary evaluation observables, cannot be
applied. The reason is that the jet-evaluation maps ${\mc
J}\on{ev}_i$, well-defined in ${\mathcal M}_{g,n}(X,\beta)$ do not
extend to the moduli space $\ol{\mathcal M}_{g,n}(X,\beta)$ of stable
maps. Therefore we obtain integrals of differential forms on
$\ol{\mathcal M}_{g,n}(X,\beta)$ with poles on the compactification
divisor. To compute the correlation functions, we need to regularize
these integrals. This is directly related to the fact that the naive
definition of the operators corresponding to the jet-evaluation
observables (in terms of the $\beta\gamma$-$bc$-fields) requires
regularization, a phenomenon familiar to us from the study of quantum
mechanical models in Part I. This regularization is not unique and
leads to a mixing of operators (and states, via the state--operator
correspondence) with their logarithmic partners.

In other words, to obtain a true operator of the non-perturbative
sigma model, we need to take a jet-evaluation observable in its
perturbative definition {\em together} with a consistent set of
regularization rules for all of its correlation functions. Changing
the regularization rules will mean adding to this operator its
logarithmic partners.

Thus, divergence of the integrals expressing the correlation functions
of the jet-evaluation observables and potential ambiguity of their
regularization are important manifestations of the logarithmic nature
of the sigma models in the infinite radius limit. In Sections
\ref{reg}--\ref{OPE} we will present explicit examples of this
regularization, which indicate the existence of a rich and interesting
structure underlying the correlation functions of the jet-evaluation
observables in the two-dimensional sigma model.

\ssec{Regularization of correlation functions}    \label{reg}

The discussion of the previous two sections implies that the naive
three-point correlation functions of jet-evaluation observables on
$\pone$ {\em diverge} and require regularization. In this section we
explain how to implement this regularization and what it means for
the operators of the sigma model.

Consider, for example, the correlation function of the evaluation
observables $\omega_1 \psi \ol\psi$, \linebreak $\omega_2 \psi
\ol\psi$, placed at the points $0,1$, and the jet-evaluation
observable $\omega_3 \psi \ol\psi \pa X \ol\pa \ol{X}$ placed at
$\infty$. Using formula \eqref{divergence}, we obtain that the
corresponding correlation function is given by the integral
\begin{equation}
q \underset{{\mathbb{P}}^{1} \times {\mathbb{P}}^{1} \times
{\mathbb{P}}^{1}}\int \prod_{{\al}=1}^{3} {\om}_{\al} (X_{\al},
{\ol{X}}_{\al}) d^{2} X_{1} \wedge d^{2}X_{2} \wedge d^{2}X_{3} \cdot
\frac{\vert X_{1} - X_{3} \vert^{2} \vert X_{2} - X_{3} \vert^{2}}
{\vert X_{1} - X_{2} \vert^{2} }
\label{3pt}
\end{equation}
(it corresponds to the instanton number $1$, for dimensional reasons,
hence the overall factor $q$). This integral diverges when when $X_{1}
\to X_{2}$. Writing $$X_2 = X + \xi/2, \qquad X_1 = X - \xi/2,$$ we
obtain that the divergent part may be approximated by the integral
\begin{equation}    \label{res}
q \underset{{\mathbb{P}}^{1} \times {\mathbb{P}}^{1} \times
U}\int {\om}_{1} (X, {\ol{X}}) {\om}_{2} (X,
{\ol{X}}) {\om}_{3} (Y, \ol{Y}) d^{2} X \wedge
d^{2} Y \ {\vert X - Y \vert}^{4} \ {d^{2}\xi \over
|\xi|^{2}},
\end{equation}
where $U$ is a small neighborhood of $0$ in the $\xi$-plane. This
integral has logarithmic divergence in $\xi$.

We will now explain why this divergence is not really surprising. In
fact, we have observed and studied similar divergences in analogous
quantum mechanical models in Part I. In those models it was the
definition of the distributions corresponding to the states that
required regularization. For instance, in the supersymmetric quantum
mechanical model on $\pone$ we have states of the form $X^n \Xb^{\nb},
n \geq 0, \nb \geq 0$, where $X$ is a coordinate on $\pone$. The
matrix element of an evaluation observable $\omega(X,\Xb) d^2 X$,
which is a differential form on $\pone$, between this state and the
co-vacuum is given by the integral
\begin{equation}    \label{qm int}
\int \omega(X,\Xb) X^n \Xb^{\nb} d^2 X.
\end{equation}
These integrals diverge in general, and we regularize them by the
corresponding {\em partie finie} (also known as the {\em
Epstein--Glaser regularization} \cite{EG}). It is obtained by
integrating over the domain $|X| < \ep^{-1}$, viewing the integral as
a function in $\ep$ and picking the constant term in this
function. This regularization is best interpreted as a way to extend
the tempered distribution corresponding to $X^n \Xb^{\nb}$ to an
ordinary distribution on $\pone$.

However, the resulting distribution is not canonical. Indeed, our
prescription involved the choice of a coordinate on $\pone$. But such
a coordinate is only unique up to a scalar. So we could choose, say,
$2\ep$ as a small parameter, instead of $\ep$. Then the answer will
change, because of the presence of logarithmic dependence in $\ep$ in
the integral over the domain $|X| < \ep^{-1}$. We will therefore pick
up an extra term, which is a multiple of the derivative of the
delta-function at $\infty$, $\pa_X^{n-1} \pa_{\Xb}^{\nb-1}
\delta^{(2)}_\infty$. This is the {\em logarithmic partner} of the
state $X^n \Xb^{\nb}$.

In other words, the distribution corresponding to $X^n \Xb^{\nb}$ is
only well-defined up to an addition of a multiple of $\pa_X^{n-1}
\pa_{\Xb}^{\nb-1} \delta^{(2)}_\infty$. We obtain that the space of
``in'' states (more precisely, its subspace consisting of differential
forms of degree $0$) is a non-trivial extension of the space of
polynomials $\C[X,\Xb]$ by the space of delta-like distributions
$\C[\pa_X,\pa_{\Xb}] \cdot \delta^{(2)}_\infty$ at $\infty \in
\pone$. Furthermore, the Hamiltonian is not diagonalizable on this
space, but has Jordan block structure, mixing $X^n \Xb^{\nb}$ and
$\pa_X^{n-1} \pa_{\Xb}^{\nb-1} \delta^{(2)}_\infty$. (Note that there
is no mixing involving the vacuum state $1$, corresponding to
$n=\nb=0$, and to the purely chiral and anti-chiral states, with
$\nb=0$ and $n=0$, respectively.)

Our Morse function on $\pone$ has only two critical points, and hence
$\pone$ decomposes into the union of two strata: $\C_0 = \pone \bs
\infty$ and the point $\infty$. They give rise to the two sectors in
the space of ``in'' states. This is why states have at most one
logarithmic partner and the maximal size of the Jordan block appearing
in the Hamiltonian is two. For manifolds of higher dimension there
will be more sectors and hence more logarithmic partners in general
(see Part I for details).

\medskip

We are now observing a similar phenomenon in the two-dimensional sigma
model. Indeed, one possible Hamiltonian interpretation of \eqref{3pt}
is as the matrix element
\begin{equation}    \label{int ham}
\langle \omega_1(X,\Xb) \psi \ol\psi\ | \ \omega_2(X,\Xb) \psi
\ol\psi\ | \ \omega_3(X,\Xb) \psi \ol\psi \pa X \ol\pa \Xb \rangle
\end{equation}
of the evaluation observable $\omega_2 \psi \ol\psi$ between the
excited state corresponding to $$\omega_3(X,\Xb) \psi \ol\psi \pa X
\ol\pa \Xb$$ and the co-vacuum state corresponding to $\omega_1(X,\Xb)
d^2 X$.  Note that in contrast to one-dimensional quantum mechanical
models, in the two-dimensional theory we now have the state--operator
correspondence. We may then interpret the same integral as the matrix
element
\begin{equation}    \label{int ham1}
\langle \omega_2(X,\Xb) \psi \ol\psi\ | \ \omega_3(X,\Xb) \psi \ol\psi
\pa X \ol\pa \Xb\ | \ \omega_1(X,\Xb) \psi \ol\psi \rangle,
\end{equation}
in which differential forms $\omega_1$ and $\omega_2$ are interpreted
as vacuum and co-vacuum states, and $\omega_3(X,\Xb) \psi \ol\psi \pa
X \ol\pa \Xb d^{2}X$ is interpreted as an operator placed at the point
$1$. However, it is the interpretation \eqref{int ham} that is closest
to the quantum mechanical matrix element \eqref{qm int}.

In light of the above discussion of the quantum mechanical models, it
should not come as a surprise that the integral \eqref{int ham} has
logarithmic divergence. It is quite similar to the logarithmic
divergence of the integral \eqref{qm int}, which, as we have seen
above, is responsible for the appearance of logarithmic partners.

Hence we follow the same strategy as in quantum mechanics and {\em
define} the matrix element \eqref{int ham} (or the matrix element
\eqref{int ham1}) as the {\em partie finie} regularization of the
integral \eqref{int ham}. In other words, we cut out the part of the
domain of integration (which is the product of three copies of
${\mathbb P}^1$) which is within the radius $\ep$ of the diagonal
$X_1=X_2$ with respect to the Fubini-Study metric. (In
principle, here we could choose an arbitrary metric; the ambiguity of
this choice is one of the reasons that the resulting regularized
integral is not canonically defined.) We then evaluate the integral as
a function of $\ep$. One can show that as a function in $\ep$ it may
be uniquely represented in the form
\begin{equation}    \label{C log}
C_0 + \sum_{i>0} C_i \ep^{-i} + C_{\log} \log \ep + o(1),
\end{equation}
where the $C_i$'s and $C_{\log}$ are some numbers (see
\cite{Hoermander}, pp. 70-71). The {\em partie finie} of the above
integral as the constant coefficient $C_0$ obtained after discarding
the terms with negative powers of $\ep$ and $\log \ep$ in the integral
\eqref{int ham} and taking the limit $\ep \to 0$.

Just as in the quantum mechanical case, this definition is
non-canonical. It depends on the choices we have made (such as the
metric on $\pone$). However, formula \eqref{3pt} shows that the
discrepancy for two different regularization schemes (with varying
$\omega_1, \omega_2$) may be represented as a multiple of the integral
\begin{equation}    \label{another}
q \underset{{\mathbb P}^{1}}\int {\omega}_{1} (X, {\ol{X}})
{\om}_{2} (X, {\ol{X}}) \cdot {}^{\ell} \omega_3(X,\Xb) \psi \ol\psi,
\end{equation}
where, by definition,
\begin{equation}
{}^{\ell}\omega(X,\Xb) =
\int_{\mathbb{P}^{1}}\ | Y - X |^{4}\ \omega(Y,\ol{Y}) d^{2}Y.
\label{groth1}
\end{equation}
Note that $\omega_3$ is a coefficient in front of a
$(2,2)$-differential on $\mathbb{P}^{1}$, and the map \Ref{groth1} is
a canonical map from the space of $(2,2)$-differentials on
$\mathbb{P}^{1}$ to the space of $(-1,-1)$-differentials, so that the
integral \eqref{another} is well-defined. The integral kernel in
\secref{groth1} is the inverse square of the absolute value squared of
the so-called prime form.

We now interpret \eqref{another} as the following matrix element:
\begin{equation}    \label{free field}
q \langle\ \omega_{1}(X,\Xb) \psi \ol\psi\ | \ \omega_{2} (X,\Xb) \psi
\ol\psi\ | \ {}^{\ell}\omega_3 \pi \ol\pi\ \rangle \ .
\end{equation}
Here we have introduced the fermionic fields $\pi, \ol\pi$ in order to
reduce the total fermionic number (which counts the degree of the
differential form on the moduli space of holomorphic maps) from $3$ of
the original matrix element \eqref{int ham} to $1$. Hence we now
integrate over the moduli space of holomorphic maps from to ${\mathbb
P}^1$ of degree $0$ (that is, the constant maps), which is
one-dimensional.

\ssec{Logarithmic partners}    \label{log partners}

The upshot of this calculation is that

\bigskip

{\em the state ${}^{\ell}\omega_3(X,\Xb) \pi \ol\pi$ is the
logarithmic partner of $\omega_3(X,\Xb) \psi \ol\psi \pa X \ol\pa
\Xb$},

\bigskip

\noindent just like $\pa_X^{n-1} \pa_{\Xb}^{\nb-1}
\delta^{(2)}_\infty$ is the logarithmic partner of the state $X^n
\Xb^{\nb}$ in quantum mechanics!  Likewise, the operator
$^{\ell}\omega_3 \pi \ol\pi(z,\zb)$ is the logarithmic partner of the
operator $\omega_3 \psi \ol\psi \pa X \ol\pa \Xb(z,\zb)$.

Here it is useful to recall the structure of the space of states of
the two-dimensional sigma model on $\pone$ from
\secref{delta-forms}. The ``big'' space of states $\wt\CH$ has a
filtration $\wt\CH_{\geq i}, i \in \Z$, where $\wt\CH_{\geq i} \simeq
\wt\CH_{\geq j}$ for all $i,j$ and $\wt\CH_{\geq i}/\wt\CH_{\geq(i+1)}
\simeq \wt\CH_i$, the space of delta-forms on the ascending manifold
$(\wt{L{\mathbb P}^1})_i$ corresponding to the $i$th preimage of the
critical set ${\mathbb P}^1 \subset L{\mathbb P}^1$ of constant
loops. We have a ``deck transformation'' map sending $\wt\CH_{m}$ to
$\wt\CH_{m+1}$ isomorphically. A physical state $\Psi$ is a vector in
$\wt\CH$ which is an eigenvector of this transformation with the
eigenvalue $q$.

The space $\wt\CH$ may be identified with a direct product
\begin{equation}    \label{dir dec}
\wt\CH = \prod_{m \in \Z} \wt\CH_m,
\end{equation}
but not canonically. Let us choose such an identification, in such a
way that the deck transformation identifies all the $\wt\CH_m$ with
each other. Then it is convenient to identify all of them with $\wt{\mc
H}_0$ and to write
$$
\wt\CH = \wt\CH_0 \otimes \C[[T,T^{-1}]],
$$
where $T$ is a formal variable. A vector in this space is then
represented by a sum
$$
\Psi = \sum_{m \in \Z} \wt\Psi_m T^m, \qquad \wt\Psi_m \in \wt\CH_0
$$
(without any convergence condition!). The deck transformation acts by
multiplication by $T$. Hence it sends this vector to
$$
\sum_{m \in \Z} \wt\Psi_{m} T^{m+1}.
$$
The eigenvector condition on $\Psi$ then becomes
\begin{equation}    \label{tau eq3}
\wt\Psi_{m} = q \ \wt\Psi_{m+1}, \qquad m \in \Z.
\end{equation}
By this condition, $\wt\Psi_0$ determines the remaining $\wt\Psi_m$,
and so the space of states becomes isomorphic to $\wt\CH_0$. This is
in fact how we defined it in \secref{delta-forms}. However, this
definition is non-canonical because to obtain it we need to choose the
direct product decomposition \eqref{dir dec} of the ``big'' space of
states $\wt\CH$, which is non-canonical.

\subsubsection{Toy model}

It might be helpful to illustrate the difference between the space
$\wt\CH_0$ and the true space of states by the following elementary
example. Let $V$ be a vector space and suppose that we are given an
extension
\begin{equation}    \label{extension}
0 \to V \to V_2 \to V \to 0.
\end{equation}
Of course, any extension of vector spaces can be split, but if they
carry additional structures (such as OPE, correlation functions, etc.,
in our case), then there may not be a splitting respecting these
structures. We can iterate this extension and construct new
extensions
$$
0 \to V_n \to V_{n+1} \to V \to 0, \qquad n>2.
$$
The space $V_n$ has a filtration $0 \subset V = V_{1} \subset V_2
\subset \ldots \subset V_{n-1} \subset V_n = V$ such that the
consecutive quotients are all isomorphic to $V$. Let us label the
filtration on $V_{2n}$ as follows: $V_{\geq i} = V_{n-i},
i=-n,\ldots,n$. Next, we take the limit of $V_{2n}$ when $n \to
\infty$ in a way compatible with this filtration. This means that to
pass from $V_{2n}$ to $V_{2n+2}$ we ``glue'' one $V$ to $V_{2n}$ as a
subspace and another $V$ as a quotient. Then the limit $V_\infty$ has
a filtration $V_{\geq i}, i \in \Z$, such that $V_{\geq i}/V_{\geq
(i+1)} \simeq V$. (This is the analogue of $\wt\CH$.) Moreover, by
construction, we have a canonical identification of $V_{\geq i}$ with
$V_{\geq j}$ for all $i,j \in \Z$. Thus, we have a canonical ``shift''
operator $S$ which maps $V_{\geq i}$ to $V_{\geq (i+1)}$
isomorphically.

Now let $\ol{V}$ be the subspace of vectors $v \in V_\infty$ which are
eigenvectors of this transformation with eigenvalue $q \in \C^\times$,
$S(v) = qv$. (This is the analogue of our space of states.) If we
choose a splitting of the exact sequence \eqref{extension}, then we
identify $V_n \simeq V^{\oplus n}$ and hence identify $V_\infty$ with
the direct product of infinitely many copies of $V$, labeled by the
integers, splitting the filtration $(V_{\geq i})$. Then the space of
eigenvectors of $S$ as above may be identified with any of these
copies of $V$, for instance, the one labeled by $0 \in \Z$. Thus, we
obtain an isomorphism $V \simeq \ol{V}$. But this isomorphism is not
canonical! Indeed, if we choose a different splitting of
\eqref{extension}, that is, a different isomorphism $V_2 \simeq V
\oplus V$, then the lift to $V \oplus V$ of a vector from the quotient
$V$ will be equal to $v + M(v)$ for some linear operator $M: V \to
V$. Therefore a vector of $\ol{V}$ which appears as $v \in V$ under
the previous identification, will appear as the vector $\wt{v} = v + q
M(v) + q^2 M^2(v) + \ldots$ with respect to the new identification.

\subsubsection{Back to the space of states of the sigma model}

The fact that the space of states cannot be canonically identified
with $\wt\CH_0$ (which, we recall, is the chiral-anti-chiral de Rham
complex of $X$) has an important consequence. When we represent an
excited state of the theory in the form $\wt\Psi_0 = \omega \psi
\ol\psi \pa X \ol\pa \ol{X}$, say, tacitly choose a direct product
decomposition of the form \eqref{dir dec}. Indeed, this expression is
well-defined as a (semi-infinite) delta-form on the ascending manifold
$(\wt{L{\mathbb P}^1})_0$. But it is not intrinsically defined as a
state in the sigma model with the target ${\mathbb P}^1$, because of
there are non-trivial extensions of this space by the space of
delta-forms of $(\wt{L{\mathbb P}^1})_1$ and other ascending manifolds
in the closure of $(\wt{L{\mathbb P}^1})_0$.

This is similar to writing an excited state in the quantum mechanical
model on ${\mathbb P}^1$ as a monomial $X^n \Xb^{\nb}$ on the big cell
$\C_0$. It is really well-defined as a state in the model on $\C_0$,
but not in the model on ${\mathbb P}^1$, where it is actually mixed
with its logarithmic partner $\pa_X^{n-1} \pa_{\Xb}^{\nb-1}
\delta^{(2)}_\infty$. (Note, however, that the vacuum states, such as
the monomial $1$ in the quantum mechanical mode, or the states
$\omega(X,\Xb) \psi \ol\psi$ in sigma model, are intrinsically
defined, as they are not mixed with anything.) The analogue of this in
the two-dimensional sigma model is the statement that the general
states corresponding to delta-forms on a particular ascending manifold
of the Morse--Bott--Novikov--Floer function, are mixed with the states
corresponding to delta-forms on other ascending manifolds which
appear in its closure.

Therefore we should not be surprised that the state with
$$
\wt\Psi_0 = \omega \psi \ol\psi \pa X \ol\pa \ol{X} \in \wt\CH_0
\subset \wt\CH,
$$
which is a delta-form on the ascending manifold $(\wt{L{\mathbb
P}^1})_0$, is mixed with the state
$$
\wt\Psi'_1 \otimes T = {}^\ell
\omega \pi \ol\pi \otimes T \in \wt\CH_0 \otimes T \subset \wt\CH.
$$
This is now a delta-form on the ascending manifold $(\wt{L{\mathbb
P}^1})_1$, which lies in the closure of $(\wt{L{\mathbb P}^1})_0$. Now
we can use the equivariance condition \eqref{tau eq3} to interpret this
state as a state in $\wt\CH_0$; namely, we identify $\wt\Psi'_1$ with
$$
\wt\Psi'_0 = q \ {}^\ell \omega \pi \ol\pi \in \wt\CH_0.
$$
Note that the appearance of $q$ here, from the equivariance
condition \eqref{tau eq3}, matches its appearance in the definition of
logarithmic partners: in formula \eqref{free field} we have to
introduce $q$ by hand, because we express the residue of a divergent
integral over the moduli space of holomorphic maps of degree $1$
(corresponding to the matrix element \eqref{int ham}) in terms of an
integral over the moduli space of holomorphic maps of degree $0$.

In this example we have only one logarithmic partner. More generally,
there may be more of them, with higher powers of $q$ (this is similar
to what happens in quantum mechanics).

\subsubsection{Strata in $\wt{L{\mathbb P}^1}$ and boundary divisors
  in $\ol{\mc M}_{\Sigma}(X,\beta)$}

We note that the structure of the ascending manifolds in
$\wt{L{\mathbb P}^1}$ exhibits an analogue of the ``bubbling
phenomenon'' discussed at the end of \secref{genus zero ex}, which was
ultimately responsible for the singularity of the integrals such as
\eqref{3pt}. To see that, recall the interpretation of the covering
$\wt{L{\mathbb P}^1}$ of $L{\mathbb P}^1$ from \secref{as qm} in which
points of $\wt{L{\mathbb P}^1}$ are realized as equivalence classes of
(continuous) maps $\wt\ga: D \to X$, where $D$ is a unit disc. In this
interpretation the ascending manifold $(\wt{L{\mathbb P}^1})_0$ is
realized as the subset of equivalence classes of {\em holomorphic
maps} $\wt\ga: D \to X$, whereas the ascending manifold in its
closure, $(\wt{L{\mathbb P}^1})_1$, may be realized as the set of
equivalence classes of holomorphic maps in which a sphere ``bubbles
out'' at the origin of the disc (see Figure 6). This is in agreement
with the description of the boundary strata in $\ol{\mc
M}_{\Sigma}(X,\beta)$ as the maps from a curve $\Sigma$ in which a
sphere ``bubbles out'' at the point $p$.

\bigskip
\begin{center}
\epsfig{file=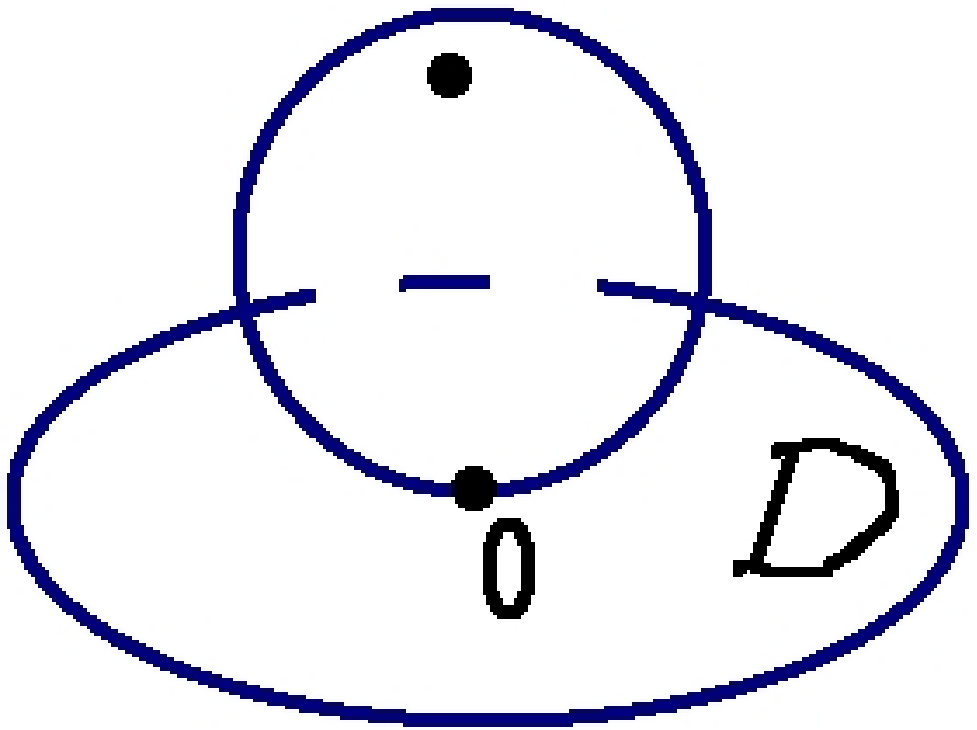,width=70mm}

\vspace*{10mm}
{\bf Figure 6. Disc with a ``bubble''.}
\end{center}

\vspace*{5mm}

The two pictures, the local one with $\wt{L{\mathbb P}^1}$, and the
global one with $\ol{\mc M}_{\Sigma}(X,\beta)$, are actually closely
related. Indeed, let us cut a small disc $D$ around the point $p$ from
$\Sigma$. Let $\ol\Sigma$ be the closure of $\Sigma \bs D$. Then $D$ and
$\ol\Sigma$ intersect along their boundary circles and general
holomorphic maps $\Sigma \to X$ may be described as loops $\BS^1 \to
X$ which are simultaneous the boundary values of holomorphic maps $D
\to X$ and $\ol\Sigma \to X$ (see Figure 7). Thus, we may identify the
moduli space ${\mc M}_{\Sigma}(X,\beta)$ of holomorphic maps $\Sigma
\to X$ (before the compactification) with the intersection of the
space of boundary values of holomorphic maps $D \to X$ and the space
of boundary values of holomorphic maps $\ol\Sigma \to X$, viewed as
subsets in $L{\mathbb P}^1$ (see \cite{GK}).

\bigskip
\begin{center}
\epsfig{file=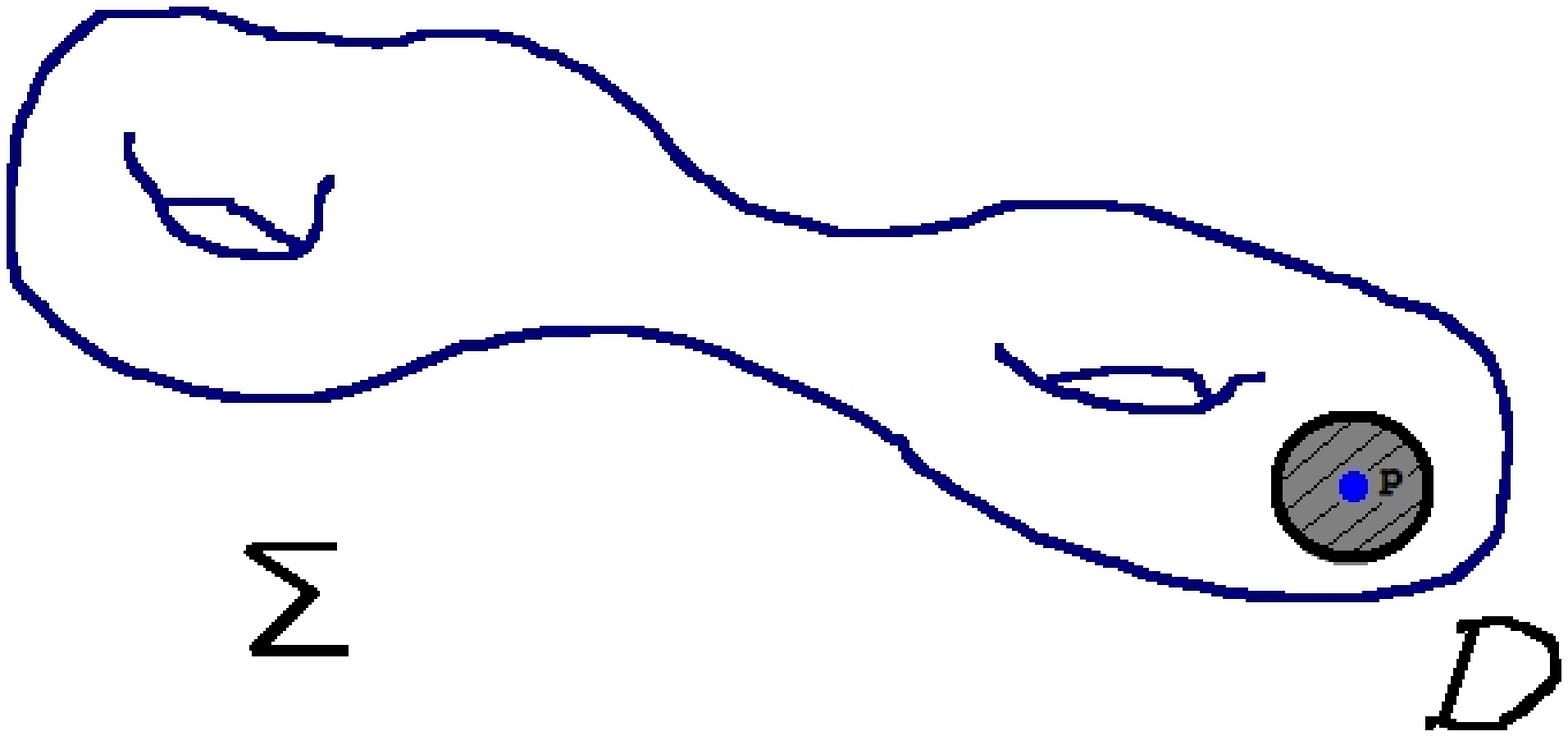,width=120mm}

\vspace*{5mm}
{\bf Figure 7. The curve $\Sigma$ as the union of a disc $D$ and
its complement $\ol\Sigma$.}
\end{center}

\bigskip

These subsets isomorphically lift to the covering $\wt{L{\mathbb
P}^1}$, with the first subset lifting to $(\wt{L{\mathbb
P}^1})_0$. The divisors in the stable map compactification $\ol{\mc
M}_{\Sigma}(X,\beta)$ of ${\mc M}_{\Sigma}(X,\beta)$ then correspond
to the intersection of the ascending manifolds $(\wt{L{\mathbb
P}^1})_n, 0 < n \leq \beta$, which lie in the closure of
$(\wt{L{\mathbb P}^1})_0$, with the space of boundary values of
holomorphic maps $\ol\Sigma \to X$. This is why we have such
parallelism between the patterns of gluing the boundary strata in
$\wt{L{\mathbb P}^1}$ and in $\ol{\mc M}_{\Sigma}(X,\beta)$.

This way the loci of singularities of the correlation functions of the
jet-evaluation observables may be directly linked to the structure of
the closures of the ascending manifolds in $\wt{L{\mathbb P}^1}$ and
hence to the non-trivial extensions in the space of operators of the
sigma model which, as we have seen above, ultimately lead to the
appearance of logarithmic partners.

\begin{remark}

Note that there is a difference between the two patterns: in the local
picture, on $\wt{L\pone}$, we do not fix the map of the bubbled sphere
to $\pone$, only its degree, whereas in the global picture, on
$\ol{\mc M}_{\Sigma}(X,\beta)$, we do fix it, up to rescalings. For
this reason the boundary strata in $\ol{\mc M}_{\Sigma}(X,\beta)$ have
codimension one, whereas the corresponding codimension of boundary
strata in $\ol{\mc M}_{\Sigma}(X,\beta)$ grows linearly with the
degree of the map from the bubbled sphere. Note also that in $\ol{\mc
M}_{\Sigma}(X,\beta)$ there are other boundary divisors corresponding
to spheres that bubble out away from the point $p$. They do not have
analogues in $\wt{L\pone}$.\footnote{We thank A. Givental for a
useful discussion of these issues.}

\end{remark}

\ssec{General case}

We now investigate the general phenomenon of logarithmic mixing of
operators. In the formulas below we indicate the passage from a given
operator to its logarithmic partner by an arrow. The first example
considered above is
$$
A(X, \ol{X})\psi\pa X \ol{\psi\pa X}
\longrightarrow q B_{A}(X, \ol{X}) {\pi}\ol{\pi},
$$
where
$$
B_{A} (X) = \int_{\pone} A (Y) |X - Y|^{4} d^{2}Y.
$$
Similarly, we obtain
$$
A(X, \ol{X})\psi\pa\psi \ol{\psi\pa\psi} \longrightarrow q B_{A}(X,
\ol{X}) p\ol{p},
$$

Consider next the case of purely bosonic operators. Suppose we have
two partitions $\la$ and $\mu$ with $|{\la}| = |\mu| = 3$. Then, using
formula \eqref{dlzzx}, we obtain
\begin{equation}    \label{atob}
C_{{\la}\mu}(X, \ol{X}) \dlx{\la}\odl{\mu} \longrightarrow
q\cdot  B_{C_{\la\mu}}{\pi}{\pa}{\pi} \ol{\pi}\ol{\pa}\ol{\pi},
\end{equation}
where the fermionic content of the log-partner is uniquely fixed by
the dimension and the ghost number (the ${\pi}{\psi}$ charge)
considerations, and
\begin{equation}
B_{C_{\la\mu}} (X) = 4 \int_{\pone} d^{2}Y \, 
C_{\la\mu}(Y)
|X- Y|^{2} ( X - Y)^{{\ell}({\la})} (\ol{X} - \ol{Y}
)^{{\ell}({\mu})}\ .
\label{logbc}
\end{equation}

The more general operator mixing for jet-evaluation observables may be
described similarly. Suppose that we wish to determine whether an
operator $A$, of perturbative conformal dimension $({\Delta},
\ol{\Delta})$ is log-mixed with operators $^{\ell}A^{(r)}$, so that
\begin{equation}
Q^{L_{0}}\ol{Q}^{\ol{L}_{0}}\left( A \right) = Q^{\Delta}
\ol{Q}^{\ol{\Delta}}\cdot \left( A + {\rm log} 
|Q|^{2} \left[ ^{\ell}A^{(1)} \right] + \ldots + {1\over k!}
\left( {\rm log} 
|Q|^{2} \right)^{k} \left[ ^{\ell}A^{(k)} \right]
  \right),
\label{qab}
\end{equation}
where we allow for multiple log-partners. Then for a test operator $B$
with the same perturbative conformal dimension we should have
\begin{equation}    \label{bootstrap}
\langle A(z,\zb) B(0) \rangle = z^{-2\Delta} \zb^{-2\ol\Delta}\left(
G_{AB} + \sum_{r=1}^k
\frac{(\log|z|^2)^{r}}{r!} G_{^{\ell}A^{(r)},B} \right),
\end{equation}
where $G_{{\tilde A}B}$ is the perturbative Zamolodchikov
metric. Therefore we could try to find $^{\ell}A^{(r)}$ by computing
the two-point functions $\langle A(z,\zb) B(0) \rangle$ and
interpreting the logarithmic terms as the two-point functions $\langle
{}^{\ell}A^{(r)}(z,\zb) B(0) \rangle$. As before, all of these
correlation functions may be computed in terms of (regularized)
integrals over the moduli spaces of holomorphic maps. This leads to a
kind of bootstrap, which in principle should enable us to compute the
logarithmic partners by a recursive procedure. The first term in this
recursion corresponds to taking the perturbative part in $\langle
{}^{\ell}A^{(r)}(z,\zb) B(0) \rangle$. Let us discuss it in more
detail in the case when $r=1$. Let us write
$$
{}^{\ell}A^{(1)} = \sum_{m>0} {}^{\ell}A^{(1)}_m q^m.
$$

It is useful to pass to the logarithmic coordinate $x = {\rm log}X,
{\psi}^{x} = {\psi}^{X} /X $ and write
\begin{align}
& x = x_{0} + \sum_{i=1}^{m} {\rm log} \left( 
\frac{1 - w_{i}^{+} z}{1-w_{i}^{-}z} \right), \cr
& {\psi} = dx_{0} + \sum_{i=1}^{m} \left( 
-\frac{dw_{i}^{+} z}{1 - w_{i}^{+} z} + 
\frac{dw_{i}^{-} z}{1 - w_{i}^{-} z} \right)
\label{instbeta}
\end{align}
in the sector with the instanton number $m$.  If two operators $A$ and
$B$ are jet-evaluation observables, that is, depend only on $x,
\psi$ and their derivatives, then the correlation function $\langle
A({\infty}) B(0) \rangle$ is given by the sum
$$
\sum_{m \in \Z_+} q^m \underset{{\mc
    M}_{\pone}(\pone,m)}\int A \wedge B,
$$
where ${\mc M}_{\pone}(\pone,m)$ is the moduli space of degree $m$
holomorphic maps from a paramet\-rized worldsheet $\pone$ to the
target $\pone$. Let us assume for simplicity that if $B(0)$ is not a
jet-evaluation observable, then $\langle A({\infty}) B(0) \rangle =
0$, so we may restrict ourselves to observables of this type.

Now, the moduli space ${\mc M}_{\pone}(\pone,m)$ is acted upon by the
group ${\C}^{\times}$ which preserves the points $0$ and $\infty$, and
the correlation function is non-trivial only if the dimensions of $A$
and $B$ coincide.  Therefore the integrand will be
${\C}^{\times}$-invariant, hence the integral will be divergent, due
to the volume of ${\C}^{\times}$. We interpret this divergence as the
$\log|z|^2$ term in formula \eqref{bootstrap}. Hence we can identify
the prefactor with the two-point function of $^{\ell}A^{(1)}_m$ and
$B$. Thus, the correlator of the log-partner $^{\ell}A^{(1)}$ of $A$
and $B$ should be equal to the sum of the integrals of the
differential forms corresponding to $A$ and $B$ over the quotients of
${\mc M}_{\pone}(\pone,m), m>0$, by $\C^\times$.

The quotient ${\mc M}_{\pone}(\pone,m)/\C^\times$ may be compactified
to the moduli space of stable maps with two marked points,
$\ol{{\CM}_{0,2}({\pone},m)}$, which is mapped by the evaluation maps
${\rm ev}_{0} \times {\rm ev}_{\infty}$ to ${\pone} \times
{\pone}$. Let us denote the pre-image of the point $(e^{x},e^{y}) \in
\pone \times \pone$ by ${\CM}(x, y;m)$. This space is a particular
compactification of the space of $m$-tuples of pairs $(w_{i}^{+},
w_{i}^{-})$, $ i = 1, \ldots , {m}$, which obey:
\begin{equation}
\prod_{i=1}^{m} \frac{w_{i}^{+}}{w_{i}^{-}} = e^{x-y}
\label{mbxy}
\end{equation}
modulo the ${\C}^{\times}$-action: 
\begin{equation}
( w_{i}^{+}, w_{i}^{-} ) \mapsto ( tw_{i}^{+}, t w_{i}^{-} ) \ , t \in
{\C}^{\times}
\label{tac}
\end{equation}
Then we compute the integral over this moduli space (using formula
\eqref{dlzzx}) and interpret the result as the perturbative
correlation function of $^\ell A^{(1)}_m$ and $B$. As the result, we
obtain the following expression for $^{\ell}A^{(1)}_m$:
\begin{equation}
^{\ell}A^{(1)}_m(x) = \sum_{B} B^\vee(x)
\int_{\pone} A(y) \int_{{\CM}(x,y;{m})} {\Omega}_{x,y} (A, B; {m})
\wedge {\ol{\Omega}_{\ol{x},\ol{y}}} (A, B;{m}),
\label{lala}
\end{equation}
where $\Omega_{x,y} (A, B; {m})$ is a meromorphic top degree form on
${\CM}(x,y;{m})$ constructed out of $A$ and $B$. We sum over the space
of all local operators using some basis $B$ and the dual basis
$B^\vee$ with respect to the perturbative Zamolodchikov metric.

For example, if (in the notation of \secref{jet-ev})
\begin{align*}
A &= A_{\la\mu}(x,\ol{x}) \dlx{\la}\odl{\mu}, \\ B &=
B_{\la'\mu'; \nu,\ka}(x, \ol{x})\dlx{\la'}\odl{\mu'}\,
{\psi}{\pa}^{{\nu}_{1}-1}{\psi}{\pa}^{{\nu}_{2}-2}{\psi}\ldots
{\ol{\psi}}{\ol{\pa}}^{{\ka}_{1}-1}\ol{\psi}\ldots\ ,
\end{align*}
then
\begin{align}
& {\Omega}_{x,y}(A, B; {m}) = \iota_{E} \frac{d^{m}w^{+} \wedge
d^{m} w^{-}}{d \, {\rm log} \left( \prod_{i} w_{i}^{+}/w_{i}^{-}
\right)} \cr & \prod_{k} \left[ \sum_{i} (w_{i}^{+})^{{\la}_{k}} -
(w_{i}^{-})^{{\la}_{k}} \right] \left[ \sum_{i}
(w_{i}^{+})^{-{\la}_{k}'} - (w_{i}^{-})^{-{\la}_{k}'} \right]\cr &
\qquad\qquad \times\, {\rm Det}
\Vert \, (w_{i}^{+})^{-{\nu}_{k} + k} \ (w_{i}^{-})^{-{\nu}_{k} + k}\,
\Vert_{1\leq i \leq m\, 1 \leq k \leq 2m},
\label{omgabb}
\end{align}
where $\iota_E$ is the contraction with the Euler vector field $E =
w^{+}_i{\pa}_{w^{+}_i} + w^{-}_i{\pa}_{w^{-}_i}$. 

Formula \Ref{omgabb} can be also derived using the free field
realization with the help of the holomortex operators ${\Psi}^{\pm}$
discussed in \secref{target pone}, along the lines of the OPE
calculation in \secref{OPE from hol}.

The above integrals may further diverge. Their regularization, in
turn, will give rise to terms with higher power of logarithms in
$\langle A(z,\zb) B(0) \rangle$, which we can use to recursively
compute $A^{(r)}_m$ with $r>1$. We will discuss this in more detail in
the follow-up paper \cite{FLN:new}.

\ssec{Operator product expansion} \label{OPE}

An important feature of quantum field theory is the operator product
expansion (OPE) which gives rise to an algebraic structure on the
space of fields. The OPE may be described in especially nice terms in
two-dimensional conformal field theories (CFT), where the expansion
may be written in terms of the rational functions of the form $(z-w)^n
(\zb-\wb)^{\nb}$. In logarithmic CFT the expansion also involves terms
with the logarithms $\log(z-w), \log(\zb-\wb)$. Our analysis of the
two-dimensional sigma models in the infinite radius limit shows that
they are logarithmic CFTs. Therefore it is natural to ask whether the
OPE in these models may be computed and logarithmic terms be observed
explicitly. In this section we will address this question in the case
of the target manifold $\pone$ (this is the simplest non-trivial
case). We will show that there are instanton corrections to the OPE
which do involve logarithmic terms. However, these corrections only
appear when we consider observables involving both chiral and
anti-chiral fields. The chiral algebra of the model (which we have
determined to be the global chiral de Rham complex in \secref{chiral})
is free of logarithms.\footnote{This is similar to the fact that in
quantum mechanical models considered in Part I the subspaces of purely
chiral and anti-chiral states have bases (such as the monomial bases
$X^n, n\geq 0$, and $\Xb^{\nb}, \nb\geq 0$, in the $\pone$ model) of
true eigenvectors of the Hamiltonians; in other words, there are no
Jordan blocks in the Hamiltonian on the chiral and anti-chiral
states.}

\subsubsection{Instanton corrections to OPE: factorization approach}

Our goal is to understand the instanton corrections to the OPE in the
sigma model on $\pone$ in the infinite radius limit. We will compute
the OPE by studying the factorization of the four-point correlation
functions in a particular channel. 

We start with the correlation function of dimension zero observables,
inserted at the points $z_{1}, z_{2}, z_{3}, z_{4}$ on the worldsheet
$\pone$, and consider the channel where $z_1 \to z_2, z_3 \to z_4$
(see Figure 8).

\bigskip
\begin{center}
\epsfig{file=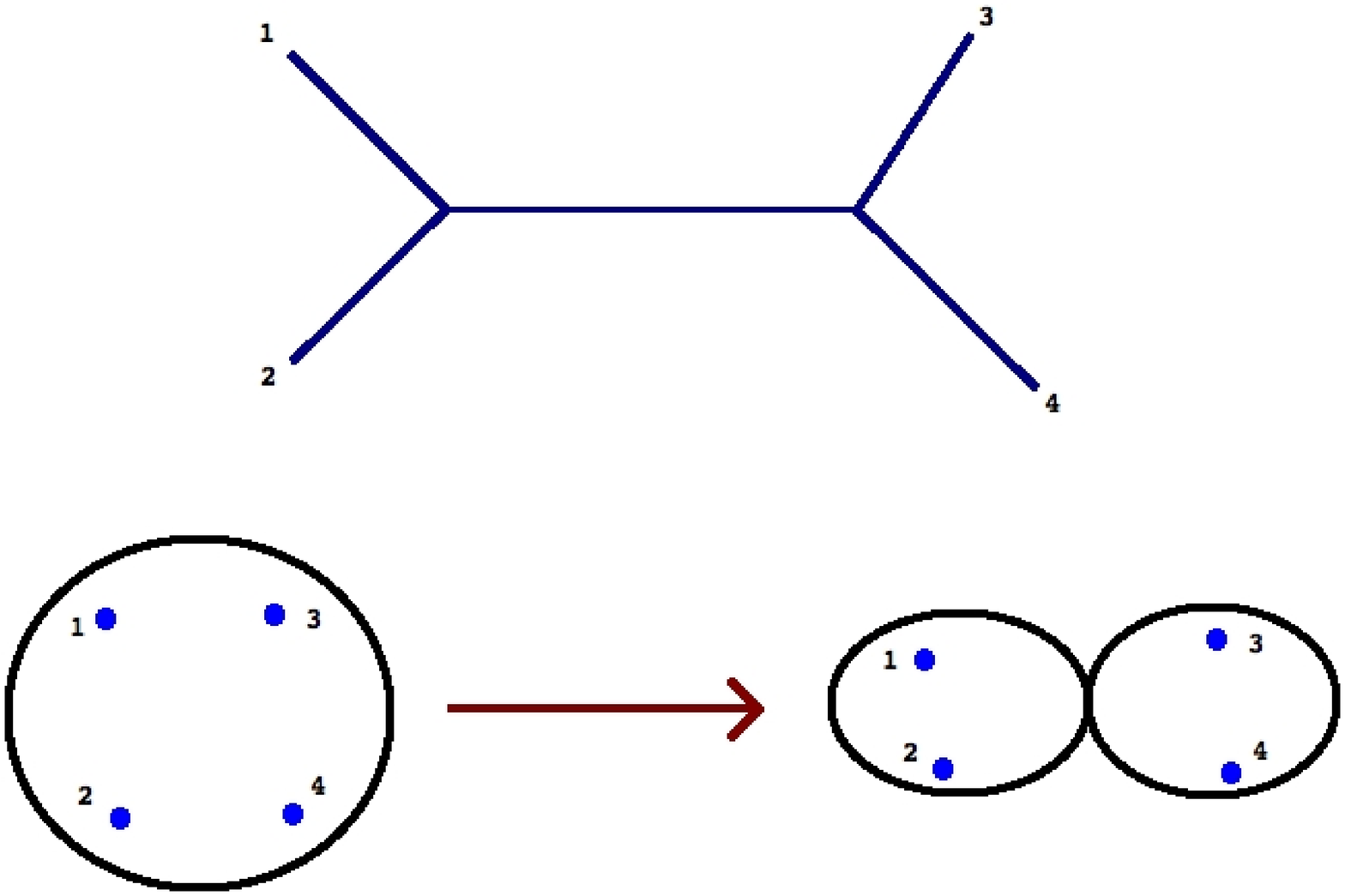,width=140mm}

\vspace*{5mm}
{\bf Figure 8. The degeneration of $\pone$ with four points that we
  consider.}
\end{center}

\bigskip

In principle, one could also try to do this by analyzing the
three-point functions in the limit when $z_1 \to z_2$. However, the
$PGL_2$-invariance of the correlation functions of dimension zero
operators eliminates any free parameters in the three-point functions,
and therefore it is difficult to analyze the corresponding limit. In
the case of four-point functions we have one free parameter which we
can use to study the limit.

Let ${\om}_{i} \in {\Omega}^{2} ({\mathbb{P}}^{1})$, $i=1,2,3$, be
three smooth two-forms on $\pone$, and $f$ a smooth function on
${\mathbb{P}}^{1})$. Consider the four-point function
\begin{equation}    \label{4pt}
\Biggl\langle {\mathcal{O}}_{{\om}_{1}}(z_{1})
{\mathcal{O}}_{{\om}_{2}}(z_{2}) {\mathcal{O}}_{{\om}_{3}}(z_{3})
{\mathcal{O}}_{f}(z_{4}) \Biggr\rangle.
\end{equation}
where $\mathcal{O}_{\om}(z) = {\om}(X(z), \ol{X}(\ol{z}))
{\psi}\ol{\psi} (z , \ol{z})$, and $\mathcal{O}_{f}(z) = f(X(z),
\ol{X}(\ol{z}))$.  The complex dimension of the moduli space of
holomorphic maps $\Phi$ from ${\mathbb P}^1$ (with at least three
marked points) to ${\mathbb P}^1$ of degree $d$ is $2d+1$. Therefore
the only non-zero contribution to this correlation function will come
from the component with $d=1$, which is isomorphic to $PGL_2$ and
which we compactify to $(\pone)^3$ by considering the values of $\Phi$
at $z_1,z_2,z_3$, denoted by $X_1,X_2,X_3$ (see \secref{genus zero
ex})\footnote{Recall from \secref{genus zero ex} that the stable map
compactification $\ol{\CM_{0,3}({\pone},1)}$ of this moduli space is
the blow-up of $(\pone)^{3}$ along the principal diagonal
$X_1=X_2=X_3$; thus, it differs from $(\pone)^{3}$ by a measure zero
subset.}. The value at $z_4$ is determined by these, and is equal to
\begin{equation}
\Phi(z_4) = {{\la} X_{1} X_{23} + X_{3} X_{12} \over X_{12} + {\la}
  X_{23} } = X_{3} + \frac{{\la}X_{23}/X_{12}}{1+{\la}
  X_{23}/X_{12}}X_{13} ,
\end{equation}
where
\begin{equation}    \label{double ratio}
{\la} = {z_{43}z_{21} \over z_{41} z_{23}}.
\end{equation}
Here and below we use the notation $z_{ij} = z_i-z_j$.

The correlation function is therefore equal to
\begin{equation}
q \int_{\pone \times \pone \times
\pone} \prod_{{\al}=1}^{3} {\om}_{\al} (X_{\al},
{\Xb}_{\al}) d^{2} X_{1} \wedge d^{2}X_{2} \wedge d^{2}X_{3} \ f
\left( {{\la} X_{1} X_{23} + X_{3} X_{12} \over X_{12} + {\la} X_{23}
} \right),
\label{fourpoint}
\end{equation}
where $q=e^{-\tau}$ is the instanton parameter. This integral
converges for smooth $f$ and ${\om}_{\al} = {\om}_{\al} ( X , {\ol{X}}
) d^{2}X$.

We now want to study the asymptotics of \Ref{fourpoint} in the limit
$z_{2} \to z_{1}$, $z_{3} \to z_{4}$. In an ordinary (non-logarithmic)
CFT we expect to have the following OPE (recall that conformal
dimensions of the evaluation observables are equal to zero):
\begin{align}    \label{opea}
\mathcal{O}_{{\om}_{1}}(z_1,\ol{z}_{1}) \
\mathcal{O}_{{\om}_{2}}(z_2,\ol{z}_{2}) &= \sum_A
A(z_1,\zb_1) z_{12}^{\Delta_A} \zb_{12}^{\ol\Delta_A}, \\
\label{opeb} \mathcal{O}_{{\om}_{3}}(z_3,\ol{z}_{3}) \
\mathcal{O}_{{f}}(z_4,{\ol{z}}_{4}) &= \sum_B B(z_4,\zb_4) z_{34}^{\Delta_B}
\zb_{34}^{\ol\Delta_B},
\end{align}
where $A(z_1,\zb_1)$ denotes a field of conformal dimensions
$(\Delta_A,\ol\Delta_A)$, and similarly for $B(z_4,\zb_4)$. (Recall
that the conformal dimension of ${\mc O}_\omega$ is $(0,0)$.)

In the ordinary CFT we would have the following
expansion of \eqref{fourpoint}:
\begin{multline}
\Biggl\langle {\mathcal{O}}_{{\om}_{1}}(z_{1})
{\mathcal{O}}_{{\om}_{2}}(z_{2}) {\mathcal{O}}_{{\om}_{3}}(z_{3})
{\mathcal{O}}_{f}(z_{4}) \Biggr\rangle \to \sum_{A, B}
z_{12}^{{\Delta}_{A}}{\ol{z}}_{12}^{{\ol{\Delta}}_{A}}
z_{34}^{{\Delta}_{B}}{\ol{z}}_{34}^{{\ol{\Delta}}_{B}}
\Biggl\langle {\mathcal{O}}_{A}(z_{1})
{\mathcal{O}}_{B}(z_{4}) \Biggr\rangle = \\ \sum_{A, B} \left(
{z_{12} \over z_{41} } \right)^{{\Delta}_{A}} \left( {
\ol{z}_{12} \over \ol{z}_{41} } \right)^{{\ol{\Delta}}_{A}}
\left( {z_{34} \over z_{41}} \right)^{{\Delta}_{B}} \left(
{{\ol{z}}_{34} \over \ol{z}_{41} } \right)^{{\ol{\Delta}}_{B}}
G_{AB}.
\label{opes}
\end{multline}
Here $G_{AB}$ is the Zamolodchikov metric, and $G_{AB} \neq 0$ only
for ${\Delta}_{A}= {\Delta}_{B}, \ol{\Delta}_{A} = \ol{\Delta}_{B}$.

Let us write
\begin{equation}
{\la} = {z_{43} z_{21} \over z_{41} ( - z_{41} + z_{43} + z_{21} )} =
- \sum_{a,b=0}^{\infty} {(a+b)! \over a! b!} \left( { z_{43} \over
z_{41}} \right)^{a+1} \left( {z_{21} \over z_{41} } \right)^{b+1}.
\label{dr}
\end{equation}
If we formally expand \Ref{fourpoint} as a power series in $\la$ near
$\la = 0$, the coefficients should give us the OPE \eqref{opea},
\eqref{opeb}.

\subsubsection{Instanton corrections to the chiral part of the
  operator product expansion}

The lowest order term in $\la$ is the term ${\la}^{0}$. It is equal to
the constant
\begin{equation}
q \int_{\pone} {\om}_{1} \int_{\pone} {\om}_{2}
\int_{\pone} f {\om}_{3},
\label{zeroo}
\end{equation}
which corresponds to 
$$
A(z_{1}, {\ol{z}}_{1}) = \left( \int_{\pone}
{\om}_{1} \int_{\pone} {\om}_{2} \right) {\bf 1}\ , \qquad
{\Delta}_{A} = 0,
$$
$$
B(z_{4}, \ol{z}_{4} ) = f ( X(z_{4}) ,
{\ol{X}}({\ol{z}}_{4})) {\om}_{3} ( X(z_{4}) , {\ol{X}}({\ol{z}}_{4}))
{\psi}(z_{4}) {\ol{\psi}}({\ol{z}_{4}}) \ , \qquad {\Delta}_{B} = 0,
$$
in \eqref{opea}, \eqref{opeb}. This is in fact the term responsible
for the quantum cohomology of $\pone$.

The terms in the OPE expansion of $\mathcal{O}_{{\om}_{1}}$ and
$\mathcal{O}_{{\om}_{2}}$ which are analytic in $\la$ near $\la=0$
come from the terms of dimension $(k,0)$ which appear in front of
${\la}^{k}$. By expanding the integrand in \Ref{fourpoint} in ${\la}$,
we obtain the following terms proportional to ${\la}^{k}$:
\begin{align}
& q\, {\la}^{k} \sum_{l}C^{(k,l)}\ \int_{\pone \times \pone \times
\pone} {\om}_{1}( X_{1}, \ol{X}_{1}) {\om}_{2} (X_{2}, \ol{X}_{2})
d^{2}X_{1} \wedge d^{2}X_{2} \cr & \qquad\qquad\qquad \times \, \left(
\frac{X_{2} - X_{3}}{X_{1} - X_{2}} \right)^{k} (X_{3} - X_{1} )^{l}\,
{\pa}_{X_{3}}^{l}f(X_{3}, \ol{X}_{3}){\om}_{3} ( X_{3}, \ol{X}_{3} )
d^{2}X_{3},
\label{lkterms}
\end{align}
where $$C_{f}^{(k,l)} = {(-1)^{k-l}(k-1)!\over l!(l-1)!(k-l)!}.$$ The
integrals in \Ref{lkterms} are conditionally convergent, for $k >
1$. We shall assume that the prescription of the integration over the
angles first is employed.  The $(k,l)$ term in \Ref{lkterms} can be
interpreted in the following way. Perturbatively, the OPE
$\mathcal{O}_{f}(z,\zb)\mathcal{O}_{\om}(0)$ we has the following
terms analytic in $z$:
\begin{align}
& \mathcal{O}_{f}(z,\zb)\mathcal{O}_{\om}(0) \sim \sum_{\mu}
z^{|{\mu}|} \mathcal{O}_{{\om}f^{({\mu},0)}}(0)\cr
& \mathcal{O}_{{\om}f^{({\mu},0)}} = {\om}(X, \Xb) \dlx{\mu} \
{\pa}^{{\ell}({\mu})}_X f (X, \Xb) 
{\psi}\ol{\psi}
\label{pertope}
\end{align}
(in the notation of \secref{jet-ev}). The operators
$\mathcal{O}_{{\om}f^{({\mu},0)}}$ are examples of the jet-evaluation
observables.  The three-point function $\langle
\mathcal{O}_{{\om}_{1}}(z_{1}) \ \mathcal{O}_{{\om}_{2}}(z_{2})\
\mathcal{O}_{{\om}f^{({\mu},0)}} (z_{3}) \rangle$ is saturated by the
one-instanton contribution, which, according to formula \eqref{dlzzx},
is given by the integral \Ref{lkterms}, where $k=|\mu|, l =
\ell(\mu)$. We interpret this as the appearance of the following terms
in the OPE:
\begin{equation}    \label{zmu}
\mathcal{O}_{{\om}_{1}}(z,\zb) \ \mathcal{O}_{{\om}_{2}}(0) \sim q
\sum_{{\mu}} z^{|{\mu}|} \mathcal{O}_{12, \mu}(0),
\end{equation}
where
\begin{align*}
\mathcal{O}_{12, \mu} &= \Wick \, C_{12, \mu}(X)
\prod_{i=1}^{{\ell}({\mu})} (-1)^{\mu_i-1}\frac{\mu_i!}{(2\mu_i-1)!}
{\pa}^{{\mu}_{i}-1}_{z}p \, \Wick + \on{terms \ with \ fermions} , \\
C_{12, \mu}(X) &= \int_{\pone\times\pone} \frac{( X-
X_{1})^{{\ell}({\mu})}(X - X_{2})^{|{\mu}|}}{(X_{1}-X_{2})^{|{\mu}|}}
{\om}_{1}(X_{1},\ol{X}_{1}) {\om}(X_{2}, \ol{X}_{2})d^{2}X_{1} \wedge
d^{2}X_{2}.
\end{align*}
Indeed, the two-point function of $\mathcal{O}_{12, \mu}$ and
$\mathcal{O}_{{\om}f^{({\mu},0)}}$ is equal to the integral
\Ref{lkterms} with $k=|\mu|, l = \ell(\mu)$. To fix the fermionic
terms in $\mathcal{O}_{12, \mu}$, we need to compute more two-point
functions of this type.

\medskip

To give an example of the fermionic terms arising in $\mathcal{O}_{12,
\mu}$, we consider the following OPE, obtained in a similar way:
\begin{equation}
\mathcal{O}_{{\om}_{1}}(z,\ol{z}) \, ({\om}_{2}(X, \ol{X}) {\pa}_{z}X
{\ol{\psi}}) (0) \sim q\, {\pi} V_{12}(X) (0) + \ldots,
\label{regope}
\end{equation}
where we have a holomorphic vector field $V_{12}\pa_X$ given by the
formula
\begin{equation}
V_{12}(X) = \int_{\pone \times \pone} \frac{(X -X_{1})(X-X_{2})}{
X_{1} - X_{2}}\, {\om}_{1}(X_{1},\ol{X}_{1}){\om}_{2}(X_{2},
\ol{X}_{2}) d^{2}X_{1} \wedge d^{2}X_{2}\, .
\label{vectf}
\end{equation}
By applying the supercharge $\mathcal{Q} = \int p(z) \psi(z) dz$ to
both sides of \Ref{regope}, we obtain that the $z$-term in \eqref{zmu},
corresponding to $\mu=(1)$, is equal to
$$
[{\mc Q},q {\pi} V_{12}(X)] = q \left(\Wick\, C_{12,(1)}(X) p\, \Wick +
\pa_X C_{12,(1)}(X) \Wick\, \psi \pi \,\Wick \right),
$$
which corresponds to the Lie derivative by the vector field
$V_{12}(X)\pa_X$ (see \secref{jet-ev}). Thus, we have
$$
\mathcal{O}_{12,(1)} = \Wick\, C_{12,(1)}(X) p\, \Wick +
\pa_X C_{12,(1)}(X) \Wick\, \psi \pi \,\Wick \, .
$$
The second term involves the fermionic combination $\Wick \psi\pi
\Wick$\, . Similar terms arise in front of higher powers of $z$ in
\eqref{zmu}.

\sssec{Logarithmic terms in the OPE}

A truly interesting term in the expansion of \eqref{fourpoint} is
\begin{equation}
{\la}\ol{\la} \approx \left\vert {z_{12} z_{34} z_{41}^{-2} }
\right\vert^{2}\, , 
\label{llb}
\end{equation}
which is equal to
\begin{multline}
\la \ol\la \underset{\pone \times \pone \times
\pone}\int \; \; \prod_{{\al}=1}^{3} {\om}_{\al} (X_{\al},
{\ol{X}}_{\al}) \ {\pa}_{X}{\ol{\pa}}_{\ol{X}} f (X_{3}, \ol{X}_{3}) \
{\vert X_{1} - X_{3} \vert^{2} \vert X_{2} - X_{3} \vert^{2} \over
\vert X_{1} - X_{2} \vert^{2} } \cdot \\ \cdot d^{2} X_{1} \wedge
d^{2}X_{2} \wedge d^{2}X_{3}.
\label{coeff}
\end{multline}

\medskip

This integral is very similar to the integral \eqref{3pt} studied
above, and, like that integral, it is also divergent. A troublesome
region is $X_{1} \approx X_{2}$. To extract the divergent part of
\Ref{coeff}, we write (similarly to the calculation in \secref{reg})
$$
X_1 = X - {\xi}/2, \qquad X_2 = X + {\xi}/2.
$$
Then the divergent part of \Ref{coeff} may be approximated by the
integral
\begin{multline}    \label{coeff1}
\la \ol\la \underset{\pone_{1} \times \pone_{3}
\times U}\int {\om}_{1} (X_{1}, {\ol{X}}_{1}) {\om}_{2} (X_{1},
{\ol{X}}_{1}) \ \left( {\om}_{3} {\pa}_{X}{\ol{\pa}}_{\ol{X}} f
\right) (X_{3}, \ol{X}_{3}) {\vert X_{1} - X_{3} \vert}^{4} \cdot \\
\cdot d^{2} X_{1} \wedge d^{2}X_{3} \wedge \ {d^{2}\xi \over |\xi
|^{2}},
\end{multline}
where $U$ is a small neighborhood of $0$ in the $\xi$-plane. The
integral \eqref{coeff1} has logarithmic divergence in $\xi$. This
means that the naive model \eqref{opea}, \eqref{opeb} for the OPE that
we had assumed above is incorrect. What is going on here?

\sssec{Expansion of integrals}

We need to pause for a moment and try to understand the analytic
phenomenon that we have just encountered. The integral
\eqref{fourpoint} is well-defined for all values of $\la$. Yet,
when we expand the integrand in $\la$ near $\la=0$, we obtain
divergent integrals. The reason is that we are trying to switch the
order in which we take two different limits: one corresponds to
expansion in Taylor series, and the other is the limit $R \to \infty$
we take by evaluating our integral over the region $|X_i| < R$ in each
of the three $\pone$. To understand this better, we consider
as a toy model, the integral
\begin{equation}    \label{toy}
\int_0^\infty \frac{dt}{(1+{\la} t)(1+t)^2}, \qquad \la \in {\R}.
\end{equation}
It converges for all non-negative values of $\la$. However, the
integrals of the coefficients in the expansion of the integrand in
$\lambda$ are divergent. The explanation is that the true expansion
of \eqref{toy} in $\lambda$ actually contains $\log \la$. Indeed, the
exact answer is
$$
- \frac{1}{1+\la} + \frac{\la \log \la}{1-\la^2}.
$$
When we try to expand it in the neighborhood of $\la=0$, we
find, in addition to a power series in $\la$ (corresponding to the
first term), a power series times $\log \la$.

On the other hand, we may first expand the integrand in \eqref{toy} in
a power series in $\la$ and then integrate the terms of this
expansion. Then we obtain integrals of the form
\begin{equation}    \label{nth term}
\int_0^\infty \frac{t^n dt}{(1+t^2)}, \qquad n \geq 0,
\end{equation}
which diverge for $n>0$, just like the integral \eqref{coeff}. We
would like to relate the {\em partie finie} regularization of the
integrals \eqref{nth term} and the logarithmic terms in the expansion
of \eqref{toy}.

Consider the following more general situation. Let $f^\la(t)$ be an
analytic function in $\la$ with the Taylor series expansion
$$
f^\la(t) = \sum_{n \geq 0} f_n(t) \la^n,
$$
Suppose that we have the following expansion:
$$
\int_0^\infty f^\la(t) dt = \sum_{n \geq 0} \left( C_n \la^n +
C_{\log,n} \la^n \log \la \right)
$$
in a small neighborhood of $\la=0$. On the other hand, we have (see
\cite{Hoermander}, pp. 70-71)
\begin{equation}    \label{ep ex again}
\int_0^{1/\ep} f_n(t) dt = C'_n \ep^0 + C'_{\log,n} \log \ep +
\sum_{i>0} C'_{i,n} \ep^{-i} + o(1).
\end{equation}
Empirical evidence from calculations with \eqref{toy} and similar
integrals suggests the following conjecture:

\begin{equation}    \label{conj}
C_n = C'_n, \qquad C_{\log,n} = C'_{\log,n}.
\end{equation}

\medskip

It might be well-known to specialists in analysis, but we were unable
to locate it in the literature.

\subsubsection{Back to the integral \eqref{coeff}.} We have found that
the $|\la|^2$-term in the expansion of this integral is
divergent. This means that our model \eqref{opea}, \eqref{opeb} for
the OPE was oversimplified, and in fact, in addition to the power
terms in $z_{ij}, \zb_{ij}$, there are logarithmic terms. Indeed, in
view of our conjecture \eqref{conj}, we should expect the term with
$|\la|^2$ as well as $|\la|^2 \log|\la|^2$. Recalling \eqref{llb}, 
we find that the expansion of $|\la|^2 \log|\la|^2$ will contain the
terms $|z_{12}|^{2} |z_{34}|^2 \log|z_{12}|^2$ as well as
$|z_{12}|^{2} |z_{34}|^2 \log|z_{34}|^2$. Note that since we consider
the limit $z_{12}, z_{34} \to 0$, with finite $z_{14}$, we are not
expanding in terms of $z_{14}$. Such an expansion will be relevant
when we consider the factorization of the four-point function
\eqref{4pt} in a different channel.

Te appearance of the term $|z_{12}|^{2} |z_{34}|^2 \log|z_{12}|^2$
means that we should include the term with $|z_{12}|^{2}
\log|z_{12}|^2$ in formula \eqref{opea}. According to the above
conjecture, to find this term, we need to introduce the ``cut-off''
$|\xi| > \ep$ in the integral \eqref{coeff}. Then the $\ep^0$-term in
the corresponding expansion of the form \eqref{ep ex again} should
give us the correlation function
$$
\langle A(z_1,\zb_1) B(z_3,\zb_3) \rangle
$$
where $A$ is the $|z_{12}|^{2}$-term in the OPE \eqref{opea} and $B$
is the $|z_{34}|^2$-term in the OPE \eqref{opea}. These are in fact
the terms in the perturbative OPEs of our operators, so we have
$$
A = {\om}_{1} (X,\Xb) {\om}_{2}(X,\Xb) {\psi}{\pa}{\psi}
{\ol{\psi}\ol{\pa}\ol{\psi}}
$$
and
\begin{equation}    \label{ABop}
B = {\om}_{3} (X,\Xb) {\pa_X\ol{\pa}_{\Xb}}f(X,\Xb)
{\pa}{X} \ol{\pa} {\ol{X}} \psi {\ol{\psi}},
\end{equation}

On the other hand, we interpret the $\log \ep$-term in the expansion
of \eqref{coeff} as the correlation function
$$
\langle {\mathbb A}(z_1,\zb_1) B(z_3,\zb_3) \rangle,
$$
where now ${\mathbb A}$ is the $|z_{12}|^{2}\log|z_{12}|^2$-term in
the modified OPE \eqref{opea}, and $B$ is given by formula
\eqref{ABop}. We reproduce this two-point function if we set
$$
{\mathbb A} = q \ {}^\ell A(X,\Xb) p \ol{p},
$$
where ${}^\ell ({\om}_{1} {\om}_{2})$ is given by formula
\eqref{groth1},
\begin{equation}
{}^{\ell}({\om}_{1} {\om}_{2})(X,\Xb) = \int_{\mathbb{P}^{1}}\ | Y - X
|^{4}\ ({\om}_{1} {\om}_{2})(Y,\ol{Y}) d^{2}Y.
\label{groth}
\end{equation}
This is precisely the logarithmic partner of $A(z,\zb)$ that we have
already found in \secref{log partners}. (This is not surprising
because the two computations, one in \secref{log partners}, involving
three-point functions, and another one here, involving four-point
functions, are essentially equivalent.)

Thus, though naively the operator $A(z,\zb)$ appears to be a primary
field of conformal dimension $(1,1)$, it is in fact a logarithmic
primary, with the log-partner ${\mathbb A}(z,\zb)$ of exact conformal
dimension $(1,1)$. This logarithmic partner appears as the coefficient
in front of the logarithmic term $|z_{12}|^{2} \log|z_{12}|$ in the
OPE \eqref{opea}.

\medskip

Likewise, the appearance of the term $|z_{12}|^{2} |z_{34}|^2
\log|z_{12}|^2$ in the integral \eqref{coeff} implies that there is a
logarithmic term $|z_{34}|^{2} \log|z_{34}|^2$ in the OPE
\eqref{opeb}. It is nothing but the logarithmic partner of the field
$B(z,\zb)$ appearing in formula \eqref{ABop},
$$
{\mathbb B} = {}^\ell (\omega \pa_X \pa_{\ol{X}}f) \pi \ol{\pi},
$$
which we had also found previously in \secref{log partners}.

\medskip

Let us summarize: we have found the following terms in the operator
product expansions (up to some numeric factors):
\begin{align} \notag
({\om}_{1}(X, {\Xb}) {\psi} {\ol{\psi}})(z,\ol{z}) \ & ({\om}_{2}(X,
{\Xb}) {\psi} {\ol{\psi}})(w,\ol{w}) \sim q \int_{{\mathbb P}^1}
\omega_1 d^2 X \int_{{\mathbb P}^1} \omega_2 d^2 X \cdot \bone \\
\label{om om} &+ ({\om}_{1} (X,\Xb) {\om}_{2}(X,\Xb) {\psi}{\pa}{\psi}
{\ol{\psi}\ol{\pa}\ol{\psi}})(w,\wb) |z-w|^2 \\ \notag &+ q \ ({}^\ell
(\omega_1 \omega_2) p \ol{p})(w,\wb) |z-w|^2 \log|z-w|^2 + \ldots ,
\end{align}
\begin{align} \notag
({\om}(X,{\Xb}) {\psi} {\ol{\psi}})(z,\ol{z}) \ & f(X,{\Xb})(w,\wb)
\sim \\ \label{om f} & ({\om}(X,\Xb) {\pa_X\ol{\pa}_{\Xb}}f(X,\Xb)
{\pa}{X} \ol{\pa} {\ol{X}} \psi {\ol{\psi}})(w,\wb) |z-w|^2 \\ \notag
+ & q \ ({}^\ell (\omega \pa_X \pa_{\ol{X}}f) \pi \ol{\pi})(w,\wb)
|z-w|^2 \log|z-w|^2 + \ldots
\end{align}
Of course, there are many other terms on the right hand sides of these
formulas, but in principle, they may all be computed in the same way
as above.

In a similar way, we compute another example, when both observables
are functions on $\pone$:
\begin{multline}
f(X,\Xb)(z,\zb) g(X,\Xb)(w,\wb) \sim (fg(X,\Xb))(w,\wb)
\\ + |z-w|^{6}\, {\rm log} |z-w|^2 \
(B_{f,g}(X,\Xb){\pi}{\pa}{\pi}\ol{\pi \pa \pi})(w,\wb) + \ldots,
\label{logope}
\end{multline}
where the coefficient function $B_{f,g}(X,\Xb)$ is given by the
integral transform:
\begin{multline}
B_{f,g}(X,\Xb) = \\ \int_{\mathbb{P}^{1}} d^{2}Y \ {\rm
Coeff}_{y^{3}{\ol{y}^{3}}} \left[ f( Y +y , \ol{Y} + \ol{y} ) g ( Y -
y , \ol{Y} - \ol{y}) \left\vert \left( Y - X \right)^{2} - y^{2}
\right\vert^{4} \right].
\label{bfgx}
\end{multline}
To compute the last line, we use the following correlation function:
\begin{multline}
\langle A(X){\psi}{\pa\psi}{\pa^{2}\psi}
\ol{{\psi}{\pa\psi}{\pa^{2}\psi}} (z) \ f(X(z_{1}))\ g(X(z_{2}))
\rangle = \\ 4\left\vert { z_{1} - z_{2} \over ( z_{1} - z ) ( z_{2} - z)}
\right\vert^6 \; \int_{\mathbb{P}^{1}\times \mathbb{P}^{1}
\times \mathbb{P}^{1}} d^{2}X_{1} d^{2}X_{2} d^{2}X_{3} \ f(X_1)
g(X_2) A(X_{3}) \ \left\vert {X_{13} X_{23} \over X_{12}^2 }
\right\vert^4.
\label{triple}
\end{multline}
Near the diagonal $X_1=X_2$ the integral in \Ref{triple} diverges in a
power-like fashion:
\begin{equation}\int {d^{2}v \over |v|^{8}} \times 
\int_{\mathbb{P}^{1}\times 
\mathbb{P}^{1}} f(X) g(X) A(Y) 
\vert  X - Y 
\vert^{8} d^2 X \, d^2 Y.
\label{triplex}
\end{equation}
(The integral \Ref{triple} also looks divergent at infinity in $X_1$,
$X_2$, but this is an illusion.) In order to extract the logarithmic
term in the OPE \eqref{logope}, we have to expand the integrand in
\Ref{triple} in $X_{12}$ to get the term with logarithmic
divergence. This gives the integral \Ref{bfgx}.

\subsubsection{Analogy with quantum mechanics}

There is a precise analogy between the behavior of the correlation
functions of the sigma models observed above and the correlation
functions in quantum mechanical models studied in Part I.

These correlation functions are best understood in terms of the moduli
spaces of stable maps and their quantum mechanical analogues,
compactified moduli spaces of gradient trajectories.

In the quantum mechanical case (on $\pone$), we considered in Section
5.2 of Part I the two-point functions of the form
\begin{equation}    \label{two-point}
_\infty\langle \wh\omega(t_1) \wh{F}(t_2) \rangle_0 = \int_{\pone} \omega
F(qz,q\zb), \qquad q = e^{t_2-t_1}.
\end{equation}
Here $\wh\omega$ and $\wh{F}$ are observables corresponding to a
two-form and a function on $\pone$, respectively. The integral is over
the moduli space of gradient trajectories connecting the point $z=0$
(``north pole'') and the point $z=\infty$ (``south pole'') on $\pone$.
(This moduli space may be compactified to $\pone$.) We consider the
expansion of \eqref{two-point} when $t_1 \gg t_2$, which corresponds
to the limit $q \to 0$.

In order to understand this limit, it is instructive to look at the
moduli space ${\mc M}_{0,\infty,2}$ of triples $(t_1,t_2,\Phi(t))$,
where $t_1 > t_2$ are points on the affine line ${\mathbb R}$, and
$\Phi(t): {\mathbb R} \to \pone$ is a gradient trajectory such that
$\Phi(-\infty) = 0, \Phi(+\infty) = \infty$, modulo the diagonal
action of $\R$,
$$
(t_1,t_2,\Phi(t)) \to (t_1+u,t_2+u,\Phi(t+u)), \qquad u \in {\mathbb R}
$$
(see Section 2.6 of Part I). This moduli space is compactified to
$\ol{\mc M}_{0,\infty,2}$ in a way similar to the stable map
compactification for the moduli space of holomorphic maps (see
\cite{Cohen}).

We have a map ${\mc M}_{0,\infty,2} \to \R_{>0}$ taking
$(t_1,t_2,\Phi(t))$ to $t_1-t_2$. Then the moduli space over which we
integrate in \eqref{two-point} is just the fiber of this map over
$t_1-t_2$.

We also compactify $\R_{>0}$, which is isomorphic to the open interval
$(0,1)$, by the closed interval $[0,1]$, adding points at $t=0$ and
$t=+\infty$. Then we have a map $\ol{\mc M}_{0,\infty,2} \to
[0,1]$. The limit of the integral \eqref{two-point} when $t_1-t_2 \to
+\infty$ (which corresponds to the point $1 \in [0,1]$) may be
described in terms of integration over the fiber of the latter map
over $1 \in [0,1]$.

The result is the identity (5.6) presented in Section 5.2 of Part
I. The expansion of the integral \eqref{two-point} as $q \to 0$ is
expressed as the sum of three terms: the first one involves the
expansion of the function $f$ in the Taylor series around the point $0
\in \pone$, the second one involves the expansion of the two-form
$\omega$ in the Taylor series around the point $\infty \in \pone$ --
these are power series in $q$ (which converge on a small disc in the
$q$-plane), and the third term is the logarithmic correction which
involves $\log(q)$.  It balances out the first two terms and depends
on the kind of regularization used in the first two terms.

\medskip

Now, the four-point function \eqref{fourpoint} may be analyzed in a
similar way. Here we have the moduli space of stable maps $\ol{\mc
M}_{0,4}(\pone,1)$, mapping to the Deligne--Mumford moduli space
$\ol{\mc M}_{0,4}$, which is isomorphic to $\pone$ via the double
ratio $\lambda$ given by formula \eqref{double ratio}. The integral
\eqref{fourpoint} is over a smooth fiber of this map at some $\la \neq
0,1,\infty$ (recall that this fiber is the blow-up of $(\pone)^3$ at
the main diagonal, so the integral is the same as over $(\pone)^3$). We
are interested in the expansion of this integral at $\la=0$. Hence it
should be described in terms of the fiber of the map $\ol{\mc
M}_{0,4}(\pone,1) \to \ol{\mc M}_{0,4}$ at $\la=0$.

This is in a complete analogy with the quantum mechanical situation
discussed above. Indeed, the expansion of \eqref{two-point} near $q=0$
is similar to the expansion of \eqref{fourpoint} near $\la=0$. The map
$\ol{\mc M}_{0,4}(\pone,1) \to \ol{\mc M}_{0,4}$ is the analogue of
the map $\ol{\mc M}_{0,\infty,2} \to [0,1]$. In the quantum mechanical
case, (regularized) integrals over the fiber of the latter map over $1
\in [0,1]$ (corresponding to $q=0$) may be used to obtain the
expansion of \eqref{two-point} leading up to the identity (5.6) of
Part I. Likewise, the expansion of \eqref{fourpoint} near $\la=0$ may
be understood in terms of (regularized) integrals over the fiber of
the map $\ol{\mc M}_{0,4}(\pone,1) \to \ol{\mc M}_{0,4}$ at $0 \in
\pone = \ol{\mc M}_{0,4}$ (which corresponds to $\la=0$). This gives
rise to an identity which is is analogous to the quantum mechanical
identity (5.6) of Part I and describes completely the OPEs ${\mc
O}_{\omega_1}(z_1) {\mc O}_{\omega_2}(z_2)$ and ${\mc
  O}_{\omega_3}(z_3) {\mc O}_f(z_4)$. We have focused above on the
logarithmic terms of these OPEs, which are obtained from the
logarithmic divergences of the integrals over the fiber at
$\la=0$. We will discuss the full identity and the full OPEs that it
describes in the follow-up paper \cite{FLN:new}.

\subsubsection{Computation of the instanton corrections using the
  holomortex operators}    \label{OPE from hol}

In \secref{target pone} we have reviewed the description \cite{AiB} of
the sigma model with the target $\pone$ as a deformation of a free
field theory by the holomortex operators. This description gives us an
alternative way for computing the correlation functions and the OPE in
the sigma model on $\pone$. Here we show how this works on the example
of the OPE of two simplest evaluation observables. Let $f$ be a smooth
function on $\pone$, and $\om$ a smooth two-form on $\pone$. They
correspond to the dimension zero observables, \be {\mathcal{O}}_{f}(z,
\ol{z}) = f (x(z), \ol{x}(\ol{z}))\ , \ {\mathcal{O}}_{\om}(z, \ol{z})
= {\om}(x(z), \ol{x}(\ol{z})) {\psi}(z) {\ol{\psi}}(\ol{z}).
\label{ocof}
\ee
Here we use the logarithmic variables $x,\xb$ on $\pone$ and the
corresponding variables $p,\psi,\pi$ and their complex conjugates.

We wish to calculate their OPE
$$
{\mathcal{O}}_{f} (z, \ol{z}) {\mathcal{O}}_{\om}(0,0),
$$
using the holomortex description of the sigma model on $\pone$.

Let us compute the one-instanton correction to this OPE and try to
reproduce the last term in \eqref{om f}, containing the
logarithm. One-instanton correction corresponds to including one
holomortex operator of each kind, $\Psi^+(w_+,\wb_+)$ and
$\Psi^-(w_-,\wb_-)$, given by formula \eqref{Psi pm}, and integrating
over their positions, $w_+$ and $w_-$ (note that according to the
definition, $\Psi_\pm$ transforms as a $(1,1)$-form on the worldsheet,
so that this integration is intrinsically defined). Thus, we need to
study the integral
\begin{equation}    \label{integral}
q \int d^{2}w_{+} d^{2}w_{-} \left( {\mathcal{O}}_{f} (z, \ol{z})
{\mathcal{O}}_{\om}(0,0)\Psi^{+}(w_{+}, \ol{w}_{+}) \Psi^{-} ( w_{-},
\ol{w}_{-} ) \right),
\end{equation}
where
$$
\Psi^\pm(z) = e^{\pm i \int^z P} \pi(z)
\ol\pi(z), \qquad P = p(w) dw + \ol{p}(\ol{w}) d\ol{w}.
$$
Inside the brackets, we take the ordinary, that is,
{\it perturbative} OPE of the operators involved in the framework of the
free field theory. (The instanton corrections come about due to the
insertion and integration of the holomortex operators.)

The expression in brackets is given by the formula
\begin{equation}    \label{expression}
f(x(z),\xb(\zb)) \omega(x(0),\ol{x}(0)) {\psi}(0) {\ol{\psi}}(0) e^{i
\int_{w_-}^{w_+} P} \pi(w_+) \ol\pi(\wb_+) \pi(w_-) \ol\pi(\wb_-).
\end{equation}
We have the OPE
$$
x(z) e^{i \int_{w_-}^{w_+} P} = \log
\left( \frac{z-w_+}{z-w_-} \right)  e^{i \int_{w_-}^{w_+} P} + \Wick
x(z) e^{i \int_{w_-}^{w_+} P} \Wick \, ,
$$
and similarly for $\xb(\zb)$. Therefore for any function
$f(x(z),\xb(\zb))$ we have
$$
f(x(z),\xb(\zb)) e^{i \int_{w_-}^{w_+} P} =
\Wick f(\wt{x}(z),\ol{\wt{x}}(\zb)) e^{i \int_{w_-}^{w_+} P} \Wick \, ,
$$
where
\begin{align*}
\wt{x}(z) &= x(z) + \log \left( \frac{z-w_+}{z-w_-} \right), \\
\ol{\wt{x}}(\zb) &= \xb(\zb) + \log \left( \frac{\zb-\wb_+}{\zb-\wb_-}
\right).
\end{align*}
The fermionic part of \eqref{expression} is given by the formula
$$
{\psi}(0) {\ol{\psi}}(0) \pi(w_+) \ol\pi(\wb_+) \pi(w_-) \ol\pi(\wb_-)
\sim \left| \frac{1}{w_+} - \frac{1}{w_-} \right|^2 \pi(0) \ol\pi(0).
$$
Thus, \eqref{expression} is equal to
$$
\Wick f(\wt{x}(z),\ol{\wt{x}}(\zb)) \omega(\wt{x}(0),\ol{\wt{x}}(0))
e^{i \int_{w_-}^{w_+} P} \Wick \, \left|
\frac{1}{w_+} - \frac{1}{w_-} \right|^2 \pi(0) \ol\pi(0).
$$
Now we need to expand $f(\wt{x}(z),\ol{\wt{x}}(\zb))$ at $z=0$. Since we
wish to reproduce the last term in \eqref{om f}, we need to take the
$|z|^2$ coefficient in the expansion, which is equal to
$$
(\pa_x \pa_{\xb} f)(\wt{x}(0),\ol{\wt{x}}(0)) \left. \frac{\pa
  \wt{x}}{\pa z} \frac{\pa \ol{\wt{x}}}{\pa \zb} \right|_{z=0} = (\pa_x
\pa_{\xb} f)(\wt{x}(0),\ol{\wt{x}}(0)) \left|
  \frac{1}{w_+} - \frac{1}{w_-} \right|^2
$$
plus terms that are less singular (in the variable $a$ introduced
below) as $w_+,w_- \to 0$, which do not contribute to the divergent
part of the integral that we are trying to calculate; hence we ignore
them in this computation.

Substituting this back into the integral \eqref{integral}, we obtain
that the naive $|z|^2$-term in the OPE is given by the integral
$$
q |z|^2 \pi(0) \ol\pi(0) \int d^{2}w_{+} d^{2}w_{-} \left|
  \frac{1}{w_+} - \frac{1}{w_-} \right|^4 \Wick
  \omega(\wt{x}(0),\ol{\wt{x}}(0)) (\pa_x \pa_{\xb}
  f)(\wt{x}(0),\ol{\wt{x}}(0)) e^{i \int_{w_-}^{w_+} P} \Wick \, .
$$
Let us set $u = e^{x(0)}, v = e^{\wt{x}(0)}$ and make the following
change of integration variables (note that $u$ is fixed in the
integral):
$$
\left( w_{+},w_{-} \right) \mapsto 
\left( a = w_+ w_-\, ,\, v = u\frac{w_+}{w_-} \right)\ .
$$
Accordingly, we now write everything in terms of the algebraic
coordinate $X=e^x$ on $\pone$ and use the corresponding fermionic
variables, which we will denote by $\psi',\ol\psi',\pi',\ol\pi'$. Note
that we have $\pi'(0) \ol\pi'(0) = \frac{1}{|u|^2} \pi(0) \ol\pi(0)$
and that $$\omega(\wt{x}(0),\ol{\wt{x}}(0)) d^2 \wt{x}(0) =
\omega(\wt{x}(0),\ol{\wt{x}}(0)) \frac{d^2 v}{|v|^2}.$$

Therefore in terms of our new variables we obtain the following
integral:
$$
q |z|^2 \pi'(0) \ol\pi'(0) \int \frac{d^2 a}{|a|^2} \; d^2 v |u-v|^4
\;   \Wick \omega(v,\ol{v}) (\pa_v \pa_{\ol{v}} f)(v,\ol{v}) e^{i
    \int_{av/u}^{au/v} P} \Wick \, .
$$

This integral diverges as $a \to 0$. As we explained before, the
$|z|^2 \log|z|^2$ term in the OPE should be equal to the prefactor in
front of the divergent term $\frac{d^2 a}{|a|^2}$. This prefactor is
equal to (note that in the limit $a \to 0$ the exponential operator
drops out)
$$
q \pi'(0) \ol\pi'(0) \int d^2 v |u-v|^4 \omega(v,\ol{v}) (\pa \ol\pa
  f)(v,\ol{v}) = q \; ({}^\ell (\omega \pa \ol\pa f) \pi' \ol\pi')(0),
$$
which coincides with the last term in formula \eqref{om f}. Thus, we
have reproduced a one-instanton logarithmic correction to the OPE
using the holomortex calculus! In a similar way one can reproduce the
logarithmic term in the OPE \eqref{om om} (for the derivation of the
first term in \eqref{om om}, see \cite{AiB}, Section 3.4).

\section{Gauged sigma models}    \label{gauged sm}

The sigma models presented above can be treated, in a certain
restricted sense, as the quantum mechanical models on the loop
space. However, the Floer function $f$, as we have discussed, is
only-well defined on a universal cover of the loop space. In addition,
it is not a Morse--Novikov function, for its critical points are not
isolated. It is, however, a Morse--Bott--Novikov function. In this
section we shall consider a deformation of the standard sigma model,
which can be described using the Morse--Novikov functions, i.e., the
functions on the covering space with isolated critical points. This
deformation is the sigma model in a background gauge field (we will
see below for which gauge group). Note that in this deformation we do
not integrate over the gauge field, so it only plays a classical
role. But at the end of the section we will also comment on the models
obtained by integrating over the gauge field.

\ssec{Quantum mechanical models}

As before, we first look at the corresponding quantum mechanical
models. Let us start with the following simple observation concerning
deformations.  In ordinary quantum mechanics on some phase space $\CP$
the first order action has the form
\begin{equation}
\int p dq - H (p,q) dt
\label{fstor}
\end{equation}
The $pdq$ part of the action is can be written, invariantly, as $\int
{\phi}^{*} d^{-1} {\omega}$, where ${\phi} : I \to {\CP}$ is the map
of the worldline to the phase space. This term in the action does not
require additional geometric structures on the worldline. The term
$\int H(p,q) dt$, which is responsible for the non-trivial dynamics of
the system, contains a one-form $dt$.  There are several ways to
interpret $dt$ in \Ref{fstor}. We can view it as the ein-bien, i.e.
defining a metric $g = dt^{2}$ on $I$, or as a gauge field $A =
dt$. In the evolution operator ${\exp}\left( - T {\hat H}\right)$, the
time interval $T$ will be interpreted as a length of the worldline in
the first approach, or as the holonomy of the gauge field in the
second.  In passing from one to two dimensions both interpretations,
with the metric and with the gauge field, prove useful. Moreover, they
become in a sense related.

Consider now the deformation problem. Suppose we wish to deform the
Hamiltonian $H(p,q)$ by adding to it a small perturbation $H \to H +
{\epsilon} h$. The action \Ref{fstor} becomes
\begin{equation}
\int pdq - H(p,q) dt - {\epsilon} h(p,q) dt.
\label{fstorii}
\end{equation}
Let us assume that the deformation forms a first class system with the
original Hamiltonian, i.e., the Poisson bracket $\{ H, h \}$ is a
linear combination of $H$ and $h$.  More generally let us assume that
the Hamiltonian of the model is a linear combination of Hamiltonians
$H^{a}$ forming a representation $R$ of a Lie algebra $\mathfrak{g} = {\rm
Lie}G$:
$$
[ {\hat H}^{{a}}, {\hat H}^{b} ] = f^{ab}_{c} {\hat H}^{c},
$$
or, classically, 
$$
\{ H^{a} , H^{b} \} = f^{ab}_{c} H^{c}.
$$
In this case it is natural to couple $H^{a}$ to a $\mathfrak{g}$-valued
gauge field:
\begin{equation}
S = \int pdq - A_{a} (t) H^{a}(p,q) dt.
\label{nonab}
\end{equation} 
The Lie algebra $\mathfrak{g}$ acts in the phase space $\CP$ by Poisson
vector fields, and we assume that this action integrates to the action
of the Lie group $G$.
 
The evolution operator can be interpreted as the element $ P {\exp}
\int_{0}^{T} A dt$ of the Lie group $G$, taken in the representation
$R$ in the Hilbert space of the model. The path integral is invariant
under the gauge transformations:
\begin{equation}
A \mapsto g A g^{-1} + g dg^{-1}\ , \ g: I \to G\ , \ {\rm s.t.} \
g(0) = g(T) = 1 \ .
\label{ggeqm}
\end{equation}
This is shown by performing a change of integration variables
$$(p(t),q(t)) \mapsto g(t) (p(t), q(t)).$$

The main example of Part I, the quantum mechanics on $\mathbb{P}^{1}$
(i.e., ${\CP} =T^{*}\mathbb{P}^{1}$) dealt with the abelian $\mathfrak{g} =
{\R}^{{2}} = {\rm Lie}{\C}^{{\times}}$. The gauge field in this case
has two components which can be normalized as: $A = ( {\al} dt,
{\beta} dt)$, the first component couples to the gradient vector field
${\half} \left( X{\pa}_{X}+ \ol{X}\pa_{\ol{X}} \right)$, the second
component couples to the Hamiltonian vector field ${i\over 2} \left(
X{\pa}_{X} - \ol{X}\pa_{\ol{X}} \right)$.  The time $T$ evolution
operator is equal to: $q^{L}\ol{q}^{\ol{L}}$, where $L = X{\pa}_{{X}},
\ol{L}=\ol{X}\pa_{\ol{X}}$, and $q = e^{- {{\al}T\over 2} + i
{{\beta}T\over 2}}, \ol{q} =e^{- {{\al}T\over 2} - i {{\beta}T\over
2}}$.  We see that the evolution operator (and correlation functions)
are invariant with respect to the transformations ${\beta} \mapsto
{\beta} + 2{\pi} {n\over T}$, for $n \in {\Z}$. The conceptual reason
for this invariance is the gauge transformation \Ref{ggeqm},
\begin{equation}
g(t) = e^{2{\pi}i n t\over T} \ .
\label{resgg}
\end{equation}
Clearly, $\al$, or, more precisely, ${\al}T$, is a gauge invariant. 

\subsection{Deformations of sigma models}

Now let us move on to the sigma model case.  First of all, we wish to
describe invariantly the analogue of the ${\al},{\beta}$-system
above. The phase space corresponding to the sigma model with the
target space $X$ and the worldsheet ${\Sigma} = {\BS}^{1}\times
{\R}^{1}$ is ${\CP} = T^{*}LX$, the cotangent bundle to the loop
space. The coordinates are $\left(X^{a}({\sigma}),
\ol{X}^{\ol{a}}({\sigma})\right)$, the $X$-valued functions on
${\BS}^{1}$, and the momenta are the one-forms $\left( P_{a
\sigma}({\sigma}) , \ol{P}_{\ol{a} \sigma} ({\sigma}) \right)$.  The
Hamiltonian is
\begin{align}    \notag
H_{{\al},{\beta}} &= {\al} \int_{{\BS}^{1}} \left( P_{a \sigma} i
{\pa}_{\sigma} X^{a} - \ol{P}_{\ol{a}{\sigma}} i {\pa}_{\sigma}
\ol{X}^{\ol{a}} \right) d{\sigma} \\ &+ {\beta} \int_{{\BS}^{1}}\left(
P_{a \sigma} {\pa}_{\sigma} X^{a} + \ol{P}_{\ol{a}{\sigma}}
{\pa}_{\sigma} \ol{X}^{\ol{a}} \right) d{\sigma}.
\label{abham}
\end{align}
Now let us write the analogue of the action \Ref{fstorii}:
\begin{multline}
S = \int_{{\R}^{1}} dt \left[ 
\int_{{\BS}^{1}} \left( P_{a \sigma} 
{\pa}_{t} X^{a}  +
\ol{P}_{\ol{a}{\sigma}}{\pa}_{t} \ol{X}^{\ol{a}} 
\right) d{\sigma}  -  H_{{\al},{\beta}} \right] \\
= \int_{{\R}^{1}\times {\BS}^{1}} 
dt d{\sigma} \left[ P_{a {\sigma}} \left( 
{\pa}_{t}X^{a} - 
i {\al} {\pa}_{\sigma} X^{a} - {\beta} 
{\pa}_{\sigma}X^{a} \right) +
\ol{P}_{\ol{a}\sigma} \left( {\pa}_{t}\ol{X}^{\ol{a}} 
+ i {\al} {\pa}_{\sigma} \ol{X}^{\ol{a}} - 
{\beta}  {\pa}_{\sigma} \ol{X}^{\ol{a}} 
\right) \right] .  \label{fstoriii}
\end{multline}
Using the complex coordinate $z = {\al} t - i \left( {\sigma} +
{\beta}t \right)$ on $\Sigma$, the metric
\begin{equation}
g = dz d\ol{z} = {\al}^{{2}} dt^{2} + \left( d{\sigma} + {\beta} dt
\right)^{2}
\label{twodm}
\end{equation}
and the redefinitions $P_{a \sigma} \to -{i \over 2}p_{a z},
\ol{P}_{\ol{a}{\sigma}}\to -{i\over 2} \ol{p}_{\ol{a}\ol{z}}$, we show
that up to the topological terms, the action \Ref{fstoriii} becomes
the bosonic part of the action \Ref{2D third action}. The parameters
${\al}, {\beta}$ have been traded for the metric on the worldsheet
${\Sigma}$.  Note that $\beta$ is both a part of the two-dimensional
metric and a component of a one-dimensional gauge field, which is
nothing but the Kaluza-Klein gauge field corresponding to the
compactification from two to one dimension.  Since we study the finite
time evolution $t \in I = [0 , T]$, one can perform the coordinate
transformation ${\sigma} \to {\sigma} + 2{\pi} {n t\over T}$, $n \in
{\Z}$, which is trivial at the endpoints. This amounts to acting on
$\beta$ via ${\beta} \mapsto {\beta} + 2{\pi}{n\over T}$, which is a
gauge transformation \Ref{resgg}.

In what follows we shall denote by $H_{{\al},{\beta}}$ the full
Hamiltonian corresponding to \Ref{2D third action}, including the
fermions.

Now suppose the target space $X$ of our sigma model is the K\"ahler
manifold with the holomorphic ${\C}^{\times}$-action, so that the
corresponding $U(1)$ subgroup acts isometrically. We can then define a
${\C}^{\times}$-action on the loop space with the isolated fixed
points. These fixed points are the constant loops which land at the
fixed points of the ${\C}^{\times}$-action in $X$.

In this case we can deform the $H_{{\al},{\beta}}$ Hamiltonian by the
one generating the ${\C}^{\times}$-action on the target space.  Let us
discuss the Lagrangian aspects of this deformation.
 
Let $v$ be the holomorphic vector field on $X$, generating a
holomorphic ${\C}^{\times}$-action. A holomorphic vector field can be
multiplied by an arbitrary complex constant $v \to {\mu}v$. When this
vector field generates ${\C}^{\times}$ we can fix its normalization by
requiring that $i \left( {v} - {\ol{v}} \right)$ generates $U(1) =
{\R}/2{\pi}{\Z}$.

Consider the two-dimensional sigma model with the target space $X$, on
the worldsheet ${\Sigma} = {\R} \times {\BS}^{1}$, with the following
action, generalizing that in \Ref{fstoriii}, \Ref{2D third action}:
\begin{multline}    \label{actionwithxi}
S = \int_{\Sigma} \left( -i p_a \left( \pa_{\ol{z}} X^{a} +
{\mu} {v}^{a} \right) - i p_{\ol{a}} \left( \pa_z \ol{X^a} +
\ol{\mu}\ol{v}^{\ol{a}} \right) + \right. \\ + \left.  i \pi_a \left(
D_{\ol{z}} \psi^{a} + {\mu} D_{b} {v}^{a} {\psi}^{b} \right) + i
\pi_{\ol{a}} \left( D_{z} \psi^{\ol{a}} + \ol{\mu} D_{\ol{b}}
{\ol{v}^{\ol{a}}} {\psi}^{\ol{b}} \right) \right) d^2 z + \int_\Sigma
\tau_{a\ol{b}} dX^a \wedge d\ol{X^b},
\end{multline}
where ${\mu} \in {\C}$ is a complex constant. The action
\Ref{actionwithxi} describes the quantum mechanics on the loop space
$LX$ with the Hamiltonian $H_{{\al},{\beta},{\gamma},{\delta}} =
H_{{\al},{\beta}} + h_{{\wt\al},{\wt\beta}}$, where the Hamiltonian
$h_{{\wt\al},{\wt\beta}}$ is given by:
\begin{multline}
h_{{\wt\al},{\wt\beta}} = {\wt\al} \int_{{\BS}^{1}}\left( P_{a \sigma}
v^{a}(X({\sigma})) + \ol{P}_{\ol{a}{\sigma}} \ol{v}^{\ol{a}} (
\ol{X}({\sigma} ) \right) d{\sigma} + \\ \qquad\qquad i{\wt\beta}
\int_{{\BS}^{1}} \left( P_{a\sigma}v^{a}(X({\sigma})) -
\ol{P}_{\ol{a}{\sigma}} \ol{v}^{\ol{a}} ( \ol{X}({\sigma}) ) \right)
d{\sigma},
\label{gdham}
\end{multline}
where ${\wt\al}, {\wt\beta} \in {\R}$, ${\mu} = {\wt\al} + i
{\wt\beta}$.

The equations of motion are the following modification of the
equations of holomorphic maps:
\begin{equation}    \label{mod hol}
\pa_{\ol{z}} X^{a} + {\mu} {v}^{a} = 0.
\end{equation}
The path integral localizes on the moduli space of solutions of these
equations, in the same way as in the ordinary sigma model.

\medskip

Let us recapitulate. The loop space $LX$ for Kahler $X$ with
$U(1)$-isometry has a natural $G={\C}^{\times} \times
{\C}^{\times}$-action. The first ${\C}^{\times}$ has been already
exploited, its $U(1)$ part is the rotation of the loops. The second
${\C}^{\times}$ translates the loops in target space using the target
space ${\C}^{\times}$-action.  Thus, the quantum mechanical
Hamiltonian \Ref{gdham} on $LX$ has four parameters, $({\al}, {\beta},
{\wt\al}, {\wt\beta})$.  The action $$S = \int P \cdot {\p}_{t} X \
dtd{\sigma} - \left( H_{{\al},{\beta}} + h_{{\wt\al},{\wt\beta}}
\right) dt,$$ given by the expression \Ref{actionwithxi}, is
well-defined on the cylinder ${\BS}^{1} \times {\R}^{1}$. However, we
wish to study sigma models on more general worldsheets.  To this
end we need to find a way to write the $v$-dependent couplings in
\Ref{actionwithxi} is a covariant way.  Since the vector field $v$
generates a symmmetry of the target space, it is natural to couple it
to a gauge field on the worldsheet. In other words, we replace
$$
\left( \pa_{\ol{z}} X^{a} + {\mu}{v}^{a} \right) 
$$
by
$$
{\nabla}_{\ol{z}}X^{a} := \pa_{\ol{z}} X^{a} + A_{\ol{z}} {v}^{a},
$$
where $A_{\ol{z}}$ is the $(0,1)$-component of the gauge field. 
Since we don't take a quotient of $X$ by the action ${\C}^{\times}$, 
the gauge field is to be viewed as a background field, just like
the worldsheet complex structure:
\begin{equation}    \label{actiongauge}
S = - \int_{\Sigma} \ d^{2}z \ \left( i p_a
\ol{\nabla}_{\ol{z}} X^{a} + i p_{\ol{a}} {\nabla}_z \ol{X^a} \right)
\ + \ {\rm fermions} \ + \int_\Sigma \tau_{a\ol{b}} dX^a \wedge
d\ol{X^b}.
\end{equation}

\subsection{Gauge group}    \label{gauge group}

What is the gauge group corresponding to the action
\Ref{actionwithxi}?  Is it ${\C}^{\times}$, $\R$, or $U(1)$?  In other
words, what are the gauge transformations of $A = A_{z} dz +
A_{\ol{z}} d\ol{z}$ leaving invariant the correlation functions in the
theory with the action \Ref{actionwithxi}?

We are dealing with the twisted supersymmetric model. That model has a
non-anomalous $U(1)_{L} \times U(1)_{R}$ symmetry,
\begin{equation}
A_{\ol{z}} \mapsto 
A_{\ol{z}} + {\ol{\p}}_{\ol{z}} f \ , \ 
A_{z} = A_{\ol{z}}^{*} \mapsto A_{z} + {\p}_{z} {\ol{f}} \ , 
\label{leftright}
\end{equation}
for complex valued $f$. The transformation \Ref{leftright} is
accompanied by the transformation $X \mapsto {\exp} ( f \cdot v) X$,
which is a $z, \ol{z}$-dependent diffeomorphism of the target space.
\footnote{Note that in the purely bosonic sigma models, which
we shall study in Part III, the ${\C}^{\times}$-gauge invariance
is broken generically by anomaly down to the $U(1)$ gauge
invariance. The issue here is the definition of the measure in
the path integral. In the bosonic model the chiral measure
is defined using a holomorphic top form which may be not invariant
under the ${\C}^{\times}$-action. Under the $U(1)$-action the
holomorphic top form is multiplied by the phase, which may cancel the
similar phase of the conjugate anti-chiral measure.}.

Thus, on a compact Riemann surface, the correlation functions in our
model are functions on the Jacobian, ${\rm Jac}({\Sigma})$, which
we identify with the quotient of the space of $(0,1)$-forms
$A_{\ol{z}}$ by the action of the group of ${\C}^{\times}$-gauge
transformations:
\begin{equation}
A_{\ol{z}} \mapsto A_{\ol{z}} + {\ol{\p}}_{\ol{z}} f
\label{ggetr}
\end{equation}
Note that \Ref{leftright} is not a ${\C}^{\times}$-gauge
transformation of the full connection $A = A_{z}dz+A_{\zb}d{\zb}$.
Indeed, the ${\C}^{\times}$-gauge transformation would transform
$A_{z} \mapsto A_{z} + {\p}_{z} f$, in addition to \Ref{ggetr}, in
variance with \Ref{leftright}. That transformation cannot be
accompanied, for general complex valued $f$, by a diffeomorphism of
the target space and does not, therefore, define a symmetry of the
theory.  In fact, by using \Ref{leftright} we can make $A$ into a
unitary flat connection. So, in the end, the sigma model is naturally
coupled to the $U(1)$-flat connection. This result seems surprising
and a little bit counterintuitive.

In order to understand it better, let us look at one example. We
take $X = {\C\mathbb{P}}^{1}$, and $\Sigma = T^{2}$, the torus which
is the quotient of the complex line by the lattice: $z \sim z +
2{\pi}i \left( m + n{\tau}\right)$, $m,n \in {\Z}$, where
\begin{equation}
{\tau} = ( {\beta} + i {\al} ) {T\over 2{\pi} } .
\label{taut}
\end{equation}
The vector field ${v}$ on $X$ is our friend $X {\pa}_{X} + \Xb
\pa_{\Xb}$.  The equations of motion read
\begin{equation}
{\pa}_{\zb}X + A_{\zb}X = 0.
\label{oncpone}
\end{equation}

We may, and will, assume that $A_{\zb } = {\mu} \in {\C}$ is a
constant. Let us perform a gauge transformation (in fact, we should
rather call it a change of variables in the path integral):
\begin{equation}
X (z, {\zb} ) \mapsto 
e^{{\ol{\bf m}} z + {\bf m} {\zb}} X ( z, {\zb} ), 
\label{abchange}
\end{equation}
accompanied by the change
${\mu} \mapsto {\mu} - {\bf m}$. 
The periodicity of the ${\C}^{\times}$-valued function requires:
\begin{align}    \notag
2{\pi}i \left( {\ol{\bf m}}-{\bf m} \right) &= 2{\pi}i m, \quad m \in
\Z, \\ 2{\pi}i \left( {\ol{\bf m}}{\tau} - {\bf m} {\ol{\tau}}\right)
&= 2{\pi} i n, \quad n \in {\Z},
\label{aib}
\end{align}
which gives
\begin{equation}
{\bf m} = \frac{n - m {\tau}}{{\tau} - \ol{\tau}}, \qquad m,n \in \Z.
\label{bblin}
\end{equation}
So we see that the moduli space of physically inequivalent
$\mu$-parameters is an elliptic curve; in particular, it is compact.
This is in striking contrast with the situation in quantum mechanics.

Indeed, the quantum mechanical analogue of the partition function of
the sigma model on the torus is the calculation of the path integral
on the circle, i.e., the trace of the evolution operator:
\begin{equation}
Z(T, {\al}, {\beta}) = {\rm Tr}\ e^{- T \left( {\al} H_{{\R}^{\times}}
+ {\beta} H_{U(1)} \right)}.
\label{uiv}
\end{equation}
We denote by $H_{{\R}^{\times}}$ the Hamiltonian corresponding to the
gradient vector field $V = - {\nabla}f$, and by $H_{U(1)}$ the
Hamiltonian corresponding to the Hamiltonian vector field $U =
{\om}^{-1}df$.  As the latter generates an action of $U(1)$, the
partition function \Ref{uiv} is invariant under the shifts ${\beta}
\to {\beta} + n$ for $n \in {\Z}$. This is analogous to the
$n$-dependent shifts in \Ref{bblin}. However, the gradient vector
field generates an action of the group ${\R}^{\times}$, therefore no
periodicity in the $T$-dependence\footnote{One should not confuse the
periodicity of the analytic continuation of $Z(T, {\al},{\beta})$
which may occur due to the integrality of the spectrum of
$H_{{\R}^{\times}}$ -- here we are talking about the real values of
$T$.} is expected.  The two-dimensional result \Ref{bblin}, however,
predicts a periodicity of a new kind: $T \to T + m$, $m \in {\Z}$.  It
is now time to explain the origin of this mysterious symmetry.

The truth of the matter is the existence of a novel symmetry of the
loop space $LX$, sometimes referred to as the {\em spectral flow}. It
is related to the $U(1)$ action on $X$. Namely, we define an action of
$\Z$ on $LX$ by the formula
\begin{equation}
x({\sigma}) \mapsto \; g_{n}(x({\sigma})) = {}^{n}x({\sigma}) = e^{in
{\sigma}}\cdot x({\sigma}),
\label{spflow}
\end{equation}
where $e^{i\vartheta}\cdot x$ denotes the action of the element
$e^{i\vartheta}\in U(1)$ on $x \in X$.

Now, let $U_{1},U_{2}$ be the vector fields on $LX$ generating the
compact subgroup $U(1)\times U(1) \subset G = \C^\times \times
\C^\times$, $U_{1}$ being the rotation of the loop, $U_{2}$ coming
from the $U(1)$ action on $X$.  Let $V_{1}, V_{2}$ denote the
generators of ${\R}^{\times}\times{\R}^{\times} \subset G$. We claim
that
$$
g_{n}^{*}V_{1} = V_{1} + n V_{2}\ , \qquad g_{n}^{*}U_{1} = U_{1} + n
U_{2}.
$$
It then follows that
\begin{equation}
g_{n}^{-1} H_{\al, \beta} g_{n} = H_{\al, \beta} + n h_{\al, \beta}
\label{abab}
\end{equation}
at the quantum level. Thus, the trace of the evolution operator has an
additional $\Z$-symmetry, leading up to two integer parameters in
\Ref{bblin}.

\medskip

As a side remark, note that we may define the space $LX_{{\vartheta}}$
of twisted loops which satisfy
\begin{equation}
x({\sigma}) \in LX_{\vartheta}\, , \ x ( {\sigma} + 2{\pi}) =
e^{i\vartheta} \cdot x({\sigma}).
\label{twlp}
\end{equation}
Then, \Ref{spflow} generalizes to the action of $\R$ on
$\cup_{\vartheta \in {\BS}^{1}} LX_{\vartheta}$, covering the standard
action of $\R$ on ${\BS}^{1} = {\R}/{\Z}$: $x({\sigma})
\mapsto \; {}^{\vartheta}x({\sigma}) = e^{i \vartheta {\sigma}} \cdot
x({\sigma})$. While we do not really need these spaces here, they
give rise to the twisted sectors in the space of states in the models
obtained by gauging away the group $U(1)$ which will be discussed in
\secref{gauging away}.

\medskip

We conclude this section with one further remark. One may wonder why
the coupling to the vector fields $V_{1}, U_{1}$ is most naturally
described using the two-dimensional metric while the coupling to the
vector fields $V_{2}, U_{2}$ is described using the two-dimensional
gauge field.  The reason is that $V_{1}, U_{1}$ are actually a part of
a much larger Lie algebra acting on the loop space, the
complexification of the Virasoro algebra (with trivial central charge
in our supersymmetric case), while $V_{2}, U_{2}$ are the generators
of the Cartan subalgebra of the affine current algebra ${\hat
u}(1)_{L} \times {\hat u}(1)_{R}$ (with the trivial level). The former
usually couples to the two-dimensional metric while the latter to the
two-dimensional gauge field.

\ssec{Singularities of the gauge fields}

We shall now show that we have to allow the gauge fields with
singularities.  Indeed, take our starting example, ${\Sigma} =
{\R}^{1} \times {\BS}^{1}$. We can interpret \Ref{actionwithxi} as
\Ref{actiongauge} with the particular gauge field $A = dt$.  If we
compactify the cylinder by adding two points, we get a two-sphere
${\BS}^{2}$, with the coordinate $z = e^{- t + i {\sigma}}$ (note that
this $z$ differs from the coordinate $z$ used earlier in this
section). The points we have added are $z = 0$ and $z = \infty$. The
gauge field
\begin{equation}
A =-2 dt =  \left(  {dz \over z} + {d\ol{z} \over \ol{z}} \right)
\label{ggfcy}
\end{equation}
has two singular points, at $z = 0$ and $z = \infty$. We can 
generalize \Ref{ggfcy} to
\begin{equation}
A = \left( {\alpha} {dz \over z} + \ol{\alpha} {d\ol{z} \over \ol{z}}
\right), \qquad \al \in \C.
\label{ggfcyl}
\end{equation}
The action \Ref{actiongauge} forces the map $X(z, \ol{z})$ to obey:
\begin{equation}
{\ol{\p}}_{\ol{z}} X^{a} + \frac{\ol{\al}}{\ol{z}} \ {v}^{a} (X) = 0.
\label{holm}
\end{equation}
The condition of non-singularity near $z = 0$ or $z=\infty$ implies
that
$$
{v}^{a}\left( X\left( z=0\right)\right) = 0, \qquad {v}^{a}\left(
X\left( z = {\infty}\right)\right) = 0.
$$
Thus the map $X(z, {\ol{z}})$ sends the points $z =0$ and $z =\infty$
to the fixed points of ${\C}^{\times}$-action on $X$. Consider the
component $\mathcal{M}_{a,b; d}$ of the moduli space of such maps, for
which $X\left( z=0\right) = a$, $X\left( z = {\infty}\right) = b$. In
addition the component of the moduli space is labeled by $d = \left[
X({\BS}^{2}) \right] \in H_{2}(X, {\z})$. The dimension of the
component $\mathcal{M}_{a,b; d}$ is equal to
\begin{equation}
{\rm dim}\mathcal{M}_{a,b; d} = {\n}_{-}(b) - {\n}_{+}(a) + \int_{d}
c_{1}(X),
\label{dimholg}
\end{equation}
where $c_{1}(X)$ is the first Chern class of $X$, and ${\n}_{\pm}$ are
the positive and negative Morse indices of the critical points
$a,b$. In order to understand \Ref{dimholg} we should interpret the
dimension of $\mathcal{M}_{a,b; d}$ as the index of
$\ol{\p}$-operator, acting on the space of linearized maps.

As we have already discussed, the points $a,b$, viewed as constant
loops landing at $a,b$, should be viewed as the critical points of
Morse--Novikov function on the loop space $LX$ obtained by deformation
of the Morse--Bott--Novikov function \eqref{f on LX},
\begin{equation}
f(\wt\ga) = \int_D \wt\ga^*(\omega_K) - \int_{\BS^1} \ga^*(H) dt,
\label{mnloop}
\end{equation}
where $H$ is the Morse function on $X$, whose gradient vector field is
$v$; it coincides with the Hamiltonian of the $U(1)$ action (here we
use the notation introduced in \secref{as qm}). This function is
well-defined on the universal cover $\wt{LX}$ of $LX$. Its critical
points are the preimages in $\wt{LX}$ of the constant maps $\ga: \BS^1
\to \pone$ landing at the critical points of $H$.

\subsection{Sigma model on $\pone$ in the background of a gauge field}
\label{Morse def}

We consider as an example of the models studied above the
supersymmetric sigma model with the target $\pone$ in the infinite
radius limit, coupled to the vector field $v = X \pa_X + \ol{X}
\pa_{\ol{X}}$, where $X$ is the algebraic coordinate on $\pone$ (as
studied in Part I) and the connection \eqref{ggfcyl}, where we will
assume that $\al$ is real, and $-1 < \al < 0$. As
discussed above, this model may be recast as a quantum mechanical
model on the loop space $\pone$ with the Morse--Bott--Novikov function
\eqref{f on LX} deformed to the Morse--Novikov function
\eqref{mnloop}. The equation of gradient trajectory is now
\eqref{holm}, which we rewrite as follows:
\begin{equation}    \label{holm1}
\pa_{\ol{z}} X + \frac{\al}{\ol{z}} X = 0.
\end{equation}

Let us describe the space of ``in'' states of this model using the
general results of \secref{spaces of delta}. Let $\wt{L\pone}_0$
(resp., $\wt{L\pone}_{\infty}$) be the ascending manifolds in
$\wt{L\pone}$ consisting of (homotopy classes of) maps $D \to \pone$
satisfying \eqref{holm1} and such that $0 \in D \mapsto 0 \in \pone$
(resp., $\infty \in \pone$). We denote by $\wt{L\pone}_{0,m}$ and
$\wt{L\pone}_{\infty,m}, m \in \Z$, their translates with respect to
the group $\pi_1(L\pone) = \Z$ acting on $\wt{L\pone}$. Let
$\wt\CH_{0,m}$ and $\wt\CH_{\infty,m}, m \in \Z$, be the spaces of
semi-infinite delta-forms supported on these ascending
manifolds. Introduce the big space of states
\begin{equation}    \label{big}
\wt\CH = \prod_{m \in \Z}
\wt\CH_{0,m} \oplus \prod_{m \in \Z} \wt\CH_{\infty,m}.
\end{equation}
The space $\CH$ of ``in'' states of our model is {\em non-canonically}
isomorphic to the space of vectors
$$
(\Psi_{0,m},\Psi_{\infty,m})_{m \in \Z} \in \wt\CH
$$
satisfying the equivariance condition \eqref{tau eq},
\begin{equation}    \label{q eq}
\wt\Psi_{0,m} = q \wt\Psi_{0,m+1}, \qquad \wt\Psi_{\infty,m} = q
\wt\Psi_{\infty,m+1}.
\end{equation}
This condition determines all $\Psi_{0,m},\Psi_{\infty,m}, m \in \Z$,
from $\Psi_{0,0},\Psi_{\infty,0}$. Therefore we may identify
$$
\CH \simeq \wt\CH_{0,0} \oplus \wt\CH_{\infty,0}.
$$
However, there are non-trivial extensions at play here, similar to
those described in \secref{log partners}, leading up to the
non-diagonalizability of the Hamiltonian and the logarithmic
mixing of operators.

Let us now describe the spaces $\wt\CH_{0,0}$ and $\wt\CH_{\infty,0}$
more precisely.

\subsubsection{The space $\wt\CH_{0,0}$}

Recall that $\wt\CH_{0,0}$ is the space of semi-infinite
delta-forms supported on the maps $D \to \pone$ satisfying
\eqref{holm1} and such that $0 \in D \mapsto 0 \in \pone$. Maps of
  this form may be written as follows:
\begin{equation}    \label{sol X}
X(z,\zb) = |z|^{-\al} \sum_{n\leq 0} \ga_n z^{-n},
\end{equation}
where the series converges on a disc of small radius (recall that $-1
< \al < 0$). Hence we obtain natural coordinates $\ga_n, n \leq 0$, on
$\wt{L\pone}_{0,0}$. The transversal coordinates are $\ga_n, n > 0$.

The space $\wt\CH_{0,0}$ may be modeled on the Fock representation of
the chiral-anti-chiral $\beta\gamma$-$bc$ system associated with the
open subset $\C_0 = \pone \bs \infty$ of $\pone$. It is generated by
the fields
$$
\beta(z) = \sum_{n \in \Z} \beta_n z^{-n-1}, \qquad \gamma(z) =
\sum_{n \in \Z} \ga_n z^{-n},
$$
$$
b(z) = \sum_{n \in \Z} b_n z^{-n-1}, \qquad c(z) = \sum_{n \in \Z} c_n
z^{-n},
$$
and their complex conjugates. We have the standard OPEs
\begin{align}    \label{OPE bg}
\gamma(z) \beta(w) &= \frac{1}{z-w} + \Wick \gamma(z) \beta(w) \Wick
\, , \\ \notag
c(z) b(w) &= \frac{1}{z-w} + \Wick c(z) b(w) \Wick \, .
\end{align}

We set $\wt{\CH}_{0,0} = {\mc F}_0 \otimes \ol{\mc F}_0$, where ${\mc
  F}_0$ is the Fock representation of the chiral $\beta\gamma$-$bc$
  system, generated by a vacuum vector $\vac$ such that
$$
\ga_n \vac = c_n \vac = 0, \quad n > 0, \qquad \beta_m \vac = b_m \vac
= 0, \quad m \geq 0.
$$
Therefore we have
$$
{\mc F}_0 = \C[\ga_n,\beta_{m}]_{n\leq 0,m<0} \otimes
\Lambda[c_n,b_{m}]_{n\leq 0,m<0} \cdot \vac.
$$
The anti-chiral counterpart $\ol{\mc F}_0$ is generated by the
vector $\ol\vac$ satisfying analogous relations with respect to
the anti-chiral generators.

\subsubsection{Digression: chiral de Rham complex of $\pone$}

Likewise, we have the Fock representation ${\mc F}_\infty$ of the
chiral $\beta\gamma$-$bc$ system associated with the open subset
$\C_\infty = \pone \bs 0$ of $\pone$. It is generated by the vacuum
vector $\wt\vac$ under the action of the Fourier coefficients of the
fields $\wt\beta(z), \wt\gamma(z), \wt{b}(z), \wt{c}(z)$, satisfying
the same relations as above, so that
$$
{\mc F}_\infty = \C[\wt\ga_n,\wt\beta_{m}]_{n \leq 0,m < 0} \otimes
\Lambda[\wt{c}_n,\wt{b}_{m}]_{n \leq 0,m < 0} \cdot \wt\vac.
$$
$\ol{\mc F}_\infty$ is its anti-chiral counterpart.

The two open sets, $\C_0$ and $\C_\infty$, overlap over
$\C^\times$. On this overlap we also have the chiral
algebra which we denote by ${\mc F}^\times$ and we have two embeddings
\begin{multline}    \label{embedding}
{\mc F}_0 = \C[\ga_n,\beta_{m}]_{n\leq 0,m<0} \otimes
\Lambda[c_n,b_{m}]_{n\leq 0,m<0} \cdot \vac \hookrightarrow \\
\C[\ga_0^{-1}] \otimes \C[\ga_n,\beta_{m}]_{n<0,m<0} \otimes
\Lambda[c_n,b_{m}]_{n\leq 0,m<0} \cdot \vac = {\mc F}^\times,
\end{multline}
\begin{multline}    \label{embedding1}
{\mc F}_0 = \C[\wt\ga_n,\wt\beta_{m}]_{n\leq 0,m<0} \otimes
\Lambda[\wt{c}_n,\wt{b}_{m}]_{n\leq 0,m<0} \cdot \vac \hookrightarrow
\\ \C[\wt\ga_0^{-1}] \otimes \C[\wt\ga_n,\wt\beta_{m}]_{n<0,m<0}
\otimes \Lambda[\wt{c}_n,\wt{b}_{m}]_{n\leq 0,m<0} \cdot \vac = {\mc
F}^\times,
\end{multline}

The general formulas \eqref{trans} give us transformation formulas
between the fields with and without tildes. We have
\begin{align*}
\wt\ga(z) &= \ga(z)^{-1} = \ga_0^{-1} \frac{1}{1 + \ga_0^{-1} \sum_{n
    \neq 0} \ga_n z^{-n}}, \qquad \wt{c}(z) = - \ga(z)^{-2} c(z), \\
\wt\beta(z) &= - \Wick \ga(z)^2 \beta(z) \Wick - 2 \ga(z) \Wick b(z)
c(z) \Wick \, , \qquad \wt{b}(z) = \ga(z)^2 b(z).
\end{align*}
These formulas give rise to an identification of two versions of ${\mc
F}^\times$, one in \eqref{embedding} and the other in its
counterpart \eqref{embedding1} with the tildes. This in turn allows us
to ``glue'' together ${\mc F}_0$ and ${\mc F}_\infty$ giving us the
usual definition of the chiral de Rham complex on $\pone$ \cite{MSV}.

The cohomologies of the chiral de Rham complex on $\pone$ (in the Cech
realization) are equal to the cohomologies of the two-step complex
\begin{equation}    \label{coho cdr}
{\mc F}_0 \oplus {\mc F}_\infty \to {\mc F}^\times,
\end{equation}
with the differential being the sum of the embeddings
\eqref{embedding} and \eqref{embedding1}.

\subsubsection{The space $\wt\CH_\infty$}

Naively, one might expect that the space $\wt\CH_\infty$ is equal to
${\mc F}_\infty \otimes \ol{\mc F}_\infty$. But equation \eqref{holm1}
written with respect to the coordinate $Y = X^{-1}$ around the point
$\infty \in \pone$ has the form
$$
\pa_{\ol{z}} Y - \frac{\al}{\ol{z}} Y = 0.
$$
A solution $Y(z,\zb)$ of this equation such that $Y(0,0) =
\infty$ has the form
$$
Y(z,\zb) = |z|^{\al} \sum_{n<0} \wt\ga_n z^{-n}.
$$
Thus, unlike the solution \eqref{sol X}, it involves only the negative
modes of $\wt\ga(z)$, not the zero mode. Therefore we obtain that 
$\wt\CH_\infty = {\mc F}^1_\infty \otimes \ol{\mc F}^1_\infty$, where
${\mc F}^1_\infty$ is generated by a vacuum vector $\wt{|1\rangle}$
satisfying
\begin{equation}    \label{vacuum one}
\wt\ga_n \wt{|1\rangle} = \wt{c}_n \wt{|1\rangle} = 0, \quad n \geq 0,
\qquad \wt\beta_m \wt{|1\rangle} = \wt{b}_m \wt{|1\rangle} = 0, \quad
m > 0.
\end{equation}
Hence we have
$$
{\mc F}^1_\infty = \C[\wt\ga_n,\wt\beta_{m}]_{n < 0,m \leq 0} \otimes
\Lambda[\wt{c}_n,\wt{b}_{m}]_{n < 0,m \leq 0} \cdot \wt{|1\rangle}.
$$
Thus, ${\mc F}^1_\infty$ is different from ${\mc F}_\infty$; in fact,
we have a short exact sequence
\begin{equation}    \label{exact seq}
0 \to {\mc F}_\infty \to {\mc F}^\times \to {\mc F}^1_\infty \to 0,
\end{equation}
so that we have
$$
{\mc F}^1_\infty \simeq {\mc F}^\times/{\mc F}_\infty.
$$
The anti-chiral counterpart $\ol{\mc F}^1_\infty$ is defined in
a similar way.

\subsubsection{Space of states}

We conclude that the space of states of our model is non-canonically
isomorphic to the direct sum
\begin{equation}    \label{dir sum}
({\mc F}_0 \otimes \ol{\mc F}_0) \oplus ({\mc F}^1_\infty \otimes
  \ol{\mc F}^1_\infty).
\end{equation}
However, the canonical structure is more complicated.

Let us look again at the ascending manifolds of the Morse--Novikov
function. There is one of them in each complex ``semi-infinite''
dimension, and the structure of their closures is the following: the
closure of $\wt{L\pone}_{0,n}$ contains $\wt{L\pone}_{\infty,n}$ (as a
complex codimension one stratum), and the closure of
$\wt{L\pone}_{\infty,n}$ contains $\wt{L\pone}_{0,n+1}$ (also in
codimension one).  This leads to non-trivial extensions between the
spaces of delta-forms supported on consecutive strata; namely, there
are extensions
$$
0 \to \wt\CH_{\infty,n} \to ? \to \wt\CH_{0,n} \to 0,
$$
$$
0 \to \wt\CH_{0,n+1} \to ? \to \wt\CH_{\infty,n} \to 0,
$$
which are obtained similarly to the the extensions of the spaces of
delta-forms in the quantum mechanical models discussed in detail in
Part I.

Thus, the structure of the big space of states $\wt\CH$ is more
complicated: instead of the direct sum decomposition \eqref{big}, we
have a space that has a canonical filtration such that the alternating
consecutive quotients are isomorphic to $\wt\CH_{0,n}$ and
$\wt\CH_{\infty,n}, n \in \Z$. The space of states of our model is
then obtained by imposing the equivariance condition \eqref{q eq}.
The non-triviality of the above extensions leads to the
non-diagonalizability of the Hamiltonian, as we will see in the next
subsection.

\medskip

We remark that the above semi-infinite stratification of $\wt{L\pone}$
(and $\wt{LX}$ for a general flag variety $X$) and the corresponding
spaces of holomorphic delta-forms have been considered in
\cite{FF:si}. These spaces are representations of the affine Kac-Moody
algebra $\ghat$ with level $0$, which are closely related to the
Wakimoto modules. In the ${\mc N} = (0,2)$ supersymmetric version the
level $0$ algebra $\ghat$ gets replaced by the $\ghat$ at the critical
level $-h^\vee$ (see \cite{FF:si,F:rev}). The corresponding models are
closely related to the geometric Langlands correspondence. This will
be discussed in Part III of this article.

\subsubsection{Action of the Hamiltonian}

We now compute the action of the Hamiltonian using as the prototype
the formulas obtained in the quantum mechanical models in Part I.

According to the results of Section 4.8 of Part I, if we choose an
identification of the space of states with the direct sum of spaces of
delta-forms on the ascending manifolds $X_\al$ of the Morse function,
then the naive Hamiltonian (which is the Lie derivative with respect
to the gradient vector field) will get a correction term given by
the formula (up to some constant factors, which we will ignore)
\begin{equation}    \label{modif}
\sum_{\al,\beta} \; \delta_{\al\beta} \otimes \ol\delta_{\al\beta}.
\end{equation}
Here the summation is over pairs of adjacent strata such that
$X_\beta$ is a codimension one stratum in the closure of $X_\al$, and
$\delta_{\al\beta}$ and $\ol\delta_{\al\beta}$ are the holomorphic and
anti-holomorphic {\em Grothendieck--Cousin} (GC) operators acting
between the spaces of holomorphic and anti-holomorphic delta-forms on
the two strata. Their definition is spelled out in Section 4.8 of Part
I.

Formula \eqref{modif} suggests that if we identify the space of states
of our model with the direct sum \eqref{dir sum}, then the naive
Hamiltonian $L_0 + \ol{L}_0$ will get a correction term equal to
\begin{equation}    \label{corr H}
\delta_{(0,0),(\infty,0)} \otimes \ol\delta_{(0,0),(\infty,0)} +
\delta_{(\infty,0),(0,1)} \otimes \ol\delta_{(\infty,0),(0,1)},
\end{equation}
where the $\delta$'s are the infinite-dimensional analogues of the GC
operators, which we will now compute explicitly.

We start with the first summand. Its chiral factor, the operator
\begin{equation}    \label{GC one}
\delta_{(0,0),(\infty,0)}: {\mc F}_0 \to {\mc F}^1_\infty,
\end{equation}
is supposed to take a holomorphic delta-form supported on the
stratum $\wt{L\pone}_{0,0}$ and extract its ``polar part'' along the
codimension one stratum $\wt{L\pone}_{\infty,0}$, as in Section 4.8 of
Part I. Therefore this operator is equal to the composition
$$
{\mc F}_0 \hookrightarrow {\mc F}^\times \twoheadrightarrow {\mc
  F}^1_\infty,
$$
where the first map is just the embedding \eqref{embedding}, and the
second map is the map in the short exact sequence \eqref{exact
seq}. According to \eqref{exact seq}, the latter map is an
operator $S_\infty: {\mc F}^\times \to {\mc F}^1_\infty$ whose kernel
is equal to ${\mc F}_\infty$. (The operator
$\ol\delta_{(0,0),(\infty,0)}$ is defined similarly.) We now need to
find the operator $S_\infty$.

\subsubsection{Friedan--Martinec--Shenker bosonization}

We will show that the operator $S_\infty$ arises naturally from
the Friedan--Martinec--Shenker (FMS) bosonization \cite{FMS}, which is
a realization of the chiral $\beta\gamma$-system in terms of scalar
fields. Recall that we are presently considering the
$\beta\gamma$-$bc$-system ${\mc F}_\infty$, in which the generating
fields have tildes. But to simplify our notation, we will omit the
tildes in the formulas below.

Introduce two chiral fields, $u(z)$ and $v(z)$ with the OPEs
\begin{align*}
u(z) u(w) &= - \log(z-w) + \Wick u(z) v(w) \Wick \, , \\
v(z) v(w) &= \log(z-w) + \Wick v(z) v(w) \Wick \, .
\end{align*}
The FMS bosonization formulas look as follows:
\begin{equation}    \label{bosonization}
\gamma(z) = e^{u(z)+v(z)}, \qquad \beta(z) =  - \Wick \pa_z v(z)
e^{-u(z)-v(z)} \Wick \, .
\end{equation}
It is easy to check that the fields $\beta(z)$ and $\gamma(z)$ satisfy
the OPE \eqref{OPE bg}.

Let us denote the (purely bosonic) $\beta\gamma$-chiral algebra by
${\mc F}_{\infty,\on{bos}}$. Let $L$ be the (Minkowski) lattice chiral
algebra $L$, generated from $\{
e^{nu+mv}, n,m \in \Z \}$ under the action of $\pa_z u(z), \pa_z
v(z)$ and their derivatives.

Formulas \eqref{bosonization} give rise to a homomorphism ${\mc
F}_{\infty,\on{bos}} \to L$. However, unlike the well-known
bosonization of fermions, this is {\em not} an
isomorphism. Nevertheless, let us invert $\gamma(z)$, i.e., consider
the bigger chiral algebra ${\mc F}_{\on{bos}}^\times$ generated by
$\gamma(z)^{\pm 1}$, $\beta(z)$ and their derivatives. Let $L_0$ be
the subalgebra of the chiral algebra $L$, generated by the
one-dimensional sublattice
$$
\{ e^{n(u+v)}, n \in \Z \} \subset \{ e^{nu+mv}, n,m \in \Z \}
$$
under the action of $ \pa_z u(z), \pa_z v(z)$ and their derivatives. 
Then
$$
{\mc F}_{\on{bos}}^\times \simeq L_0
$$
(see \cite{FF:weil} for the proof). In particular, ${\mc
  F}_{\on{bos}}^\times$ and ${\mc F}_{\infty,\on{bos}}$ act on $L$.

Now let $L_1 \subset L$ be the subspace of $L$ generated by $\{
e^{nu+(n+1)v}, n \in \Z \}$. Then the vector $e^v \in L_1$ satisfies
the relations \eqref{vacuum one}. Thus, we obtain an embedding ${\mc
F}^1_{\infty,\on{bos}} \hookrightarrow L_1$.

Consider the {\em screening operator}
$$
S = \int e^{v(z)} dz: L_0 \to L_1.
$$
It was proved in \cite{FF:weil} that
\begin{align*}
\on{Ker} S &= {\mc
F}_{\infty,\on{bos}} \subset {\mc F}_{\on{bos}}^\times = L_0, \\
\on{Im} S &= {\mc F}_{\infty,\on{bos}}^1 \subset L_1.
\end{align*}

Now let us add the fermionic $bc$-system ${\mc F}_{\on{ferm}}$
generated by $b(z)$ and $c(z)$ and consider the full chiral
$\beta\gamma$-$bc$-system ${\mc F}_\infty$ as above. We have
$$
{\mc F}^\times = {\mc F}_{\on{bos}}^\times \otimes {\mc
  F}_{\on{ferm}} = L_0 \otimes {\mc F}_{\on{ferm}}, \qquad
  {\mc F}_\infty = {\mc F}_{\infty,\on{bos}} \otimes {\mc
    F}_{\on{ferm}}.
$$

Consider the operator
$$
S_\infty = \int e^{v(z)} dz \otimes 1: L_0 \otimes {\mc
F}_{\on{ferm}} \to L_1 \otimes {\mc F}_{\on{ferm}}.
$$
Then (note that the fermionic parts in ${\mc F}_\infty$ and ${\mc
  F}^1_\infty$ are the same)
\begin{align*}
\on{Ker} S_\infty & = {\mc F}_{\infty,\on{bos}} \otimes {\mc
  F}_{\on{ferm}} = {\mc F}_{\infty}, \\ \on{Im} S_\infty &= {\mc
  F}_{\infty,\on{bos}}^1 \otimes {\mc F}_{\on{ferm}} = {\mc
  F}^1_\infty.
\end{align*}
Therefore we may, and will, view it as an operator acting from ${\mc
F}^\times$ to ${\mc F}^1_\infty$. Thus, we have found an operator
${\mc F}^\times \to {\mc F}^1_\infty$ whose kernel is equal to ${\mc
F}_\infty$. Composing it with the embedding ${\mc F}_0 \hookrightarrow
{\mc F}^\times$, we obtain the sought-after GC operator \eqref{GC
one}. We conclude that {\em the GC operator is equal to the FMS
screening operator}!

\subsubsection{Mystery of the FMS bosonization revealed}

The FMS formulas have given us what we were looking for, but what is
the geometric meaning of these formulas? To answer this question, let
us look at the chiral algebra ${\mc F}^\times$ from the point of view
of the logarithmic coordinate $x$ on $\C^\times$ such that $\ga =
e^x$. We will denote the fields of the corresponding chiral de Rham
complex by $x(z), p(z), \psi(z), \pi(z)$, and similarly for the
anti-chiral analogue. We have the OPE formulas \eqref{OPEfrom},
$$
x(z) p(w) = \frac{i}{z-w} + \Wick x(z) p(w) \Wick \, , \qquad 
\psi(z) \pi(w) = \frac{i}{z-w} + \Wick \psi(z) \pi(w) \Wick \, .
$$

Using the general transformation formulas \eqref{trans} under the
action of changes of variables, we find that
\begin{align*}
\gamma(z) &= e^{x(z)}, \qquad \beta(z) = -i \left( \Wick p(z)
e^{-x(z)} \Wick + \Wick \psi(z) \pi(z) e^{-x(z)} \Wick \right) , \\
c(z) &= e^{x(z)} \psi(z), \qquad b(z) = -i e^{-x(z)} \pi(z),
\end{align*}

Next, we make the $T$--duality transform of \cite{AiB}, Section 2.3,
$$
p(z) \mapsto \pa_z U(z),
$$
where $U(z)$ has the following OPE with $x(w)$:
$$
U(z) x(w) = - i \log(z-w) + \Wick U(z) x(w) \Wick \, .
$$
Then the above formulas become
\begin{align*}
\gamma(z) &= e^{x(z)}, \qquad \beta(z) = -i \left( \Wick \pa_z U(z)
e^{-x(z)} \Wick - \Wick \psi(z) \pi(z) e^{-x(z)} \Wick \right) , \\
c(z) &= e^{x(z)} \psi(z), \qquad b(z) = -i e^{-x(z)} \pi(z).
\end{align*}

Finally, observe that if we introduce an additional field $\phi(z)$ with
the OPE
$$
\phi(z) \phi(w) = \frac{1}{z-w} + \Wick \phi(z) \phi(w) \Wick \, ,
$$
and bosonize the fermions $\psi(z), \pi(z)$ by the usual formulas
$$
\psi(z) = e^{\phi(z)}, \qquad \pi(z) = i e^{-\phi(z)}, \qquad \pa_z
\phi(z) = -i \Wick \psi(z) \pi(z) \Wick \, ,
$$
then we can rewrite the above formulas as follows:
\begin{align*}
\gamma(z) &= e^{x(z)}, \qquad
\beta(z) = \Wick \pa_z(-i U(z) + \phi(z))e^{-x(z)} \Wick \, , \\
c(z) &= e^{x(z)+\phi(z)} , \qquad b(z) = e^{-x(z)-\phi(z)}.
\end{align*}
In particular, if we set $u(z) = x(z) - i U(z) + \phi(z), v(z) = -
\phi(z) + i U(z)$, then we recover the FMS formulas
\eqref{bosonization}.

Thus, the FMS formulas can be explained by extending the
$\beta\gamma$-system to its supersymmetric version and using the change
of variables $\gamma = e^x$ in the corresponding chiral de Rham
complex, followed by the T--duality of \cite{AiB} (see also \cite{Bor}
for a closely related computation).

\subsubsection{FMS screenings and holomortex operators}

Now we can write the screening operator $S_\infty$ in terms of the
variables $x(z), p(z), \psi(z), \pi(z)$. We find that
$$
e^v = e^{-\phi+iU} = -i \pi e^{i \int p dz}.
$$
Actually, we have worked above with the chiral de Rham complex on the
open subset $\C_0 \subset \pone$, with the coordinate $\ga = e^x$. But
we need to make our computation on the open subset $\C_\infty$ with
the coordinate $\wt\ga = \ga^{-1} = e^{-x}$. Therefore we need to
replace $p \mapsto -p, \pi \to -\pi$ in the formulas, so that
$$
e^v = i \pi e^{-i \int p dw}.
$$
Finally, we find that
$$
S_\infty = i \int \pi e^{-i \int p dw} dz.
$$
This formula expresses the GC operator $\delta_{(0,0),(\infty,0)}$ of
formula \eqref{GC one} in terms of the logarithmic coordinate on
$\pone$.

Likewise, we find that the anti-chiral operator
$\ol\delta_{(0,0),(\infty,0)}$ is given by the formula $-i \int \ol\pi
e^{-i \int \ol{p} d\wb} d\zb$. Hence the first correction term to the
Hamiltonian is
$$
\delta_{(0,0),(\infty,0)} \otimes \ol\delta_{(0,0),(\infty,0)} = - 
\int \pi(z) e^{-i \int p dw} dz \otimes \int \ol\pi(\zb)
e^{-i \int \ol{p} d\wb} d\zb.
$$
We interpret the latter integral as the integral
\begin{multline}    \label{cor1}
- \pf \; \pi(z) \ol\pi(\zb) e^{-i \int (p(w) dw +
  \ol{p} d\wb)} |z|^2 d\sigma := \\ - \left( \int_{|z|=\ep} \pi(z)
  \ol\pi(\zb) e^{-i \int (p(w) dw + \ol{p} d\wb)} |z|^2 d\sigma
  \right)_{\ep^0},
\end{multline}
where $\sigma$ is the phase of $z$, regularized via the familiar
Epstein--Glazer procedure. Namely, we retain the $\ep^0$-term in the
expansion of this integral as a function of $\ep$.

Deformation of the Hamiltonian by this operator corresponds to the
deformation of the action by the integral  
$$
\int \pi(z) \ol\pi(\zb) e^{-i \int (p(w) dw + \ol{p} d\wb)} d^2 z =
\int \Psi^-(z,\zb) d^2 z,
$$
which is one of the two holomortex operators introduced in
\cite{AiB} and in \secref{target pone}.

\subsubsection{The second holomortex}

We claim that the second correction term $\delta_{(\infty,0),(0,1)}
\otimes \ol\delta_{(\infty,0),(0,1)}$ to the Hamiltonian (see
\eqref{corr H}) corresponds to the second holomortex operator.

The chiral factor $\delta_{(\infty,0),(0,1)}$ is supposed to take a
holomorphic delta-form supported on the stratum
$\wt{L\pone}_{\infty,0}$ and extract its ``polar part'' along the
codimension one stratum $\wt{L\pone}_{0,1}$. The result is a
holomorphic delta-form on $\wt{L\pone}_{0,1}$. The operator
$\ol\delta_{(\infty,0),(0,1)}$ is its complex conjugate. Thus we
obtain an operator $\wt\CH_{(\infty,0)} \to \wt\CH_{(0,1)}$. By the
$q$-equivariance condition \eqref{q eq}, $\wt\CH_{(0,1)}$ is
identified with $\wt\CH_{(0,0)}$ up to multiplication by
$q$. Therefore we obtain an operator
$$
\wt\CH_{(\infty,0)} \to \wt\CH_{(0,0)}.
$$
To compute this operator, we observe that we have a basic symmetry in
our problem, reversing the two critical points $0$ and $\infty$. It
acts by $\ga \mapsto \ga^{-1} = \wt\ga$, or $x \mapsto -x$. The
operator $\delta_{(\infty,0),(0,1)} \otimes
\ol\delta_{(\infty,0),(0,1)}$ may be obtained from
$\delta_{(0,0),(\infty,0)} \otimes \ol\delta_{(0,0),(\infty,0)}$ by
applying this change of variables, and multiplying by $q$ (because of
the equivariance condition). Thus, we find that
$$
\delta_{(0,0),(\infty,1)} \otimes \ol\delta_{(0,0),(\infty,1)} = - 
q \int \pi(z) e^{i \int p dw} dz \otimes \int \ol\pi(\zb)
e^{i \int \ol{p} d\wb} d\zb,
$$
which we again interpret as the integral
\begin{equation}    \label{cor2}
- q \pf \; \pi(z) \ol\pi(\zb) e^{i \int (p(w) dw +
  \ol{p} d\wb)} |z|^2 d\sigma,
\end{equation}
regularized, as above, via the Epstein--Glaser procedure.

\medskip

We conclude that the naive Hamiltonian acquires the correction equal
to the sum of \eqref{cor1} and \eqref{cor2}. This corresponds to the
deformation of the action of the free theory with the target
$\C^\times$ by the holomortex operators
$$
\int \pi(z) \ol\pi(\zb) \left( e^{-i \int (p(w) dw + \ol{p}
  d\wb)} + e^{i \int (p(w) dw + \ol{p} d\wb)} \right) d^2 z =
\int (\Psi^-(z,\zb) + q \Psi^+(z,\zb)) d^2 z.
$$
By rescaling the different summands in the space of states, we can
rewrite the integrand as $q_1 \Psi^+(z,\zb) + q_2 \Psi^-(z,\zb)$,
where $q_1 q_2 = q$. If we choose $q_1=q_2=q^{1/2}$, then we
obtain the action of the sigma model on $\pone$ found in \cite{AiB}
and reviewed above in \secref{target pone}.

\medskip

Let us summarize. We have adapted the quantum mechanical formulas from
Part I to compute the correction terms to the Hamiltonian of the sigma
model on $\pone$ coupled to the vector field $v$. These correction
terms are given by the infinite-dimensional analogues of the
Grothendieck--Cousin operators corresponding to adjacent semi-infinite
cells in $\wt{L\pone}$. We have found that they may be expressed as
the screenings arising in the FMS bosonization, which, in turn,
coincide with the holomortex operators from \cite{AiB} and
\secref{target pone}. Thus, we obtain that the Hamiltonian of our
model is equal to the deformation of the Hamiltonian of the free
theory on $\C^\times$ by the holomortex operators, as expected.

\subsubsection{Cohomology of the supercharges}

We can also use the quantum mechanical formulas from Part I, Section
4.9, to derive the action of the supercharges of the sigma model on
$\pone$. The total supercharge splits as the sum of two terms $Q +
\ol{Q}$, where (up to a factor of $-i$)
\begin{align*}
Q &= \int \psi p dz + \pf \; \left( q e^{i \int P} -
  e^{- i \int P} \right) \ol\pi d\ol{z}, \\
\ol{Q} &= \int \ol\psi \ol{p} d\ol{z} + \pf \;
\left( q e^{i \int P} - e^{- i \int P} \right) \pi dz.
\end{align*}
These are the supercharges found in \cite{AiB}. The cohomology of $Q +
\ol{Q}$ is equal to the quantum cohomology of $\pone$, that is, $H^0 =
H^2 = \C$ and $H^i = 0$ for $i \neq 0,2$. In \cite{AiB} we have also
computed the cohomology of the right-moving supercharge $\ol{Q}$ and
found that we obtain the same answer (with the non-trivial cohomology
occurring in degrees $0$ and $1$ with respect to the grading
associated with $\ol{Q}$).

Finally, let us compute the cohomology of the perturbative version of
$\ol{Q}$, that is, when we set $q=0$. Then the term $q {\pf} \; e^{i
\int P} \pi dz$ will disappear. The cohomology of the resulting
complex reduces to the cohomology of the two-step complex
$$
{\mc F}^\times \overset{S_\infty}\longrightarrow {\mc F}^1_\infty.
$$
Because of the exact sequence \eqref{exact seq}, this complex has the
same cohomology as the complex \eqref{coho cdr}, which is the Cech
complex computing the cohomology of the chiral de Rham complex of
$\pone$. Thus, we conclude that the cohomology of the right-moving
supercharge $\ol{Q}$ {\em in the perturbative regime} is isomorphic to
the chiral de Rham complex of $\pone$, in agreement with the
prediction of \cite{Kapustin,Witten:cdo}. (The same result was
obtained in \cite{AiB} in a different way. See also
\cite{Bor,Feigin,MS,GMS1} for related work.) When we include the
instantons, we turn on a non-zero parameter $q$ and hence obtain an
extra term $q {\pf} \; e^{i \int P} \pi dz$ in the
differential. Because of this, the cohomology shrinks to the quantum
cohomology of $\pone$.

Similar results may be obtained for other toric
varieties, along the lines of the above analysis and \cite{AiB}.

In Part III of this paper we will consider the ${\mc N} = (0,2)$
supersymmetric sigma models in the infinite radius limit. (We have
studied the corresponding quantum mechanical models in Part I, Section
6.4.) For the target manifold $\pone$, we will find, by methods
similar to the ones used above, that cohomology of the (right-moving)
supercharge is, perturbatively, equal to the cohomology of the chiral
algebra of differential operators of $\pone$, the purely bosonic
version of the chiral de Rham complex \cite{GMS}. This is in agreement
with \cite{Witten:cdo}. When we include the instantons, the cohomology
becomes identically zero, so this model has spontaneously broken
supersymmetry. This agrees with the results of
\cite{Witten:private,Tan} obtained by other methods. In Part III we
will also obtain similar results for the ${\mc N} = (0,2)$
supersymmetric sigma model on the flag manifolds of simple Lie groups,
which, like the model on $\pone$, possess affine Kac--Moody algebra
symmetry of critical level $k=-h^\vee$ \cite{FF:si,F:rev}.

\ssec{Gauging the Lie group symmetry}    \label{gauging away}

Let $X$ be a Kahler manifold with the isometric action of a Lie group
$G$.  We can gauge the infinite radius limit of the sigma model on $X$
described above to obtain a sigma model on $X/G$ (in the infinite
radius limit).  To this end we enlarge the set of fields by adding the
gauge multiplet: $(A, {\Psi}, {\phi}, {\ol\phi}, {\eta}, {\chi}, H
)$, on which the $\mathcal{Q}$-operator acts as follows:
\begin{align*}
{\CQ} A &= {\Psi} \ , \ {\CQ}{\Psi} = d_{A}{\phi}\ , \ 
{\CQ} {\phi} = 0 \\
{\CQ} {\ol\phi} &= {\eta}\ , \ {\CQ}{\eta} = [ {\phi}, {\ol\phi} ] \\
{\CQ} {\chi} &= H \ , \ {\CQ} H = [ {\phi} , {\chi} ].
\end{align*}
We then write the gauged sigma model action as follows:
\begin{align*}
S = S_{0} + {\CQ} \cdot &\left( \int_{\Sigma}
 {\tr} \left( {\chi} \cdot \left(
F_{A} + {\mu}( x, {\xb} ) {\om} \right) \right)
+ {\tr} \left( {\Psi} \star d_{A}{\ol\phi} \right)
+ {\tr} \left( {\eta} [ {\phi}, {\ol\phi} ] \right) \right.
\\
&+ {\pi}_{iw} \left( 
{\pb}_{\wb} x^{i} + A_{\wb}^{a} V_{a}^{i}(x) \right)
+ {\ol{\pi}}_{\ib\wb} \left( 
{\pa}_{w} {\xb}^{\ib} + A_{w}^{a} V_{a}^{\ib}({\xb}) \right)
\\
&+ \left. G^{i\jb}\left( {\pi}_{iw} p_{\jb\wb} + p_{iw} {\pi}_{\jb\wb} 
+ {\Gamma}\on{-terms}\right) + {\tr} \left( {\chi} H \right) \right).
\label{lagr1}
\end{align*}

The study of these models is beyond the scope of this paper (another
model with gauge symmetry, the four-dimensional Yang--Mills theory,
will be discussed in the next section). We note, however, that they
may be analyzed following the same methods that we have used in the
study of the ordinary sigma models in the infinite radius limit, in
the preceding sections. In particular, the space of states of this
model may be described as certain extensions of spaces of delta-forms
on semi-infinite strata in the universal coverings of the loop space,
as well as the twisted loop spaces, introduced in \secref{gauge
group}.

The correlation functions are given by integrals over the moduli
spaces of twisted maps $\Sigma \to X$ corresponding to holomorphic
principal $G$-bundles ${\mc P}$ over $\Sigma$ (that is, holomorphic
sections of the associated bundle ${\mc P} \underset{G}\times X$),
satisfying a stability condition. In the case when the observables
represent equivariant cohomology classes, we obtain gauge theory
analogues of the Gromov--Witten invariants. (Another approach to these
invariants, via $K$--theory, is presented in \cite{FTT}.) These
comprise the topological sector of the gauged sigma model. More
general correlation functions, such as the correlation functions of
gauge theory analogues of the jet-evaluation observables introduced
above, will be represented by divergent integrals over these moduli
spaces. Their regularization will involve the logarithmic mixing of
the type discussed in \secref{log mixing}.

\section{Four-dimensional gauge theory}    \label{ym}

In this section we briefly discuss the four-dimensional analogue of
our constructions. The natural venue for the instanton physics is the
place where they were originally found, namely the four dimensional
gauge theory \cite{polyakov}. The approach of this paper, namely, the
reduction to the supersymmetric quantum mechanics, the weak coupling
limit after some redefinition of the wave-functions, and the extraction
of the spectrum of the resulting Hamiltonian from the correlation
functions, which typically reduce to finite-dimensional integrals,
works in the four-dimensional case as well. However, there is yet
another interesting twist of the story: the so-called equivariant
Morse theory.

The plan of this section is the following. First, we discuss the
gauged supersymmetric quantum mechanics. The analogue of the $\ol\tau
\to \infty$ limit in this theory can be performed in several ways, and
the one which is most relevant for the four-dimensional gauge theory
will be reviewed.  Then we briefly introduce the ${\CN}=2$ twisted
superfields, and write the Lagrangian of the theory. In the limit
$\ol\tau \to \infty$ the path integral becomes the sum of integrals
over finite-dimensional moduli spaces of instantons, i.e., solutions
to the anti-self-duality equations on the curvature of the gauge
field. We discuss the analogue of the evaluation observables in this
theory and conclude with the example of the instanton correlation
function in the case of $G = SU(2)$ gauge theory, at the instanton
charge one.  We find the logarithmic dependence of the correlator on
the positions of the operators, and conclude that the theory is a
logarithmic conformal theory in four dimensions.

\ssec{Self-dual Yang--Mills theory}

The obvious analogue of our ``weak coupling with instantons'' limit in
four dimensions is the so-called self-dual Yang--Mills theory.  The
action of (ordinary, i.e., non-supersymmetric) Yang--Mills theory, on a
Riemannian manifold ${\bf M}^{4}$, i.e., the action of Euclidean
theory, has the form:
\begin{equation}
S_{\rm YM} = \frac{1}{4 g^{2}} \int_{{\bf M}^{4}} {\tr} F
\wedge \star F  -
\frac{i \vartheta}{2\pi}  \int_{{\bf M}^{4}}{\tr} F \wedge F.
\label{stym}
\end{equation}
Using the decomposition
$$
F = F^{+} + F^{-}, \qquad \star F^{\pm} = \pm F^{\pm}
$$
of the curvature two-form, we can rewrite \Ref{stym} as
\begin{equation}
         S_{\rm YM} = \frac{i}{4{\pi}} \left( {\tau} \ {\tr}
\Vert F^{-} \Vert^2 - {\ol\tau}\ \Vert
F^{+} \Vert^2 \right),
\label{stymm}
\end{equation}
where
$$
{\tau} = {{\vartheta}\over 2\pi} + {4{\pi} i \over g^2}
$$
is the complexified coupling. Our limit consists of taking ${\ol\tau}
\to \infty$ while keeping $\tau$ finite. In this limit the action
\Ref{stymm} is not very useful. Instead, a first order action is more
adequate:
\begin{equation}
S_{\rm sdYM} = - i\int_{{\bf M}^{4}}
{\tr} \ H^{+} \wedge F + \frac{\tau}{4\pi}\ {\tr} F \wedge F.
\label{sdym}
\end{equation}
At this point we need to fix our normalization of the Killing form
${\tr}$ on the Lie algebra $\mf g$ of the Lie group $G$. We normalize
it so that the instanton charge
$$
- \frac{1}{8{\pi}^{2}} \int_{{\bf M}^{4}} \ {\tr} F \wedge F
$$
assumes integer values, and is a non-negative integer on the
anti-self-dual connections.

\subsubsection{Supersymmetric Yang--Mills theory}

Let us now discuss the ${\CN}=2$ supersymmetric Yang--Mills theory
\cite{W:tft}. We shall consider, as in the case of the sigma models,
the twisted supersymmetry. This is done in order to have a naturally
defined measure in the path integral. In addition to the gauge field
$A$, there are the fermionic one-form $\psi$, self-dual two-form
$\chi^{+}$, the fermionic scalar $\eta$, and the pair of bosonic
scalar fields $\phi$ and $\ol{\phi}$.  All fields transform in the
adjoint representation of the gauge group $G$. In the standard
super-Yang--Mills theory instead of the integer spin fields $\psi,
\chi^{+}, \eta$ one has a pair of Weyl fermions, ${\lambda}_{{\al}
i}$, and their conjugates $\ol{\lambda}_{{\dot \al} i}$, ${\alpha},
{\dot \alpha}, i = 1,2$. The bosonic symmetry group of the physical
theory is the group $Spin(4) \times SU(2)_{I}$. The twisting consists
of embedding $Spin(4)$ into this group is a non-standard way, under
which ${\ol{\lambda}}_{{\dot \alpha} i} = {\psi}_{\mu}$,
${\lambda}_{\alpha i} = {\chi}^{+} \oplus {\eta}$. In the standard
theory the fields $\phi$ and $\ol{\phi}$ are complex conjugates of
each other. In the applications of the twisted theory to the Donaldson
theory it is much more natural to view $\phi$ and $\ol{\phi}$ as
independent fields \cite{WittenCohFt}. In the modern language this is
a reflection of the fact that the mathematically better defined are
the $I$-models, as opposed to the $B$-models which look
more natural physically (see the discussion of this in the case of
two-dimensional sigma models in \cite{AiB}).

In order to proceed we need to know the normalizations of various
terms in the action of super-Yang--Mills theory. One way to fix the
normalization is to view the theory as the dimensional reduction of
the six-dimensional minimal supersymmetric gauge theory. In six
dimensions the action reads, schematically:
$$ S_{\rm 6d SYM} = {1\over 4g_{6}^{2}} \int \ {\tr} \left( F \wedge
\star_{6} F + {\ol\lambda} {\dirac}_{A} {\lambda} \right).
$$
Upon reduction, one gets the four-dimensional action:
\begin{align} \notag
S_{\rm 4d SYM} &= {1\over 4g^{2}} \int \ {\tr} \left( F \wedge \star F
+ D_{A} {\phi} \wedge \star D_{A}{\ol\phi} + {\rm vol}_{g} \ [
{\phi}, {\ol\phi} ]^{2} \right. \\ \label{fullfdac}
&\qquad\qquad\qquad + {\chi}^{+} \wedge D^{+}_{A} {\psi} + {\eta}
\wedge \star D_{A}^{*} {\psi} \\ \notag &\qquad\qquad\qquad +
\left. {\phi} [{\chi}^{+}, {\chi}^{+}] + {\phi} [ {\eta}, {\eta} ]
{\rm vol}_{\rm g} + {\ol\phi} [ {\psi}, \star \psi ] \right) \\
\notag &\qquad\qquad\qquad - { i {\vartheta}\over 8{\pi}^{2}} \int \
{\tr} F \wedge F,
\end{align}
where
$$
{\rm vol}_{\rm g}  = {\rm g}^{\half} d^{4}x = \star 1
$$
is the metric volume-form. The last term in \Ref{fullfdac}, the one
with the theta-angle, can be turned on in four dimensions without
breaking Lorentz invariance. In six dimensions this term requires an
introduction of a background two-form (it is dual to the axion scalar
in four dimensions). Introduce the auxiliary self-dual bosonic
two-form field $H^{+}$, and rewrite the gauge kinetic term:
\begin{multline}
S_{\rm 4d SYM} = \\ {1\over g^{2}} \ {\CQ} \cdot \int \ {\tr} \left(
-i {\chi}^{+} \wedge F + \ {\chi}^{+} \wedge H^{+} + \frac{1}{4}{\psi}
\wedge \star D_{A}{\ol\phi} + \frac{1}{4}{\rm vol}_{\rm g} \ {\eta} [
{\phi}, {\ol\phi} ] \right) - { i {\tau}\over 4{\pi}}
\int \ {\tr} F \wedge F.
\label{fssfdac}
\end{multline}
Here we have used the topological supercharge $\CQ$, i.e., the
supercharge ${\varepsilon}^{{\al}i} Q_{{\al}i}$ which becomes a scalar
upon the twisting. It acts as follows:
\begin{align} \notag
{\CQ} A &= {\psi}\ , \ {\CQ} {\psi} = D_{A}{\phi}\\ \label{qch} {\CQ}
{\chi}^{+} &= H^{+}\ , \ {\CQ} H^{+} = [ {\phi} , {\chi}^{+}] \\
\notag {\CQ} {\ol{\phi}} &= {\eta}\ , \ {\CQ} {\eta} = [ {\phi},
{\ol{\phi}}] \ , \ {\CQ} {\phi} = 0.
\end{align}

\subsubsection{The weak coupling limit\ ${\ol{\tau}} \to \infty$.}

As \Ref{fssfdac} stands, it is not suited for considering our limit
$g \to 0$ with $\tau$ fixed. However, let us perform the following simple
field redefinition:
\begin{equation}
\left( \ol{\phi} , {\eta}, H^{+}, {\chi}^{+} \right) \mapsto \left(
g^{2}\ \ol{\phi} , g^{2}\ {\eta}, g^{2}\ H^{+}, g^{2}\ {\chi}^{+}
\right)
\label{redeta}
\end{equation}
which keeps \Ref{qch} intact. Now we are in position to take the $g
\to 0$ limit. Indeed, with \Ref{redeta} the action \Ref{fssfdac}
splits as:
\begin{equation}
S_{\rm 4d SYM} = S_{\rm ssdYM} + g^{2} \ {\CQ} \int {\tr} \left(
{\chi}^{+} \wedge H^{+} + {\eta} [ {\phi}, \ol{\phi} ]\ {\rm vol}_{\rm
g} \right),
\label{split}
\end{equation} where $S_{\rm ssdYM}$ is the action of the limit theory:
\begin{multline}
\label{sdsym}
S_{\rm ssdYM} = \\ \int\ {\tr} \left( -i H^{+} \wedge F + i {\chi}^{+}
\wedge D_{A}^{+} {\psi} + {\eta} \wedge \star D^{*}_{A} {\psi} + D_{A}
{\phi} \wedge \star D_{A} {\ol\phi} + [ {\psi} , \star {\psi} ]
{\ol\phi} \right) - \frac{i {\tau}}{4\pi} \int\ {\tr} F \wedge F.
\end{multline}
The action \Ref{split} makes sense on any four-manifold ${\bf
M}^{4}$. The twist made sure that on any ${\bf M}^{4}$ this theory has
at least one fermionic symmetry, generated by $\CQ$.  The usual
feature of such a theory is the $\CQ$-exactness of the stress-energy
tensor, which follows from the fact that all the metric dependence of
\Ref{split} is contained in the $\CQ$-exact terms in the Lagrangian.

The conformal invariance of \Ref{sdsym} (even that of \Ref{split}, at
the classical level) is much less appreciated. To achieve it, let us
assume that ${\phi}$ is a scalar, degree zero field, while $\ol{\phi}$
and $\eta$ are half-densities, i.e., transform as section of ${\rm
vol}_{\rm g}^{\frac{1}{2}}$, under the coordinate transformations.
Then \Ref{sdsym} can be rewritten, with explicit metric dependence,
as (we use the standard notation ${\rm g} = {\rm det}({\rm g}_{\mu\nu})$):
\begin{align} \notag
S_{\rm ssdYM} &= {\CQ} \int\ -i {\rm g}^{\mu\mu'} {\rm g}^{\nu\nu'}
{\rm g}^{\half} {\tr} \left( {\chi}^{+}_{\mu\nu} F_{\mu'\nu'} \right)
+ {\rm g}^{\m\nu}{\rm g}^{\frac{1}{4}} {\tr} \left( {\psi}_{\nu}
D_{\m} \ol{\phi} \right) \\ \label{sdsymcft} &-\frac{1}{4} {\CQ} \int\
{\rm g}^{\mu\nu}{\rm g}^{\frac{1}{4}} {\tr} \left( {\psi}_{\mu}
{\ol{\phi}} \right) {\pa}_{\nu} {\rm log}( {\rm g} ) \\ \notag &-\frac{i
{\tau}}{4\pi} \int_{{\bf M}^{4}} {\tr} F \wedge F,
\end{align}
where $D_{\mu} = {\pa}_{\mu} + [ A_{\mu}, \cdot ]$.

{}The path integral in the theory \Ref{sdym} localizes onto the
anti-self-dual gauge field configurations:
\begin{equation}
F_{A}^{+} = 0.
\label{asd}
\end{equation}
We now wish to apply our techniques to the case of gauge theory. To
this end we need to reformulate the theory as the quantum mechanics of
the same type we encountered before. It turns out that in addition to
the complication of the configuration space being non-simply
connected, we have an additional feature -- {\em equivariance}. We
shall spend some time discussing the equivariant versions of Morse
theory and the supersymmetric quantum mechanics.

\subsubsection{Four-dimensional gauge theory as quantum mechanics}

We can interpret the four-dimensional gauge theory as supersymmetric
gauged quantum mechanics \cite{Atiyah,W:tft}. Let us consider the
four manifold of the form ${\bf M}^{4} = {\R} \times M^3$, where $M^3$
is a compact three-dimensional manifold. Let $t$ denote the coordinate
along the $\R$ factor. Let us assume the four-dimensional metric to be
of the product form:
\begin{equation}
g_{\mu\nu}dx^{\mu}dx^{\nu} = dt^{2} +  h_{ij}({\vec x})dx^idx^j
\label{fdmet}
\end{equation}
where $\vec x = ( x^{1}, x^{2}, x^{3}) = ( x^{i})$. The
four-dimensional gauge field splits as:
$$
A = A_{t}dt + a = A_{t} dt + a_{i} dx^{i}
$$
Let $B_{a} = \star_{3} F_a$ be the three-dimensional magnetic
field, a one-form on $M^{3}$, valued in the adjoint bundle.  
In the gauge $A_{t} =0$ the anti-self-duality equation \Ref{asd}
$$
F_{A}^{+} = 0 \Leftrightarrow {\dot a} + B_{a} = 0
$$
can be interpreted as the gradient flow, with respect to the "Morse
function" $f$ given by the {\em Chern--Simons functional}
\begin{equation}
f = - {\half} \int_{M^{3}} {\tr} \left( a d  a + \frac{2}{3} a^{3}
\right),
\label{cs}
\end{equation}
if the metric on the space $\CA$ of gauge fields on $M^3$ is defined
with the help of the metric $h_{ij}$:
\begin{equation}
h_{aa} = \int_{M^{3}} {\tr} \ {\delta} a \wedge \star_{3} {\delta} a
  = \int \ d^{3}x \ h^{\half} h^{ij} \ {\tr}  {\delta}a_{i} {\delta}a_{j} 
\label{trdm}
\end{equation}
This metric and the functional \Ref{cs} are invariant under the gauge
transformations from the group ${\CG}_{0}$, the component of identity
of the full gauge group. Thus $f$ can be viewed as a Morse (or, if
${\pi}_{1}(M^{3}) \neq 0$, a Morse--Bott) function on the space
${\hat\CX} = {\CA}/{\CG}_{0}$ of the gauge equivalence classes of
connections on a principal $G$-bundle $\CP$ on $M^3$, together with a
choice of a path, up to homotopy, connecting the connection with the
trivial one.

The actual symmetry group of the gauge theory is $\CG$, and the actual
configuration space is ${\CX} = {\hat\CX}/{\Gamma}$,
${\Gamma} =  {\pi}_{0}({\CG}) = {\CG}/{\CG}_{0}$. For example,
for $M^{3} = {\mathbb S}^{3}$, and simple $G$, ${\Gamma} = {\Z}$.

An important subtlety is related to the singularities of $\CX$ and
$\hat\CX$. These arise because the gauge group does not act freely on
$\CA$.  For example, the trivial connection $a = 0$ is left fixed by
the group $G$ of constant gauge transformations, while generic $a$ has
a trivial stabilizer. In addition, there is the whole zoo of
connections which have a stabilizer $H$, which is anywhere between $G$
and the trivial one.

Gauge theory is supposed to give us a {\em definition} of the quantum
mechanics on the space with singularities of this form. Of course,
such a singularity may or may not be a serious issue. Consider, for
example, quantum mechanics of a free particle on a group manifold
$G$. Now let us impose as a gauge symmetry the adjoint action of
$G$. At the level of wave-functions this is a very simple selection
rule: only the $Ad(G)$-invariant functions on $G$ should be kept, in
other words, only the characters of the irreducible representations of
$G$. Now, if we look at the quotient space $X = G/ Ad(G) = T/W$, it
has singularities, and the point $g=1$, the point with the maximal
stabilizer, is one of them. The vicinity of the point $a = 0$ in $\CX$
is somewhat similar to the vicinity of the point $g=1$ in $X$.

A safe way to avoid dealing with the singular quotients is to discuss
the instanton equations in the gauge-covariant way. The equations
\Ref{asd}, written in the form of the evolution equations, read:
\begin{equation}
{\dot a} = D_{a} A_{t} + B_{a}
\label{asdb}
\end{equation} 
We shall now discuss the general setup where the equations like \Ref{asdb}
naturally appear.

\subsubsection{Equivariant integration on the space of paths}

We now study supersymmetric quantum mechanics on a smooth manifold
$X$, possibly, infinite-dimensional. We assume that $X$ is endowed
with $G$-invariant function $f$, which has the Morse property in the
directions transversal to $G$-orbits (so that it gives rise to an
ordinary Morse function on $X/G$ if this quotient exists and is
smooth). The corresponding gradient vector field $v^{\m} = h^{\m\nu}
{\p}_{\nu}f$ commutes with the action of $G$. We define a
generalization of the gradient trajectory: given a map $A_{t} : {\R}
\to {\mf g}$, consider the equation
\begin{equation}
{\dot x}^{\m} (t) = v^{\m} (x(t))+ {\CV}^{\m}_{a}(x(t)) A^{a}_{t}(t)
\label{eqyigr}
\end{equation}
A solution of this equation gives rise to an ordinary gradient
trajectory on $X/G$, if it exists. In general, we use this definition
as a replacement of the notion of gradient trajectory.
 
The space ${\CX}$ of pairs $( x(t), A_t (t))$ is acted upon by the
group ${\CG} = {\rm Maps}( {\R} , G)$:
$$
g: \left( x(t), A_{t}(t) \right) \mapsto \left( g(t) \cdot x(t), \
g^{-1}(t) {\p}_{t} g (t)+ g^{-1}(t) A_{t}(t) g(t) \right)
$$
Ideally, we would like to divide ${\CX}$ by $\CG$, however, the
critical points may have stabilizers. A safer route is to use the
equivariant integration theory developed below.

\medskip

We arrive at the quantum mechanical model with the following
structure.  The configuration space of the theory is the space
${\Pi}TX \times {\Pi}T{\mf g} \times {\mf g}$. The functions on
this space are the differential forms on $X \times \mf g$, taking
values in the space of function on the Lie algebra $\mf g$. The
coordinates on the Lie algebra $\mf g$ are denoted by $\phi, \ol
\phi$, and the coordinates on $\Pi T \mf g$ are $( \ol \phi,
\eta)$, $\eta = d{\ol \phi}$.

The supersymmetry generator $\CQ$ acts as follows:
\begin{align} \notag
{\CQ} x^{\m} &= {\psi}^{\m} , \qquad {\CQ} {\psi}^{\m} =
{\CV}^{\m}({\phi}) = {\phi}^{a} {\CV}^{\m}_{a} \\ \label{eq:susyeq}
{\CQ} {\ol\phi} &= {\eta} , \qquad {\CQ} {\eta} = [ {\phi} , {\ol
\phi} ] \\ \notag {\CQ} {\pi}_{\m} &= p_{\m} , \qquad {\CQ} p_{\m} =
{\p}_{\m} {\CV}^{\nu} ({\phi}) {\pi}_{\nu}.
\end{align}
This generator squares to the infinitesimal gauge transformation of
$G$, generated by $\phi$:
$$
{\CQ}^{2} = {\CL}_{{\CV}({\phi})} + ad ( {\phi}).
$$
In order to ensure the nilpotent nature of $\CQ$, one imposes
the condition of the $G$-invariance:
$$
{\CH} = \left( {\Omega}^{\bullet} ( X \times {\mf g} ) \otimes {\rm
Fun}({\mf g}) \right)^{G}.
$$
In the Lagrangian approach this invariance is enforced with the help of
the Lagrange multiplier $A_{t}$ which can be interpreted as
a one-dimensional gauge field  taking values in $\mf g$. Its superpartner
${\Psi}_{t}$ obeys:
$$
{\CQ} A_{t} = {\Psi}_{t} , \qquad {\CQ} {\Psi}_{t} = D_{t} {\phi} =
{\p}_{t} {\phi} + [ A_{t} , {\phi} ].
$$
The Lagrangian of a first order theory (or, rather, ${\ol \tau} = \infty$ 
theory) looks as follows:
\begin{equation}
L  = {\CQ} \left(  - i {\pi}_{\m} \left( {\dot x}^{\m} - {\CV}^{\m} (
A_{t}) - v^{\m} \right)
+ {\tr} \left( {\Psi}_{t} D_{t} {\ol\phi} \right)  +
h_{\m\nu} {\CV}^{\m}({\ol \phi}) {\psi}^{\nu} \right).
\label{eq:lagreq}
\end{equation}
Note that the Mathai-Quillen interpretation of the Lagrangian
\Ref{eq:lagreq} can be deduced from the lectures \cite{Moore} on the
so-called {\it projection form}.

{}The finite radius theory differs from \Ref{eq:lagreq} by
\begin{equation}
{\Delta}L = \frac{1}{\la} {\CQ} \left(  h^{\m\nu} {\pi}_{\m}
p^{\prime}_{\nu} + {\tr} \left( [ {\phi}, {\ol\phi}] {\eta} \right)
\right).
\label{eq:deffr}
     \end{equation}

\ssec{Gauged quantum mechanics}

At this stage it is perhaps useful to recall a few facts about the
gauged supersymmetric quantum mechanics and the $\ol\tau\to \infty$
limit in this context. The available reviews consider the
zero-dimensional quantum field theories, i.e., integrals over the
quotient spaces, e.g., \cite{Moore}. We need to discuss a quantum
mechanical version, corresponding to one-dimensional quantum field
theories.

So let us  consider the following more general situation.
Let $X$ be compact smooth manifold with the compact simple Lie group
$G$ action. Let ${\mf g} = \on{Lie}(G)$ and  ${\CV} : {\mf g} \to Vect
(X)$ be the corresponding
homomorphism of the Lie algebras.  Let $( x^{\m}, {\psi}^{\m})$ denote
local bosonic and fermionic coordinates on ${\Pi}TX$.
Let $h = h_{\m\nu}dx^{\m}dx^{\nu}$ be a $G$-invariant metric on
$X$.
Sometimes it is convenient to introduce a basis ${\bf t}_{a}$ on
$\mf g$, and the corresponding structure constants:
$$
[ {\bf t}_{a} , {\bf t}_{b} ] = f_{ab}^{c} {\bf t}_{c}
$$
We shall assume this basis to be orthonormal with respect to the
Killing form, which we denote by ${\tr}$. We shall fix the
normalization of ${\tr}$ later.

\subsubsection{Equivariant cohomology.}

We first discuss a finite-dimensional integral. Suppose we wish to
integrate differential forms over the quotient $X/G$, assuming it
exists. The forms on $X/G$ are ${\Pi}TG$-invariant forms on $X$. In
other words, these are $G$-invariant, horizontal forms on $X$,
sometimes also called the {\it basic} forms:
\begin{equation}
{\varpi} \in {\Omega}_{\rm basic}^{\bullet}(X) \Leftrightarrow
\ {\CL}_{{\CV}({\xi})} {\varpi} = 0 , \ \iota_{{\CV}({\xi})} {\varpi}
= 0 \qquad \forall {\xi} \in {\mf g}
\label{basic}
\end{equation}
If \Ref{basic} holds, then ${\varpi}$ is a pull-back of some
differential form ${\varpi}' \in {\Omega}^{\bullet}(X/G)$, of the same
degree.  Sometimes the quotient $X/G$ does not exist, because the
group $G$ may be acting with fixed points, etc. In these circumstances
one should use another model of the de Rham complex of $X/G$. There
are, in fact, several well-known models. For example, one can use the
Weil or Cartan models of {\it equivariant cohomology} of $X$.  We
start by reviewing the Weil model. Instead of dividing $X$ by $G$, and
restricting the de Rham complex of $X$, one multiplies it by an
acyclic complex, which models the de Rham complex of $EG$, a
contractible space, on which $G$ acts freely, and then imposes the
condition of being basic, i.e.,  being a pull-back from $(X \times EG)/
G$, the latter quotient (by the diagonal action of $G$) always being
well-defined.  The de Rham complex of $EG$ is modeled by the Weil
algebra ${\CW}^{\bullet} ({\mf g})$ of $\mf g$, the space of
functions on ${\Pi}T{\mf g}$. The unusual feature of this space is
that the odd coordinates $c^{a}$ are viewed as one-forms (this is not
really unusual), but the even, bosonic, coordinates ${\phi}^{a}$, are
viewed as two-forms. In other words, ${\Pi}T{\mf g}$ is viewed not
as a supermanifold, but rather as a graded manifold. The differential
(which eventually migrates to our supercharge ${\CQ}$) acts on
\begin{equation}
{\Omega}^{\bullet}(X) \otimes {\CW}^{\bullet}({\mf g})
\label{weilcm}
\end{equation}
as follows:
\begin{align} \notag
{\CQ} &= d_{X}  +  {\dl}, \\
\label{weildiff}
{\dl} c &= {\phi} - {\half} [ c , c]\, , \
{\dl} {\phi} = [ {\phi}, c]. \notag
\end{align}
The cohomology of ${\dl}$ on ${\CW}^{\bullet}({\mf g})$ is
one-dimensional (in degree $0$), as it should be since it represents
the de Rham complex of the contractible space $EG$.

{}The action of the group ${\Pi}TG$ on ${\Omega}^{\bullet}(X) \otimes
{\CW}^{\bullet}({\mf g})$ is generated by the operators $\iota_{a}$
and ${\CL}_{a} = \{ {\CQ} , \iota_{a} \}$:
\begin{equation}
\iota_{a} = \iota_{{\CV}_{a}} + \frac{{\p}}{{\p}c^{a}} \ , \
{\CL}_{a} = {\CL}_{{\CV}_{a}} + f_{ab}^{d} \left( {\phi}^{d}
{{\p}\over{{\p}{\phi}^{b}}}  + c^{d} {{\p}\over{{\p}c^{b}}} \right).
\label{contra}
\end{equation}
Of course, we are not interested in ${\Omega}^{\bullet}(X) \otimes
{\CW}^{\bullet}({\mf g})$, we need the space of basic forms
$$
W^{\bullet}_{G}(X) = \left( {\Omega}^{\bullet}(X) \otimes
{\CW}^{\bullet}({\mf g})\right)_{\rm basic},
$$
which is defined as the subcomplex in ${\Omega}^{\bullet}(X) \otimes
{\CW}^{\bullet}({\mf g})$, annihilated by $\iota_{a}, {\CL}_{a}$.
The useful observation of J.~Kalkman is that the first order
differential equations $\iota_{a}{\varpi} ( c, {\phi} )  = 0$ can be
solved by:
$$
{\varpi} ( c, {\phi} ) = e^{- c^{a} \iota_{{\CV}_{a}}} {\al}({\phi}),
$$
where
$$
{\al} \in {\Omega}^{\bullet} (X) \otimes {\rm Fun}({\mf g}), \qquad
{\al} ( {\phi} ) \in {\Omega}^{\bullet}(X).
$$
The remaining equations ${\CL}_{a}{\varpi}( c, {\phi} ) = 0$ translate
to the conditions of $G$-invariance on ${\al}$:
$$
g^{*} {\al} ( {\phi} ) = {\al} ( g{\phi} g^{-1}) , \qquad \forall g
\in G.
$$
Thus, the de Rham complex of $X_{G} = \left( X \times EG \right) / G$
is modeled
on
$$
{\Omega}_{G}^{\bullet}(X) = \left( {\Omega}^{\bullet}(X) \otimes
{\rm Fun}({\mf g}) \right)^{G}
$$
with the differential $\CQ$, which acts on ${\al}$ as follows:
\begin{equation}
{\CQ}{\al} ( {\phi} ) = d_{X} {\al}({\phi}) + 
\iota_{{\CV}({\phi})}{\al}({\phi})
\label{diffeq}
\end{equation}
When $X/G$ exists, the space $X_{G}$ is a fiber bundle over $X/G$ with
the contractible fiber $EG$. Whether $X/G$ exists or not,
the space $X_{G}$ is always a fiber bundle over $BG$, with the fiber
$X$. In this way, one sees explicitly the structure of the
$H^{\bullet}(BG) = (S^{\bullet}{\mf g}^*)^{G}$-module, which is also
clear from the Cartan description.

\subsubsection{Equivariant integration.}

Given a form ${\al} \in {\Omega}_{G}^{\bullet}(X)$, in the case of
free $G$-action, we can ask the following natural question: how does
one produce a basic form ${\beta} \in {\Omega}^{\bullet}_{\rm
basic}(X)$?  If $G$ acts freely, then one can find the so-called
connection forms ${\Theta}^{a} \in {\Omega}^{1}(X)$, i.e., the forms
which obey
\begin{equation}
\iota_{{\CV}_{a}} {\Theta}^{b} = {\dl}^{b}_{a}, \qquad
{\CL}_{{\CV}_{a}} {\Theta}^{b} = - f_{ac}^{b} {\Theta}^{c}.
\label{conn}
\end{equation}
The connection form is defined up to a shift by a section of $
{\Omega}^{1} ( X/G) \underset{G}\times {\mf g}$. The connection
$\Theta$ defines the curvature two-forms
$$
F = d{\Theta} + {\half} [ {\Theta}, {\Theta} ] \ , \ F^{a} =
d{\Theta}^{a} + {\half} f^{a}_{bc} {\Theta}^{b} \wedge {\Theta}^{c},
$$
which are automatically horizontal, $\iota_{{\CV}_{a}} F^{b} = 0$. As
far as the $G$-action is concerned, the connection and curvature obey:
$$
g^{*} F = g^{-1} F g \ , \ g^{*} {\Theta} = g^{-1} {\Theta} g.
$$
Given ${\al} \in {\Omega}_{G}^{\bullet}(X)$, i.e., a $G$-equivariant
map from $\mf g$ to the differential forms on $X$, ${\al}({\phi}) \in
{\Omega}^{\bullet}(X) $, we define:
\begin{equation}
   {\beta} = e^{- {\Theta}^{a} \iota_{{\CV}_{a}} } {\al} \left( - F
   \right).
   \label{horz}
   \end{equation}
   The formula \Ref{horz} defines obviously a $G$-invariant form.  It
   is also not difficult to convince oneself that $\beta$ defined by
   \Ref{horz} is horizontal, $\iota_{{\CV}_{a}}{\beta} = 0$. One
   should pay attention to the ordering of the exponential, since the
   operations of contracting with ${\CV}^{a}$ and multiplication by
   ${\Theta}^{b}$ do not commute, thanks to \Ref{conn}. Finally, note
   that $d{\beta} = 0$ if and only if ${\CQ}{\al}= 0$.

Given ${\beta} \in {\Omega}^{k}(X)_{\rm basic}$, $k = {\rm dim}(X/G)$,
one should compute an integral
   $$
   I ({\beta}) = \int_{X/G} {\beta}
   $$
The problem is to compute $I({\beta})$ in terms of ${\al} \in
{\Omega}^{\bullet}_{G}(X)$. This is done using the {\it projection
form} \cite{Moore}.

As the result, we get the following form ${\Gamma}_{\al} \in
{\Omega}^{\bullet + {\rm dim}G} (X)$:
\begin{equation}
{\Gamma}_{\al} = \frac{1}{{\rm Vol}(G)} \int d{\phi} d{\ol\phi}
d{\eta} \ e^{- {\eta}_{a} {\Theta}^{a} - {\ol\phi}_{a} \left(
{\phi}^{a} + F^{a} \right) } \ {\al}({\phi}).
\label{alphi}
\end{equation}
One checks that if $X/G$ exists, then $I(\beta) = \int_X \Gamma_\al$.
The observation of \cite{Moore} is that the form ${\Gamma}_{\al}$ is
not changed (in \cite{Moore} the attention was only paid to the
$\CQ$-cohomology, but the argument can be recycled to show the actual
form independence) if the canonically normalized connection \Ref{conn}
is replaced by a general $\mf g$-valued one form $\Sigma = \Sigma_a
{\bf t}^a$ on $X$ obeying
\begin{equation}
\iota_{{\CV}_{a}} {\Sigma}_{b} = H_{ab} \ , \
{\CL}_{{\CV}_{a}} {\Sigma}_{b} = - f_{ab}^{c} {\Sigma}_{c},
\label{ioth}
\end{equation}
where at every point $x \in X$ the matrix $H_{ab}$ is a non-degenerate
$G$-invariant, ${\CL}_{{\CV}_{c}} H_{ab} = 0$, bilinear pairing on
${\mf g}_{x}$, the tangent space to the $G$-orbit, passing through
the point $x$. The formula \Ref{alphi} gets modified to
\begin{equation}
{\Gamma}_{\al} = \frac{1}{{\rm Vol}(G)} \int \ d{\phi} d{\ol\phi}
d{\eta} \ e^{- {\eta}^{a}{\Sigma}_{a} + {\ol\phi}^{a} \left(
{\phi}^{b} H_{ab} + d{\Sigma}_{a} \right) } \ {\al}({\phi}).
\label{alphii}
\end{equation}
The relation to the previous formalism is achieved by writing
${\Theta}^{a} H_{ab} = {\Sigma}_{b}$, and by changing the variables:
${\eta} \to {\eta} + [{\ol\phi} ,{\Theta}]$.  Finally, one may study
the one-parametric family of projection forms depending on a parameter
$\la$ (these forms {\it do care} about the representatives, only for
$\CQ$-closed forms $\al$ is the $\la$-dependence unobservable)
\begin{multline}
{\Gamma}_{\al}(\la) = \\ \frac{1}{{\rm Vol}(G)} \int_{{\mf g} \times
{\Pi}T{\mf g}} d{\phi} d{\ol\phi} d{\eta} \ e^{- i
{\eta}^{a}{\Sigma}_{a} + i {\ol\phi}^{a} \left( {\phi}^{b} H_{ab} +
d{\Sigma}_{a} \right) - \frac{1}{\la} {\tr} \left( [ {\eta}, {\eta} ]
{\phi}\right) - \frac{1}{\la} {\tr} [{\phi}, \ol{\phi}]^{2} } \
{\al}({\phi}).
\label{alphiii}
\end{multline}
In the limit $\la \to \infty$ we get our original form
${\Gamma}_{\al}$. In the limit $\la \to 0$ one gets an integral
realization of the relation between the $G$-equivariant and the
$W$-invariant part of the $T$-equivariant cohomology, where $T$ and
$W$ are the maximal torus and the Weyl group of $G$,
respectively. Indeed, we can use the $G$-invariance to reduce the
integral over $\phi$ to the integral over ${\mf t}/W$, ${\mf t} =
\on{Lie}(T)$. When $\la$ is sent to $0$, the $\frac{1}{\la} {\tr}
[{\phi}, {\ol{\phi}}]^2$ term in \Ref{alphiii} dominates, and the
integral over ${\mf g}/{\mf t}$ part of $\ol{\phi}$ and $\eta$ becomes
essentially Gaussian (outside of the discriminant in ${\mf t}/W$):
\begin{equation}
{\Gamma}_{\al}(0) = {1\over{{\rm Vol}(T) \vert W \vert}}
\int_{{\mf t} \times {\Pi}T{\mf t}} \ d{\phi} d{\ol\phi} d{\eta} \
{\De}^{2}({\phi}) \
e^{- i {\eta}^{k}{\Sigma}_{k} + i {\ol\phi}^{k} \left( {\phi}^{j} H_{kj}
+ d{\Sigma}_{k} \right) } \ {\al}({\phi}),
\label{alphiv}
\end{equation}
where
$$
{\De}^{2}({\phi}) = \prod_{{\rm roots \ of } \ {\mf g}}
\langle {\rm root} , {\phi} \rangle.
$$
We close the section by giving some examples of equivariant forms.
Let $( X, {\om}) $ be a symplectic manifold with the $G$-action
preserving the symplectic form ${\om}$. Suppose that the $G$-action is
Hamiltonian and one can define the moment map ${\mu}: X \to {\mf
g}^{*}$, such that
$$
\iota_{{\CV}_{a}} {\om} = - d {\mu}_{a}.
$$
Then ${\Omega}({\phi}) = {\om} + {\phi}^{a} {\mu}_{a}$, as well as any
functions of ${\Omega}({\phi})$ are equivariantly closed.  Another
source of equivariant forms comes from $G$-equivariant vector bundles
${\CE}$ over $X$. Let us denote the fiber of $\CE$ over a point $x \in
X$ by ${\CE}_{x}$. The $G$-action on $X$ lifts to the action on
$\CE$. In particular, $g: {\CE}_{x} \to {\CE}_{g \cdot x}$.  Given a
connection $\nabla$ on $\CE$, i.e., a way to parallel transport the
vectors in ${\CE}_{x}$, the action of $G$ on $\CE$ can be decomposed
into the parallel transport from $x$ to $g\cdot x$ and the action of
$G$ on the vector space ${\CE}_{g\cdot x}$: for any $\xi \in {\mf g}$,
and ${\psi}_{x} \in {\CE}_{x}$, we have:
\begin{equation}
{\xi} \cdot {\psi}_{x} = {\nabla}_{{\CV}({\xi})} {\psi}_{x} +
R({\xi}) \cdot {\psi}_{x}.
\label{xipsi}
\end{equation}

Let $F_{\nabla} = {\nabla}^{2}$ denote the curvature two-form.  The
bundle $\CE$ defines the so-called equivariant Euler class ${\rm
Euler}_{\CE} \in H_{G}^{{\rm rk}{\CE}}(X)$. This cohomology class has
many useful representatives. A one-parametric family of such
representatives is constructed given a section $s: X \to
{\CE}$. Schematically, it is given by the Mathai-Quillen form
\begin{equation}
{\rm Euler}_{\CE} (u) = \int_{{\Pi}T{\CE}_{x}^{*}} e^{i {p \cdot s +
{\pi} \cdot {\nabla} s} - \frac{1}{2} u \langle p , p \rangle -
\frac{1}{2} u \langle {\pi} , \left(  {\Gamma}_{{\CV}({\phi})} +
R({\phi}) + F_{\nabla} \right) \cdot {\pi} \rangle - \frac{1}{2} u
\langle {\pi} , {\Gamma}_{\mu} dx^{\mu} p \rangle},
\label{eqeuler}
\end{equation}
where ${\Gamma}_{{\CV}({\phi})} = \iota_{{\CV}({\phi})} {\Gamma}$, and
${\Gamma}$ is a connection one-form for $\nabla$ on $\CE$.  The
differential ${\CQ}$ is made to act on the auxiliary fields $({\pi},
p)$, which are the fermionic and bosonic coordinates on
${\Pi}{\CE}_{x}^{*}$ and ${\CE}_{x}^{*}$ respectively, as follows:
\begin{align} \notag
{\CQ} {\pi}_{e} &= p_{e} - {\Gamma}_{e \mu}^{h} {\pi}_{h} dx^{\mu}, \\
\notag {\CQ} p_{e} &= R({\phi})_{e}^{h} {\pi}_{h} - {\Gamma}_{e
\mu}^{h} p_{h} dx^{\mu} + \frac{1}{2} F_{e \mu\nu}^{h} {\pi}_{h}
dx^{\mu} \wedge dx^{\nu} - {\Gamma}_{e \mu}^{h} {\CV}^{\mu}({\phi})
{\pi}_{h}.
\end{align}

Finally, the simplest examples of the equivariant (and also
equivariantly closed) forms are invariant functions on ${\mf g}$, e.g.,
invariant polynomials $P({\phi}) \in \C[{\mf g}]^{G} = \C[{\mf
t}]^{W}$.

\subsubsection{The first glimpses of quantum mechanics}
\label{seq:glimps}

So far we have discussed the integrals of differential forms, or
equivariant differential forms. We have also mentioned the equivariant
differential ${\CQ}$.

There are several points of view on $\CQ$ and the equivariant forms.
One viewpoint treats the latter as the functions on some super (or
graded) manifold,  where $\CQ$ acts as the odd
vector field, obeying the non-trivial (for odd fields) {\em master}
equation
$$
\{ {\CQ} ,   {\CQ} \} = 0.
$$
Another point of view identifies the equivariant forms with the
wave-functions of a supersymmetric quantum mechanics. In this approach
$\CQ$ is viewed as an operator acting in the space of states of the
quantum mechanical system. One also needs to define a conjugate
operator ${\CQ}^{*}$. For example, for $$ {\CQ} = {\psi}^{\mu}\left(
p_{\mu} + {\p}_{\mu}f \right) + {\chi}_{\mu}{\CV}^{\mu}_{a}{\phi}^{a}
+ \on{tr} \left( {\eta} \frac{\pa}{\pa \ol\phi} +
[{\phi}, \ol{\phi}] \frac{\pa}{\pa \eta} \right)$$ we get:
$$
{\CQ}^{*} = {\chi}_{\mu} h^{\mu\nu} \left( p_{\nu} + {\p}_{\nu}f
\right) + h_{\mu\nu} {\psi}^{\mu} {\CV}^{\nu}_{a} {\ol\phi}^{a} +
{\tr} \left( \frac{\pa}{\pa \eta} \frac{\pa}{\pa \phi} + [ {\phi},
  \ol{\phi}] {\eta} \right).
$$
Note that the definition of $\CQ^*$ requires a metric on the Lie
algebra and on $X$, whereas the definition of $\CQ$ is
metric-independent.

\subsubsection{Hamiltonian interpretation of gauged quantum
  mechanics.}

{}We can now elaborate on the points of the section \ref{seq:glimps}
and discuss the Hamiltonian interpretation of the theory
\Ref{eq:lagreq}.

Note that the quantum mechanics with action \Ref{eq:deffr}
contains, as a subsector, the quantum mechanics on the Lie algebra
$\mf g$. In fact, there are two versions of the theory.  In one,
which is mostly adopted in the conventional physical applications, the
variables $\phi$ and $\ol{\phi}$ are treated as complex conjugates.
In this case we are dealing with some kind of anharmonic oscillator on
the complexification ${\mf g}_{\C}$. In the second approach, which
is more directly related to the Mathai-Quillen form and equivariant
cohomology, the fields $\phi$ and $\ol{\phi}$ are independent real Lie
algebra valued variables. In this case one is dealing with the
indefinite oscillator on $(\dim \g,\dim \g)$ signature space $\mf g
\oplus \mf g$.

In either case the Hamiltonian of the resulting model looks as follows:
\begin{equation}
H = {\CL}_{v} - {\tr} \frac{{\p}^{2}}{{\p} {\phi}{\p}{\ol{\phi}}} +
{\ol{\phi}}^{b} \left( H_{ab} {\phi}^{a} +  d{\Sigma}_{b} \wedge +
f_{b}^{ac} \frac{{\p}^{2}}{{\p}{\eta}^{a} {\p}{\eta}^{c}} \right) +
{\eta}^{b} {\Sigma}_{b} \wedge
\label{combham}
\end{equation}
where
$$
H_{ab} = h_{\m\nu} {\CV}^{\m}_{a} {\CV}^{\nu}_{b} \ ,  \
{\Sigma}_{a} = h_{\m\nu} {\CV}^{\m}_{a} dx^{\nu}
$$
and we have dropped the term
\begin{equation}
{\delta}H = \frac{1}{\la} {\tr} [{\phi}, \ol{\phi}]^2 +
\frac{1}{\la} \ {\tr} \left( {\eta} [ {\phi}, {\eta}] \right)
\label{etat}
\end{equation}
with which we may play in various ways. The simplest model appears to
be with the term \Ref{etat} dropped, i.e., with $\la = \infty$. In
this case the fields $\phi$ and $\ol{\phi}$ enter at most linearly and
can be integrated out. If, instead, we send $\la$ to $0$, then the
anharmonic oscillator potential will force the Lie algebra $\mf g$
variables to be confined near a maximal torus $\mf t$, just like in
the finite-dimensional example \Ref{alphiv}.

The standard approach to solving the model with the Hamiltonian
\Ref{combham} would be to use the Born-Oppenheimer approximation. It
consists in first solving the harmonic (in the absence of \Ref{etat})
oscillator
in $\phi$ and $\ol\phi$, whose frequencies are given by the eigenvalues
of the induced metric $H_{ab} (x)$. If the group $G$ acts on
$X$ without fixed points, then the frequencies never go down to zero,
and one can approximate the wave-functions by the ground state
wave-functions in $\phi$, $\ol\phi$ directions, times some forms
on the $X$ space.

However, in the vicinity of a fixed point of the $G$ the metric
$H_{ab}(x)$ becomes degenerate. Then the Born-Oppenheimer
approximation can no longer be applied, and the theory becomes more
intricate. In particular, a new "branch" develops, where the wave
function does not exponentially decay when $\phi$ goes to infinity.

We shall not  discuss the transition between branches.
Some of the remarks on this problem (in the context of supersymmetric
models with higher degree of supersymmetry) can be found in
\cite{WittenHiggs}.

\subsubsection{Example of the group manifold} \label{seq:group}
Consider the example $X = G$ where the group $G$ acts by the right 
multiplication.

The space of states of this model, viewed as the gauged quantum
mechanics, is the space of differential forms on $G$ tensored with the
space of functions of $\phi$ and with the differential forms on
another copy of $\mf g$:
$$
{\Psi} ( g, dg, {\phi}, {\ol{\phi}}, d{\ol{\phi}}).
$$
It is customary to denote $d\ol{\phi}$ by $\eta$, and it is convenient
to denote by $\psi$ the left-invariant form $g^{-1}dg$.
Then the equivariant differential acts as follows:
\begin{align} \notag
{\CQ} g & = g \psi \ , \\
\notag
{\CQ} \psi & = {\phi} - {\half} [\psi,
\psi], \\ \label{eqdiff} {\CQ} {\phi} & = 0, \\ \notag {\CQ} {\ol{\phi}} & =
{\eta}, \\ \notag {\CQ} {\eta} & = [ {\phi}, \ol{\phi} ].
\end{align}
By passing to the $G$-invariant variables
\begin{align} \notag
A &= dg g^{-1}\ , \qquad A^{*} = g {\eta} g^{-1} + [ A, g {\ol{\phi}}
g^{-1}], \\ {\Phi} &= g {\phi} g^{-1} - {\half} [A, A] \ , \qquad
{\Phi}^{*} = g \ol{\phi} g^{-1},
\label{ginv}
\end{align}
we map $\CQ$ to the following simple differential operator:
\begin{equation}
{\CQ} = {\Phi} \frac{\p}{{\p}A} + A^{*} \frac{\p}{{\p}{\Phi}^{*}}.
\label{qop}
\end{equation}

\sssec{Observables in gauge theory}

We now go back to the problem of our interest: the four-dimensional
supersymmetric twisted gauge theory. The first question which we
should address is what are the analogues of the evaluation observables
in the gauge theory. In the case of the twisted supersymmetric sigma
model the evaluation observables ${\CO}(x, \psi)$ have many nice
properties: i) they corresponded to the dimension zero operators; ii)
the set of these operators is non-trivially acted upon by the
supercharge ${\CQ}$; iii) their definition did not require any short
distance regularization perturbatively. If the condition i) is relaxed
one gets the jet-evaluation observables. Finally, these observables
have a clear geometric interpretation.

In gauge theory we must demand that the observables be
gauge-invariant. The gauge theory observables, analogous to the
operators ${\CO}(x,{\psi})$ in sigma model are the local
gauge-invariant functionals ${\CO}(A, {\psi})$. These operators rarely
obey the property i). However, they do obey ii) and
iii). Geometrically, these correspond to the ${\CG}_{{\bf
M}^{4}}$-invariant differential forms on ${\CA}_{{\bf M}^{4}}^{+}$,
the space of anti-self-dual gauge fields. Indeed, the integral over
$H^{+}$ field enforces the anti-self-duality condition, $F^{+}_{A}=0$,
and the integral over the gauge equivalence classes of $A$ becomes an
integral over the moduli space of instantons
\begin{equation}
\CM = {\CA}_{{\bf M}^{4}}^{+}/{\CG}_{{\bf M}^{4}} \ ,
\label{instmod}
\end{equation}
where ${\CA}_{{\bf M}^{4}}^{+}$ is the space of anti-self dual
connections on ${\bf M}^{4}$ (satisfying equation \eqref{asd}) and
${\CG}_{{\bf M}^{4}}$ is the group of gauge
transformations. Then the integration over $\chi^{+}, \eta$ makes
$\psi$ a one-form on $\CM$, valued in one-forms on ${\bf M}^{4}$ and
satisfying the equation
\begin{equation}
D_{A}^{+} {\psi} = 0  \ , \qquad D_{A}^{*}{\psi} = 0 \ . \label{psieq}
\end{equation}
The last equation $D_{A}^{*}{\psi}=0$ can be interpreted as a
gauge-fixing for the fermionic gauge symmetry
${\delta}_{\varepsilon}{\psi} = D_{A} {\varepsilon}$.  In classifying
the observables we can replace the condition of horizontality by the
cohomology of the corresponding BRST operator, which acts as follows:
\begin{equation}
{\delta}{\psi} = D_{A}{\phi}.
\label{psibrst}
\end{equation}
Thus, we define the gauge-evaluation observables as the cohomology of
the operator \eqref{psibrst} acting on the gauge-invariant functionals
${\CO}(A, {\psi}, {\phi})$.

After the functional integration over $\ol\phi$, the field $\phi$
becomes a curvature two-form
\begin{equation}
\label{unicurv}
\phi = \frac{1}{{\De}_A} [ {\psi} , \star \psi ],
\end{equation}
where ${\De}_{A} = D_{A} D_{A}^{*} + D_{A}^{*} D_{A}$ is the
gauge-covariant Laplacian (here acting on zero-forms).

\subsubsection{Universal connection}

According to \cite{WittenCohFt}, it is convenient to think of
$\phi, \psi, F_{A}$ as of the three components of the universal
curvature:
\begin{equation}
{\bf F} = {\phi} + {\psi} + F_{A}
\label{unicurvi}
\end{equation}
of the universal connection on the universal principal $G$-bundle over
${\CM} \times {\bf M}^{4}$.
Let us review the construction of this connection.

Let $A(m) = A_{\mu} ( x, m) dx^{\mu}$, $m \in {\CM}$, $x \in {\bf
M}^{4}$ be a family of anti-self-dual connections on ${\bf M}^{4}$,
defined over some open domain $U \subset {\CM}$, so that for each $m
\in U$, $F_{A(m)}^{+}=0$ and the gauge equivalence class obeys $[A(m)]
= m$. Now let us study the $m$-dependence. Clearly, the derivative
$\frac{\pa}{\pa m^{i}} A(m)$ obeys the linearized instanton equation,
$$
D_{A}^{+}\frac{\pa A(m)}{\pa m^{i}}  = 0 \ .
$$
Therefore it can be decomposed:
\begin{equation}
\frac{\pa A(m)}{\pa m^{i}} = {\psi}_{i} (m) + D_{A(m)}
{\varepsilon}_{i} (m),
\label{deco}
\end{equation}
where the set of one-forms $\psi_{i}(m)$, $ i = 1, \ldots, {\rm
dim}{\CM} $, obeys $$ D_{A(m)}^{+}{\psi}_{i} (m) = 0 , \
D_{A(m)}^{*}{\psi}_{i}(m) = 0 \ .
$$
More precisely, the set $( \psi_{i} ) $ forms a basis in $H^{1}_{m}$,
which coincides with $T_{m}{\CM}$ in a nice situation, where both
$H^{0}_{m}$ and $H^{2}_{m}$ vanish. Assuming that this is the case, $
( \varepsilon_{i} ) $ in \Ref{deco} is the set of compensating gauge
transformations,
$$
{\varepsilon}_{i}(m) = {\varepsilon}_{i} ( x, m) , \qquad  $$
which we can actually compute from \Ref{deco} that
$$
{\varepsilon}_{i}(m) = {1\over{{\De}_{A(m)}}} \ D_{A(m)}^{*} \frac{\pa 
A(m)}{\pa m^{i}}.
$$
In components:
\begin{align} \notag
{\psi}_{i} (m) &= {\psi}_{i \mu}(x, m) dx^{\mu} \\ \label{univconn}
D_{\mu}{\psi}_{i\nu} &- D_{\nu} {\psi}_{i \mu} + {\half}
{\epsilon}_{\mu\nu\mu'\nu'} {\rm g}^{\mu'\mu''}{\rm g}^{\nu\nu''} {\rm
g}^{\half} \ D_{\mu''}{\psi}_{i\nu''} = 0 \\ \notag D_{\mu} & \left(
{\rm g}^{\mu\nu} {\rm g}^{\half}\ {\psi}_{i\nu} \right) = 0.
\end{align}
By combining $A(m)$ and ${\varepsilon}(m)$,
\begin{equation}
{\bf A} = A(m) + {\varepsilon}_{i}(m) dm^{i},
\label{univconni}
\end{equation}
we arrive at \Ref{unicurvi} with
\begin{align}    \label{unicurvii}
{\bf F} &= {\bf d} {\bf A} + {\half} [ {\bf A}, {\bf A}] = {\phi} +
{\psi} + F_{A(m)} \\ \notag {\psi} &= {\psi}_{i \mu} dm^{i} \wedge
dx^{\mu} \\ \notag {\phi} &= {\half} {\phi}_{ij} dm^{i} \wedge dm^{j}
\\ \notag {\phi}_{ij} &= \frac{\pa}{\pa m^{i}} {\varepsilon}_{j} -
\frac{\pa}{\pa m^{j}} {\varepsilon}_{i} + [ {\varepsilon}_{i},
{\varepsilon}_{j} ],
\end{align}
and \Ref{unicurv} follows:
$$
{\De}_{A(m)}{\phi}_{ij} = \frac{\pa}{\pa m^{[i}} D_{A(m)}^{*}
\frac{\pa A(m)}{\pa m^{j]}} + [ D_{A(m)}^{*} \frac{\pa A(m)}{\pa
m^{[i}} , {\epsilon}_{j]}] - D_{A(m)}^{*} [ \frac{\pa A(m)}{\pa
m^{[i}} , {\epsilon}_{j]}] - $$
$$
[\frac{\pa A(m)}{\pa m^{[i}} , \star D_{A(m)} {\epsilon}_{j]}] + [
D_{A(m)} {\epsilon}_{i} , \star D_{A(m)} {\epsilon}_{j}] = [{\psi}_{i}
, \star {\psi}_{j}] \ ,
$$
as claimed.

The universal connection one-form $\bf A$ was defined over $U \subset
\CM$, starting with a section $A(m) : U \to {\CA}_{{\bf
M}^{4}}^{+}$. Now, we may have chosen a different section:
$$
A(m)' = A(m)^{g(m)} = g(m) A(m) g^{-1}(m) + g(m) dg^{-1}(m)
$$
which is related to $A(m)$ by the gauge transformation $g(m)$. It will
change the connection one-form ${\bf A}$ by the corresponding gauge
transformation.  Hence the invariant observables, like the
differential forms ${\tr} {\bf F}^{l}$, are well-defined forms on
${\CM} \times {\bf M}^{4}$.

\subsubsection{Deformation complex, finite $g^2$, and the gauge theory}

The fermion kinetic term of the theory in the limit \Ref{sdsym} is
directly related to the Atiyah--Hitchin--Singer (AHS) complex, which is
the instanton version of the deformation complex corresponding to the
general moduli problem. Recall that the AHS complex is built given a
solution $A$ to the instanton equations, $F_{A}^{+} = 0$. Then:
\begin{equation}
0 \longrightarrow
{\Omega}^{0}({\bf M}^{4}) \otimes_{\CP} {\mf g} 
\longrightarrow^{\hspace*{-5mm} D_{A}}
{\Omega}^{1}({\bf M}^{4}) \otimes_{\CP} {\mf g} 
\longrightarrow^{\hspace*{-5mm} D_{A}^{+}}
{\Omega}^{2,+}({\bf M}^{4}) \otimes_{\CP} {\mf g}
\longrightarrow 0
\label{ahs}
\end{equation}
The sequence \Ref{ahs} is indeed a complex, as $D_{A}^{+} \circ D_{A}
= F_{A}^{+} = 0$. The cohomology of the AHS complex will be denoted by
$H^{i}_{[A]}$. The AHS complex and its cohomology in particular
characterize the vicinity of the point $[A]$ in the moduli space
${\CM}$. First of all, if the only non-vanishing cohomology group of
the AHS complex is $H^{1}_{[A]}$, then the moduli space $\CM$ is
smooth, and its tangent space at $[A]$ coincides with $H^1_{[A]}$. If,
on the other hand, the zeroth cohomology $H^0_{[A]}$ of \Ref{ahs} is
non-vanishing, then the moduli space is singular, as the point $[A]$
has a non-trivial stabilizer, $G_{[A]}$, whose Lie algebra is
isomorphic to $H^0_{[A]}$. Finally, the group $H^2_{[A]}$ is an
obstruction to the smoothness of $\CM$. More precisely it is
responsible for the possibility to extend the first order deformation
of $[A]$, labeled by $H^1_{[A]}$, to the second order
deformation. Indeed, suppose the first order deformation $a_{[1]}$,
which is a solution to
$$
D_{A}^{+}a_{[1]} = 0
$$
up to the infinitesimal gauge transformation $a_{[1]} \sim a_{[1]} +
D_{A} {\varphi}_{[1]}$, is given. Then the second order deformation
$a_{[2]}$ has to be such that
$$
D_{A}^{+} a_{[2]} + {\half} [ a_{[1]}, a_{[1]}] = 0
$$ and it is defined up to the following transformations:
$$
a_{[2]} \sim a_{[2]} + D_{A} {\varphi}_{[2]} + [a_{[1]},
{\varphi}_{[1]}] + {\half} [ D_{A} {\varphi}_{[1]} ,{\varphi}_{[1]}].
$$
Such $a_{[2]}$ can be found if and only if the image of $[a_{[1]},
a_{[1]}]$ in $H^2_{[A]}$ is zero. The map
$$
K = [\ \cdot \ ,^{\hspace*{-1.5mm} \wedge} \ \cdot \ ]: H^1_{[A]}
\times H^1_{[A]} \longrightarrow H^2_{[A]}
$$
is called the {\it Kuranishi map}. There exists a theory which
identifies the germ ${\CM}_{[A]}$ of the moduli space near the point
$[A]$ with the quotient of the preimage $K^{-1}(0) \subset
H^{1}_{[A]}$ of zero in $H^{1}_{[A]}$ by the stabilizer $G_{[A]}$,
${\CM}_{[A]} \approx K^{-1}(0)/G_{[A]}$.

The $g^2 \to 0$ limit of the gauge theory, as the infinite radius
limit of the sigma model, has some subtleties, related to the
existence of unwanted zero modes. In our derivation of the reduction
of the path integral to the integral over the space ${\CM}$ of
instantons, we assumed that the integral over the fermionic fields
$\chi^{+}$ and $\eta$ gave us the equations \Ref{psieq}. In general,
however, it may happen that the conjugate operators
$$
D_{A}^{+,*} \oplus D_{A} : \left( {\Omega}^{2,+} \oplus {\Omega}^{0}
\right) \otimes_{\CP} {\mf g} \to {\Omega}^{1} \otimes_{\CP} {\mf
g}
$$
have non-vanishing kernel, which is clearly isomorphic to $H^{2}_{m}
\oplus H^{0}_{m}$ (in this case the space of solutions to the
equations \Ref{psieq} has the kernel which is strictly larger then the
virtual tangent space $T_{m}{\CM}$). Then the integral over these zero
modes of $\chi^{+}$ and $\eta$ has to be regularized. Also, the
fermionic zero modes $\chi_{0}^{+}, {\eta}_{0}$ are accompanied by the
bosonic zero modes $H_{0}^{+}, {\ol{\phi}}_{0}$, which might produce
infinities unless properly regularized.

Finally, the formula \Ref{unicurvi} for the field $\phi$ now has to be
modified. Indeed, since $D_{A}$ has zero modes on scalars, the
solution to the equation $$ {\De}_{A} {\phi} = [ {\psi} , \star
{\psi}] $$ now has to be written as
$$
{\phi} = {\phi}_{0} + \frac{1}{{\De}_{A}} [ {\psi}, \star \psi ], $$
where ${\phi}_{0}$ solves $D_{A} {\phi}_{0} = 0$.

Let us now turn on the coupling constant $g^2$, but only on the zero
modes $H_{0}^{+}, {\chi}_{0}^{+}, {\ol{\phi}}_{0}, {\eta}_{0}$:
\begin{equation}
{\De}S = g^{2} \ {\CQ} \int {\tr} \left( {\chi}_{0}^{+} \wedge
H_{0}^{+} + {\eta}_{0} [ \ol{\phi}_{0}, \phi_{0} ] \right),
\label{correct}
\end{equation}
where the zero modes are the solutions of the equations
\begin{align} \notag
D_{A} {\ol{\phi}}_{0} &= 0, \qquad D_{A}^{+, *}H_{0}^{+} = 0 \\
D_{A} {\eta}_{0} &= 0, \qquad D_{A}^{+, *}{\chi}_{0}^{+} = 0.
\label{zmds}
\end{align}
There is one more subtlety related to the fact that we now treat the
fields $\ol{\phi}, \eta$ as half-densities, but we shall ignore it.

Note that the supercharge $\CQ$ preserves the space of zero modes:
\begin{align} \notag
{\CQ} \ol{\phi}_{0} &= {\eta}_{0} \ , \qquad {\CQ} {\eta}_{0} = [
{\phi}_{0} , \ol{\phi}_{0} ] \\ {\CQ} {\chi}_{0}^{+} &= H_{0}^{+} \ ,
\qquad {\CQ} H_{0}^{+} = [ {\phi}_{0} , {\chi}_{0}^{+}].
\label{zrmds}
\end{align}
Thus the contribution of the vicinity of the point $m \in {\CM}$ to
the correlation function of some "evaluation observables" will be
given by the integral
\begin{multline} \int_{H^{2}_{m}} {\rm d}H_{0}^{+}{\rm
    d}{\chi}^{+}_{0} \ e^{ \int \
i {\tr}\left( H^{+}_{0} [a_{[1]}, a_{[1]}]\right) - i {\tr} \left(
{\chi}^{+}_{0} [ a_{[1]} , {\psi}_{[1]}] \right) - g^{2}
{\tr}H_{0}^{+} H_{0}^{+} - g^{2} {\tr} {\chi}^{+}_{0} [ {\phi}_{0} ,
{\chi}^{+}_{0}] } \\ \cdot \int_{H^{0}_{m}} {\rm d}{\ol\phi}_{0}
{\rm d}{\eta}_{0} \ e^{ \int {\tr} {\eta}_{0} [a_{[1]} , \star
\psi_{[1]} ] - g^{2} {\tr} \left( {\eta}_{0} [ {\phi}_{0}, {\eta}_{0}]
\right) - g^{2} {\tr} [ {\phi}_{0}, \ol{\phi}_{0} ]^2 },
  \label{obstr}
\end{multline}
which we should view as the differential form on $H^{1}_{m}$
by decomposing the corresponding zero modes:
$$ a_{[1]} = \sum_{i=1}^{{\rm rk}H^{1}_{m}} {\psi}_{i} \ {\mf
m}^{i}\ , \qquad {\psi}_{[1]} = \sum_{i=1}^{{\rm rk} H^{1}_{m}}
{\psi}_{i} \ {\rm d}{\mf m}^{i}
$$
The differential form \Ref{obstr} is the smooth (for $ g^2 > 0$)
representative of the Poincare dual to the zero locus of the kuranishi
map. The $\phi_0$ dependence signifies the $G_{m}$-equivariant nature
of the differential form, while the integral over $H^{0}_{m}$ in
\Ref{obstr} gives the projection form, which is suited to define the
integration theory over the quotient $K^{-1}(0)/G_{m}$.  Now, it seems
that even here we don't need to take $g^2 > 0$, as even in the $g^2
\to 0$ limit we get something reasonable: the Poincare dual gets
represented by the delta form ${\delta}(K({\mf m}))$, and the
projection form gives what we expect it to give.  In fact, it depends
on the degeneracy of the kuranishi map. In the nice situations, where
it is sufficiently non-degenerate, the $g^2 = 0$ expression
\Ref{obstr} defines a well-defined differential form on
$H^{1}_{m}$. However this is not always the case, we give an example
momentarily.

The usefulness of having $g^2 > 0$ at this point is the
possibility of working with another representative of the Poincare
dual of the zero locus of $K$; namely, the Euler class of the
corresponding obstruction bundle $\cup_{m \in \CM} H^{2}_{m}$ written
in terms of the curvatures of the corresponding metric
connections. This representation arises in the large $g^2$ limit.

In many applications involving the correlation functions of the BPS
observables both representatives are equally good, and sometimes one
is simpler to work with than another. In our story, we shall try to
work with $g^2 = 0$ for as long as it is possible.

\begin{remark}
In the case of two-dimensional sigma model coupled to the
two-dimensional topological gravity the discussion similar to the one
we just had, is quite important in figuring out the contribution of
the degree zero maps of higher genus Riemann surfaces. The analogous
obstruction bundle is related, in this case, to the Hodge bundle on
the moduli space of Riemann surfaces.  Taking into account its
contribution, i.e., the integrals of its Chern classes, to the
Gromov-Witten invariants plays important r\^ole in the modern
topological string theory.\qed
\end{remark}

Likewise, the trivial connection, $A=0$, which is unavoidable in any
gauge theory in the sector with the trivial 't Hooft fluxes, comes
with the whole host of non-trivial cohomology of the AHS complex:
$$
\qquad\qquad H^{i}_{[0]} = H^{i}({\bf M}^{4}, {\R}) \otimes {\mf g},
\quad i = 0,1,
$$
$$
H^{2}_{[0]} = H^{2,+}({\bf M}^{4}, {\R}) \otimes {\mf g}.
$$
Moreover, in this case the stabilizer $G_{0} = G$ coincides with the
group $G$ itself, while the kuranishi map is given by
$$
K ({\mf m})  = f_{ab}^{c} {\mf m}^{a k} {\mf m}^{b l} {\om}_{p} 
\int_{{\bf M}^{4}} {\om}^{p} \wedge {\al}_{k} \wedge {\al}_{l},
$$
where $({\om}_{p})_{p=1,\ldots , b_{2}^{+}}$ is the basis in the space
of self-dual harmonic two-forms, and $({\al}_{l})_{l=1,\ldots ,
b_{1}}$ is the basis in the space of harmonic one-forms.

Now, to give an example of the necessity of $g^2 > 0$ regularization,
consider the case of a simply-connected manifold ${\bf M}^{4}$.  In
this case the kuranishi map corresponding to $A = 0$ is identically
zero, so the expansion of the $H^{+}F^{+} + {\chi}^{+}D_{A}{\psi}$
term does not give us anything interesting. The integrals \Ref{obstr}
can be evaluated, to produce
$$
\left( {\rm Det}_{\mf g}^{\prime} {\rm ad}{\phi}_{0}
\right)^{b_{2}^{+} + b_{0}},
$$
where the $b_{2}^{+}$ in the exponent comes from the $H_{0}^{+},
{\chi}^{+}_{0}$ integral, while the origin of the $b_{0}$ contribution
is the measure on the ${\phi}_{0}$. The integral over $\eta_0$,
${\ol{\phi}}_{0}$ gives the determinants which cancel each other. The
unfortunate subtlety is the prime in the determinant. The components
of $H_{0}^{+}, {\chi}_{0}^{+},{\eta}_{0}, \ol{\phi}_{0}$ which commute
with ${\phi}_{0}$ do not enter the $g^2$-dependent part of
\Ref{obstr}. These modes should be taken care of separately. This and
further subtleties of the {\it stack} nature of the moduli space $\CM$
are beyond the scope of this paper.

\subsection{Gauge theory and logarithms}

We shall now discuss the main problem of our interest -- the
appearance of logarithms in the correlation functions the
four-dimensional gauge theory.

\subsubsection{Evaluation observables in gauge theory}

The correlation functions of the "evaluation
observables", which we now take to be the gauge-invariant functionals
of $A, {\psi}$ and ${\phi}$ (they correspond to the equivariant forms
on $\CA$, the space of four-dimensional connections), reduce, in the
$\ol{\tau}\to \infty$ limit, to the integrals over the moduli space of
gauge instantons. We shall need some description of these moduli
spaces. Since we wish to demonstrate the logarithmic nature of our
theory, it suffices to consider the simplest case, the charge one
instantons, for the gauge group $G = SU(2)$.

\subsubsection{From correlation functions to matrix elements.} First
of all, we need to set up our calculation in such a way so as to be
able to interpret it quantum mechanically. It is convenient to take as
a four-manifold the four-sphere ${\bf M}^{4} = {\mathbb S}^{4}$, which
we can view as the one-point compactification of ${\R}^{4} = {\C}^{2}
= {\mathbb H}^{1}$, a quaternionic line.  Let us use the quaternionic
notation: ${\bv} = v^{0} + v^{1} {\bi} + v^{2} {\bj} + v^{3} {\bk}
\in {\mathbb H}^{1}$, or, equivalently, the $2 \times 2$ matrix
notation:
\begin{equation}
{\bv} = \begin{pmatrix} v^{0} + i v^{1} & v^{2} -  i v^{3} \\
- v^{2} - i v^{3} & v^{0} - i v^{1} \end{pmatrix} =
\begin{pmatrix} w^{1} & {\wb}^{2} \\ - w^{2} & {\wb}^{1}
\end{pmatrix} \  .
\label{vwbas}
\end{equation}
where $(w^{1}, w^{2}) \in {\C}^{2}$ represents the point $(v^{0},
v^{1}, v^{2}, v^{3}) \in {\R}^{4}$, upon some identification ${\R}^{4}
\simeq {\C}^{2}$. The standard round metric on ${\mathbb S}^{4}$,
$$
ds^2 = \frac{d{\bv}^2 }{(1 + \vert {\bv} \vert^2 )^2}
$$
is conformal to the metric on ${\R} \times {\mathbb S}^{3}$, with the
points ${\bv} = 0$ and ${\bv} = \infty$ deleted, $$ ds^2 = {1\over
4{\rm cosh}^2 (t)} \left( dt^2 + d{\Omega}_{3}^{2} \right),
$$ where $t = {\rm log} \vert {\bv} \vert$.

The path integral of the gauge theory on ${\mathbb S}^{4}$ can be
interpreted as the vacuum matrix element in the quantum mechanics,
where the space of states is obtained by quantizing the gauge fields
on ${\mathbb S}^{3}$. The Hamiltonian of the theory is the generator
of the $t$ translations, $H = {\pa}_{t}$, which is the dilatation
operator in the $\bv$ coordinates.

We claimed that the path integral on ${\mathbb S}^{4}$ computes a
vacuum matrix element. The precise vacuum states depend on the type of
operators which are inserted at the points ${\bv} = 0$ and ${\bv} =
\infty$, because these are the only points points fixed by the
dilatation operator.

\subsubsection{ADHM construction}

The moduli space of charge one instantons on ${\mathbb S}^{4}$ is
well-understood. For our quantum mechanical purposes we should
actually consider the space of instantons which are located neither at
${\bv} = 0$ nor at ${\bv} = \infty$. For simplicity we take $G=SU(2)$
in what follows.

The instanton moduli are $m = ( {\bf b} , {\bf m} = {\rho} g_{2})$,
and $( {\bf b} , {\bf m})$ is identified with $ ( {\bf b} , - {\bf
m})$, so that ${\CM}_{2,1} \simeq {\R}^{4} \times \left( {\R}^{4}\bs
\{ 0 \} \right) / {\Z}_{2}$.

The charge one $SU(2)$ instantons on ${\mathbb S}^{4}$ are constructed
with the help of a family of operators
$$
{\mc D}^{+} = \begin{pmatrix} \ b_0 - z_0 & b_1 - z_1 & I \\ - {\ol
b}_{1} + {\ol z}_{1}  & {\ol b}_{0} - {\ol z}_{0} & J^{\dagger}
\end{pmatrix}
$$
parametrized by the points $z \in {\mathbb C}^{2} \approx {\mathbb
R}^{4} = {\mathbb S}^{4} - \{ \infty \}$.
The ADHM equations imply that
$$
{\mc D}^{+} = \begin{pmatrix} r g_{1} & {\rho} g_{2} \end{pmatrix},
$$
where ${\rm det} g_{1} = {\rm det} g_{2} = 1$,
$$
g_{1}^{\dagger} g_{1} = g_{2}^{\dagger} g_{2} = {\bf 1}_{2 \times 2} ,
\qquad r^2 = \Vert b - z \Vert^2 , \qquad {\rho}^2 = I I^{\dagger} =
J^{\dagger}J.
$$
The "observation point" ${\bf z} = z_0 + z_1{\bf j} = {\bf b} - r
g_{1}$.

The gauge field is written in terms of the normalized solution
${\Psi}$ to the equation
$$
{\mc D}^{+} {\Psi} = 0, \qquad {\Psi}^{\dagger} {\Psi} = 1_{2 \times 2}
$$
as
\begin{equation}
A = {\Psi}^{\dagger} d {\Psi}.
\label{univconnadhm}
\end{equation}
We point out an important feature of the ADHM construction. If we let
$d$ in formula \Ref{univconnadhm} act not only on $z$ but also on the
moduli, then $A$ becomes a universal connection, that is, a connection
on the universal bundle over
$$
{\CM} \times {\mathbb S}^{4},
$$
whose restriction onto a fiber $m \times {\mathbb S}^{4}$ of
projection to $\CM$ gives the corresponding connection on ${\mathbb
S}^{4}$.  In our case
$$
{\Psi} = \begin{pmatrix} \ \ {\rho} g_{1}^{\dagger} \\ - r
g_{2}^{\dagger} \end{pmatrix} \frac{1}{\sqrt{ r^{2} + {\rho}^{2}}}
$$
and
$$
{\bf A} = x {\theta}_{1} + ( 1- x) {\theta}_2,
$$
where ${\theta}_{i} = g_{i} d g_{i}^{\dagger}$, and
$$
x = \frac{{\rho}^{2}}{r^2 + {\rho}^2}.
$$
The universal curvature ${\bf F} = d{\bf A} + {\bf A}^2$ and the
corresponding ${\tr}\, {\bf F}^2$ invariant is given by
\begin{equation}
{\tr}\, {\bf F}^2 = d \left[ x^2 \left( 3 - 2 x \right) \right] \wedge
{\tr} \ {\theta}_{12}^3,
\label{instch}
\end{equation}
where
$$
{\theta}_{12} = ( g_{2}^{\dagger}g_{1} )
d ( g_{1}^{\dagger} g_{2} ).
$$
We can rewrite \Ref{instch} in a more suggestive form, using
quaternions:
\begin{equation}
{\tr}\, {\bf F}^2 = \frac{d^4 {\bv}_{z}}{( 1 + | {\bv}_{z} |^2 )^4},
\label{quat}
\end{equation}
where
$$
{\bv}_{z} = \frac{1}{\rho} g_{2}^{\dagger} \cdot ( {\bf b} - {\bf z}
).
$$
The geometry behind formula \Ref{quat} can be found in
\cite{atiyah}.

\subsubsection{Gauge theory calculation}

The following correlation function is a good candidate for exhibiting
the logarithmic structure of the four-dimensional gauge theory:
\begin{equation}
C _{{\CO}{\CO}{\CS}} ( x_{1}, x_{2} ; x_{3}) =
\langle {\CO} (x_{1})  {\CO} (x_{2}) {\CS} (x_{3})  {\bf GF} ({\bf
  \infty})\rangle \ ,
\label{thfp}
\end{equation}
where
$$
{\CO} ( x ) \ = \  {\tr}\, {\phi}^{2} (x) \quad , \quad
{\CS} ( x ) \ = \ {\tr}\, F_{\mu\nu} F^{\mu\nu} (x),
$$
and
$$
{\bf GF} ({\bf y}) \sim \prod_{a}
{\delta}({\phi}^{a}({\bf y})){\delta}({\ol{\phi}}^{a}({\bf
y})){\eta}^{a}({\bf y}) \times \prod_{a} {\delta}({\la}^{a}({\bf
y})){\ol{c}}^{a}({\bf y})c^{a}({\bf y})
$$
is the operator which fixes the gauge transformations at the point
${\bf y}$ to be trivial (it is a suitably regularized product of the
Faddeev-Popov ghosts and the delta functions ${\delta}({\phi}^{a})$,
all taken at one point in ${\bf M}^{4}$).

{}The correlator \Ref{thfp} is saturated by the charge one
instanton. Indeed, on the moduli space ${\CM}_{2,1}$ of charge $1$
instantons with the gauge group $SU(2)$ the operators ${\tr}{\phi}^2$
become four-forms, and ${\tr}F_{\mu\nu}^{2}$ -- the density of the
topological charge -- a function. The dimension of the moduli space
${\CM}_{N,k}^{\infty}$ of charge $k$ instantons with the gauge group
$SU(N)$, considered up to the gauge transformations, equal to the
identity at one point (e.g., ${\bf y} = {\bf \infty}$), is equal to $4
N k$.

Using the expressions \Ref{quat} for the universal curvature
invariants, computed with the help of the ADHM construction, we reduce
\Ref{thfp} to the following integral:
\begin{equation}
C_{{\CO}{\CO}{\CS}} ({\bf x}_{1}, {\bf x}_{2} ; {\bf x}_{3}) =
\frac{1}{\vert {\bf x}_{12} \vert^{4}} \ {\CC} \left( { {\bf
{\xb}}_{12} \cdot ( {\bf x}_{13} + {\bf x}_{23} ) \over \vert {\bf
x}_{12} \vert^{2}} \right),
\label{oninscal}
\end{equation}
where 
\begin{equation} 
{\CC} ( {\bq}) =
\int_{{\R}^{4} \times {\R}^{4}} 
\frac{ d^{4}{\bv}_{1} }
{( 1 + | {\bv}_{1} |^2 )^4 }
\frac{ d^{4}{\bv}_{2} }
{ ( 1 + | {\bv}_{2} |^2 )^4}
\frac{  \
\vert {\bv}_{-} \vert^{4}}
{( 1 + |
{\bv}_{+} - {\bv}_{-} \cdot {\bq} |^2 )^4}
\label{cqinst}
\end{equation}
with
$$
{\bv}_{\pm} = \frac{{\bv}_{1} \pm {\bv}_{2}}{2},
$$
and we use quaternionic notation.

Formula \Ref{oninscal} is very suggestive in that it resembles the
two-dimensional holomortex formalism (see \secref{target pone} and
\secref{OPE from hol}). Indeed, it looks like the correlation function
of two holomortex operators, inserted at the points ${\bv}_{1}$ and
${\bv}_{2}$. This suggests that the four-dimensional Yang-Mills theory
may be defined as a deformation of a much simpler model by analogues
of the two-dimensional holomortex operators. The precise definition of
these operators is beyond the scope of the present paper. We will only
remark that to observe this holomortex structure of the correlation
functions it is important to go beyond the topological sector of the
Yang-Mills theory.

The integral \Ref{oninscal} is invariant under the $SU(2)$ rotations,
${\bq} \mapsto u {\bq} {\ub}$, $u {\ub} = 1$. Thus, we can assume that
${\bq} = q \in {\C}$, or, in matrix form,
$$
{\bq} = \begin{pmatrix} q & 0 \\ 0 & {\qb} \end{pmatrix} \ .
$$
Without any detailed calculations it is clear that ${\CC}(q)$ has
singularities when $q \to -1, +1, \infty$. These limits correspond to
the situations, where the operator ${\CS}$ hits one of the operators
$\CO$, or goes away to infinity.

The correlation function 
\Ref{oninscal} can also be rewritten in the following symmetric form:
\begin{align}
C_{{\CO}{\CO}{\CS}} ( {\bf x}_{1}, {\bf x}_{2} ; {\bf x}_{3}) &= \vert
{\bf x}_{12}\vert^{4} \ {\Gamma} \left( {\bf x}_{12}^{2} ,{\bf
x}_{13}^{2}, {\bf x}_{23}^{2} \right),
\label{oninstcf}
\\ \notag \\ \notag {\Gamma} ( a_{12}, a_{13}, a_{23} ) &\ =
\int_{{\R}^{3}_{\geq 0}} \frac{ d^{3}t \left( t_{1}t_{2}t_{3}
\right)^{3} e^{-(t_{1}+t_{2}+t_{3})}} {( t_{1}t_{2} a_{12} +
t_{1}t_{3} a_{13} + t_{2}t_{3}a_{23} )^4 }.
\end{align}
Note that the expression in the denominator in 
\Ref{oninstcf} can be rewritten using
$$
\frac{t_{1}t_{2} |{\bf x}_{12}|^{2} + t_{1}t_{3} |{\bf x}_{13}|^{2}
+ t_{2}t_{3}|{\bf x}_{23}|^{2}}{t_{1}+t_{2}+t_{3}} = \sum_{i=1}^{3}
t_{i} \, | {\bf x}_{i} - {\bf x}_{t} |^{2},
$$
where
$$
{\bf x}_{t} = \frac{ t_{1}{\bf x}_{1}+ t_{2}{\bf x}_{2}+ t_{3}{\bf
    x}_{3}}{t_{1}+t_{2}+t_{3}}.
$$
Therefore the integral \Ref{oninstcf} can be written as the integral
over the plane triangle with the vertices ${\bf x}_{1}, {\bf x}_{2},
{\bf x}_{3}$ (see Figure 9).

\bigskip
\begin{center}
\epsfig{file=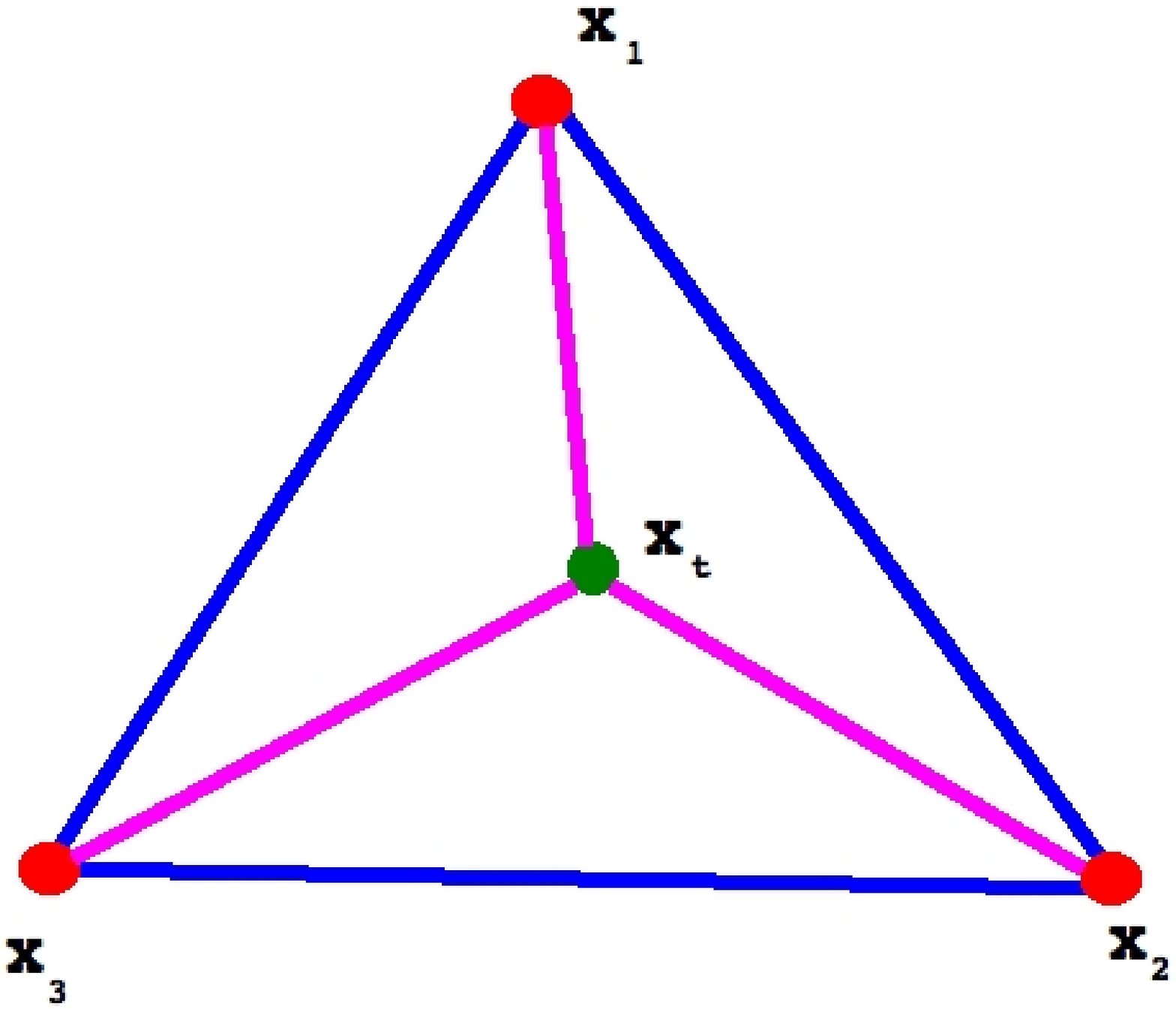,width=80mm}

{\bf Figure 9. The triangle.}
\end{center}

\bigskip

Let us introduce the kinematic variables
\begin{equation}
s_{2} = |{\bf x}_{12}|/|{\bf x}_{23}| \ , 
\ s_{3} = |{\bf x}_{13}|/|{\bf x}_{23}|,
\label{sses}
\end{equation}
which obey the usual triangle inequalities
$$
| 1 - s_{2} | \leq s_{3} \leq 1 + s_{2}.
$$
The integrals \Ref{oninscal},\Ref{oninstcf} can be reduced to the
one-dimensional integral
\begin{equation}
{\CC}({\bf q})\ = 
\int_{-\infty}^{+\infty} {\rm d}{\vartheta}\, {\rho} \left(
s_{2}^{2}+s_{3}^{2} + 2s_{2}s_{3}\, {\rm cosh}{\vartheta} \right) \ ,
\label{gammaone}
\end{equation}
where the function
\begin{equation}
{\rho}(m) = \frac{m^{2}+6m+3}{(m-1)^{6}}{\rm log}(m) - \frac{10m^{2} +
  19m +1}{3m(m-1)^{5}}\label{rhof}
\end{equation}
interpolates between 
$$
{\rho} (m) = \frac{1}{60} - \frac{4}{105} (m-1) +
\frac{5}{84}(m-1)^{2} + \ldots $$ for $m \to 1$, and
$$
{\rho}(m) = \left( {\rm log}(m) - \frac{10}{3} \right) \frac{1}{m^{4}} + 
12 \left( {\rm log}(m) - \frac{23}{12} \right) \frac{1}{m^{5}} + \ldots
$$
for $m \to \infty$. The argument of the function ${\rho}$ in
\Ref{gammaone} is bounded below by
$$
m_{\rm min} = \left( s_{2} + s_{3} \right)^{2} \geq 1
$$
(the equality is achieved only when the point ${\bf x}_{3}$ is on the
line connecting ${\bf x}_{1}$ and ${\bf x}_{2}$). Therefore the
function ${\rho}$ is finite for all values of $\vartheta$, and the
integral \Ref{gammaone} converges for all ${\bf x}_{i}$, $i=1,2,3$.

\medskip

Now we interpret the correlation function \Ref{thfp} as the
vacuum matrix element
\begin{equation}
\langle {\rm vac} \vert\ {\CO}\, e^{i {\varphi} J}\, e^{-T H}\,
{\CS}  \  \vert {\rm vac}_{\CO} \rangle.
\label{thfpii}
\end{equation}
Here the Hamiltonian $H$ generates the radial evolution, with
the position ${\bf x}_{1}$ of one of the operators ${\CO}$ being the
origin, $J$ is one of the $SO(4)$ generators
$$
J = \begin{pmatrix} 1 & 0 \\ 0 & -1 \end{pmatrix} \ ,
$$
in the notation of \Ref{vwbas}, 
and ${\varphi}$ is the angular distance between the location ${\bf
z}$ of ${\CS}$ and that of the second ${\CO}$, ${\bf x}_{2}$, when
viewed from the point ${\bf x}_{1}$:
$$
e^{i {\varphi}} = \frac{1-q}{\vert 1 - q \vert} = \frac{\ol{{\bf
x}}_{12} \cdot {\bf x}_{32} }{\vert
{\bf x}_{12} \vert \vert {\bf x}_{32} \vert}\ .
$$
Finally, the "time" $T$ is related to ${\bf x}_{1}, {\bf x}_{2}$ and
${\bf x}_{3}$ as follows:
$$
e^{T} = \frac{\vert {\bf x}_{12} \vert}{\vert {\bf x}_{32} \vert}  =
s_{2}\ .
$$
The precise analytic expression for the correlation function
\Ref{oninstcf} is rather complicated.  The function ${\rho}(m)$
decays sufficiently fast for large $m \gg m_{\rm max} \sim
4$. Therefore the typical range of $\vartheta$ which contributes to
the integral \Ref{gammaone} is $\sim {\rm log} \left(
\frac{C}{s_{2}s_{3}}\right) $ where $C$ is a numerical constant of the
order one. This is why we expect \Ref{gammaone} to contain only the
simple logarithms ${\rm log}(s_{2})$, ${\rm log}(s_{3})$. This is a
natural result for the one-instanton correlation function, as we have
argued in the case of quantum mechanics.

The upshot of this calculation is that the logarithms do appear in the
correlation functions of this model. This signifies the logarithmic
nature of the four-dimensional conformal theory which we obtain in the
$\ol{\tau} \to \infty$ limit of the twisted ${\CN}=2$
super-Yang--Mills theory.

\subsubsection{Operator product expansion and
logarithmic partners.}

The next natural step in our program is analogous to the computation
of the instanton corrections to the OPE in two-dimensional sigma
models in \secref{OPE}. Let us investigate the asymptotics of the
correlation function ${\CC}_{{\CO}{\CO}{\CS}}({\bf x}_{1}, {\bf x}_{2}
; {\bf x}_{3} )$ in the limit ${\bf x}_{2} \to {\bf x}_{1}$. The naive
asymptotics of the integral \Ref{cqinst} as ${\bf q} \to \infty$ is a
logarithmically divergent integral
$$
{\CC}({\bf q}) \sim \frac{1}{|{\bf q}|^{4}} 
\int_{{\R}^{4} \times {\R}^{4}} 
\frac{ d^{4}{\bv}_{1} }
{( 1 + | {\bv}_{1} |^2 )^4 }
\frac{ d^{4}{\bv}_{2} }
{ ( 1 + | {\bv}_{2} |^2 )^4}
\frac{1}{
\vert {\bv}_{-} \vert^{4}},
$$
which is actually cut off at $|{\bv}_{-}| \sim |{\bf q}|^{-1}$, so that
$$
{\CC}_{{\CO}{\CO}{\CS}}({\bf x}_{1}, {\bf x}_{2} ; {\bf x}_{3}) \sim
\frac{1}{|{\bf x}_{23}|^{4}} {\rm log} \, \frac{| {\bf x}_{12}|}{|{\bf
    x}_{23}|} \ , \qquad {\bf x}_{2} \to {\bf x}_{1}.
$$

We interpret this as the one-instanton correction to the OPE:
$$
{\tr}{\phi}^{2} ({\bf x}_{1}) \, 
{\tr}{\phi}^{2} ({\bf x}_{2}) \sim \Lambda_{\on{QCD}} \;
{\rm log} | {\bf x}_{12} | \, {\tr}(H^{+})^{2} ({\bf x}_{2}) + \ldots 
$$
This formula is analogous to the logarithmic terms in formulas
\eqref{om om} and \eqref{om f} obtained in sigma models. Continuing
along these lines, we can find the analogues of the jet-evaluation
observables in the four-dimensional gauge theory and observe the
logarithmic mixing of these observables. We plan to study
this in more detail in a follow-up paper.

\section{Conclusions}
\label{conclusions}

In this paper we have studied (twisted) supersymmetric two-dimensional sigma
models and four-dimensional gauge theories in the ${\ol{\tau}} \to
\infty$ limit. We have used the quantum mechanical models considered
in Part I as a prototype. A special feature of these models is that
the path integral localizes on the finite-dimensional moduli spaces of
instanton configurations (holomorphic maps in two dimensions and
anti-self-dual connections in four dimensions). This gives us good
control of the correlation functions and enables us to describe rather
explicitly the spaces of states and the spectra of these
models. However, to do this we must go beyond the topological sector
and consider observables that are not annihilated by the supercharge.

\medskip

In two dimensions, we identify a large class of observables of this
type which we call {\em jet-evaluation observables}. These are
differential forms on the jet space of the target manifold. They
generalize the familiar evaluation observables in that they depend not
only on the value of a holomorphic map in the target manifold, but
also on its derivatives. Their correlation functions are given by
integrals over the moduli spaces of holomorphic maps, generalizing the
Gromov--Witten invariants. However, in contrast to the Gromov--Witten
invariants, these integrals generally diverge on the boundary divisors
of the moduli spaces of holomorphic maps and need to be
regularized. Their regularization is not canonical, reflecting what we
call {\em logarithmic mixing} of operators (and states) of the sigma
model. This is analogous to the appearance of the logarithmic
structures in the quantum mechanical models which we studied in Part
I. We have presented here many explicit examples of the correlation
functions, OPE and logarithmic mixing in two-dimensional sigma models,
particularly, in the case of the target manifold $\pone$. We have also
revisited the holomortex description of the latter model introduced in
\cite{AiB} and shown that it may be used to effectively reproduce
these results.

Our conclusion is that the twisted ${\CN}=(2,2)$ supersymmetric sigma
models are logarithmic conformal field theories, with central charge
$c=0$. Such logarithmic CFTs have been extensively studied in the
literature recently, see, e.g.,
\cite{Gurarie,Kogan,GL,Schomerus,DF,MR} and references therein. These
models are quite interesting for both theoretical reasons and for
their applications to condensed matter. Here we consider a new class
of models of this type which have many attractive features: they are
defined geometrically (as sigma models) and their correlation
functions are computed explicitly as (regularized) integrals over
moduli spaces of holomorphic maps. We hope that further understanding
of these models will be beneficial for the investigation of
logarithmic CFTs as well as their physical applications.

\medskip

In Part III of this paper we will study sigma models with less
supersymmetry, such as the ${\mc N} = (0,2)$ and purely bosonic sigma
models. These models are more difficult to analyze because the
definition of the measure in the path integral becomes more
problematic, which leads to various anomalies that we avoid in the
${\CN}=(2,2)$ models. In particular, these models are not conformally
invariant unless the target manifold is Calabi-Yau. Nevertheless, they
may still possess non-trivial chiral algebras of symmetries, such as
an affine Kac--Moody algebra of critical level $k=-h^\vee$ in the case
when the target manifold is a flag manifold of a simple Lie
groups. The latter models have applications to the geometric Langlands
correspondence. In addition, the models with the target manifold
$\pone$ (and probably other flag manifolds as well) admit an analogue
of the holomortex description, similar to the one discussed above.

\medskip

We have also studied the twisted four-dimensional supersymmetric
Yang--Mills theory. The $\ol\tau \to \infty$ limit of this model may
be studied along the same lines as above, with the added complication
that we need to take into account equivariance with respect to the
gauge transformations. However, much of the same structure that we
have observed in one- and two-dimensional models also appears in four
dimensions. In particular, we have computed some sample correlation
functions (in the one instanton sector for the group $SU(2)$) which
exhibit the same logarithmic behavior that we have seen in lower
dimensions. These correlation functions are given by integrals over
the moduli spaces of anti-self-dual connections, which are described
explicitly by the ADHM construction. By careful analysis of these
correlation functions we obtain logarithmic terms in the OPE of the
analogues of jet-evaluation observables. Thus, we find the same kind
of logarithmic mixing of operators that we have seen in
two-dimensional sigma models. We believe that further investigation of
these phenomena will lead to better understanding of the
four-dimensional gauge theory beyond its topological sector.

\end{document}